\documentclass[preprintnumbers,superscriptaddress,nofootinbib,aps,prd,floatfix]{revtex4}
\usepackage{amsmath,multirow,graphicx,tabularx,epsfig}
\usepackage{color}
\usepackage{hyperref}
\usepackage{pgf}
\usepackage{tikz}
\usetikzlibrary{arrows,automata}
\usetikzlibrary{positioning}

\tikzset{
    state/.style={
           rectangle,
           draw=black, thin,
           minimum height=2em,
           inner sep=2pt,
           text centered,
           },
}

\newcommand\one{\leavevmode\hbox{\small1\normalsize\kern-.33em1}}

\newcommand{\lag}{\mathcal{L}}

\newcommand{\qqquad}{\qquad \qquad}
\newcommand{\qqqquad}{\qquad \qquad \qquad}

\newcommand{\gev}{{\ensuremath\rm GeV}}

\def\slashchar#1{\setbox0=\hbox{$#1$}           % set a box for #1
   \dimen0=\wd0                                 % and get its size
   \setbox1=\hbox{/} \dimen1=\wd1               % get size of /
   \ifdim\dimen0>\dimen1                        % #1 is bigger
      \rlap{\hbox to \dimen0{\hfil/\hfil}}      % so center / in box
      #1                                        % and print #1
   \else                                        % / is bigger
      \rlap{\hbox to \dimen1{\hfil$#1$\hfil}}   % so center #1
      /                                         % and print /
   \fi}

\def\eg{{\sl e.g.} \,}
\def\ie{{\sl i.e.} \,}

\begin{document}

\title{Higgs Quantum Numbers in Weak Boson Fusion}

\author{Christoph Englert}
\affiliation{IPPP, Department of Physics, Durham University, United Kingdom}

\author{Dorival Gon\c{c}alves-Netto}
\affiliation{Institut f\"ur Theoretische Physik, Universit\"at Heidelberg, Germany}

\author{Kentarou Mawatari}
\affiliation{Theoretische Natuurkunde and IIHE/ELEM, Vrije
  Universiteit Brussel\\ and International Solvay Institutes,
  Brussels, Belgium}

\author{Tilman Plehn}
\affiliation{Institut f\"ur Theoretische Physik, Universit\"at Heidelberg, Germany}

\preprint{IPPP/12/90, DCPT/12/180}

\begin{abstract}
  Recently, the ATLAS and CMS experiments have reported the discovery
  of a Higgs like resonance at the LHC. The next analysis step will
  include the determination of its spin and CP quantum numbers or the
  form of its interaction Lagrangian channel-by-channel. We show how
  weak-boson-fusion Higgs production and associated $ZH$ production
  can be used to separate different spin and CP states.
\end{abstract}

\maketitle

\tableofcontents
\newpage

%%%%%%%%%%%%%%%%%%%%%%%%%%%%%%%%%%%%%%%%%%%%%%%%%%%%%%%%%%%%%%%%%%%%%%%%
\section{Introduction}
\label{sec:intro}

Recently, ATLAS and CMS have reported the discovery of a
Higgs-like~\cite{higgs} resonance with a mass around
126~GeV~\cite{atlas,cms}. Decays to $\gamma \gamma$, $ZZ^*$, $WW^*$,
and recently $\tau \tau$~\cite{hcp,cms_spin} have been established, with
coupling strengths consistent with the Standard
Model~\cite{fits_th1,fits_th2,fits_ex}. While this `Higgs discovery'
is a great triumph for experimental and theoretical high energy
physics, the detailed study of this new state will require many years
of work at the LHC and possibly a future linear collider.\bigskip

The ultimate goal of such studies will be a term-by-term and
channel-by-channel confirmation of the structure of the Higgs
Lagrangian from data. If we assume that the observed resonance is
indeed responsible for electroweak symmetry breaking the leading
renormalized operators describing the couplings to all massive
fermions and gauge bosons are fixed. In an effective theory approach
higher-dimensional operators will induce additional (anomalous) Higgs
couplings, which can be constrained from
measurements~\cite{oscar}. Some of these higher-dimensional operators
are responsible for the most striking LHC observables --- Higgs
production in gluon fusion and Higgs decays to photons. Both of these
operators have mass dimension six before and five after electroweak
symmetry breaking, but they are not suppressed by a large mass
scale. In the presence of Yukawa couplings both, fermion and massive
gauge boson loops induce higher-dimensional operators suppressed by
powers of $v$~\cite{lecture}.

This structure of higher-dimensional operators suggests that we should
have a careful look at the structure of operators responsible for the
observation of the new `Higgs'
resonance~\cite{original,kentarou}. Only after the operator basis is
fixed we can determine the corresponding couplings. In that sense the
current coupling extractions~\cite{fits_th1,fits_th2,fits_ex} rely on
strong assumptions about the structure of the Higgs Lagrangian or the
Higgs couplings.\bigskip

The structure of the Higgs Lagrangian is strongly linked to, but not
entirely equivalent to the spin and quantum numbers of the heavy
resonance. An example is the tensor structure of a coupling of a
CP-even scalar to two $W$ bosons, which can be constructed using the
gauge field or the field strength tensor. Only the first can generate
a mass for the $W$ boson and unitarize the longitudinal Goldstone
boson scattering.  It is well known how to distinguish between
different coupling structures, using the
Cabibbo--Maksymowicz--Dell'Aquila--Nelson~\cite{nelson,zerwas} angles
in fully reconstructed $X \to ZZ$
decays~\cite{melnikov,lookalikes,cms_spin,jochum,spin_fatjets,spin_fatjets2}. We
will show in this paper that an equivalent way of determining the
Higgs coupling structure can be based on weak-boson-fusion (WBF)
events~\cite{original,kentarou,klamke,schumi,matt}. The two methods nicely
complement each other, covering a wide range of production and decay
channels~\cite{wernerb}. In addition, we will show how the angular $H
\to ZZ$ analysis can be applied to associated $ZH$ production with a
fully reconstructed decay $H \to b\bar{b}$~\cite{bdrs}.

In contrast to these two complete methods based on a full set of
angular correlations, the determination of coupling structures for
example from $X\to \gamma\gamma$ events is somewhat
limited~\cite{spinmeasus}. The same channel makes it unlikely that the
set of `Higgs' measurements is due to a spin-1 resonance, which cannot
decay to two neutral gauge bosons because of the Landau--Yang
theorem~\cite{landau_yang}. Nevertheless, we will see that there is a
case for a dedicated channel-by-channel LHC study, carefully keeping
track of implicit assumptions for example when testing spin-2 models
with and without form factor corrections.\bigskip

In this paper we develop a complete strategy to extract the structure
of the `Higgs' operators in weak boson fusion. We start by constructing
the appropriate observables and linking them to the $X \to ZZ$
angles. We then define the operator basis best suited for weak boson
fusion, which is slightly different from the basis used for Higgs
decays. In the second part of the paper we study jet-jet and jet-$X$
correlations. We show that the set of them is sufficient to probe our
set of coupling operators and compare the individual distinguishing
power for the most popular coupling structures. Finally, we add brief
comment on the angular correlation in associated $ZX$ production, with
a hadronic decay $X \to b \bar{b}$.

%%%%%%%%%%%%%%%%%%%%%%%%%%%%%%%%%%%%%%%%%%%%%%%%%%%%%%%%%%%%%%%%%%%%%%%%
\section{Kinematics}
\label{sec:setup}

Before we can develop a detailed strategy on how to extract the form
of the Higgs interaction Lagrangian at the LHC we need to define our
theoretical framework, our set of observables, and their relation for
example to the well-known $H \to ZZ$ analysis. The latter is usually
formulated in terms of a set of angles, where the angle between the
two $Z$ decay planes is one of the most promising observables. It is
proto-typical for linking the measurement of quantum numbers or
Lagrangian operators to angular correlations, leaving any kind of rate
measurement for an unbiased determination of the corresponding
coupling strength. The same approach we suggest for an angular
analysis of weak boson fusion Higgs events.  As we will see below,
weak boson fusion will offer one additional handle in form of
transverse momentum or energy spectra, but its analysis requires some
care in its interpretation.

%%%%%%%%%%%%%%%%%%%%%%%%%%%%%%%%%%%%%%%%%%%%%%%%%%%%%%%%%%%%%%%%%%%%%%%%
\subsection{Flipped Nelson}
\label{sec:angles}

%-------------------------------------------------------
\begin{figure}[t]
 \includegraphics[width=0.4\textwidth]{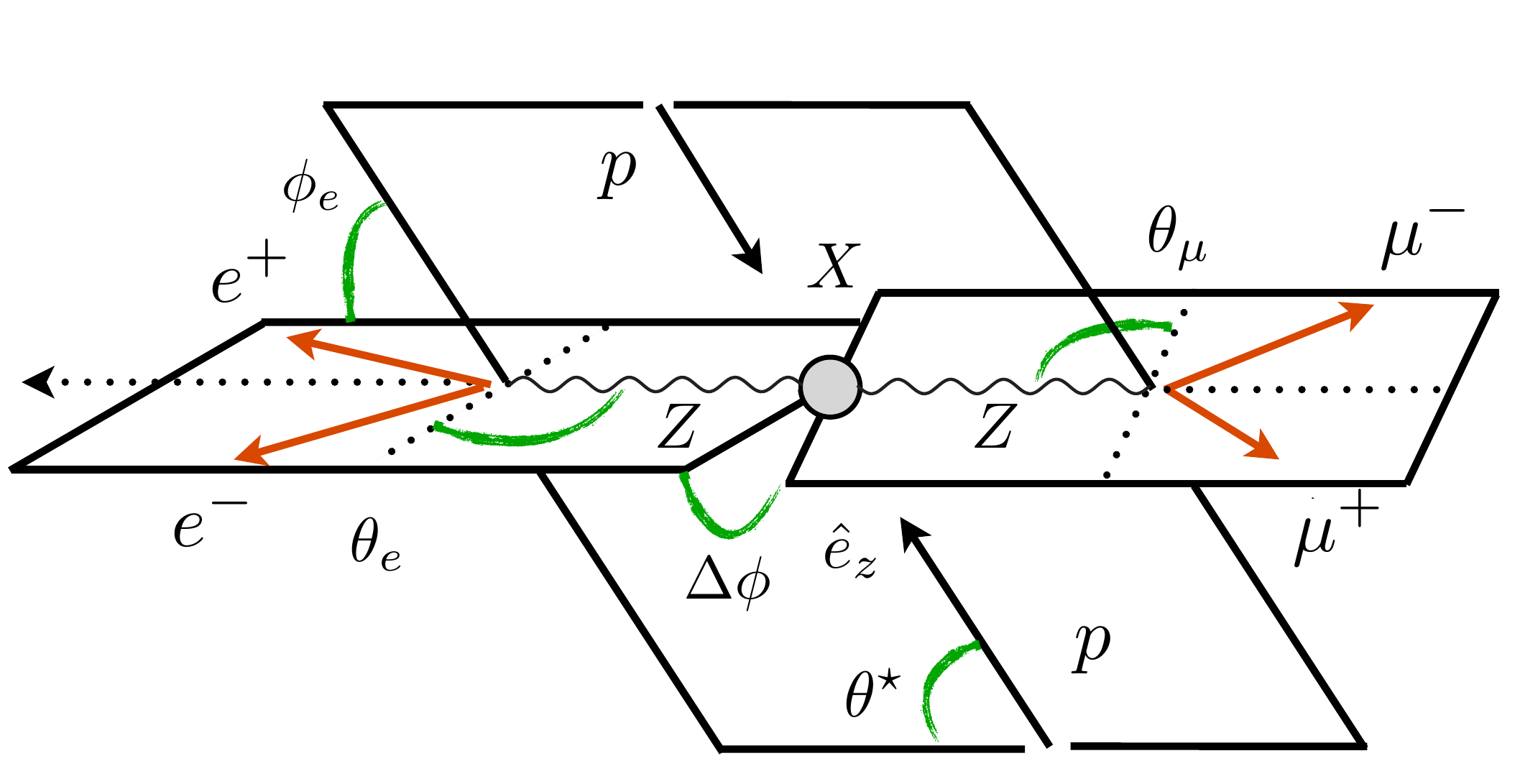}
 \hspace*{0.1\textwidth}
 \includegraphics[width=0.28\textwidth]{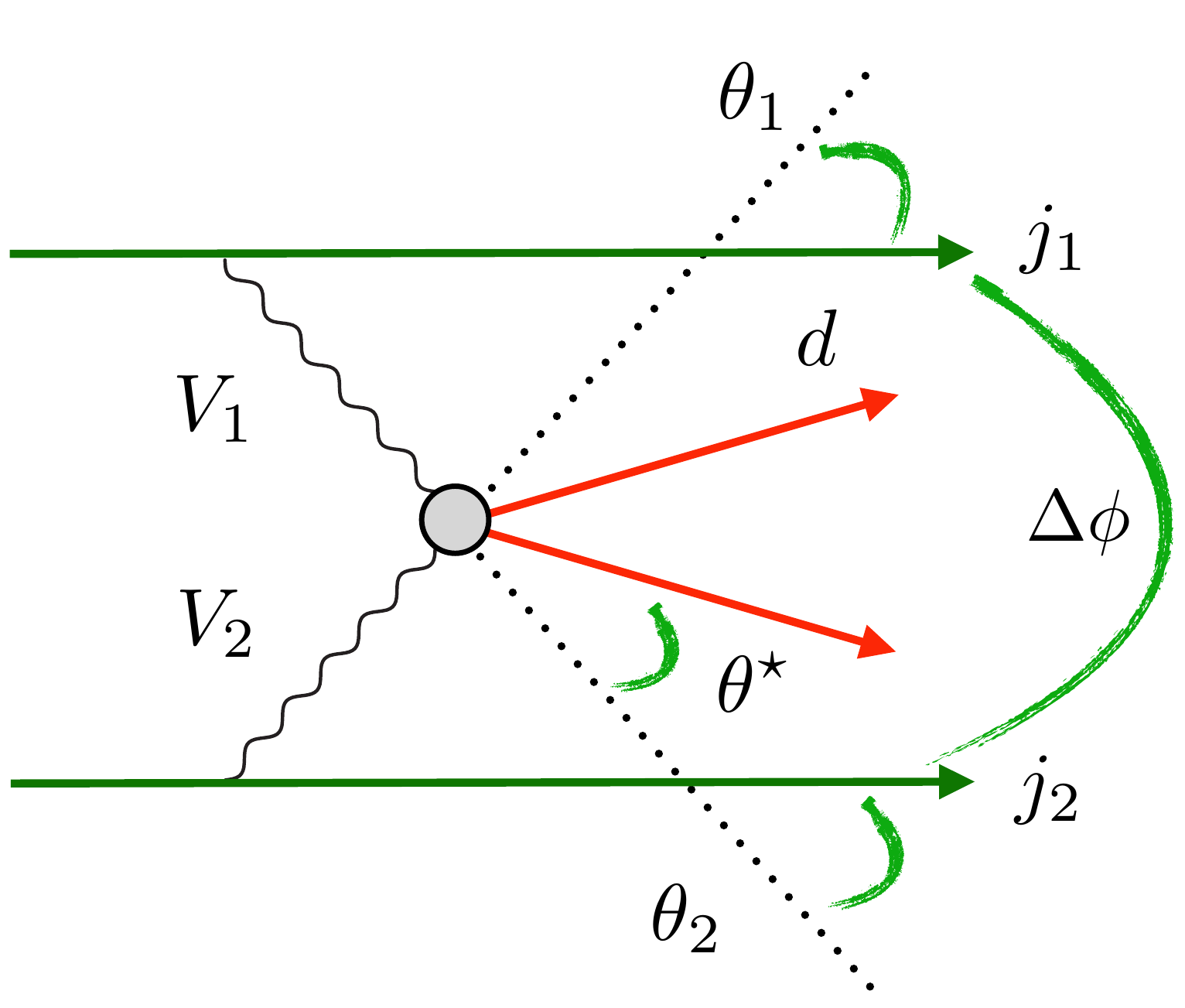}
\caption{Kinematic setup for the angular analysis of $H \to ZZ$ events
  (left) and Higgs events in WBF production (right). All angles are
  defined in Eq.\eqref{eq:angles_zz} and
  Eq.\eqref{eq:angles_wbf}.}
\label{fig:angles} 
\end{figure}
%-------------------------------------------------------

The `traditional' observables to measure the coupling structure 
of a massive state decaying to two weak gauge bosons 
are the Cabibbo--Maksymowicz--Dell'Aquila--Nelson~\cite{nelson,zerwas} angles. 
The kinematics for
the decay $X\to ZZ\to 4\ell$ is illustrated in the left panel of
Fig.~\ref{fig:angles}.  The four $Z$ decay momenta coming from a heavy
Higgs-like state $X$ are given by
\begin{alignat}{5}
p_X = p_{Z_e}+p_{Z_\mu} \; ,
\qqquad 
p_{Z_e} = p_{e^-}+p_{e^+} \; ,
\qqquad 
p_{Z_\mu} = p_{\mu^-}+p_{\mu^+} \; .
\end{alignat}
For each of these momenta and the beam direction we define unit
three-momenta $\hat{p}_i$ in the $X$ rest frame and in the two
$Z_{e,\mu}$ rest frames. Note that the one of the two $Z$ bosons will
be far of its mass shell, \ie $p^2 \ll m_Z^2$, but this does not pose
a problem for the boost into its reference frame. The set of
observable spin and CP angles are then defined
\begin{alignat}{5}
&\cos \theta_e =   
  \hat{p}_{e^-} \cdot\hat{p}_{Z_\mu} \Big|_{Z_e} 
&\qquad 
&\cos \theta_\mu =   
  \hat{p}_{\mu^-} \cdot\hat{p}_{Z_e} \Big|_{Z_\mu} 
\qqqquad 
\cos \theta^* = 
  \hat{p}_{Z_e} \cdot \hat{p}_\text{beam} \Big|_X  \notag \\
&\cos \phi_e = 
  (\hat{p}_\text{beam} \times \hat{p}_{Z_\mu}) \cdot (\hat{p}_{Z_\mu} \times \hat{p}_{e^-}) \Big|_{Z_e} 
  &\qquad
&\cos \Delta \phi = 
  (\hat{p}_{e^-} \times \hat{p}_{e^+}) \cdot (\hat{p}_{\mu^-} \times \hat{p}_{\mu^+}) \Big|_X \; .
\label{eq:angles_zz}
\end{alignat}
The index at the end of each relation indicates the rest frame in
which the angles are defined.
In the notation in Ref.~\cite{melnikov} this corresponds to $\phi_e
\to \Phi_1 $ and $ \Delta \phi \to \Phi$.  An important feature is
that the reconstruction of the angles defined in
Eq.\eqref{eq:angles_zz} requires a full reconstruction of the `Higgs'
decay at all stages. It does not require both $Z$ bosons to be
on-shell as long as we can boost into a well-defined center-of-mass
frame of the two decay leptons.  In spite of the suggestive notation
the angles $\phi$ and $\theta$ do not stand for opening angles and not
azimuthal or polar angles.\bigskip

%-------------------------------------------------------
\begin{figure}[t]
 \includegraphics[width=0.3\textwidth,clip]{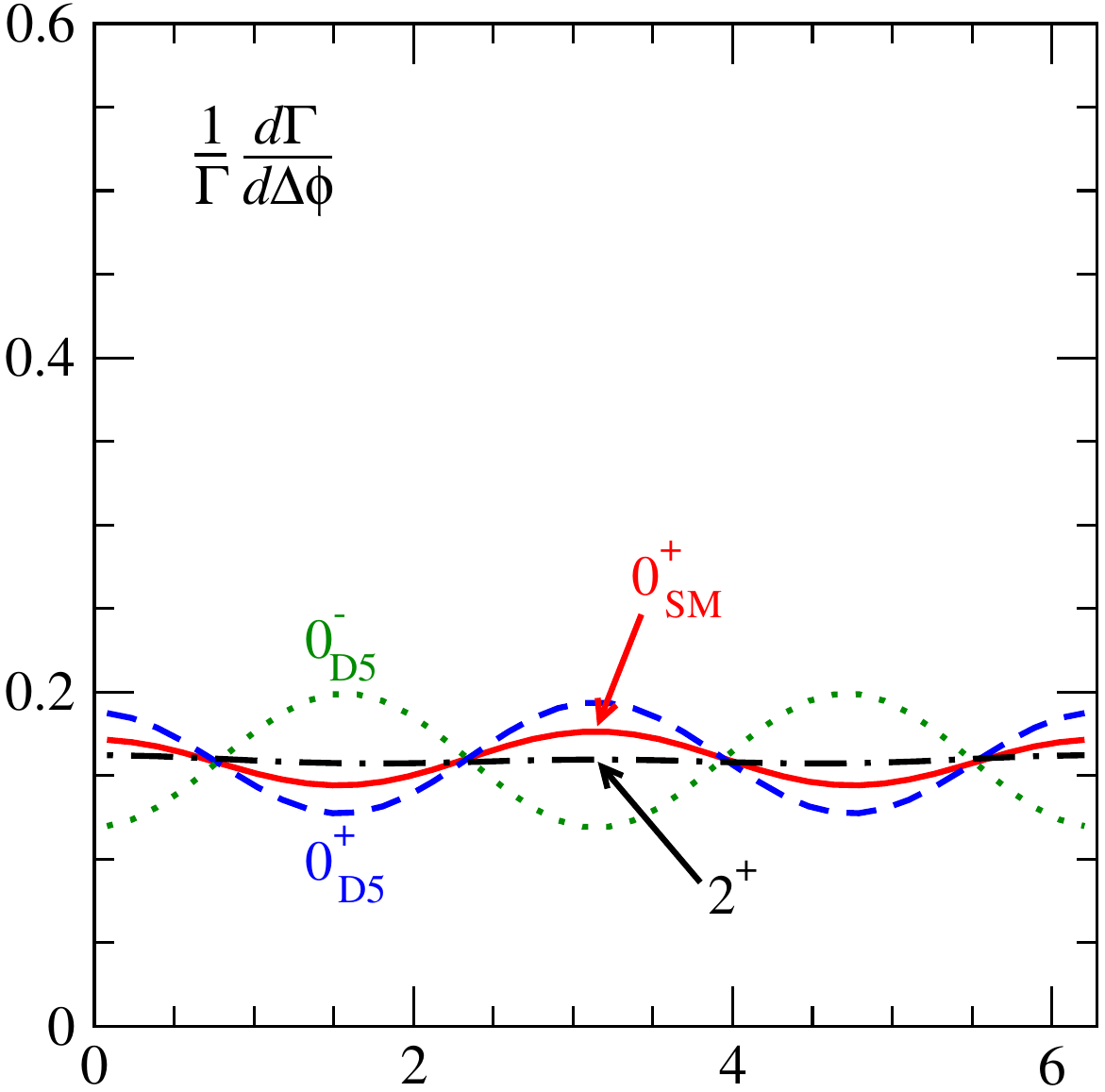}
 \hfill
 \includegraphics[width=0.3\textwidth,clip]{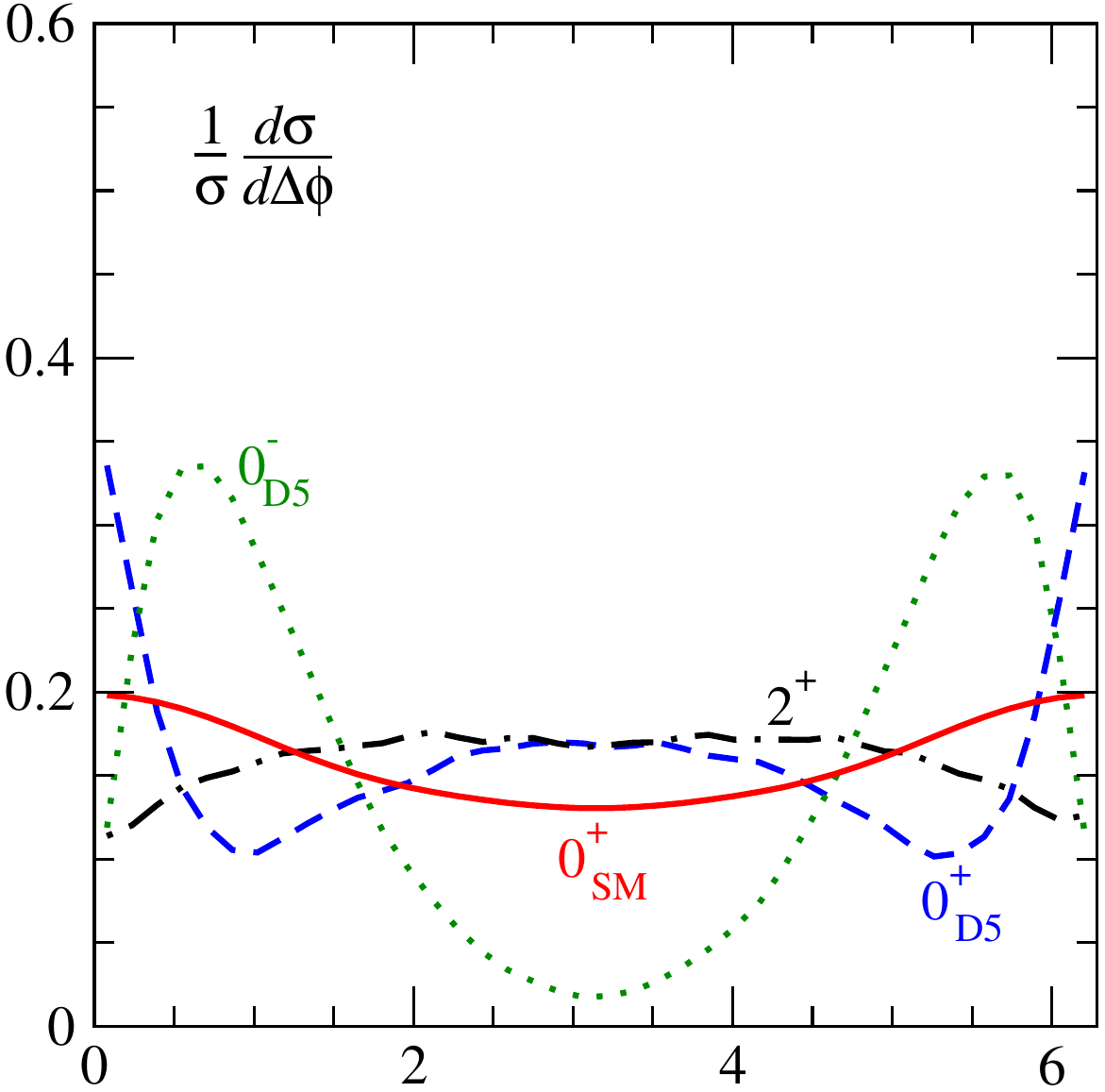}
 \hfill
 \includegraphics[width=0.3\textwidth,clip]{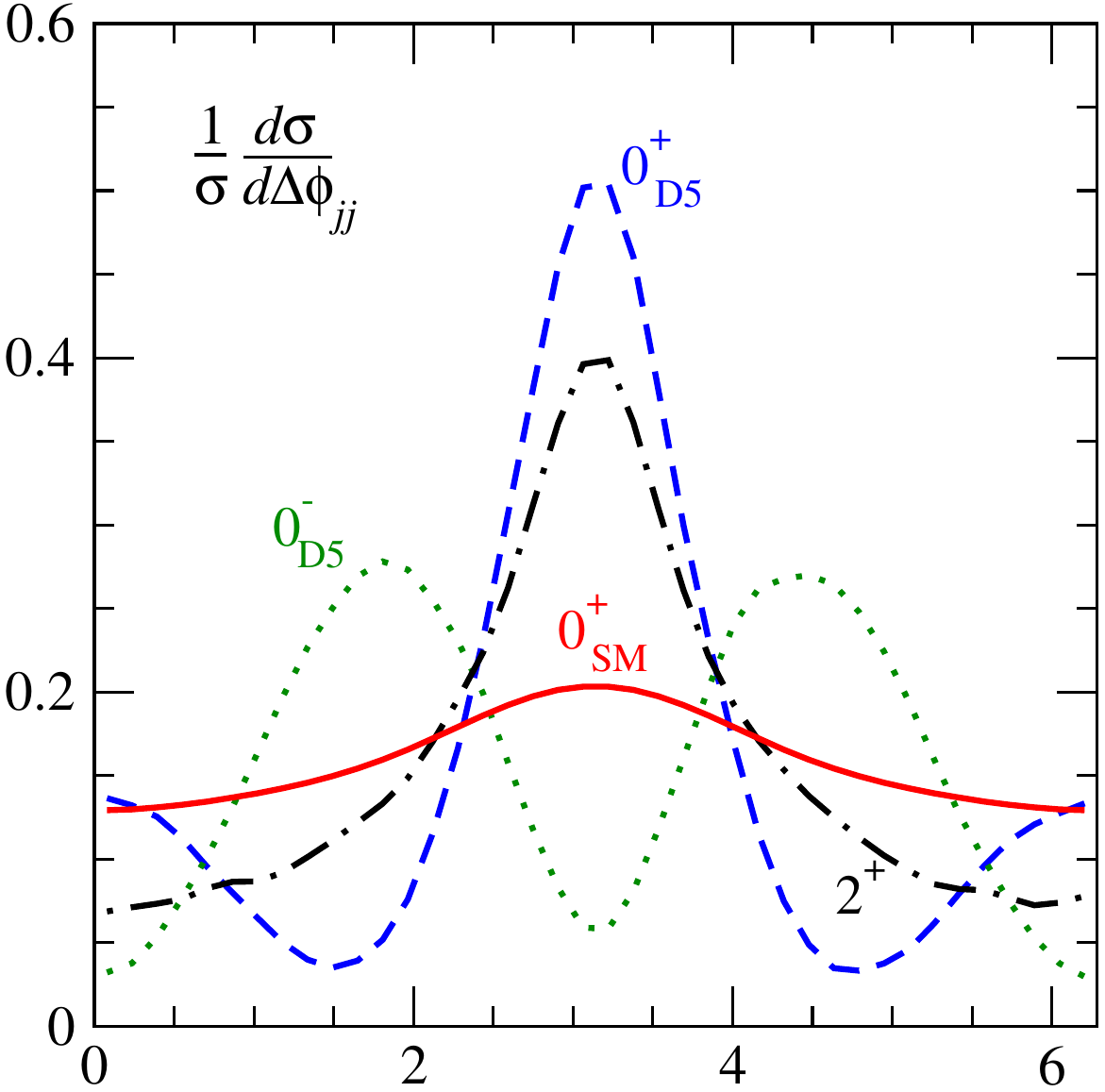}
\caption{Normalized $\Delta\phi$ distributions for $X\to ZZ$ events
  (left), WBF production in the Breit frame (center), and WBF
  production in the laboratory frame (right). We show the SM operator
  $X V_\mu V^\mu$ (solid), the CP-even dimension-5 operator 
  $X V_{\mu\nu} V^{\mu \nu}$ (dashed), the CP-odd dimension-5 operator 
  $X V_{\mu\nu}\widetilde{V}^{\mu \nu}$ (dotted), and an example for a
  spin-2 coupling (dashed-dotted). The operators are defined in
  Secs.~\ref{sec:spinzero} and \ref{sec:spintwo}.}
\label{fig:dphi} 
\end{figure}
%-------------------------------------------------------

As a first illustration we show the $\Delta\phi$ dependences for the
process $pp \rightarrow X \rightarrow ZZ \rightarrow
(e^+e^-)(\mu^+\mu^-)$ in the left panel of Fig.~\ref{fig:dphi}.  Our
hypotheses are the three allowed scalar $XZZ$ couplings structures to
mass dimension six (or five after symmetry
breaking)~\cite{kaoru_dieter,original} and a spin-2
operator~\cite{kentarou}. The corresponding operators are spelled out
in Sec.~\ref{sec:spinzero} and \ref{sec:spintwo}.

For the Standard Model coupling we expect this distribution to have a mild
modulation, which would vanish for large Higgs masses.  In contrast,
there are clear modulations in $\Delta\phi$ with a phase shift between
CP-even and CP-odd dimension-5 operators, which can be easily
understood from kinematics~\cite{original,kentarou}.\bigskip

The $X\to ZZ\to 4\ell$ topology and the weak-boson-fusion `Higgs'
production topology
\begin{alignat}{5}
q_1 q_2 \to j_1 j_2 \, (X \to d \bar{d} )
\label{eq:top_wbf}
\end{alignat}
are linked by a crossing symmetry. The labeling of the incoming and
outgoing partons as incoming quarks $q_{1,2}$ and outgoing jets
$j_{1,2}$ is only meant to allow for a definition of the angles
independently of the partonic sub-processes. The `Higgs' decay products
can be $d = \tau, W, Z, \gamma$, depending on the channel we are
looking at~\cite{wbf_tau,wbf_w,wbf_gamma}. For those observables which
require a full momentum reconstruction of $p_d$ the list of useful
Higgs decay channels is reduced.

Our aim is to generalize the angular basis of Eq.\eqref{eq:angles_zz}
to weak boson fusion, guided by the obvious crossing symmetry.  When
moving one of the final state partons to the initial state we
replace time-like $Z$ propagators with space-like $V=W,Z$ propagators
in the $t$-channel. In this situation we know that the corresponding
Breit frames are the appropriate reference frames. It is defined as
the reference frame where the momentum of the $t$-channel particle $V$
is completely space-like and can be reached through a conventional
boost. Similar to the $X\to ZZ$ case, a full reconstruction of the
$X$ decay is required.  With this caveat and fixing the directions of
all the external momenta as shown in Fig.~\ref{fig:angles} we define a
modified version of the five angles in Eq.\eqref{eq:angles_zz},
\begin{alignat}{5}
&\cos \theta_1 =   
   \hat{p}_{j_1} \cdot\hat{p}_{V_2} \Big|_{V_1 \text{Breit}} 
&\qquad 
&\cos \theta_2 =   
   \hat{p}_{j_2} \cdot\hat{p}_{V_1} \Big|_{V_2 \text{Breit}} 
\qqqquad 
\cos \theta^* = 
  \hat{p}_{V_1} \cdot \hat{p}_{d} \Big|_X  \notag \\
&\cos \phi_1 = 
  (\hat{p}_{V_2} \times \hat{p}_{d}) \cdot (\hat{p}_{V_2} \times \hat{p}_{j_1}) \Big|_{V_1 \text{Breit}} 
  &\qquad
&\cos \Delta \phi = 
  (\hat{p}_{q_1} \times \hat{p}_{j_1}) \cdot (\hat{p}_{q_2} \times \hat{p}_{j_2}) \Big|_X \; .
\label{eq:angles_wbf}
\end{alignat}
Again, we define the angle between the two planes of the Breit frames
as $\Delta \phi$. We will see below that this angle is
similar to the azimuthal difference of two jets in the laboratory
frame, $\Delta\phi_{jj}$, even though in Eq.\eqref{eq:angles_wbf}
$\phi$ and $\theta$ do not imply azimuthal or polar angles. It should
be noted that using these conventions we can define the angle
 $\Phi_+\equiv 2\phi_1 +\Delta \phi $  which typically lead to modulations
for spin-2 resonances. In the notation of the Ref.~\cite{kentarou} this corresponds
to $\Phi_+ \to \Phi_{12}$.\bigskip

Since we cannot measure the charge of the four fermions involved in
the WBF topology, we need an alternative criterion to map the leptonic
observables in Eq.\eqref{eq:angles_zz} to an LHC production process.
We break the external fermion degeneracy by imposing that each
incoming quark $q_{1,2}$ largely keeps its direction when the
scattering process turns into a tagging jet $j_{1,2}$. This defines
the weak boson 3-momenta
\begin{alignat}{5}
  \vec{p}_{V_{1,2}} = \vec{p}_{q_{1,2}}-\vec{p}_{j_{1,2}} \; ,
 \end{alignat}
where $q_{1,2}$ are the incoming partons moving in the beam direction,
$\hat{p}_{q_{1,2}} = \pm \hat{e}_\text{beam}$. 

Technically, the determination of the incoming particle momenta
$q_{1,2}$ is an issue at hadron colliders (and ill-defined in
perturbative QCD). We start by reconstructing the final state and map
it onto a set of partonic momenta. In the absence of additional jet
radiation this defines the center-of-mass system for the incoming
quarks, related to the laboratory frame by a longitudinal boost. The
incoming parton momenta are parameterized as $p_{q_1}=(E_1,0,0,E_1)$
and $p_{q_2}=(E_2,0,0,-E_2)$, which we can invert to express $E_1$ and
$E_2$ in terms of the summed energy and longitudinal momentum entries
of the final state particles. The boost from the laboratory frame to
the center-of-mass system can be approximated by reconstructing the
the events' pseudorapidity from the detector geometry and massless
calorimeter hits. This should be understood as a prescription to
obtain a set of well-defined (leading order, parton-like) four-momenta
rather than reconstructing the actual initial state.\bigskip

Among the angles listed in Eq.\eqref{eq:angles_wbf}  
$\Delta \phi$ does not require a reconstruction of
the Higgs candidate. It is constructed only from the four external
partons, two incoming and two outgoing. Comparing
Eq.\eqref{eq:angles_wbf} and Eq.\eqref{eq:angles_zz} we see that it
corresponds to the angle between the two $Z$ decay planes in $X \to
ZZ$ decays. In the central panel of Fig.~\ref{fig:dphi} we show it for
illustration, testing the same three spin-0 hypotheses as in the left
panel and properly defined in Sec.~\ref{sec:spinzero}. The Standard
Model expectation is again relatively flat, with an additional
residual dependence introduced the interference of longitudinal and
transverse amplitudes and by kinematic cuts.  The two dimension-5
operators show a distinct modulation, in complete analogy to the
corresponding measurement in $X \to ZZ$ decays, but with a larger
amplitude of the modulation.\bigskip

%-------------------------------------------------------
\begin{figure}[t]
  \includegraphics[width=0.3\textwidth]{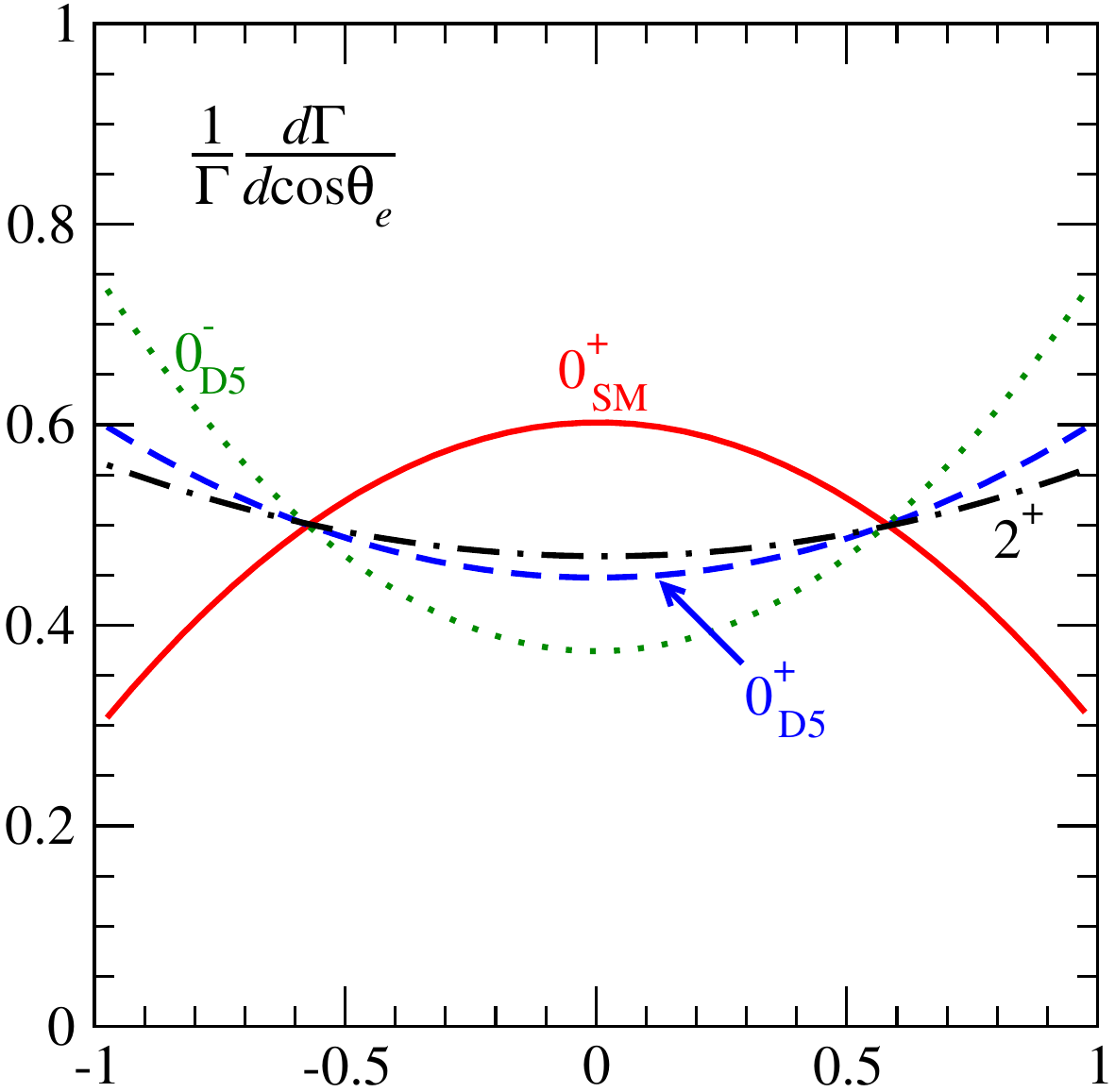}
  \hfill
  \includegraphics[width=0.29\textwidth]{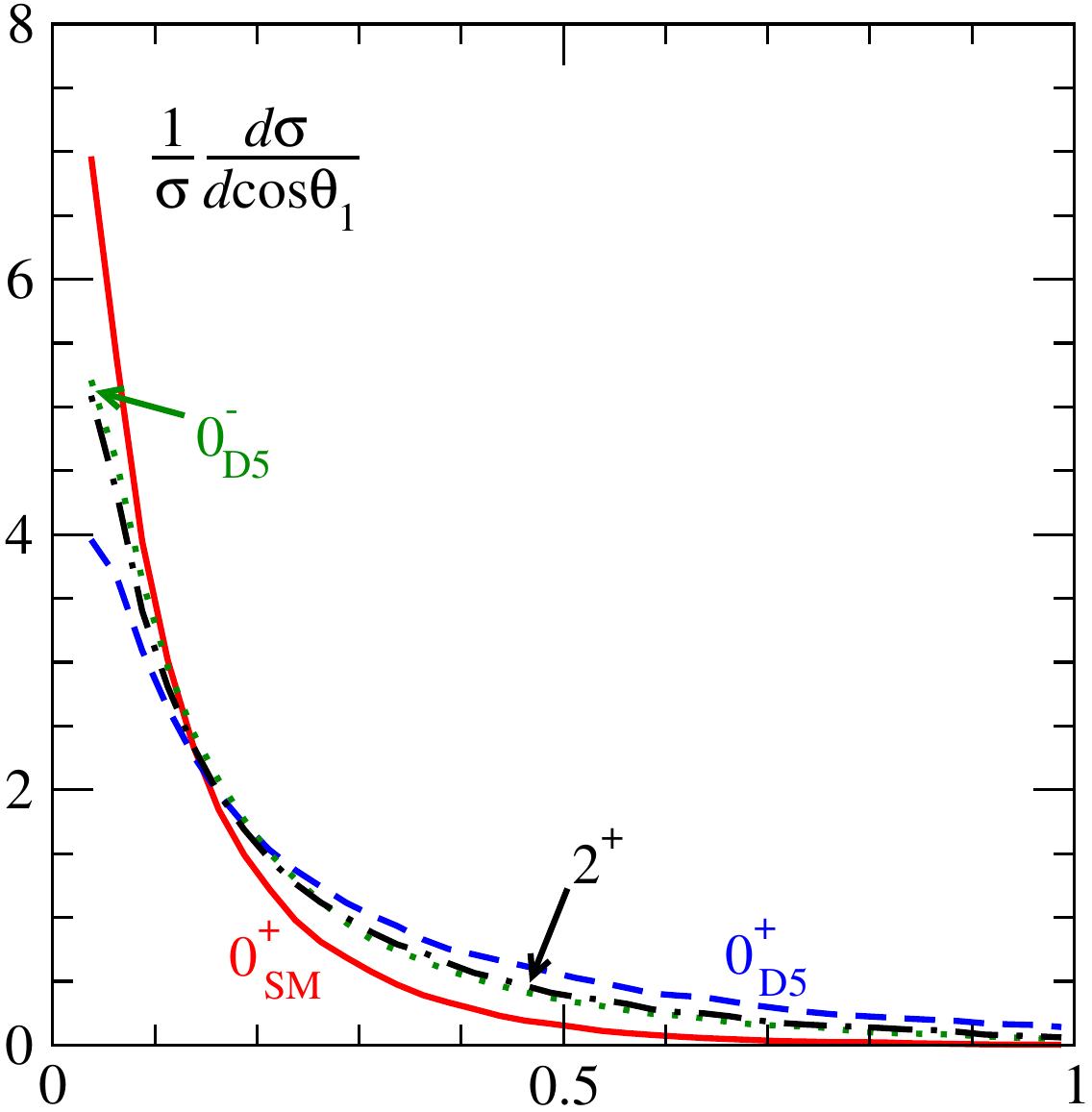}
  \hfill
  \includegraphics[width=0.3\textwidth,clip]{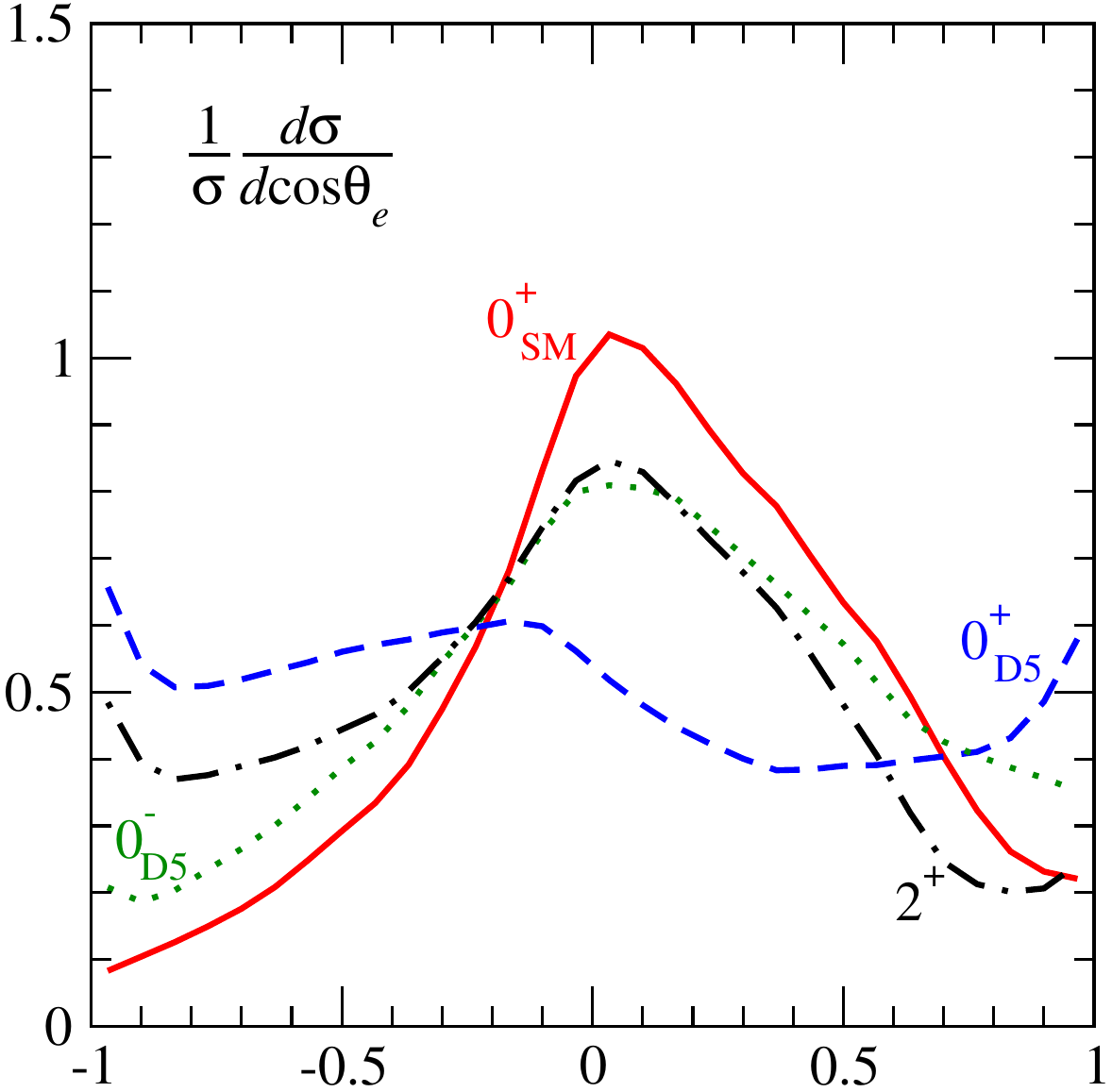}
\caption{Normalized $\cos\theta$ distributions for $X\to ZZ$ events
  (left), WBF production in the Breit frame (center), and WBF
  production after analytic continuation (right).  Again, we show $X
  V_\mu V^\mu$ (solid), $X V_{\mu \nu} V^{\mu \nu}$ (dashed), $X
  V_{\mu\nu}\widetilde{V}^{\mu \nu}$ (dotted), and the spin-2 example
  (dashed-dotted), as defined in Secs.~\ref{sec:spinzero} and
  \ref{sec:spintwo}.}
\label{fig:costheta}
\end{figure}
%-------------------------------------------------------

A second key observable in the angular analysis of the $X \to ZZ$
topology is the helicity angle $\theta_{e,\mu}$. In the left panel of
Fig.~\ref{fig:costheta} we show how it discriminates between a
spin-0 and spin-2 resonance~\cite{melnikov}.  The spin-2 Lagrangian is
defined in Sec.~\ref{sec:spintwo}.  
For the crossed WBF topology the
center panel shows the Breit frame version, defined in
Eq.\eqref{eq:angles_wbf}.  It is only of limited use to differentiate
between the different hypotheses.  

The translation of the $H \to ZZ$ angles into the Breit frame basis of
Eq.\eqref{eq:angles_wbf} is not the only possible generalization. To
avoid a boost into the Breit frame we can alternatively construct a
crossed set of momenta simply mapping the $pp\to Xjj$ kinematics to
the $X \to 4\ell$ kinematics. This way we can show the distributions
in terms of the angles defined in Eq.\eqref{eq:angles_zz}: the WBF
topology is related to the decay $X \to ZZ\to 4\ell$ via the crossing
symmetry $p_{q_i}\rightarrow -p_{q_i}$. This crossing we can perform
by analytically continuing the WBF initial state momenta to final
state momenta, rendering $p_V$ time-like.  We illustrate this
alternative mapping to space-like momenta $p_V$ in the right panel of
Fig.~\ref{fig:costheta}, including basic acceptance cuts. We see that
this observable is more distinctive than its Breit frame counterpart,
but not as clear as the decay angle correlation.\bigskip

%-------------------------------------------------------
\begin{figure}[t]
 \includegraphics[width=0.33\textwidth]{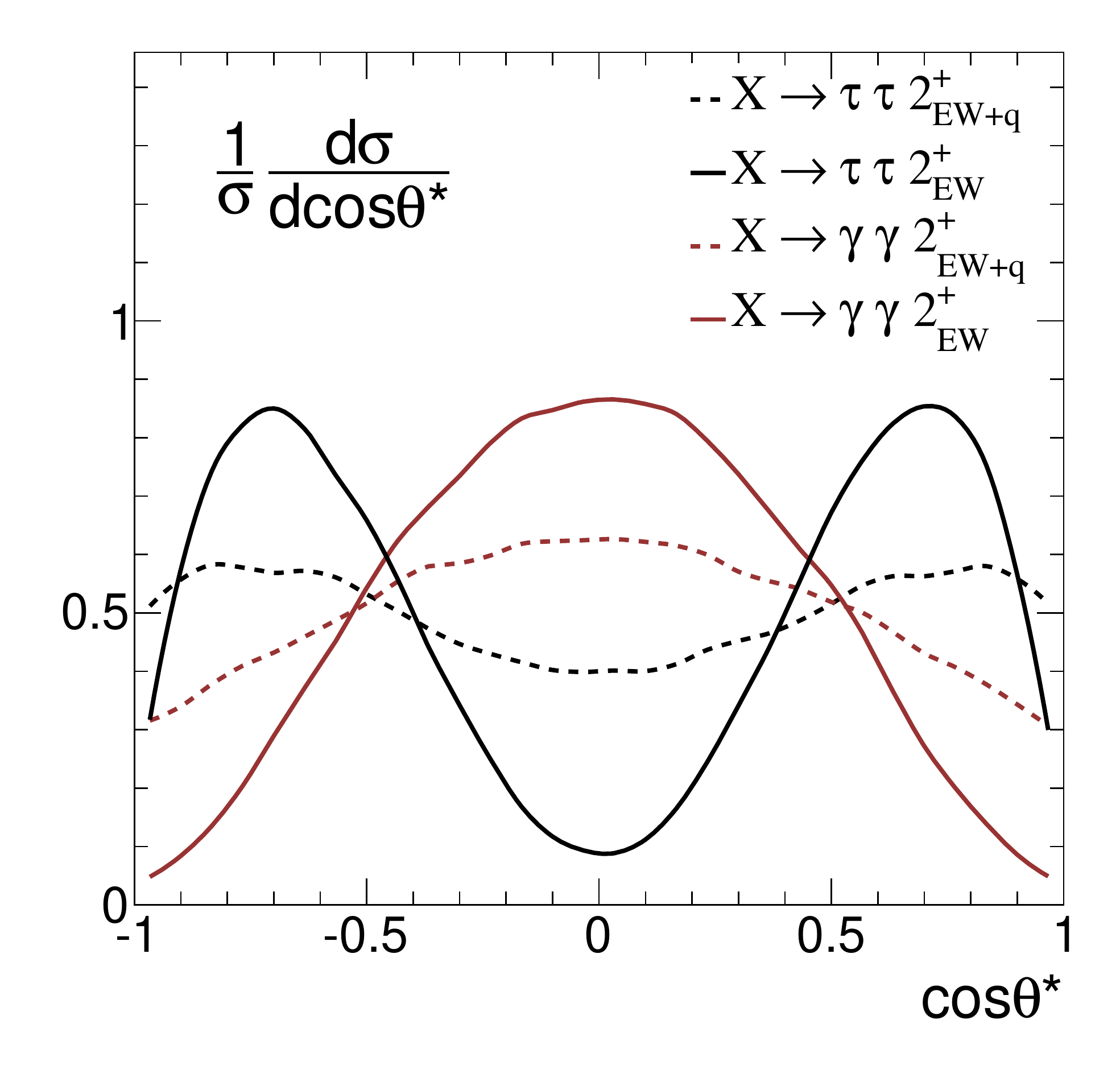}
 \hspace{0.1\textwidth}
 \includegraphics[width=0.33\textwidth]{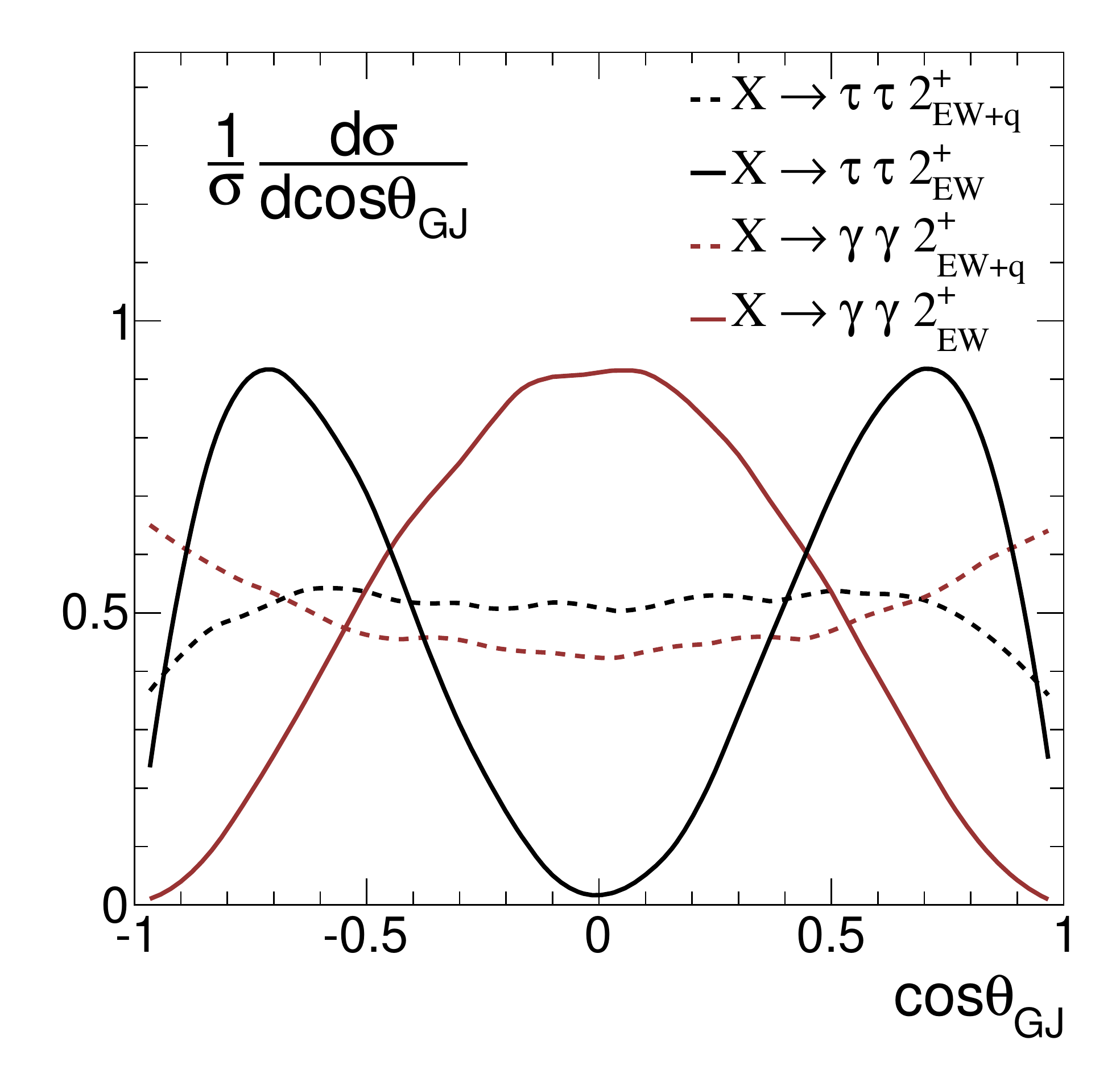}
\caption{Normalized $\cos\theta^*$ and $\cos\theta_{GJ}$ distributions for a
  spin-2 resonance coupling only to weak bosons or coupling to
  quarks as well. We show the $X$ decay to tau
  leptons and to photons.}
\label{fig:gottfried}
\end{figure}
%-------------------------------------------------------

The different observations discussed in this section gives us
confidence that a detailed study of angular correlation in WBF events
should serve as a powerful supplement to the study of $X \to ZZ$
decays.

Additional observables have been suggested for a dedicated study of a
spin-2 resonance. One of them is the sum of the two $V_{1,2}$ Breit
frame angles $\Phi_+$~\cite{kentarou}. Another is the
so-called Gottfried--Jackson angle
$\theta_{GJ}$~\cite{mrauch,gottfried}. It is the angle between the
momentum of the resonance $X$ in the laboratory frame and one of the
$X$ decay products in the $X$ rest frame (somewhat in analogy of the
top helicity angle).  Compared to the angular basis of
Eq.\eqref{eq:angles_wbf} it is very closely related to $\theta^*$, as
confirmed in Fig.~\ref{fig:gottfried}. For spin-0 and spin-1 both
distributions show a flat behavior, as shown in the Appendix. For
spin-2 the two angles are essentially equivalent.  It will turn out
that the angle $\theta^*$ appears to be more stable with respect to UV
completions or the presence of form factors appearing in the spin-2
theory, so we will focus on $\theta^*$ in the following.

%%%%%%%%%%%%%%%%%%%%%%%%%%%%%%%%%%%%%%%%%%%%%%%%%%%%%%%%%%%%%%%%%%%%%%%%
\subsection{Hadron collider observables}
\label{sec:hadcoll}

While the kinematic observables introduced in Eq.\eqref{eq:angles_wbf}
are well suited to capture the properties of a Higgs candidate and its
coupling to gauge bosons, they are not particularly well suited to
cope with QCD effects at hadron colliders. The reason is that the
reconstruction of the $V_{1,2}$ Breit frames assumes a complete
reconstruction of the hard scattering process. This assumption is not
realistic in the presence of QCD jet radiation on the one hand and
underlying event and pile-up on the other hand.\bigskip

As a rough guide line we can look at the opening angle between the two
$Z$ decay planes in $X \to ZZ$ decays again. For the WBF production
process it turns into $\cos \Delta \phi = (\hat{p}_{q_1} \times
\hat{p}_{j_1}) \cdot (\hat{p}_{q_2} \times \hat{p}_{j_2})$, evaluated
in the Higgs candidate's rest frame. In the laboratory frame each
vector $(\hat{p}_q \times \hat{p}_j)$ by definition resides in the
azimuthal plane. In the Higgs candidate's rest frame this changes, but
the impact of a modest transverse boost on the large longitudinal
parton momenta should be small as well. If we compute $\Delta \phi$ in
the laboratory frame instead, it turns into the azimuthal angle
between two vectors, each orthogonal to one of the tagging jet
direction. This is equivalent to the azimuthal angle between the two
tagging jets $\Delta \phi_{jj}$. In the right panel of
Fig.~\ref{fig:dphi} we show this azimuthal opening angle for the WBF
production of four Higgs candidates, with the scalar tree-level and
dimension-5 $XVV$ couplings or a spin-2 coupling. As expected, the
Standard Model operator predicts an essentially flat behavior, while
the other two show a distinctly different
modulation~\cite{original,schumi}, arguably the strongest of the three
approaches compared in this figure.\bigskip

Encouraged by this observation we define a new set of (essentially)
angular correlations for the WBF Higgs production process
\begin{alignat}{5}
q_1 q_2 \to j_1 j_2 \, (X \to d \bar{d} ) \; .
\label{eq:top_wbf2}
\end{alignat}
It simply uses the rapidity differences and azimuthal opening angles
between the tagging jets, the Higgs candidate (if reconstructed), and
its decay products:
\begin{alignat}{5}
\left\{ \Delta \eta_{mn}, \Delta \phi_{mn} \right\} 
\qqquad \text{for} \quad m,n = j_{1,2}, X, d, \bar{d} \; .
\label{eq:deltaetaphi}
\end{alignat}
The two tagging jets $j_{1,2}$ now have to be ordered, so we defined
$j_1$ to be harder of the two.
Should there be additional jets in
the event we choose the hardest two jets which qualify as tagging jets
following the basic cuts defined in Sec.~\ref{sec:analysis}.

Obviously, this set of observables over-constrains the kinematics in
the laboratory frame. The dimensionality of the phase space is eight,
which includes only five independent angles. While the collider
observables of Eq.\eqref{eq:deltaetaphi} might not be the most
obvious choice for testing spin and CP properties of a heavy resonance
$X$, the case of $\Delta \phi_{jj}$ shows their potential, and they
are well-defined experimental quantities at hadron colliders.\bigskip

One additional observable of interest is the boost which relates the
lab frame to the `Higgs' and the partonic $Xjj$ center-of-mass frames.
It is related to the kind of parton in the initial state,
distinguishing valence quarks from sea quarks or gluons. While it is
possible to for example construct asymmetries in the two frames, we do
not find clear links to the coupling structure of a massive particle
in weak boson fusion.\bigskip

Finally, at least for a cross check it might be useful to add
information which is not encoded in the rapidity differences or angles
in Eq.\eqref{eq:deltaetaphi}. One such observable is the transverse
momentum of the tagging jet, where the splitting probability
$P_{T,L}(x)$ is known to look different for transverse and
longitudinally polarized gauge bosons~\cite{liantao,lecture}
\begin{alignat}{5}
P_T(x,p_T) &\propto 
                 \frac{1+(1-x)^2}{2x} \;
                 \frac{p_T^2}{(p_T^2 + (1-x) \, m_W^2)^2} 
          &\quad \to \quad&
                 \frac{1+(1-x)^2}{2x} \;
                 \frac{1}{p_T^2} 
                 \notag \\
P_L(x,p_T) &\propto 
                 \frac{1-x}{x} \;
                 \frac{m_W^2}{(p_T^2 + (1-x) \, m_W^2)^2} 
          &\quad \to \quad&
                 \frac{1-x}{x} \;
                 \frac{m_W^2}{p_T^4} \; .
\label{eq:ptj}
\end{alignat}
Looking at large transverse momenta $p_T\gg m_V$ the radiation of
longitudinal $V$ bosons falls off sharper than the radiation of
transverse $V$ bosons.  Even more distinctively, for spin-1 or spin-2
particle production we know that these states do not unitarize the
partonic processes. This difference can have a huge effect on the
transverse momenta of the tagging jets~\cite{samesign,wolfgang},
depending on the additional unitarization procedure.

%%%%%%%%%%%%%%%%%%%%%%%%%%%%%%%%%%%%%%%%%%%%%%%%%%%%%%%%%%%%%%%%%%%%%%%%
\subsection{Contaminating sub-processes}
\label{sec:gf}

The main advantage of the leptonic decay $H \to ZZ \to 4\ell$ is that
after some basic kinematic cuts the backgrounds are negligible. This
holds for non-Higgs events, for example from continuum $ZZ$
production, but also for signal events which might have different
topology or different intermediate states.\footnote{A potentially
  dangerous contribution would be from Higgs decays to $Z\gamma$,
  where for a dimension-5 $HZZ$ coupling we would not expect any
  hierarchy between the effective $g_{HZZ}$ and $g_{HZ\gamma}$
  couplings.}\bigskip

Once we switch to WBF Higgs production the non-Higgs
backgrounds are still small, $S/B = \mathcal{O}(1)$, but not entirely
negligible. The angular correlations for example of the $Z \to \tau
\tau$ background process is well understood, which means that they can
be taken into account in the final
analysis~\cite{original}. Additional partonic sub-process contributing
to WBF Higgs production are
\begin{alignat}{5}
q \bar{q} \to q \bar{q} H, gg H
\qqquad
q g \to qg H
\qqquad
\bar{q} g \to \bar{q} g H
\qqquad
g g \to q\bar{q} H, gg H \; ,
\label{eq:gf}
\end{alignat}
all mediated by the effective $Hgg$ coupling, \ie with cross sections
of the order $\sigma \propto \alpha_s^2 g_{Hgg}^2$. {\sl Per se},
these rates are not small. After the basic acceptance cuts, shown in
Sec.~\ref{sec:analysis}, they can exceed the WBF signal. However, to
remove the QCD-induced $Zjj$ background at order $\alpha \alpha_s^2$
we need to apply a stiff cut on the invariant mass of the two tagging
jets $m_{jj} > 600 - 800$~GeV. This cut reduces the contamination of
gluon-fusion Higgs events in the WBF sample to a small
perturbation.\bigskip

%-------------------------------------------------------
\begin{figure}[t]
  \includegraphics[width=0.5\textwidth]{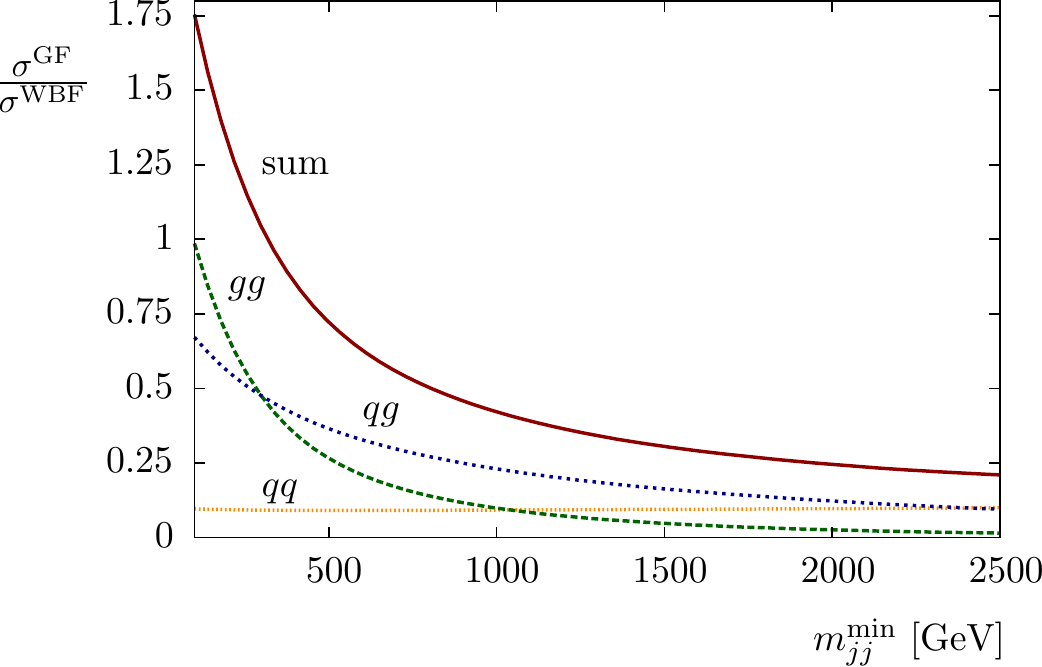}
  \caption{Contamination of WBF Higgs events with gluon-fusion events
    as a function of $m_{jj}^\text{min}$. Note that the unphysical
    separation of the partonic processes contributing to the order
    $\alpha_s^2 g_{Hgg}^2$ is only possible at leading order and scale
    dependent. The NLO rates are based on {\sc
      Vbfnlo}~\cite{vbfnlo,klamke}.}
\label{fig:gf_mjj}
\end{figure}
%-------------------------------------------------------

In Fig.~\ref{fig:gf_mjj} we show the relative size of the Higgs
production rates in gluon fusion and in weak boson fusion. Indeed, for
$m_{jj}^\text{min} \sim 250$ the two production modes still contribute
with similar rates. When we increasing $m_{jj}^\text{min}$ towards
1~TeV the combined gluon fusion contribution rapidly falls
off. Splitting it into the different contributions of order
$\alpha_s^2 g_{Hgg}^2$, as given by the partonic sub-processes, we see
that the originally dominant $gg$ initial state drops below 20\% of
the WBF rate around $m_{jj}^\text{min} \sim 500$~GeV. The mixed
quark-gluon initial state is the most dangerous, because it is almost
as large as the gluon fusion rate, but it drops only very slowly with
a remaining $30\%$ correction to the weak boson fusion rate for
realistic $m_{jj}^\text{min}$ values.  In contrast, the purely
quark-induced sub-process gives a constant correction around 10\% and
cannot be reduced choosing larger $m_{jj}^\text{min}$ values.

In combination, after applying tagging jet acceptance cuts and the
usual $m_{jj}^\text{min}$ cut we roughly expect one third of the Higgs
signal events to arise from gluon fusion. However, in a realistic
analysis we need a minijet veto to further reduce the $t\bar{t}$ and
QCD $Z$+jets backgrounds. The original (still largely valid) estimates
for the survival probabilities against such a veto range around 90\%
for the WBF signal and 30\% for the QCD $Z$+jets
background~\cite{wbf_w,wbf_tau}. Gluon fusion Higgs production will be
even more strongly suppressed, with a recently estimated jet veto
survival probability below 20\%~\cite{jetveto,manchester}. This means
that after applying such a jet veto the gluon fusion contamination
will drop from 50\% to a negligible 10\%. For $m_{jj} > 1$~TeV it
ranges around 3\%.  The interference between gluon fusion and WBF
diagrams can be safely neglected, essentially independently of the
cuts~\cite{hjjint}. In our detailed analysis we therefore omit all
gluon-induced operators and show the corresponding distributions only
in the Appendix.\bigskip

%-------------------------------------------------------
\begin{figure}[t]
 \includegraphics[width=0.32\textwidth]{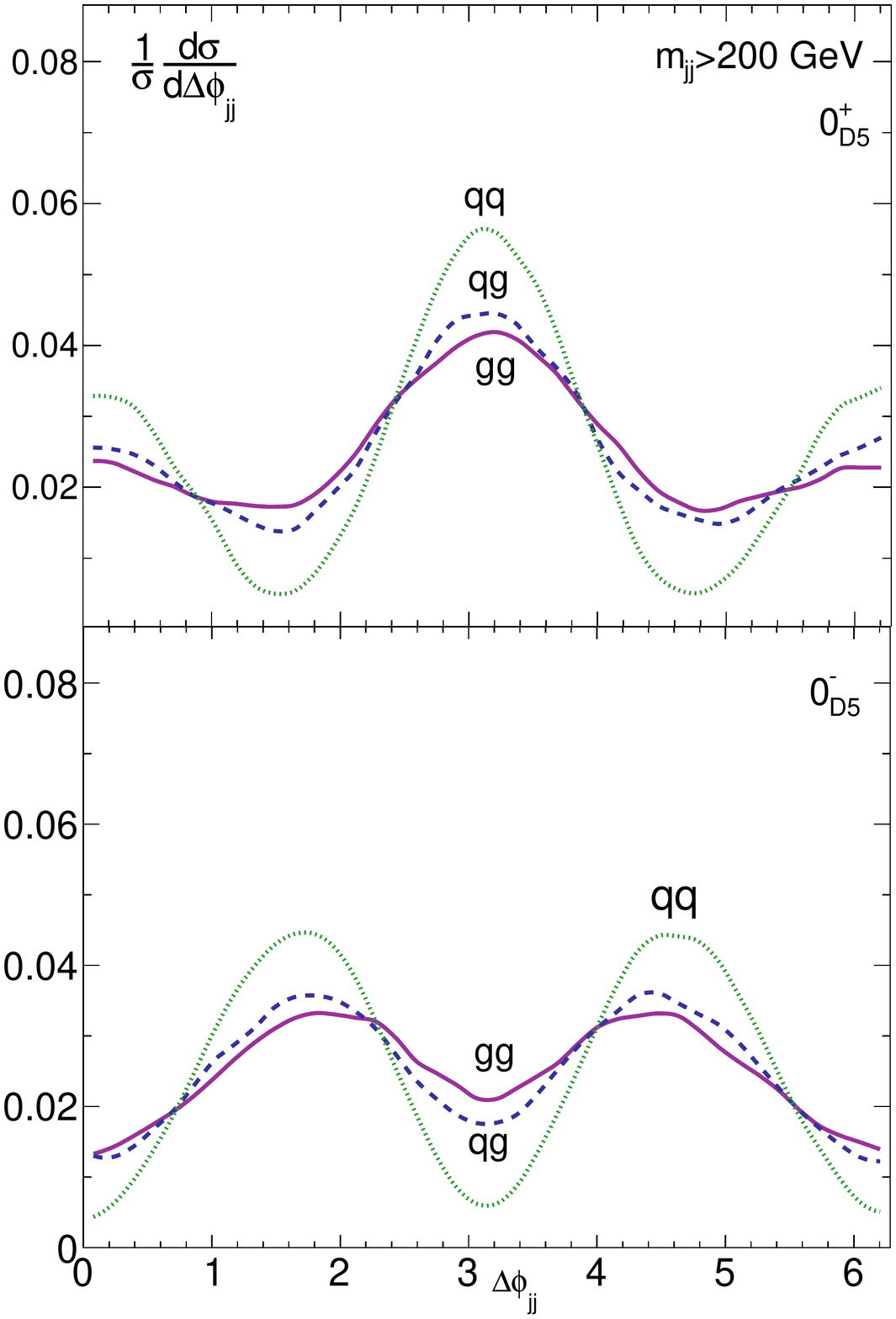}
 \hfill
 \includegraphics[width=0.32\textwidth]{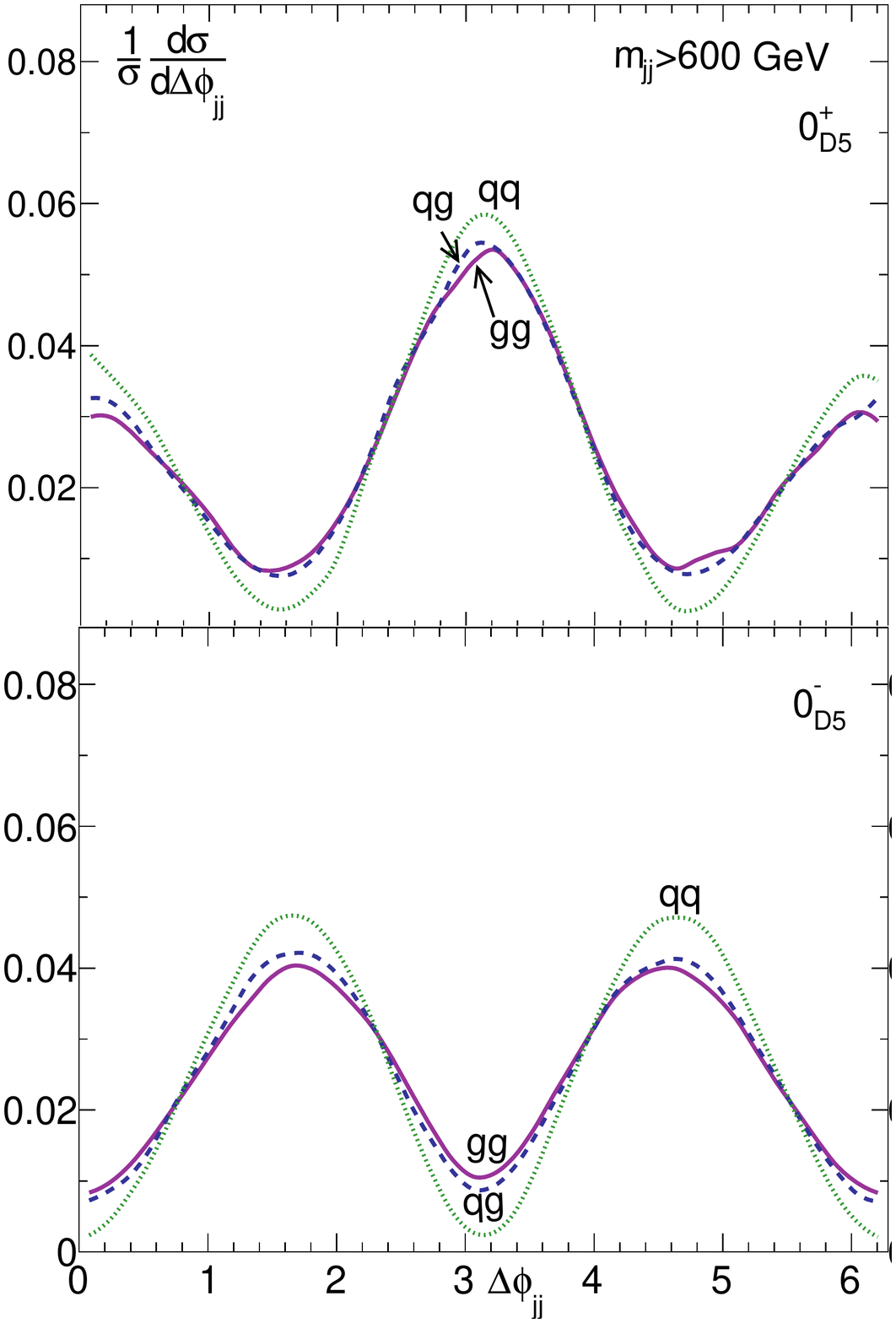}
 \hfill
 \includegraphics[width=0.32\textwidth]{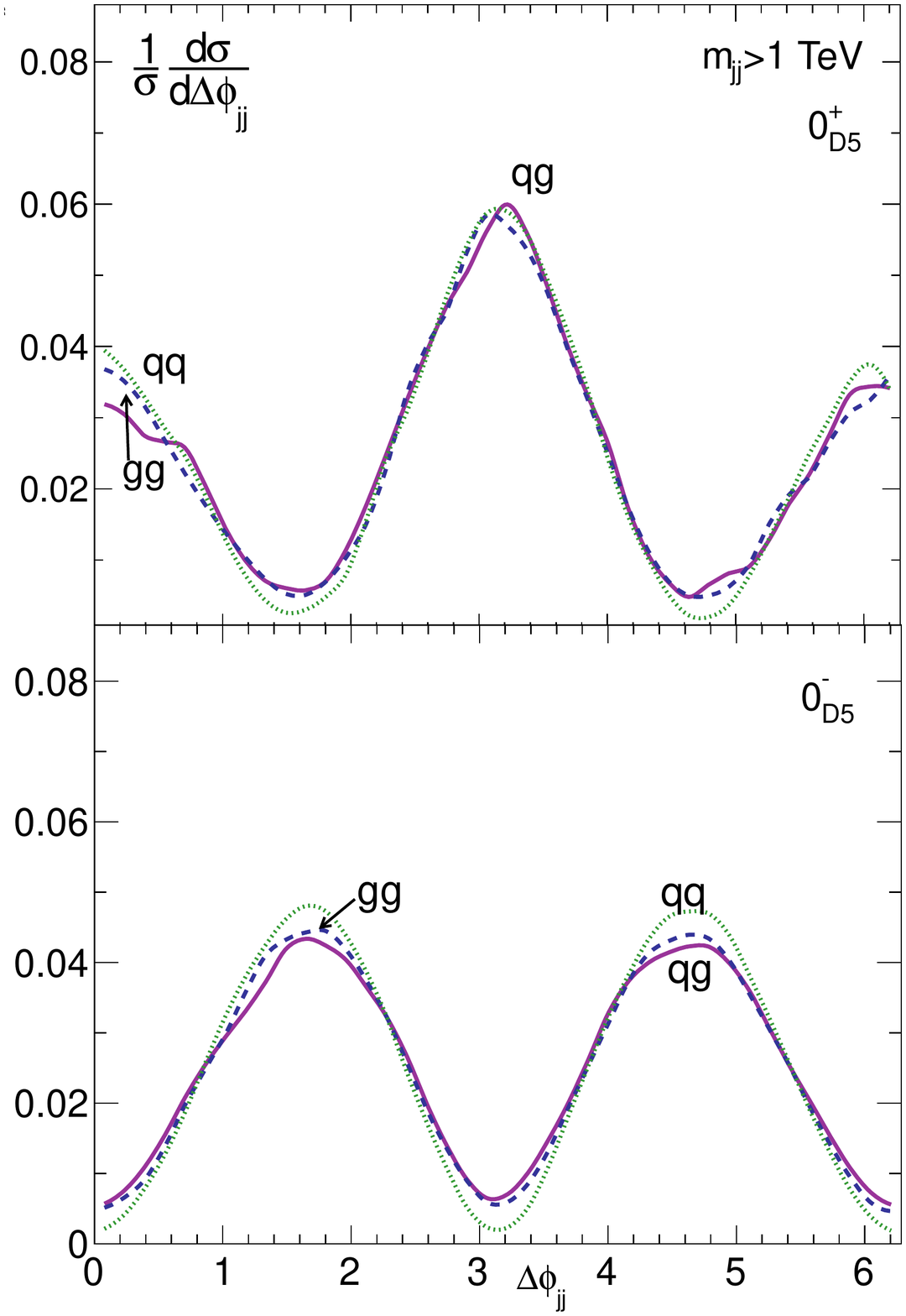}
\caption{Normalized $\Delta\phi_{jj}$ distributions for Higgs
  production in gluon fusion, separated by partonic channels to
  leading order. The CP even Higgs is represented on the top figures
  and the CP odd on the bottom ones. The $m_{jj}^\text{min}$ cut is
  increased from 200~GeV (left) to 600~GeV (central) and 1~TeV
  (right).}
\label{fig:gf_phi}
\end{figure}
%-------------------------------------------------------

In addition to the fact that the WBF Higgs event sample will only have
a small contribution from gluon fusion, there is no problem in
carrying a percentage of gluon fusion event through the Higgs coupling
analysis. For example, for a scalar Higgs boson the CP quantum number
entirely fixes the form of the effective $Hgg$ coupling structure. As
an example, in Fig.~\ref{fig:gf_phi} we show the $\Delta\phi_{jj}$
distribution for gluon fusion Higgs production, split into the three
partonic sub-processes listed in Eq.\eqref{eq:gf}. The partonic
sub-processes show a very similar behavior, essentially independent of
the detailed cuts. Once we require at least $m_{jj}^\text{min} =
600$~GeV the curves are completely degenerate. The effective
Higgs-gluon coupling in the Standard Model has the same structure as
the CP-even dimension-5 operator for weak gauge bosons, $H V_{\mu \nu}
V^{\mu \nu}$, which explains the quantitative agreement of these
results with the right panel of Fig~\ref{fig:dphi}. The
$m_{jj}^\text{min}$ cut forces all partonic sub-channels into similar
kinematic configurations, while the Higgs production is governed by
the same coupling to mostly $t$-channel gluons.  If there should be a
good reason to include more gluon-fusion Higgs events in this coupling
analysis we can expect stable results in the combination of gluon
fusion and weak boson fusion, without any complications from different
partonic sub-processes governed by the parton densities.\bigskip

Finally, an additional complication in WBF is the fact that we cannot
determine the charge of the incoming and outgoing fermions, \ie we
cannot distinguish neutral from charged current contributions. As long
as we stick to $SU(2)$-invariant operators this does not pose a
problem, but there exist Higgs-candidate operators which are forbidden
by the Bose symmetry between two $Z$ bosons and allowed for a $W^+
W^-$ pair. In Sec.~\ref{sec:spinone} we will come across them. While
we can distinguish photon-induced contributions based on the
transverse momentum spectrum of the tagging jets, $WW$ and $ZZ$
induced WBF events are indistinguishable, with a typical relative size
$1:4$, dominated by $WW$ fusion.

%%%%%%%%%%%%%%%%%%%%%%%%%%%%%%%%%%%%%%%%%%%%%%%%%%%%%%%%%%%%%%%%%%%%%%%%
\section{Lagrangian}
\label{sec:lag}

Measuring the Higgs coupling structure in weak boson fusion is not a
theory-independent measurement. Instead, we have to implement all
different hypotheses in a Monte Carlo event generator, \ie {\sc
  Madgraph}~\cite{madgraph}. The different predictions then have to be
compared to ATLAS and CMS data. In Section~\ref{sec:analysis} we will
limit ourselves to signal-only results, noting that in WBF Higgs
searches we expect a favorable signal-to-background ratio. Moreover,
we will only show a limited number of particularly promising
distributions to show that we can indeed distinguish all available
scenarios for a heavy resonance $X$. A complete list of all simulated
distributions can be found in the Appendix.

%%%%%%%%%%%%%%%%%%%%%%%%%%%%%%%%%%%%%%%%%%%%%%%%%%%%%%%%%%%%%%%%%%%%%%%%
\subsection{Spin zero}
\label{sec:spinzero}

The different hypotheses for the nature of a heavy resonance $X$ can
be easily ordered by its spin.  First, we parameterize the relevant
interactions of a CP-even or CP-odd scalar $X= H, A$ with gauge bosons
in an effective, gauge-invariant
Lagrangian~\cite{original,kaoru_dieter,looka2},
\begin{alignat}{5}
\lag_0 &=
  g_1^{(0)} \,H V_\mu V^\mu 
  -\frac{g_2^{(0)}}{4} \; H \, V_{\mu\nu}V^{\mu\nu}
  -\frac{g_3^{(0)}}{4} \; A \, V_{\mu\nu}\widetilde V^{\mu\nu} 
  -\frac{g_4^{(0)}}{4} \; H \, G_{\mu\nu}G^{\mu\nu}
  -\frac{g_5^{(0)}}{4} \; A \, G_{\mu\nu}\widetilde G^{\mu\nu} \; ,
\label{eq:spinzero}
\end{alignat}
where $V=W,Z$ are the electroweak gauge bosons and $G$ is the gluon.
The coupling to the photon can be described in complete analogy, but
it will not be relevant for our analysis of LHC production channels.
For all gauge fields the corresponding $V^{\mu\nu}$ indicates the
field strength tensor and $\widetilde V^{\mu\nu}$ is its dual.  Some
of the couplings $g_i^{(0)}$ defined this way have mass dimension, \ie
we do not distinguish between order-one coupling factors and mass
suppression. This simplified notation is appropriate, since throughout
our analysis we assume equal rates for all coupling operators.  In the
Standard Model the couplings are $g_1^{(0)}=2M_V^2/v$ and
$g_4^{(0)}=\alpha_sg_{Htt}/3\pi v$ (in the heavy top
limit)~\cite{lecture}.  The additional higher-dimensional coupling
$g_2^{(0)}$ is finite but negligible in the Standard Model.\bigskip

In an effective theory approach the couplings $g_2^{(0)} - g_5^{(0)}$,
which carry mass dimension $1/\Lambda$, can be generated through
loops. We omit the additional factors $1/\Lambda$ throughout our
discussion because it complicates the notation and will be of no
relevance to our discussion of angular correlations. Nevertheless, we
expect a hierarchy in the size of scalar `Higgs' couplings. In
general, the far off-shell processes $H\rightarrow ZZ \rightarrow 4
\ell$ and $H\rightarrow WW \rightarrow \ell^+\ell^-\nu\bar{\nu}$ come
with an additional suppression factor from the leptonic branching
ratios. In contrast, there is no such suppression for the loop-induced
decay $H\rightarrow \gamma\gamma$. In absence of the Standard Model
coupling $g_1^{(0)}$ the LHC measurements require $g_2^{(0,ZZ)}
\approx g_2^{(0,WW)} \approx 1000 g_2^{(0,\gamma\gamma)}$.  The
non-diagonal coupling $g_2^{(0,Z\gamma)}$ is similarly suppressed in
the absence of an on-shell $H\rightarrow Z\gamma$ signal. However, as
long as anomaly-free $SU(2)$ multiplets run inside the loops all 
induced couplings tend to be of similar size. This implies that a
substantial fine tuning would be needed to accommodate a
$g_2^{(0)}$-based explanation of the ATLAS and CMS results.\bigskip

While such an argument based on relative rates is convincing in an
effective theory framework with an essentially known particle content,
the analysis of angular correlations is model independent. It does not
assume any perturbative field theory and hence applies to all possible
fundamental explanations of the `Higgs' sector of the Standard
Model. The list of scalar hypotheses which we test later is summarized
in Tab.~\ref{tab:model}.

%%%%%%%%%%%%%%%%%%%%%%%%%%%%%%%%%%%%%%%%%%%%%%%%%%%%%%%%%%%%%%%%%%%%%%%%
\subsection{Spin one}
\label{sec:spinone}

The general Lagrangian describing the interaction between a neutral
CP-even or CP-odd gauge boson $X = Y^{(e)}, Y^{(o)}$ and the
electroweak gauge bosons is
\begin{alignat}{5}
\lag_1 &= 
    i g_1^{(1)} (W^+_{\mu\nu}W^{- \mu} - W^-_{\mu\nu}W^{+\mu}) \; Y^{(e)\nu} 
   + i g_2^{(1)} W^+_\mu W^-_\nu \; Y^{(e)\mu\nu}  
\notag \\
  &+ g_3^{(1)} \epsilon^{\mu\nu\rho\sigma}(W^+_\mu  \overleftrightarrow{\partial}_\rho W^-_\nu)Y^{(e)}_\sigma   
   + ig_4^{(1)} W^+_{\sigma\mu}W^{- \mu\nu} Y^{(e)\sigma}_{\nu} 
\notag \\
  &- g_5^{(1)} W^+_\mu W^-_\nu(\partial^\mu Y^{(o)\nu} + \partial^\nu Y^{(o)\mu}) 
   + ig_6^{(1)} W^+_\mu W^-_\nu \widetilde{Y}^{(o){\mu \nu}}  
   + ig_7^{(1)} W^+_{\sigma\mu} W^{-\mu\nu} \widetilde{Y}^{(o)\sigma}_\nu
\notag \\
  &+ g_8^{(1)} \epsilon^{\mu\nu\rho\sigma} Y^{(e)}_\mu  Z_\nu (\partial_\rho Z_\sigma)  
   + g_9^{(1)} Y^{(o)}_\mu  (\partial_\nu Z^\mu )Z^\nu \;.
\label{eq:spinone}
 %   + g_{V\tau} \bar\tau \gamma_\mu ( g_V - g_A \gamma^5)\tau V^\mu                      %coupling with fermions      
\end{alignat}
It includes $g_1^{(1)}-g_7^{(1)}$ describing the coupling to $W$
bosons~\cite{kaoru_dieter} and the reduced set $g_8^{(1)}-g_9^{(1)}$
for the $Z$ couplings~\cite{ly_crap}, constrained by the
Landau-Yang theorem~\cite{landau_yang}. All terms in this Lagrangian
have mass dimension four, except for $g_4^{(1)}$ and
$g_7^{(1)}$. Compared to the usual $X \to ZZ$ analysis this WBF
Lagrangian is much more complex, so we limit our number of hypotheses
to one coupling each for CP-even and CP-odd states to $W$ and $Z$
bosons,
\begin{alignat}{7}
 &1^+_W: \quad &g^{(1)}_1 = g^{(1)}_2&\neq 0 
\qqqquad 
  1^-_W: \quad &g^{(1)}_5 = g^{(1)}_6&\neq 0
\notag \\
 &1^+_Z: \quad &g^{(1)}_8 &\neq 0
 \qqqquad 
  1^-_Z: \quad &g^{(1)}_9 &\neq 0
\end{alignat}
The interactions of $Y^{(e)}$ with $W$ bosons has the same form as for
a Kaluza-Klein $Z$ boson, \ie it is a re-scaled version of the $WWZ$
coupling in the Standard Model.  For the CP-odd $Y^{(o)}$ we resort to
a special combination of dimension-4 operators only. The interactions
with $Z$ bosons is the usual complete
basis~\cite{zerwas,melnikov,lookalikes}. These four scenarios are
summarized in Tab.~\ref{tab:model}. We will see that the angular
correlations in WBF production of a $1^\pm_W$ and $1^\pm_Z$ can be
different. The reason is that in our choice of operators the two are
at best partially identical, because the list of allowed coupling
structures of a neutral resonance to the neutral $Z$ is considerably
more constrained.\bigskip

When extending our analysis of the WBF production process to include
decays of the heavy neutral resonance for example into a pair of
$\tau$ leptons we need to assume a structure of its coupling to
fermions.  For simplicity, we assume the same vector and axial-vector
structure as the $Z$ boson,
\begin{alignat}{5}
 g_V= -\frac{1}{2} + 2 \sin^2 \theta_w 
 \qqqquad g_A=- \frac{1}{2} \;,
\label{eq:spinone_fermions}
\end{alignat}
again re-scaled by a constant factor. Moreover, we only couple the new
vector to the third generation, to avoid modifications of the WBF
production process.

%-------------------------------------------------------
\begin{table}[t]
\begin{small} 
\begin{tabular}{l|ll|l}
 \hline
  & initial state \quad & couplings &  \\
 \hline
 $0_\text{SM}^+$ & $qq$ & $g_1^{(0)}$ 
  & SM Higgs scalar (D3 coupling to $W,Z$) \\
 $0_\text{D5}^+$ & $qq$ & $g_2^{(0)}$ 
  & scalar (D5 coupling to $W,Z$) \\
 $0_\text{D5}^-$ & $qq$ & $g_3^{(0)}$ 
  & pseudo-scalar (D5 coupling to $W,Z$) \\
 $0_\text{D5g}^+$ & $qq, qg, gg$ & $g_4^{(0)}$ 
  & scalar (D5 coupling to gluons)\\
 $0_\text{D5g}^-$ & $qq, qg, gg$ & $g_5^{(0)}$ 
  & pseudo-scalar (D5 coupling to gluons)\\ \hline
 $1^-_W$ & $qq$ & $g_5^{(1)}=g_6^{(1)}$ 
  & D4 couplings to $W$ \\
 $1^-_Z$ & $qq$ & $g_9^{(1)}$ 
  & vector coupling to $Z$ \\ 
 $1^+_W$ & $qq$ & $g_1^{(1)}=g_2^{(1)}$ 
  & D4 couplings to $W$ \\
 $1^+_Z$ & $qq$ & $g_8^{(1)}$
  & axial-vector coupling to $Z$ \\ \hline
 $2_\text{EW}^+$ &  $qq$  & $g_1^{(2)}$ 
  & tensor coupling to EW gauge bosons \\
 $2_\text{EW+q}^+$ & $qq, qg, gg$ & $g_1^{(2)}=g_3^{(2)}$ 
  & tensor coupling to EW gauge bosons and fermions\\
 $2_\text{QCD}^+$ & $qq, qg, gg$ & $g_2^{(2)}$ 
  & tensor coupling to gluons \\
 $2_{}^+$ & $qq, qg, gg$ & $g_1^{(2)}=g_2^{(2)}=g_3^{(2)}$ 
  & graviton-like tensor \\
 \hline
\end{tabular} 
\end{small}
\caption{Properties of the heavy resonance models discussed in this
  paper. The structure of the coupling to gluons we only list for
  completeness reasons, in the analysis we do not have to take these
  operators into account. Instead, we show them in the Appendix.}
\label{tab:model}
\end{table}
%-------------------------------------------------------

%%%%%%%%%%%%%%%%%%%%%%%%%%%%%%%%%%%%%%%%%%%%%%%%%%%%%%%%%%%%%%%%%%%%%%%%
\subsection{Spin two}
\label{sec:spintwo}

There is a number of commonly used ways to parameterize the
interaction of a spin-2 boson~\cite{melnikov}. As usually, we choose
graviton-inspired scenarios to contrast with the scalar Higgs
boson. More precisely, we consider a massive graviton resonance, which
couples to the SM particles through their energy-momentum tensors,
\begin{alignat}{5}
\lag_2 =
  -g_1^{(2)} \; G_{\mu\nu}T^{\mu\nu}_V
  -g_2^{(2)} \; G_{\mu\nu}T^{\mu\nu}_G
  -g_3^{(2)} \; G_{\mu\nu}T^{\mu\nu}_f \; ,
  \label{eq:spintwo}
\end{alignat}
where $G_{\mu\nu}$ is the spin-2 resonance and $T^{\mu\nu}_{V,G,f}$ is
the energy-momentum tensor of the electroweak gauge bosons, the gluon,
and the different fermions. For instance, the tensor for the $Z$ boson
is given by~\cite{Hagiwara:2008jb}
\begin{alignat}{5}
 T_Z^{\mu\nu} &= 
      -g^{\mu\nu}
      \left[ -\frac{1}{4}Z_{\rho\sigma}Z^{\rho\sigma}
             +\frac{m^2_Z}{2}Z_\rho Z^\rho
      \right] 
      -Z_\rho^\mu Z^{\nu\rho} +m^2_Z Z^\mu Z^\nu \notag \\
\text{with} \quad
 Z_{\mu\nu}
  &= \partial_\mu Z_\nu-\partial_\nu Z_\mu
    +ig_w \cos\theta_w
     \left( W^+_\mu W^-_\nu - W^+_\nu W^-_\mu \right) \; .
\end{alignat}
While conventional graviton excitations have universal couplings
strengths $1/\Lambda$ to fermions and gauge fields, where $\Lambda$ is
the scale parameter of the theory, for
our test we define four scenarios motivated on the WBF topology
\begin{alignat}{5}
 &2^+_\text{EW}:  && g_1^{(2)}= \frac{1}{\Lambda} \qqquad g_2^{(2)}=g_3^{(2)}=0\;, \notag \\
 &2^+_\text{EW+q}:\quad &&g_1^{(2)}=g_3^{(2)}=\frac{1}{\Lambda} \qqquad g_2^{(2)}=0\;, \notag \\
 &2^+_\text{QCD}: &&g_2^{(2)}= \frac{1}{\Lambda} \qqquad g_1^{(2)}=g_3^{(2)}=0\;, \notag \\
 &2^+: && g_1^{(2)}=g_2^{(2)}=g_3^{(2)}= \frac{1}{\Lambda} \;.
\label{eq:spin2op}
\end{alignat}
A universally-coupled spin-2 particle is heavily constrained by
Tevatron data~\cite{unigr}, but we ignore these constraints in our
discussion limited to the Higgs sector. All hypotheses shown in
Tab.~\ref{tab:model} are not meant as a valid ultraviolet completion
of the Standard Model without a Higgs boson. All they need to serve as
are general alternative scenarios for the interpretation of a
Higgs-like resonance in weak boson fusion.
Similar to the spin-1 case, the couplings to leptons are assumed to
include decays of the spin-2 resonance even for $2^+_{\rm EW}$ and 
$2^{+}_{\rm QCD}$.

%%%%%%%%%%%%%%%%%%%%%%%%%%%%%%%%%%%%%%%%%%%%%%%%%%%%%%%%%%%%%%%%%%%%%%%%
\section{Analysis}
\label{sec:analysis}

The study presented in this section is not meant to produce an
optimized  Higgs couplings analysis using the
weak-boson-fusion production mechanism. Instead, we study the behavior
of the kinematic variables described in Sec.~\ref{sec:hadcoll} for the
different `Higgs' candidates. For this purpose, we identify the most
promising variables to separate the different Higgs spins and
couplings defined in Secs.~\ref{sec:spinzero}-\ref{sec:spintwo}. Based
on this most promising subset of observables we will show how the
different hypotheses on the nature and the couplings of the heavy
state $X$ can be distinguished. In Sec.~\ref{sec:stats} we apply a
more sophisticated statistical analysis to this problem, but
eventually we will leave it to the LHC experiments to determine the
quantitative impact of our approach.\bigskip

%-------------------------------------------------------
\begin{table}[b!]
\begin{small} \begin{tabular}{c || c  c  c  c  c  c   c  c  c  c | c c}
 \hline
 \multirow{2}{*}{cuts} &  
 \multicolumn{10}{c|}{signal} & 
 \multicolumn{2}{c}{background $\tau \tau jj$}  \\
   & $0^+_\text{SM}$ & $0^+_\text{D5}$ & $0^-_\text{D5}$ &  $0^+_\text{D5g}$ &
	       $0^-_\text{D5g}$ & 
     $1^+_{W}$ & $1^-_{W}$ & $1^+_{Z}$ & $1^-_{Z}$ & $2^+$ & 
     EW & QCD  \\
  \hline
$m_{jj}$ in Eq.\eqref{eq:wbf_cuts1}              
& 0.442 & 0.241 & 0.290 & 0.072 & 0.070 & 0.318 & 0.420 & 0.291 & 0.419 & 0.161 & 0.211 & 0.027\\  
$\Delta \eta_{jj}$ in Eq.\eqref{eq:wbf_cuts2}              
& 0.389 & 0.068 & 0.108 & 0.053 & 0.053 & 0.189 & 0.242 & 0.134 & 0.220 & 0.096 & 0.124 & 0.014\\  
\hline
\end{tabular} \end{small}
\caption{Signal and background cross sections for the different
  `Higgs' coupling operators, normalized to the acceptance cuts
  Eq.\eqref{eq:general_cuts}.}
\label{tab:cuts}
\end{table}
%-------------------------------------------------------

For our numerical analysis we implement the interactions Lagrangians
Eqs.\eqref{eq:spinzero}, \eqref{eq:spinone} and \eqref{eq:spintwo}
into {\sc FeynRules}~\cite{feynrules}, which provides a UFO model
file~\cite{ufo} that we implement into {\sc
  MadGraph5}~\cite{madgraph}. All simulations for this section are
done at parton level for an LHC energy of 14~TeV. To study the impact
of the acceptance cuts on our observables we choose three possible set
of acceptance cuts:
\begin{enumerate}
\item We start with a minimal set of cuts applied on the outgoing
  partons
\begin{alignat}{5}
 p_{T,j} &\ge 20~\gev & \qqqquad
 \Delta R_{jj}&\ge 0.6 &  \qqqquad
 |\eta_j|& \le 5\;.
 \label{eq:general_cuts}
\end{alignat}

\item To suppress backgrounds and a possible contamination with
  gluon-fusion events we also include
\begin{alignat}{5}
 m_{jj} \ge 600~\gev \;.
 \label{eq:wbf_cuts1}
 \end{alignat}

\item As is common in WBF analyses we require a rapidity gap
 between the two tagging jets
\begin{alignat}{5}
  \Delta \eta_{jj} \ge 4.2 \qqqquad \eta_{j_1} \times \eta_{j_2} < 0 \;. 
 \label{eq:wbf_cuts2}
\end{alignat}
\end{enumerate}
Note that while we simulate the decay $X \to
\tau\tau$~\cite{wbf_tau,original} as a toy channel for including the
$X$ momentum and the momentum of the decay product we do not apply any
cuts on the decay products. In Tab.~\ref{tab:cuts} we present the
signal and background ratio suppression after cutting on $m_{jj}$,
Eq.\eqref{eq:wbf_cuts1}, and after requiring the rapidity gap between
the two tagging jets, Eq.\eqref{eq:wbf_cuts2}. The signal and the
backgrounds are normalized to the rates after the acceptance cuts of
Eq.\eqref{eq:general_cuts}. We see that the background suppression and
the reduced contamination from gluon-fusion are delivered by the
$m_{jj}$ cut.  In the following we will advocate that the cuts from
Eq.\eqref{eq:wbf_cuts2} should be avoided because they reduce the
number of observables available for the analysis of the Higgs coupling
structure. For the remaining part of this paper we will only apply
Eqs.\eqref{eq:general_cuts} and \eqref{eq:wbf_cuts1}.\bigskip

This reduced set of kinematic cuts has an effect on the distinguishing
power of the observable $\Delta\phi_{jj}$. Forcing the two tagging jet
into a more collinear configuration enhances the modulation for
example distinguishing the three scalar coupling
structures~\cite{original,klamke}. However, we will see that $\Delta
\eta_{jj}$ itself is a powerful observable distinguishing coupling
structures, so this effect can better be exploited through a
two-dimensional $\Delta \phi_{jj}$ vs $\Delta \eta_{jj}$ analysis.

%%%%%%%%%%%%%%%%%%%%%%%%%%%%%%%%%%%%%%%%%%%%%%%%%%%%%%%%%%%%%%%%%%%%%%%%
\subsection{Tagging jet kinematics}
\label{sec:kin_jj}

%-------------------------------------------------------
\begin{figure}[b]
 \includegraphics[width=0.3\textwidth]{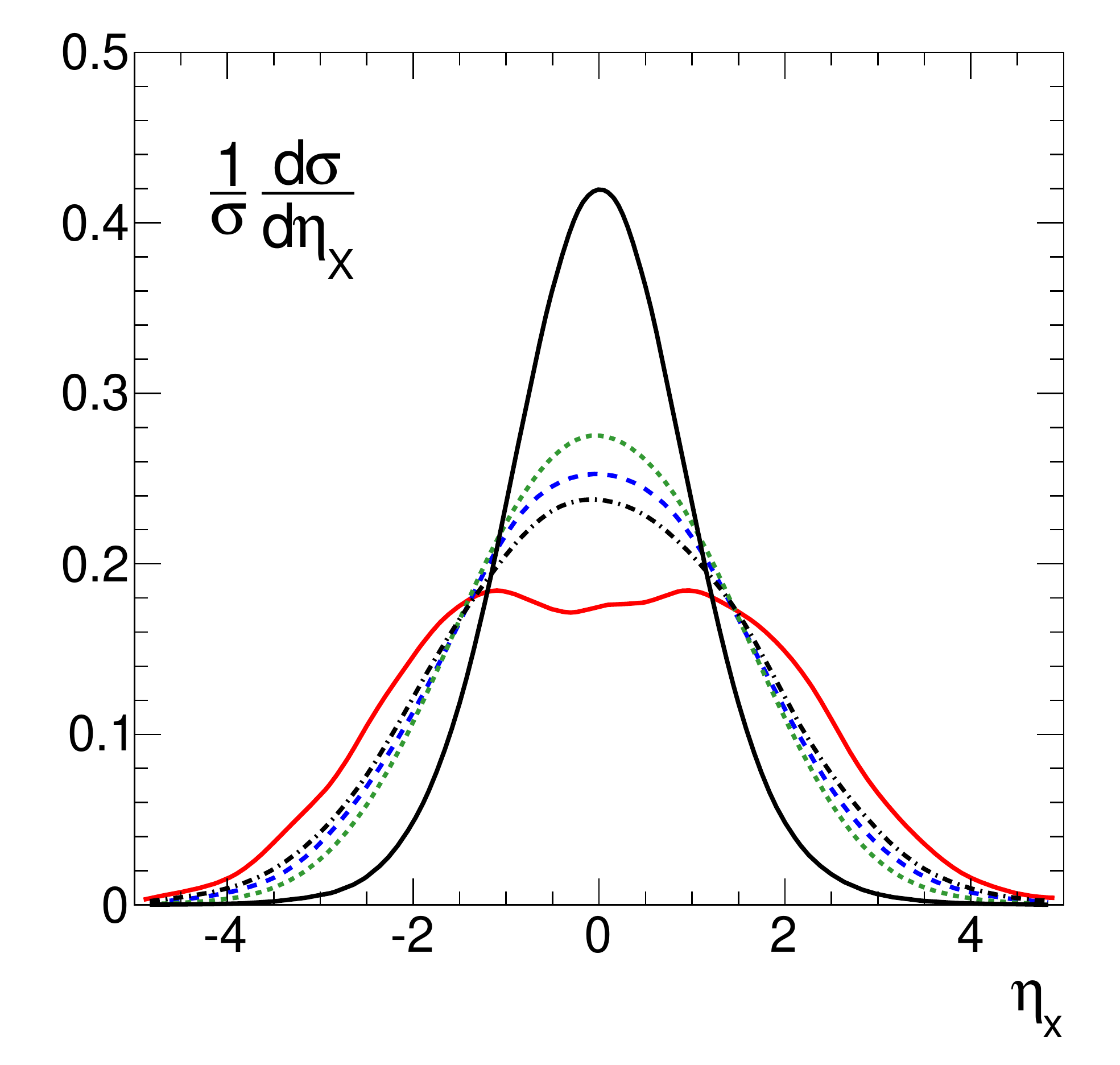}
 \hfill
 \includegraphics[width=0.3\textwidth]{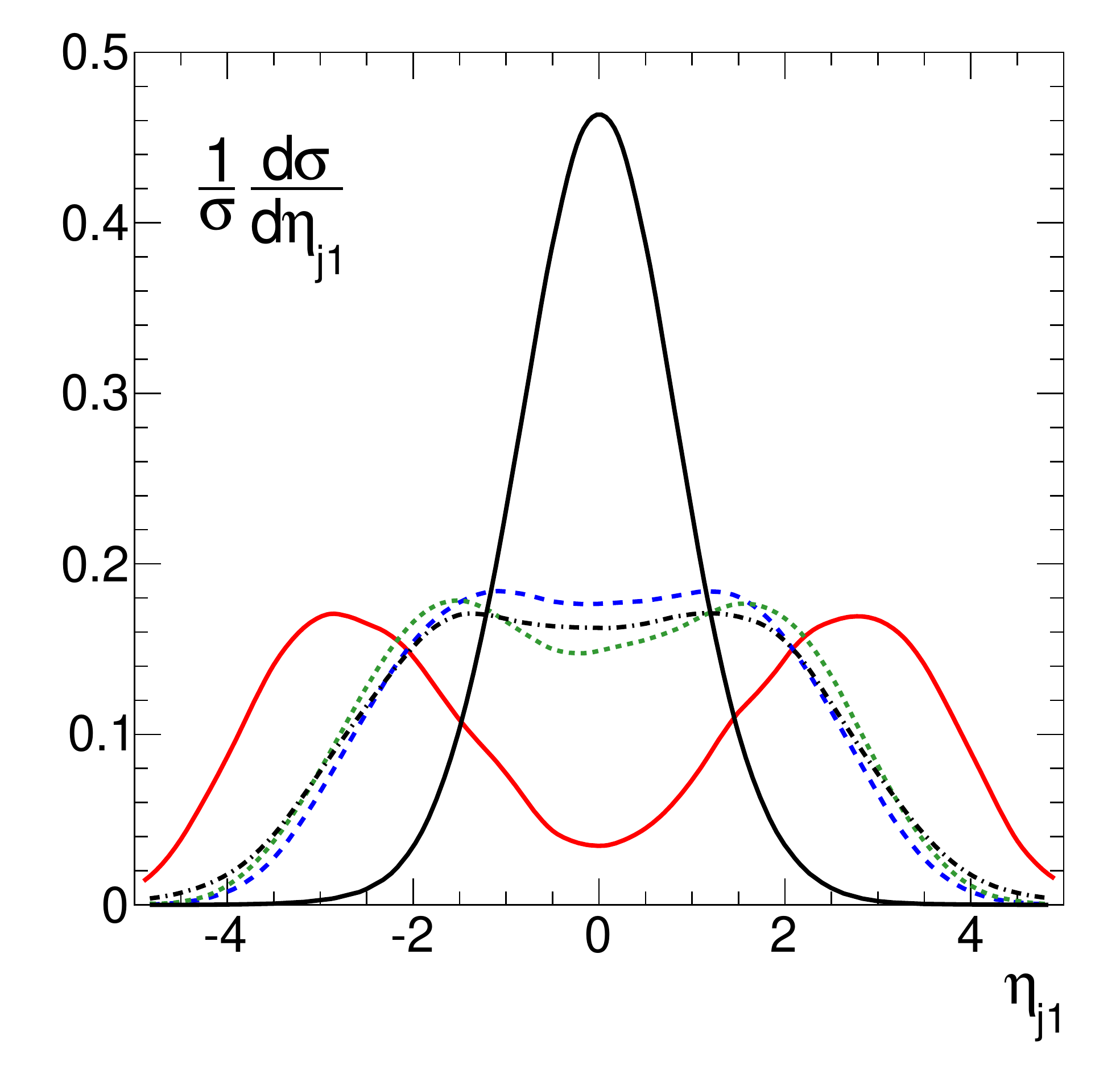}
 \hfill
 \includegraphics[width=0.3\textwidth]{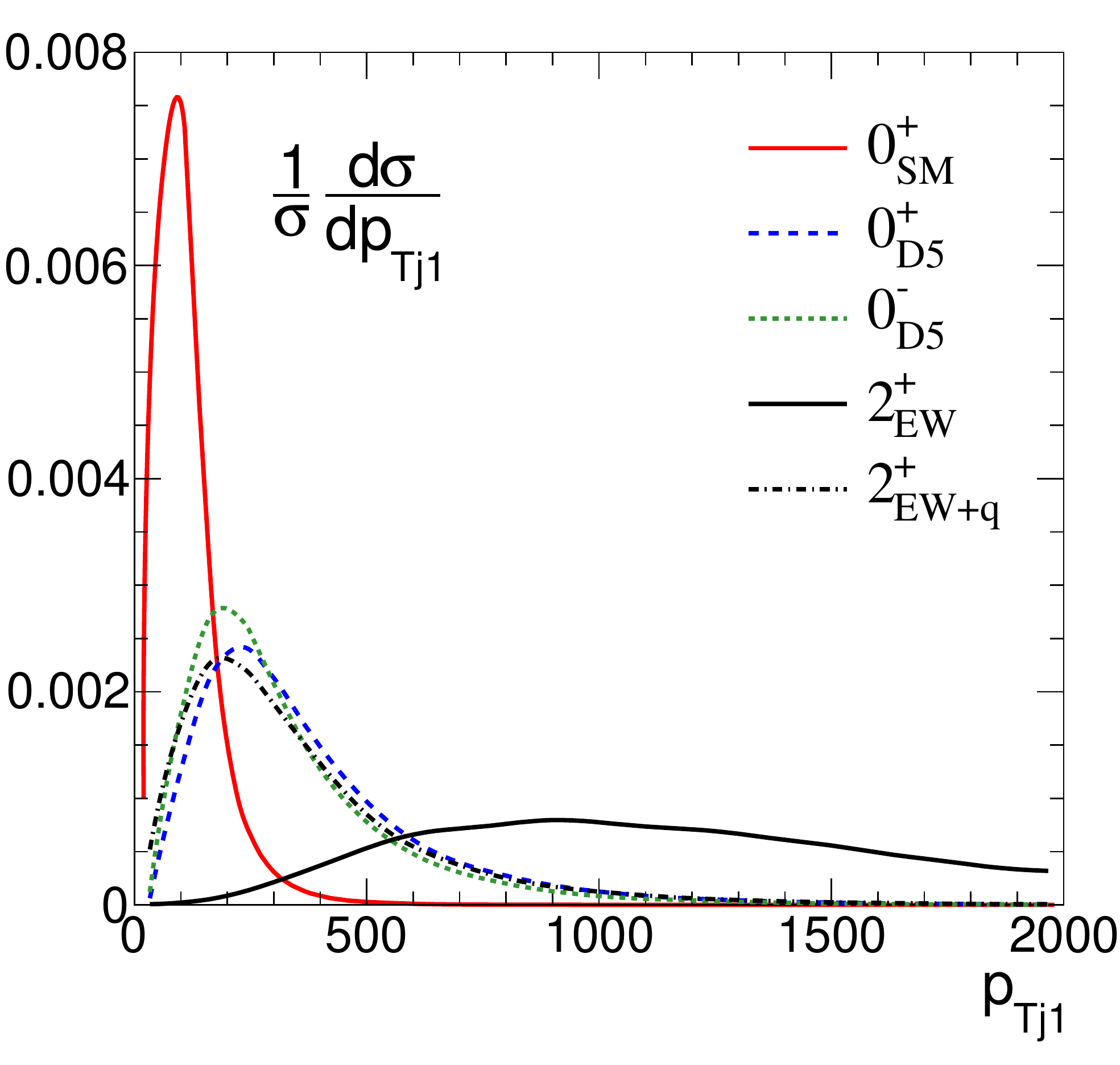}
\caption{Rapidity of the resonance $X$ (left), rapidity of the leading
  tagging jet (center), and transverse momentum of the leading tagging
  jet (right) for the different spin-0 and spin-2 hypotheses after the
  cuts of Eqs.\eqref{eq:general_cuts} and \eqref{eq:wbf_cuts1}.  The
  hypotheses $0_\text{SM}^+$ (red), $0_\text{D5}^+$ (blue dashed),
  $0_\text{D5}^-$ (green dashed), $2^+_\text{EW}$ (black), and 
  $2^+_\text{EW+q}$ (black dashed) are defined in
  Tab.~\ref{tab:model}.}
\label{fig:kin_basics}
\end{figure}
%-------------------------------------------------------

From the WBF analyses~\cite{wbf_w,wbf_tau} we know what the kinematic
features of the two tagging jets and the heavy SM Higgs boson are:
both tagging jets are distinctively forward and there is only little
color activity between them~\cite{jetveto}, except for final state
radiation off the centrally produced Higgs boson. All of this we show
in Fig.~\ref{fig:kin_basics}. The corresponding distributions for the
coupling structures $0^+_\text{D5}$ and $0^-_\text{D5}$ are very
similar, with a slight difference due to the absence of the
longitudinal amplitude for the $0^-$ coupling. In the spin-2 case with
couplings only to weak gauge bosons, $2^+_\text{EW}$, the tagging jets
are more central. Also including graviton emission from quark lines,
$2^+_\text{EW+q}$, shifts the tagging jets into the forward
region. This effect is a first indication that for a spin-2
interpretation of the Higgs-like resonance the details of the
underlying model have a significant effect on essentially all
observables. A complete set of reference distributions is shown in
Fig.~\ref{fig:app1} in the Appendix.\bigskip

Another serious issue arises when we compare the transverse momentum
distributions of the tagging jets (or the recoiling heavy
resonance). The momentum dependence of the spin-2 coupling generates
$p_{T,j}$ spectra extending beyond a TeV. However, consistent models
of a spin-2 resonance will include an additional form factor to cut
off this tail and render all $p_T$ distributions more similar to the
scalar case~\cite{wolfgang}. Including distributions with an energy
dimension, like the transverse momentum of the tagging jets and the
central resonance will only test different implementations of such a
form factor which cannot be derived from first principles.

The same behavior can be observed in WBF production of heavy $W'$
bosons, for example in little Higgs models~\cite{samesign}. Just
introducing additional heavy gauge bosons appears to predict very
large momenta of the tagging jets. However, for heavy gauge bosons we
can also check a consistent theory including heavy fermion partners
and indeed find $p_{T,j}$ distributions very similar to the known
Higgs case. To test the stability of or results with respect to the UV
completion of the spin-2 model we also show results requiring the
tagging jets to fulfill
\begin{alignat}{5}
p_{T,j} < p_T^\text{max} = 100~\gev \; .
\label{eq:ptmax}
\end{alignat}
\bigskip

%-------------------------------------------------------
\begin{figure}[!t]
 \hspace*{0.5cm}  spin-0
 \hspace*{3.25cm} spin-1
 \hspace*{3.25cm} spin-2 
 \hspace*{3.25cm} spin-2($p_T^\text{max}$)\\
 \includegraphics[width=0.24\textwidth]{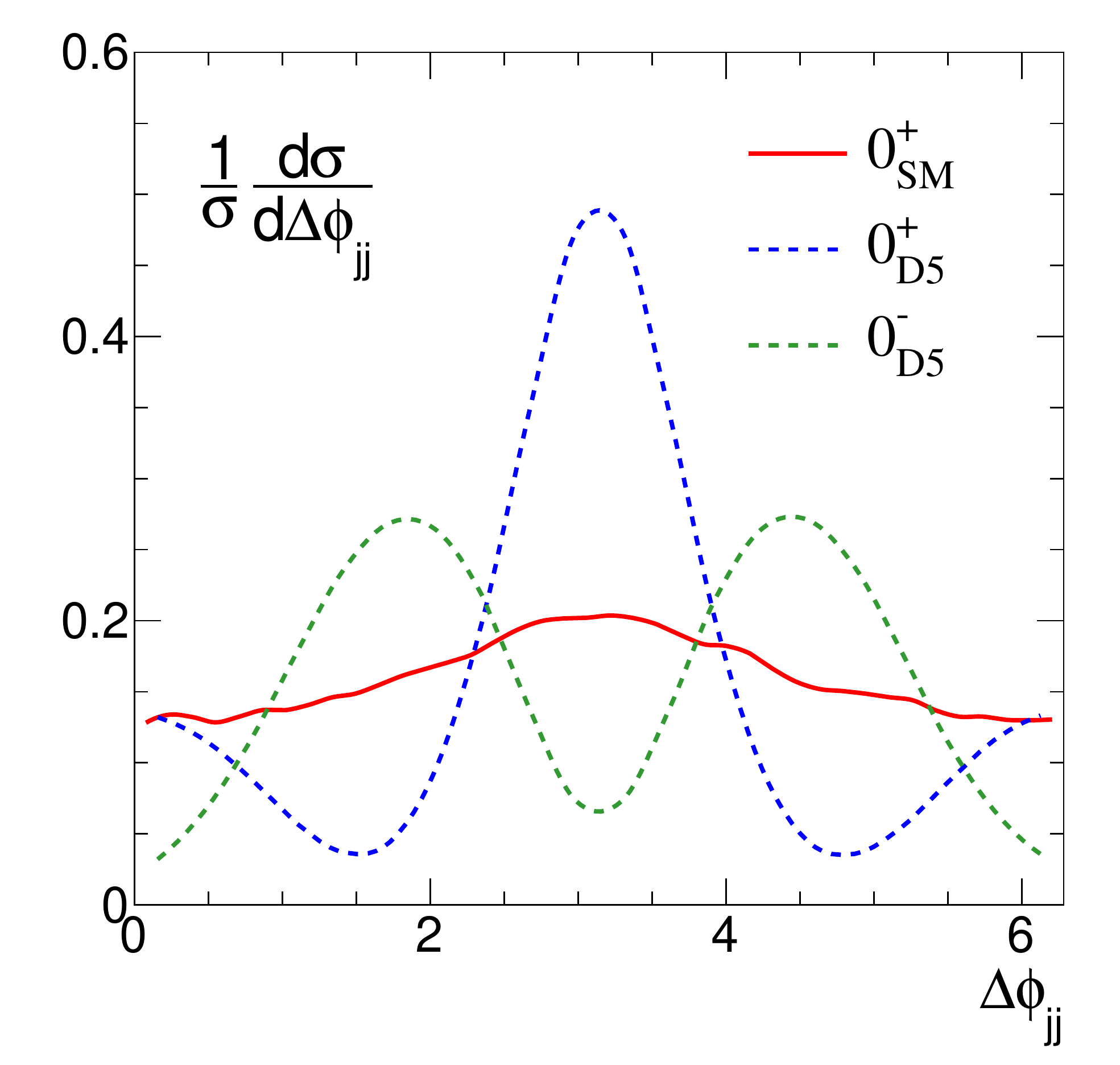}
 \hfill
 \includegraphics[width=0.24\textwidth]{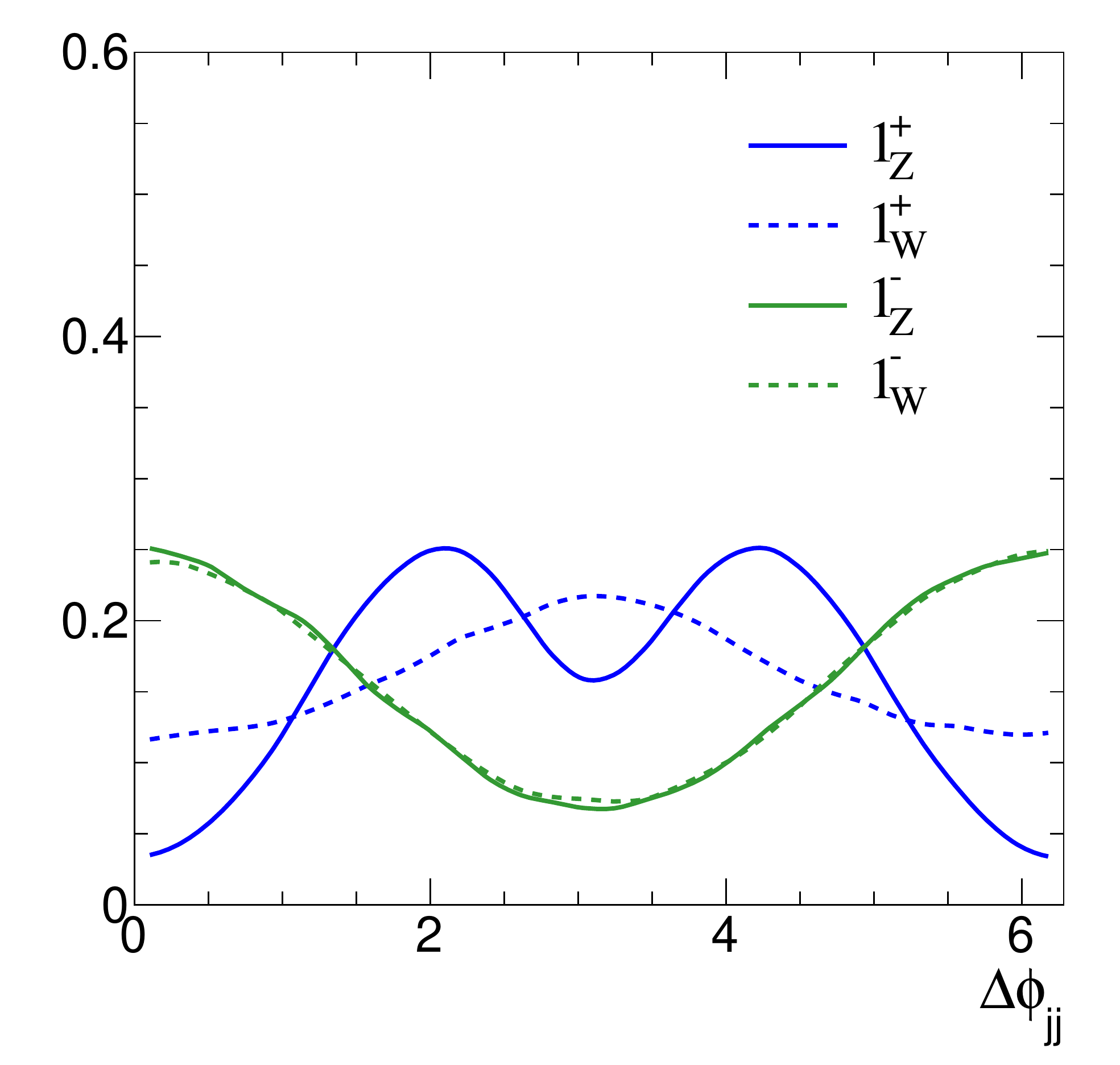}
 \hfill
 \includegraphics[width=0.24\textwidth]{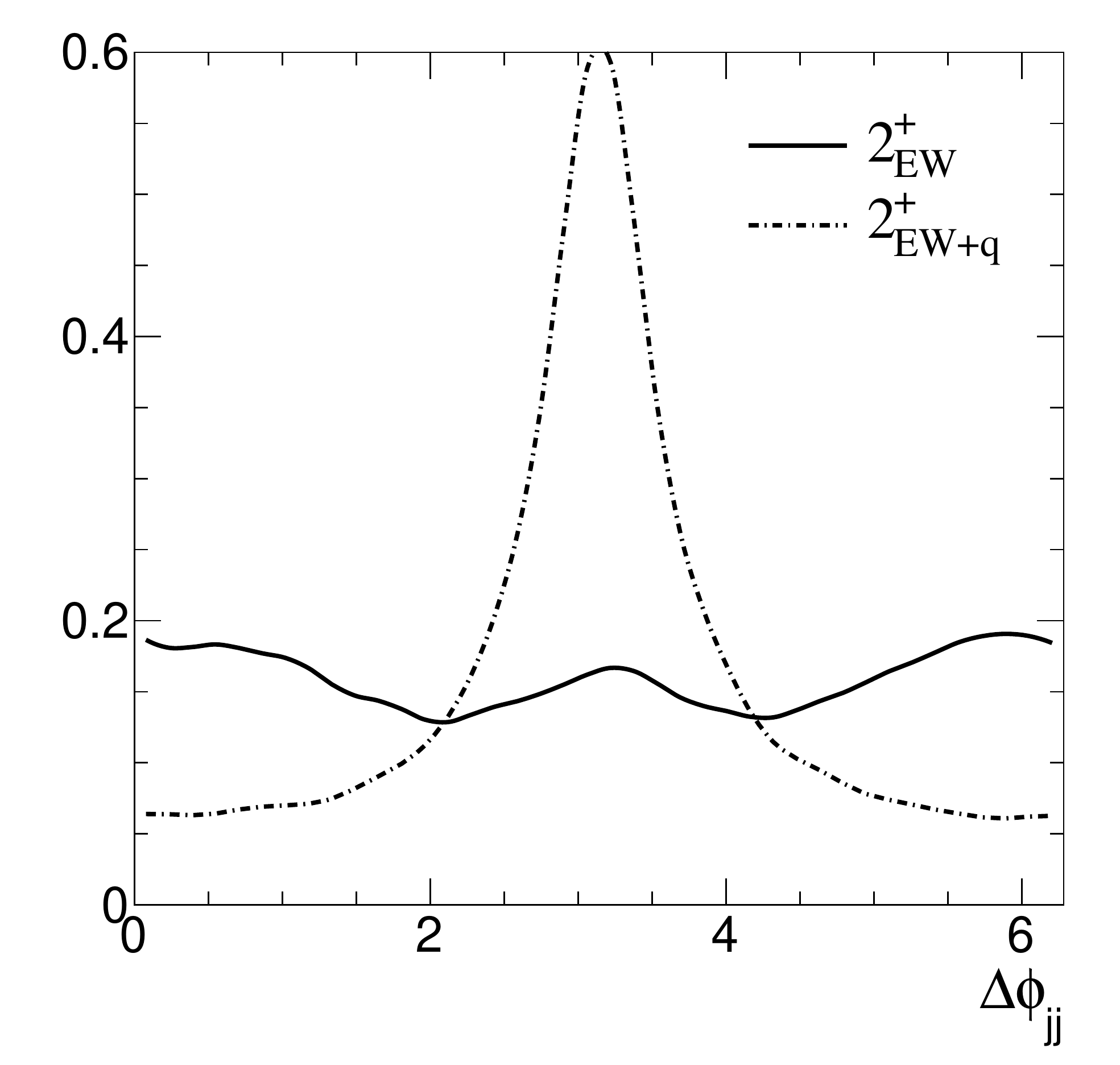}
 \hfill
 \includegraphics[width=0.24\textwidth]{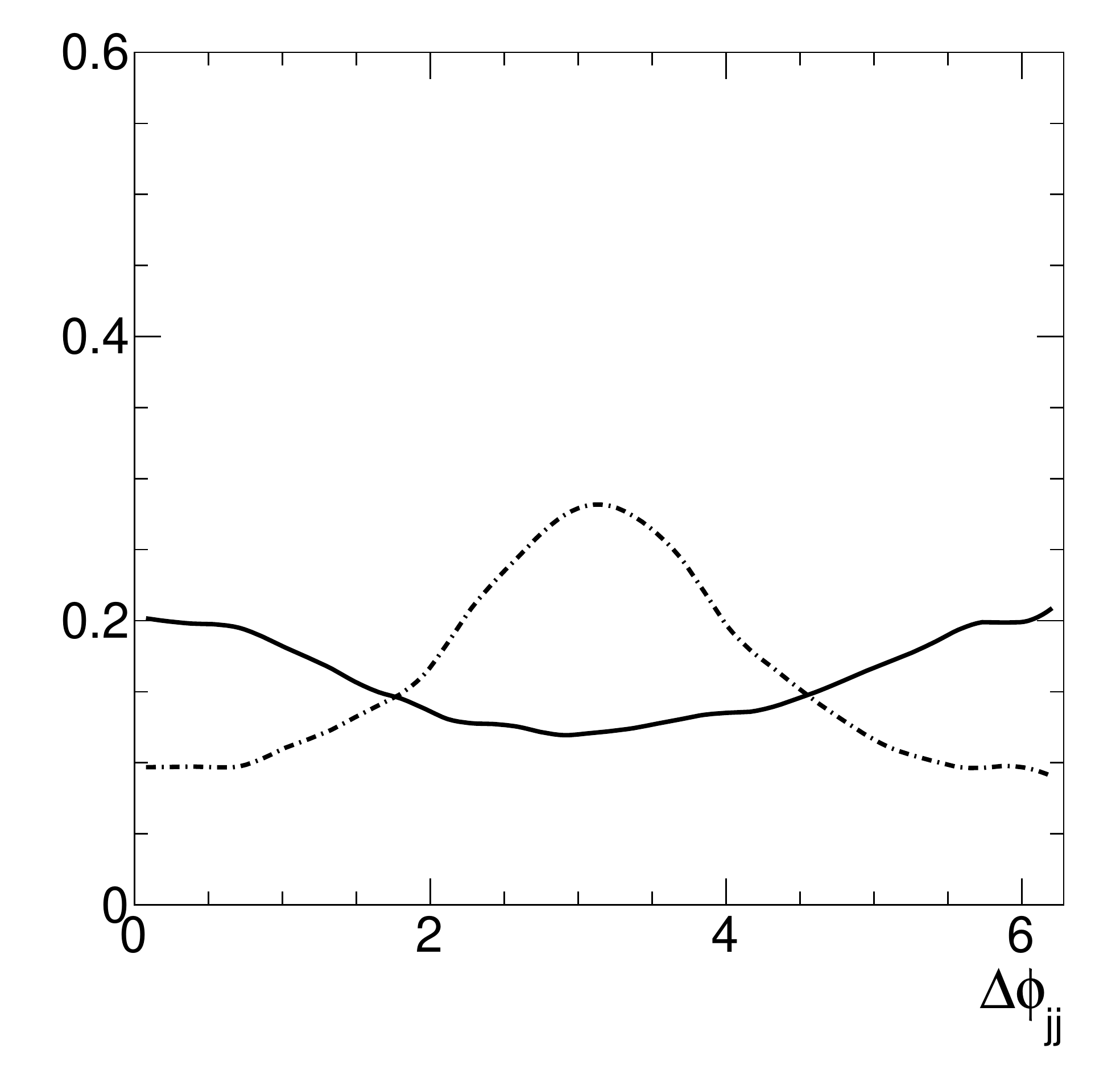} \\
 \includegraphics[width=0.24\textwidth]{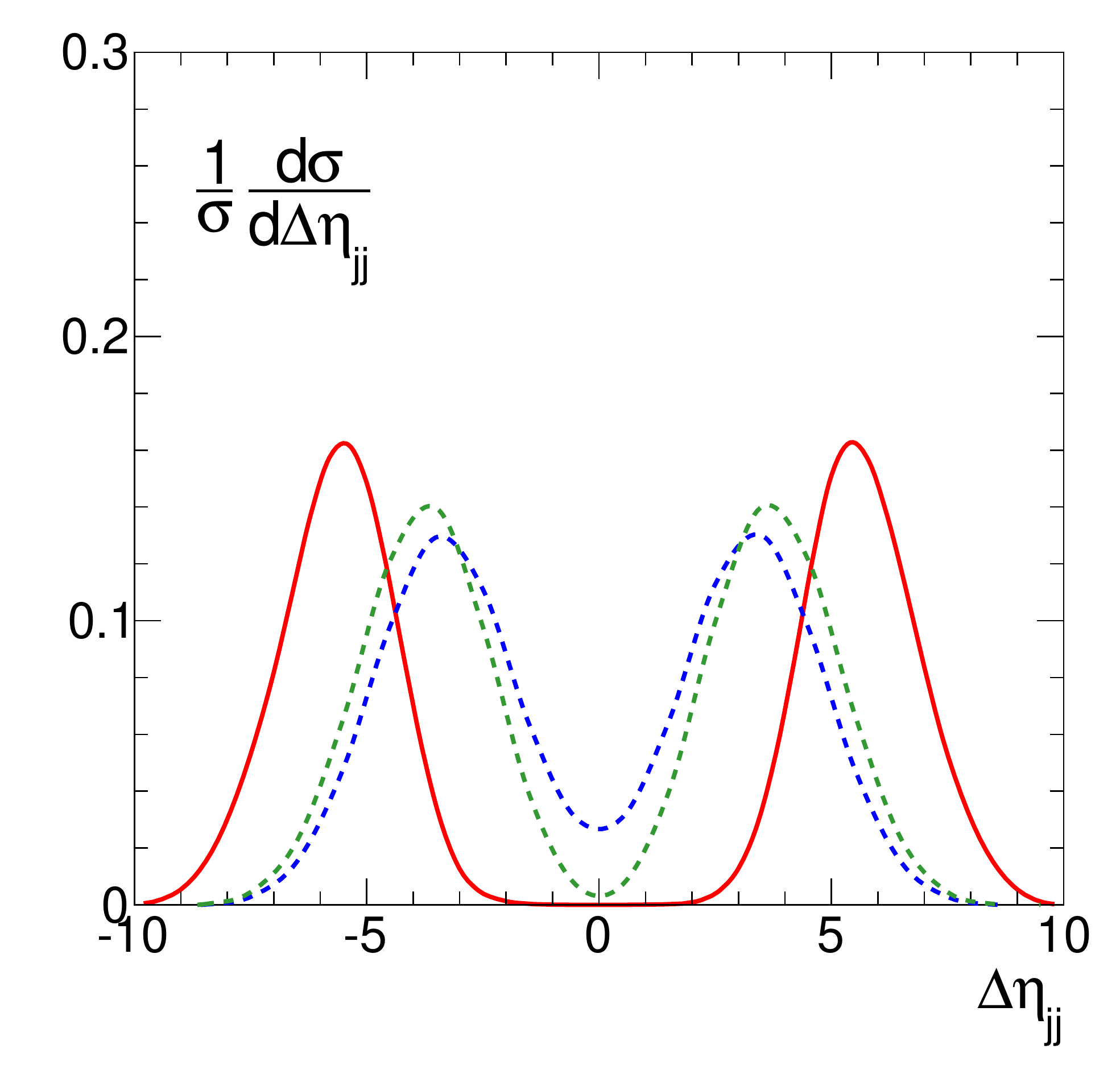}
 \hfill
 \includegraphics[width=0.24\textwidth]{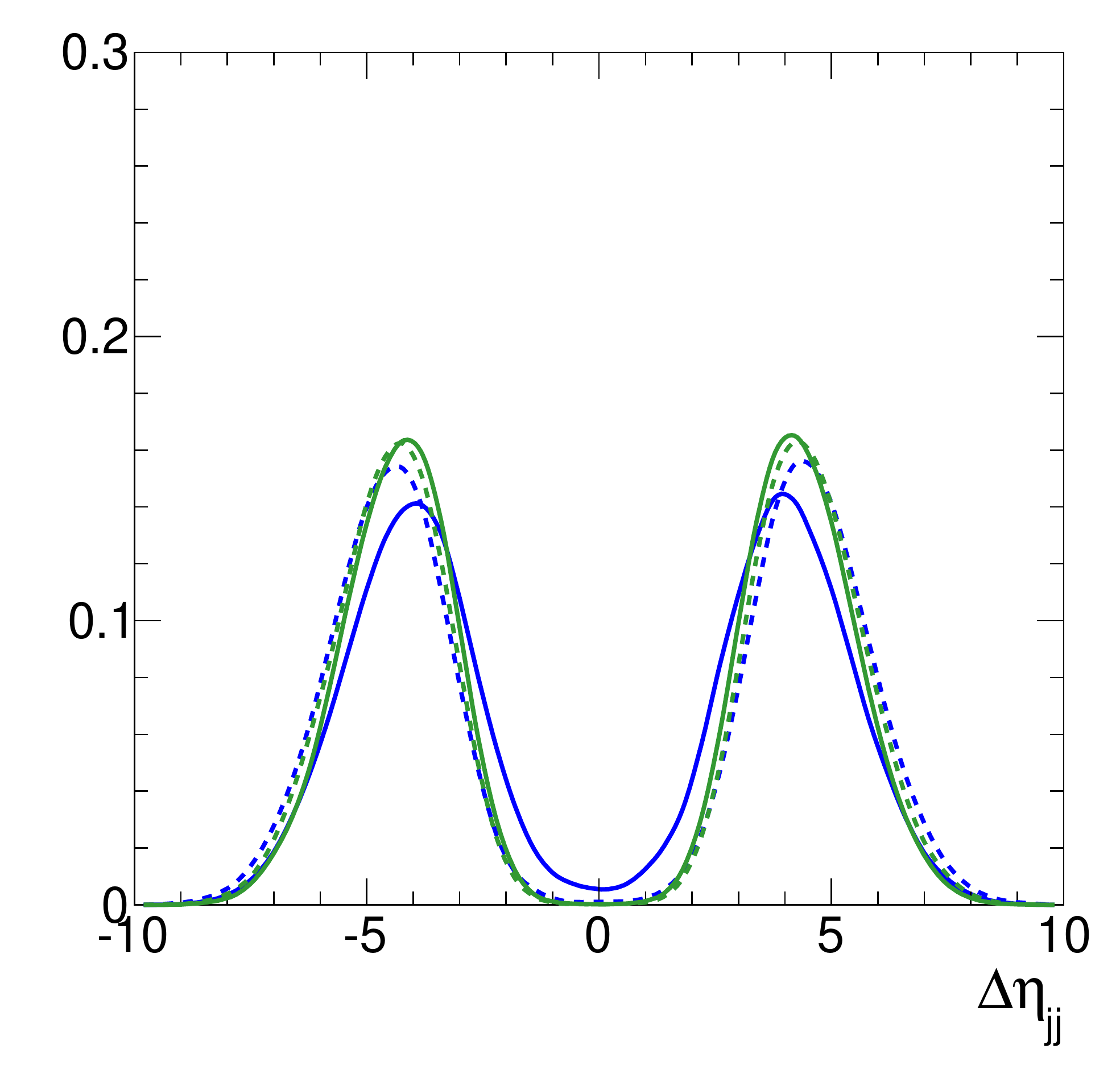}
 \hfill
 \includegraphics[width=0.24\textwidth]{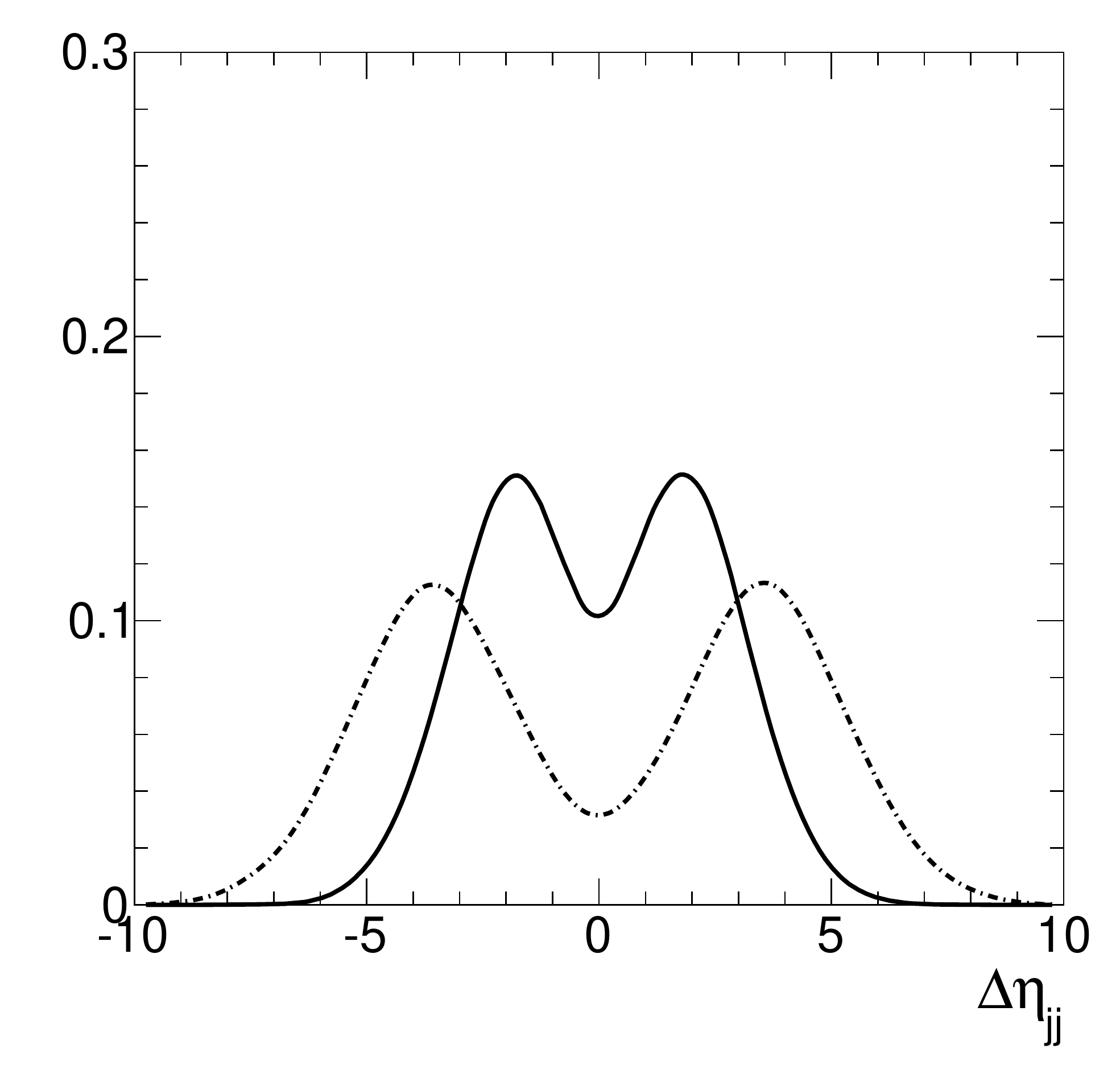}
 \hfill
 \includegraphics[width=0.24\textwidth]{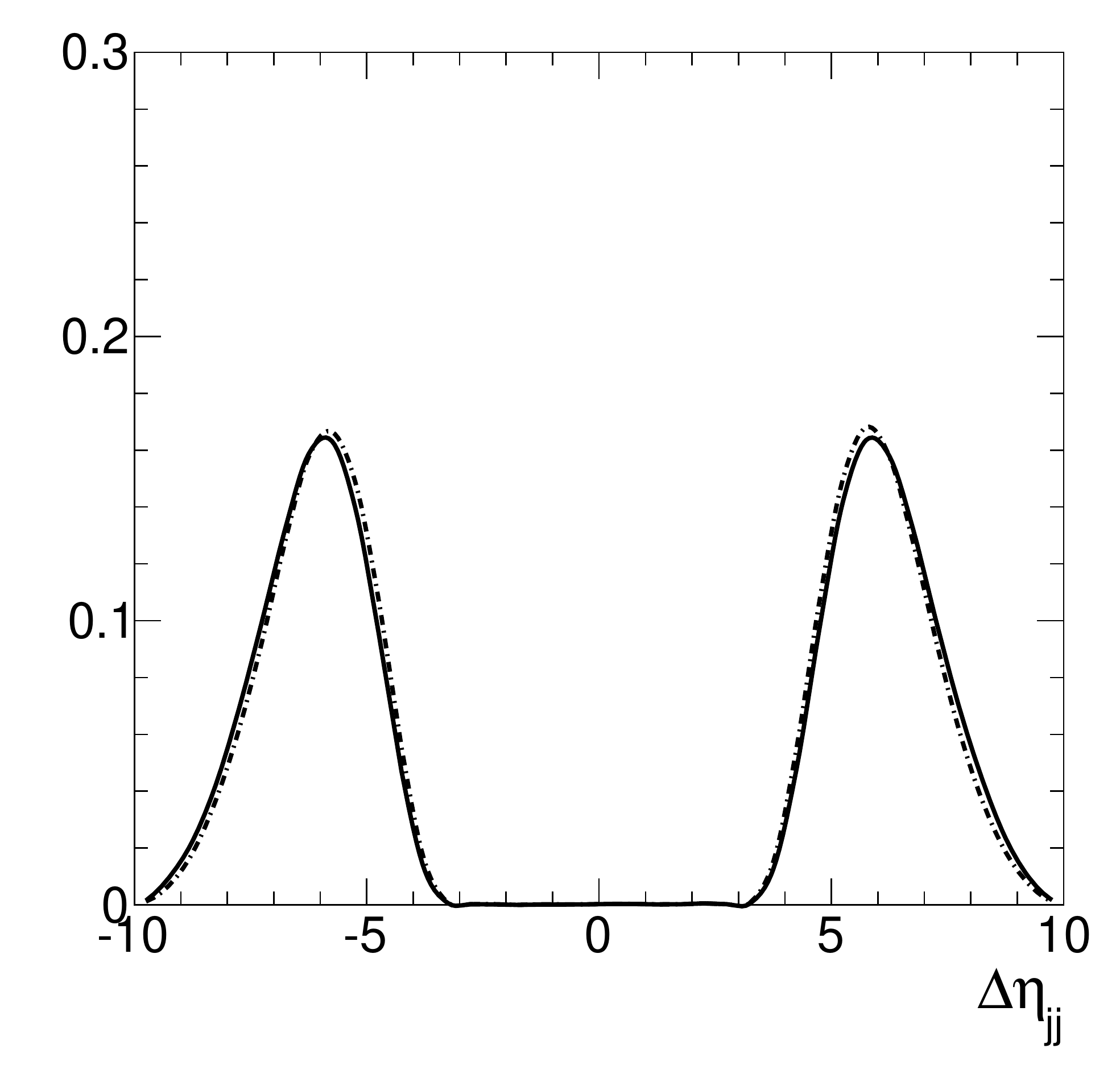} 
 \caption{Normalized correlations between the two tagging jets in WBF
   production of a heavy resonance $X$. We show the difference in the
   azimuthal angle $\Delta \phi_{jj}$ and the rapidity difference
   $\Delta \eta_{jj}$. The spin-0, spin-1, spin-2 interaction
   structures are described in Tab.~\ref{tab:model}: $0_\text{SM}^+$
   (red), $0_\text{D5}^+$ (blue dashed), $0_\text{D5}^-$ (green
   dotted); $1^+_Z$ (blue), $1^+_W$ (blue dashed), $1^-_Z$ (green),
   $1^-_W$ (green dashed); $2^+_\text{EW}$ (black), $2^+_\text{EW+q}$
   (black dashed). For the spin-2 case we also show the distributions
   after requiring $p_T^\text{max} = 100$~GeV (right panels).}
\label{fig:kin_jj}
\end{figure}
%-------------------------------------------------------

The starting point of our analysis of `Higgs' coupling structures in
weak boson fusion is the azimuthal angle between the two tagging
jets~\cite{original}. Its relation to alternative observables we
discuss in Sec.~\ref{sec:hadcoll} and specifically in
Fig.~\ref{fig:dphi}, indicating that it should be well suited to
distinguish different spin-0 and spin-2 hypotheses.

In the upper panels of Fig.~\ref{fig:kin_jj} we show the $\Delta
\phi_{jj}$ distributions for the nine different couplings listed in
Tab.~\ref{tab:model}. There are essentially four different patterns in
the left three panels: a flat $\Delta \phi_{jj}$ behavior
($0^+_\text{SM},1^+_W$,$2^+_\text{EW}$), a back-to-back peak at
$\Delta \phi_{jj} \sim \pi$ ($0^+_\text{D5},2^+_\text{EW+q}$), a preferred angle
around $\pi/2$ ($0^-_\text{D5},1^+_Z$), and preferably aligned tagging
jets $\Delta \phi_{jj} \sim 0$ ($1^-_{W,Z}$). Due to the absence of
the longitudinal amplitude for the $0^-_\text{D5}$ operator we know
that its distribution follows a $1-\cos2\Delta\phi_{jj}$
modulation. On the other hand, $0^+_\text{D5}$ follows a $\cos
\Delta\phi_{jj}$ shape from the interference between the transverse
and longitudinal amplitudes~\cite{kentarou}.

The issue with the proper definition of the spin-2 couplings also
appears in this azimuthal observable. First, the distributions without
($2^+_\text{EW}$) and with ($2^+_\text{EW+q}$) a coupling to quarks
have little in common. Secondly, removing the high-energy tale of the
jet momenta requiring Eq.\eqref{eq:ptmax} strongly affects the full
spin-2 distribution $2^+_\text{EW+q}$, changing it from distinctively
peaked at $\Delta \phi_{jj} = \pi$ to essentially identical to the
Standard Model Higgs scalar $0^+_\text{SM}$. This is a general feature
which we can trace back to the helicities probed by the spin-2
amplitude: once we limit it to relatively small energies of external
particles it probes exactly the same helicity structure as the spin-0
Standard Model operator~\cite{kentarou}.\bigskip

In the lower panels of Fig.~\ref{fig:kin_jj} we show the rapidity
difference $\Delta \eta_{jj}$ between the two tagging jets.  In the
standard WBF analyses we require $\Delta \eta_{jj} > 4.2$, cutting
away most of the spin-1 and spin-2 events and keeping mostly the
Standard Model Higgs $0^+_\text{SM}$ events. Skipping this cut we see
that there are three distinct groups of curves, with maxima around
$\Delta \eta_{jj} \sim 2$ ($2^+_\text{EW}$), $\Delta \eta_{jj} \sim 4$
($0^\pm_\text{D5}, 1^\pm_V, 2^+_\text{EW+q}$), and $\Delta \eta_{jj}
\sim 5.5$ ($0^+_\text{SM}$). Asking for tagging jets with limited
energy, Eq.\eqref{eq:ptmax}, turns both spin-2 distributions into an
exact copy of the $0^+_\text{SM}$ predictions. In this case only two
distinctly different patterns survive, SM-Higgs-like and slightly less
forward.

These rapidity differences are strongly correlated with the tagging
jet rapidities shows in Fig.~\ref{fig:kin_basics}.  Only very little
information remains in the additional observation of $\eta_{j,1}$
once we exploit $\Delta \eta_{jj}$. This is an effect of the WBF
ansatz we are using for all resonances hypotheses --- if any of the
processes under consideration included a mixed quark-gluon initial
state the boost from the laboratory frame to the center-of-mass frame
of the two tagging jets might well turn out useful.\bigskip

The main advantage of the two observables discussed in this section is
that they do not require the reconstruction of the heavy resonance. In
Sec.~\ref{sec:stats} we will compare the two observables $\Delta
\phi_{jj}$ and $\Delta \eta_{jj}$ in their distinguishing power
testing some spin-0 and spin-2 coupling structures. A complete set of
distributions can be found in Fig.~\ref{fig:app2a} in the Appendix. As
expected, the study of decay independent jet correlations is a very
useful starting point for a `Higgs' coupling
analysis~\cite{original}. However, the predictions from a spin-2 model
have to be taken with a grain of salt, because they are not stable
with respect to (slight) modifications of the underlying
model.

%%%%%%%%%%%%%%%%%%%%%%%%%%%%%%%%%%%%%%%%%%%%%%%%%%%%%%%%%%%%%%%%%%%%%%%%
\subsection{Higgs-jet correlations}
\label{sec:kin_jx}

%-------------------------------------------------------
\begin{figure}[b!]
 \hspace*{0.5cm}  spin-0
 \hspace*{3.25cm} spin-1
 \hspace*{3.25cm} spin-2 
 \hspace*{3.25cm} spin-2($p_T^\text{max}$)\\
 \includegraphics[width=0.24\textwidth]{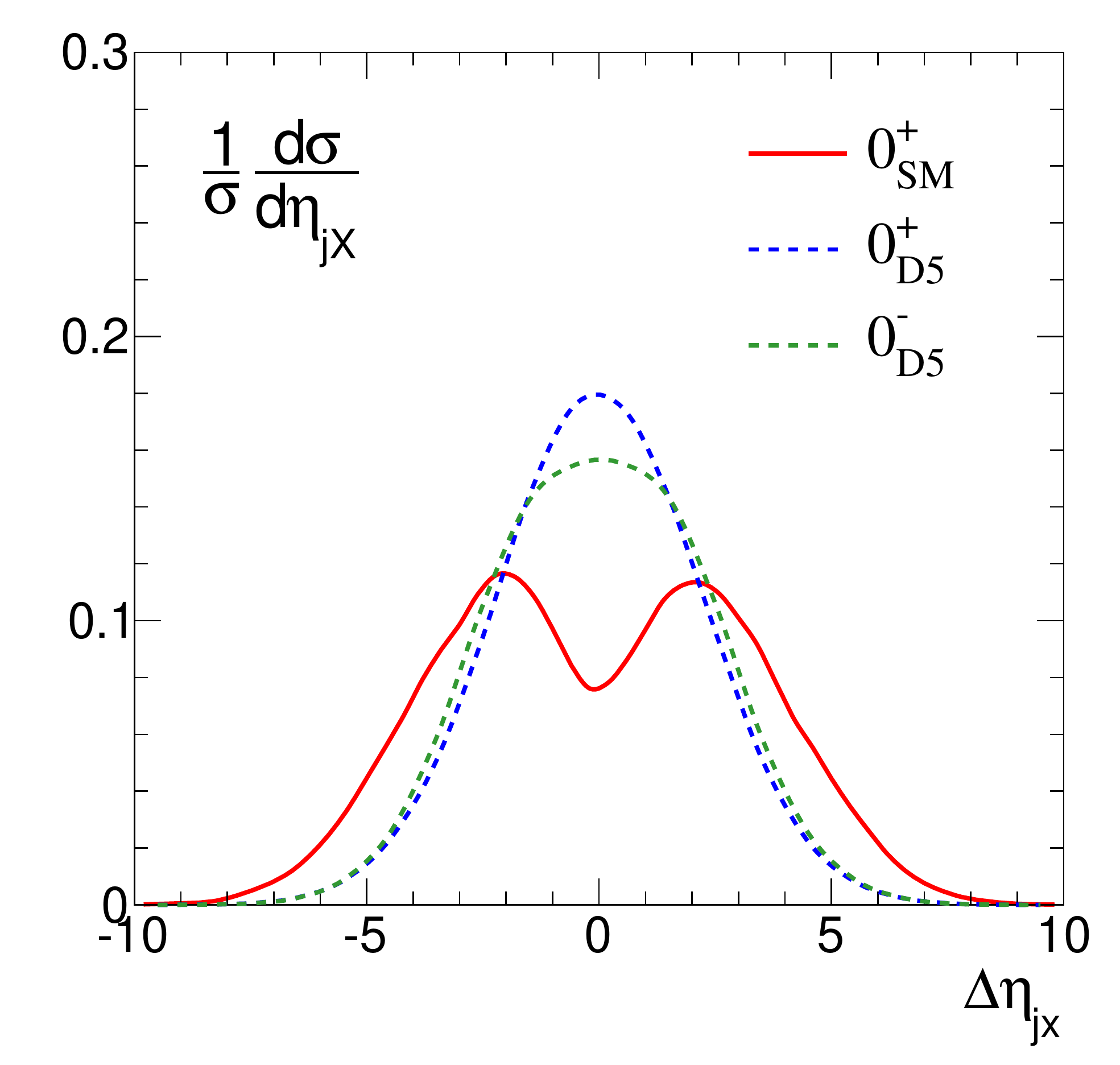}
 \hfill
 \includegraphics[width=0.24\textwidth]{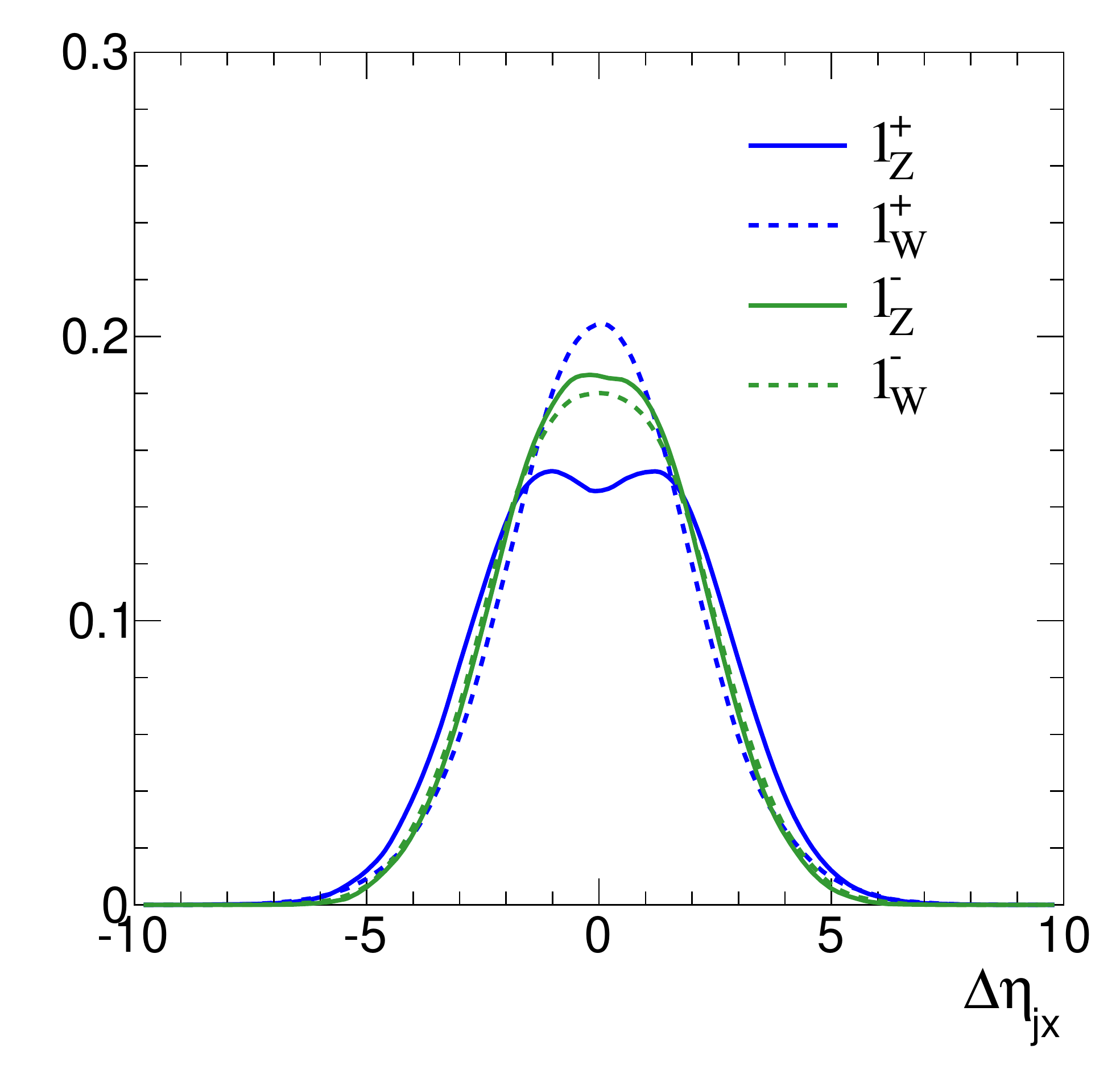}
 \hfill
 \includegraphics[width=0.24\textwidth]{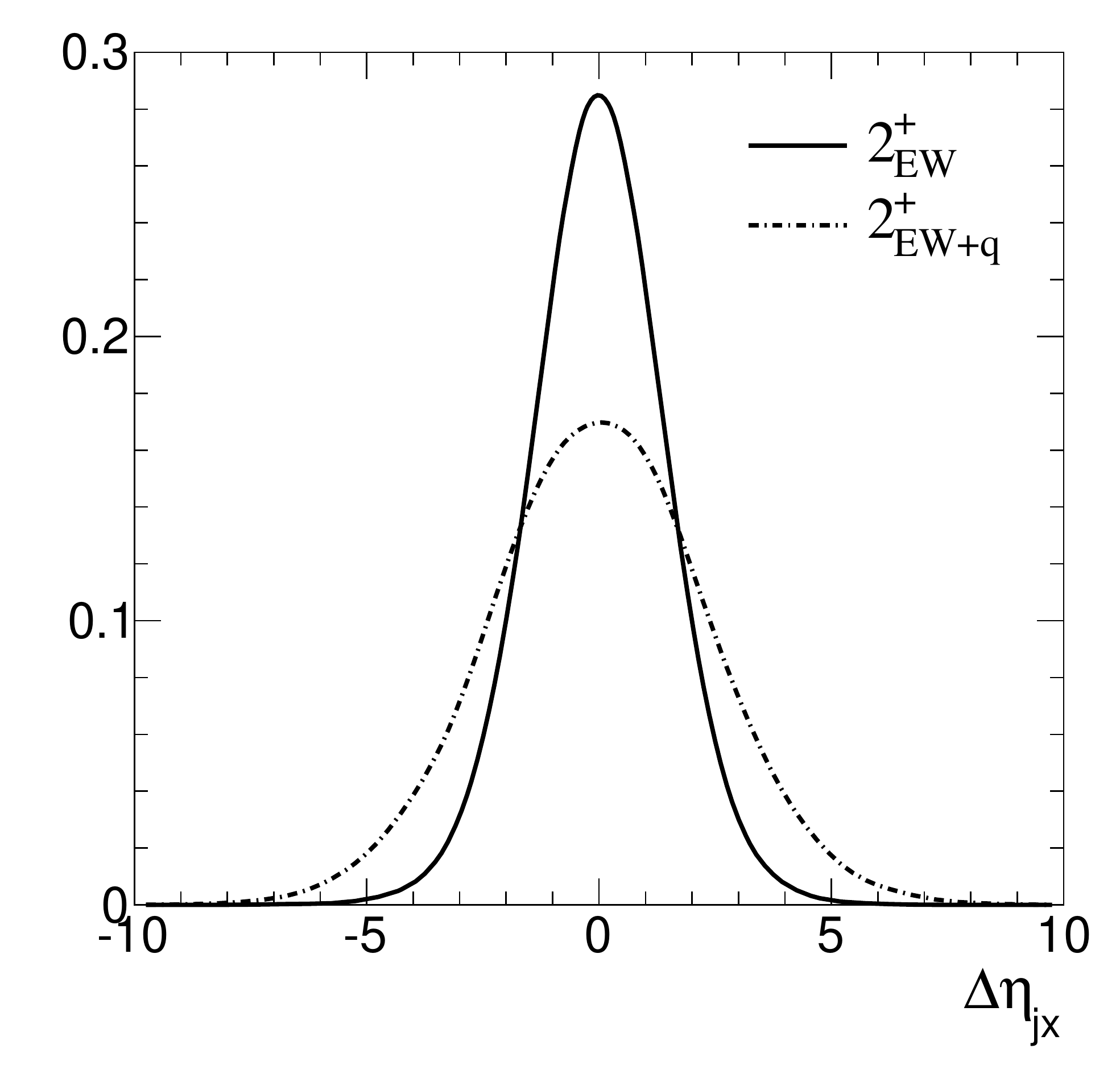}
 \hfill
 \includegraphics[width=0.24\textwidth]{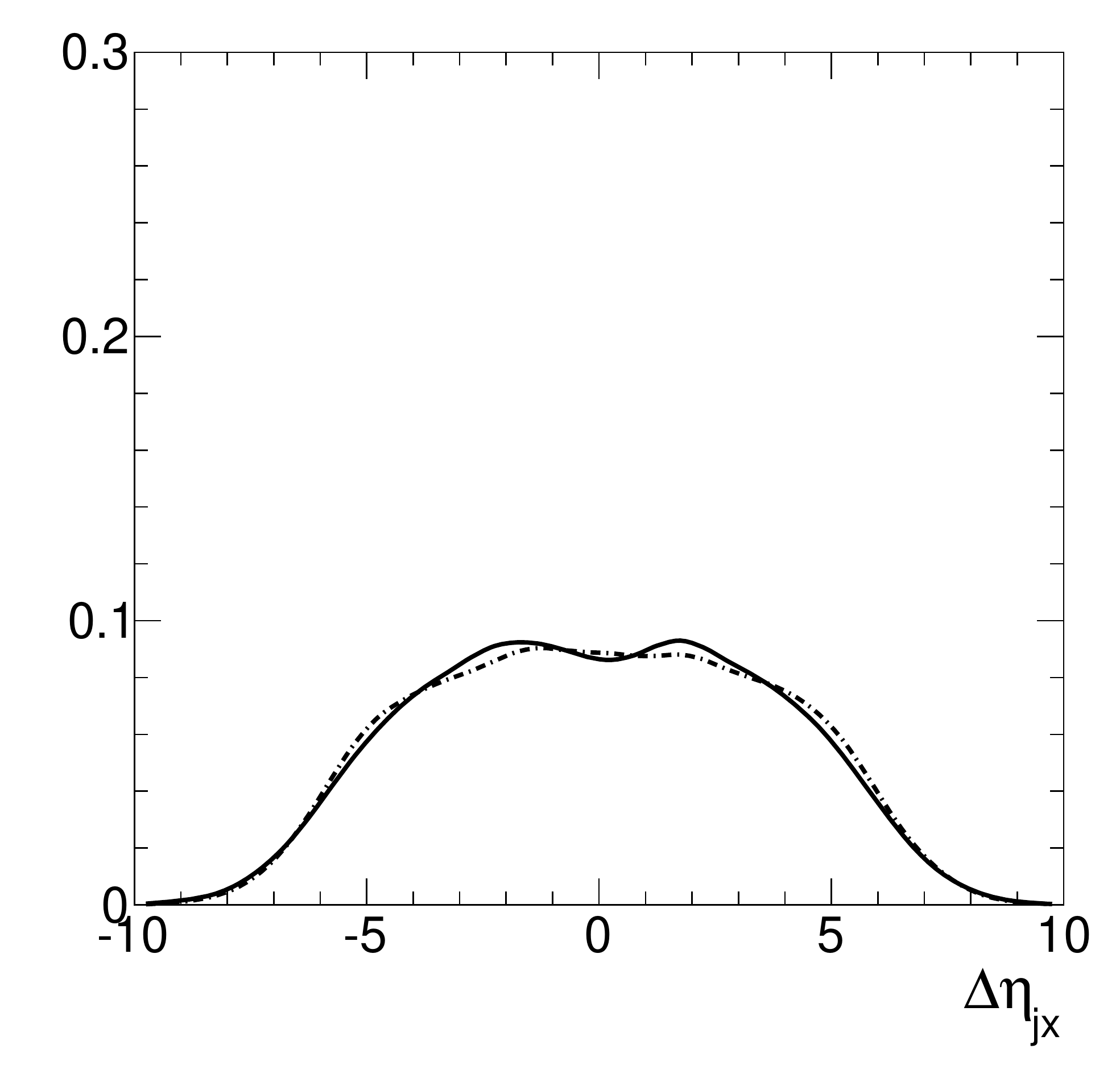}\\
 \includegraphics[width=0.24\textwidth]{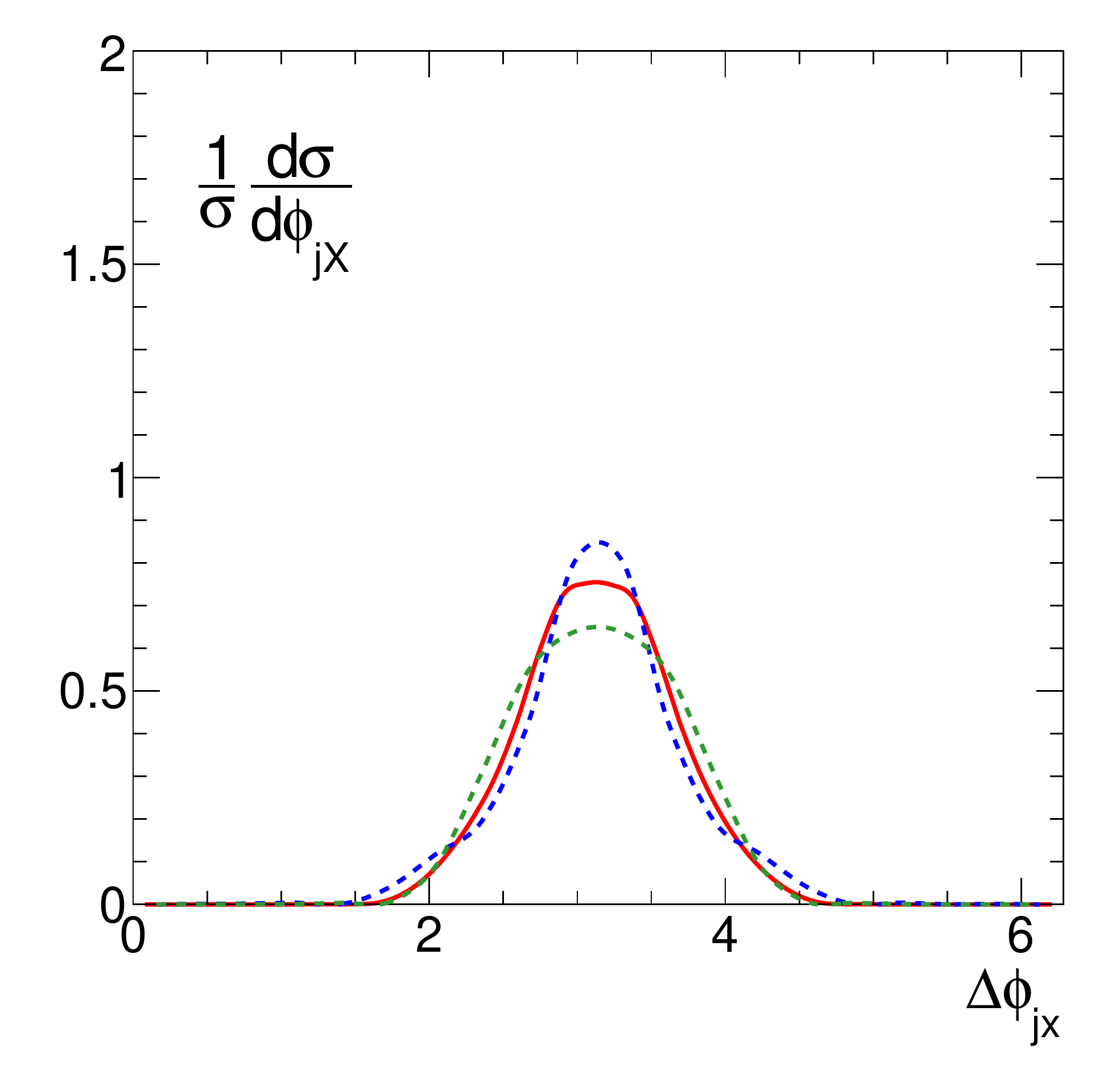}
 \hfill
 \includegraphics[width=0.24\textwidth]{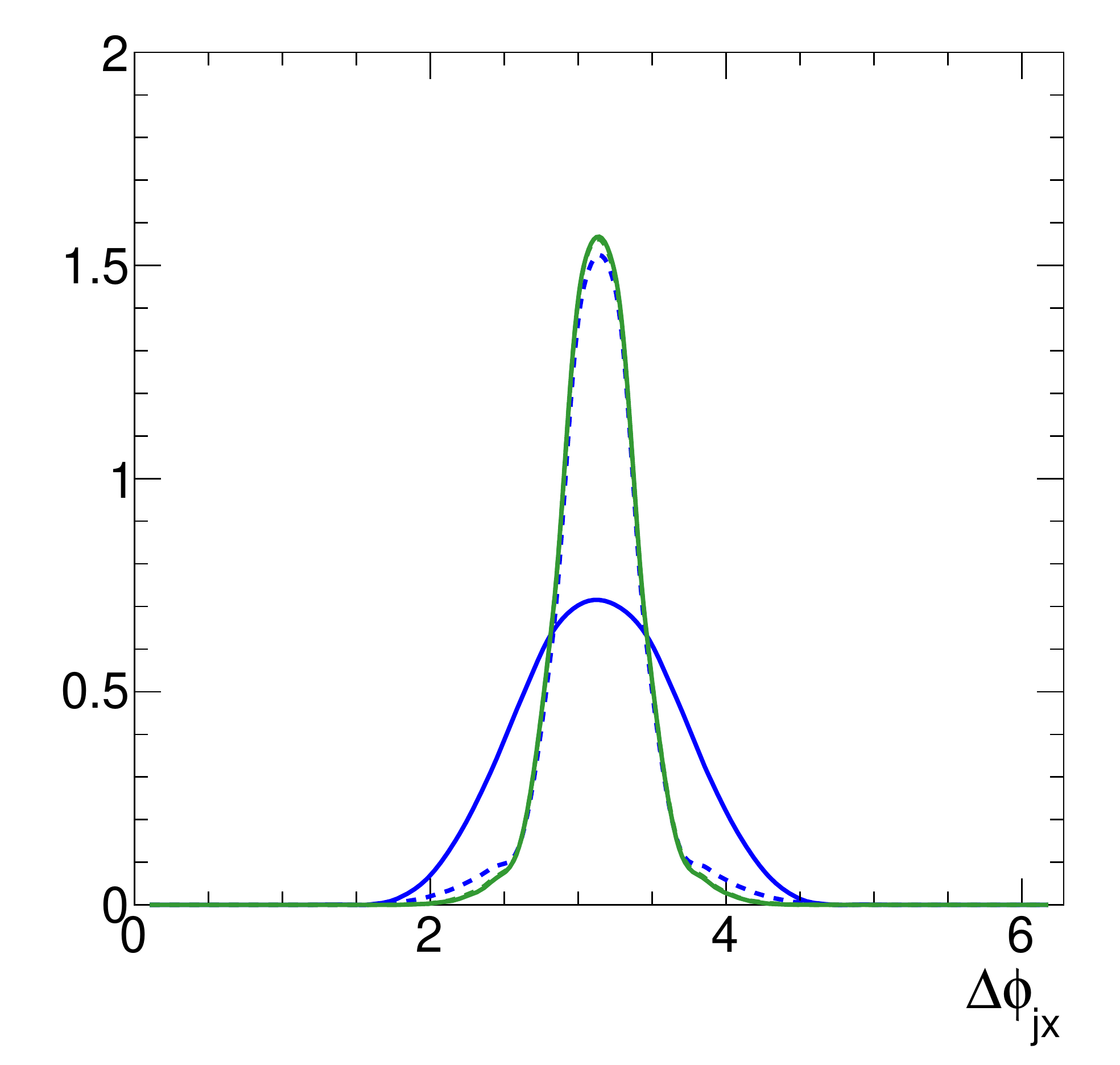}
 \hfill
 \includegraphics[width=0.24\textwidth]{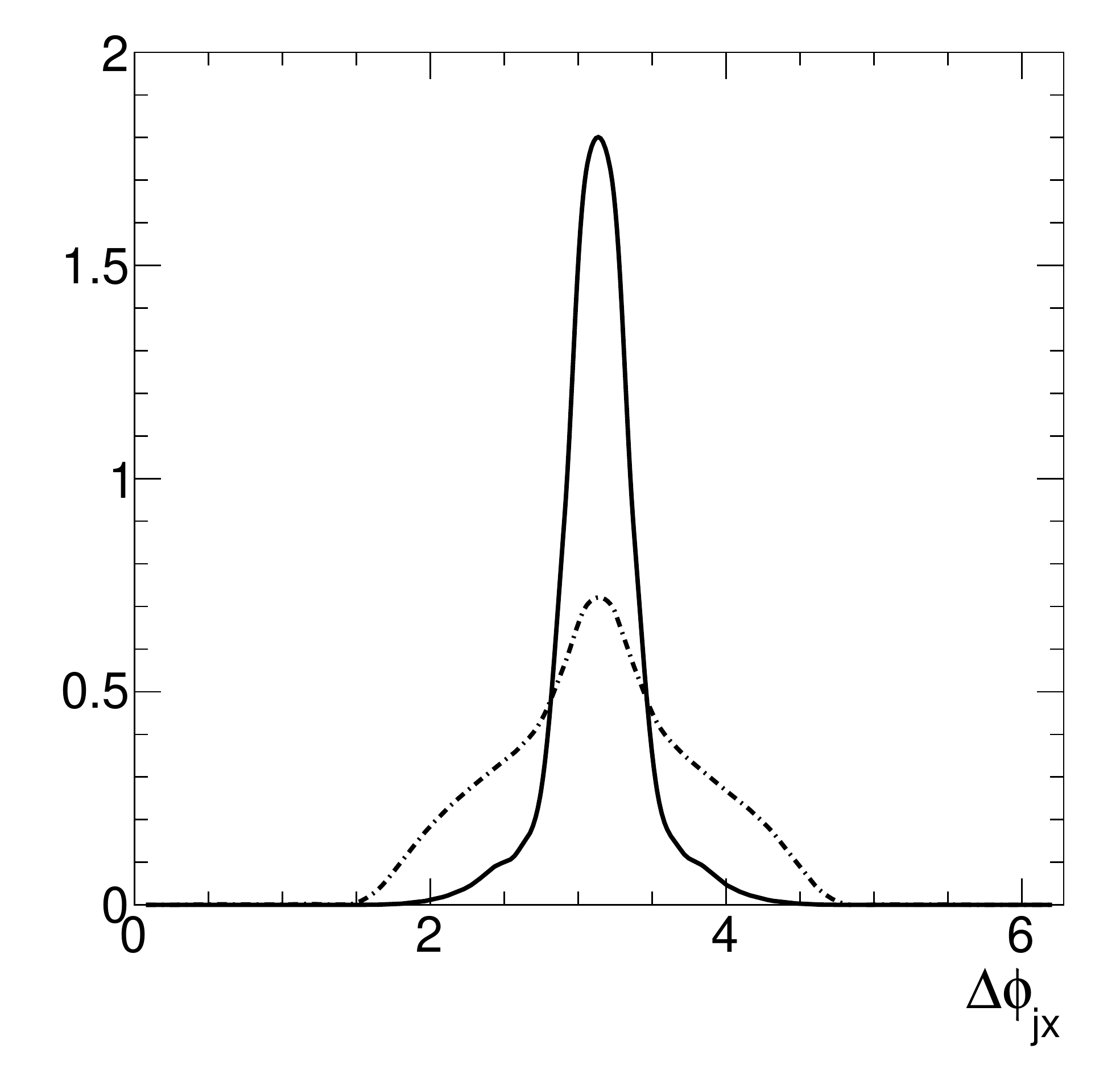}
 \hfill
 \includegraphics[width=0.24\textwidth]{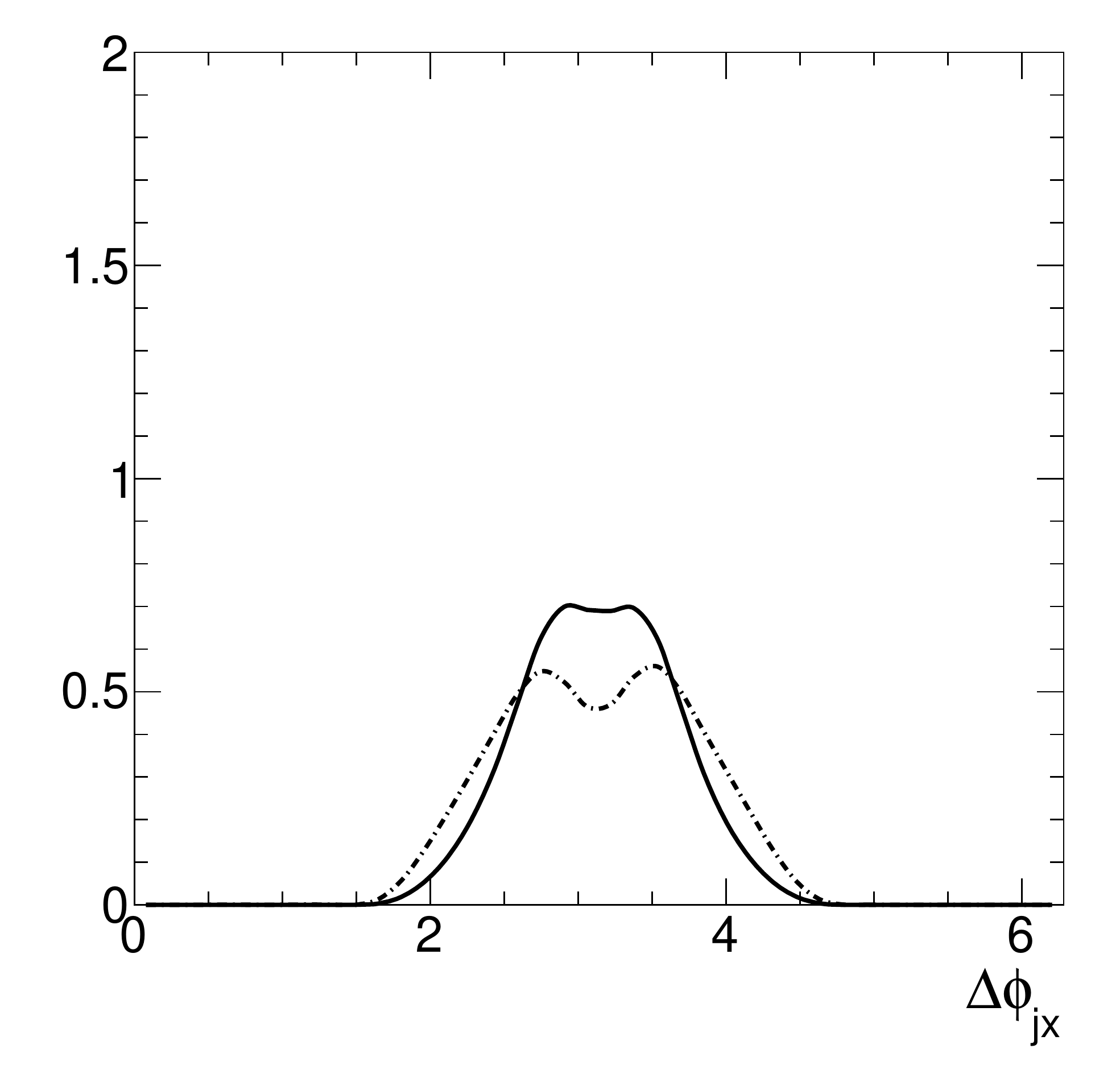}
  \caption{Normalized correlations between the leading tagging jet and
   the heavy resonance.  We show the difference in the rapidity
 difference $\Delta\eta_{jX}$ as well as the azimuthal angle
   $\Delta \phi_{jX}$. The spin-0, spin-1, spin-2 interaction structures are
   described in Sec.~\ref{sec:spinzero}-\ref{sec:spintwo}.}
\label{fig:kin_jx}
\end{figure}
%-------------------------------------------------------

Adding information on the `Higgs' momentum requires the reconstruction
of the heavy resonance. This means that the most promising channel for
such an analysis would be the decay $X \to \gamma \gamma$, possible
supplemented by the approximate reconstruction of $X \to \tau \tau$
decays. The latter requires additional assumptions about the couplings
of the heavy resonance to fermions.  This is most obvious in the
spin-1 case, where in Eq.\eqref{eq:spinone_fermions} we assume that
the heavy neutral resonance couples to the third generation in
complete analogy to a Standard Model $Z$ boson. 

This kind of assumption has for a long time been a problem in the
interpretation of $Z'$ searches, so it would be beneficial if we could
apply an analysis of the coupling structure without considering
fermionic couplings. Similarly, for WBF Higgs production the spin-2
states coupling to the energy-momentum tensor makes specific
assumptions about the production vertex and about the decay vertex. It
would be preferable if these two assumptions could be kept separate. 

A way of including significantly more information than just the
jet-jet correlations discussed in the previous section and at the same
time making minimal assumptions about the decay of the heavy resonance
is to study angular correlations between the tagging jets and the
reconstructed resonance $X$. In this second step of the angular
analysis we will focus on the observables $\Delta \eta_{jX}$ and
$\Delta \phi_{jX}$, where $j$ denotes the tagging jet with the larger
transverse momentum. This ordering of the tagging jets induces a
difference compared to the Breit-frame analysis described in
Sec.\ref{sec:angles}, where the tagging jets are ordered by
rapidity.\bigskip

In the upper panels of Fig.\ref{fig:kin_jx} we show the rapidity
difference between the reconstructed heavy resonance and the leading
tagging jet. As expected from the tagging jet distributions shown in
Sec.~\ref{sec:kin_jj} the Standard Model Higgs $0^+_\text{SM}$
is peaked at large
differences $\Delta \eta_{jX} \sim 2$.
The unitarized spin-2 resonance has a
characteristically flat behavior. The $\Delta \eta_{jX}$ distributions
for all spin-1 coupling structures as well as the dimension-5 scalars
are only marginally distinctive.

The second set of panels shows the azimuthal distance between the
leading tagging jet and the heavy resonance. The Standard Model
Higgs and the cut-off $2^+_\text{EW}$ structures predict the same
moderate back-to-back topology. For spin-1 the $1^-_V$ and $1^+_W$
distributions are much more strongly peaked in the same back-to-back
kinematics.\bigskip

Comparing the jet-$X$ correlations shown in Fig.~\ref{fig:kin_jx} with
the jet-jet correlations from Fig.~\ref{fig:kin_jj} clearly shows that
in weak boson fusion the spin and coupling information is largely
encoded in the tagging jets, not in the angular correlations of the
heavy resonance~\cite{original}, with the notable exception of a
unitarized spin-2 model~\cite{kentarou}.

%%%%%%%%%%%%%%%%%%%%%%%%%%%%%%%%%%%%%%%%%%%%%%%%%%%%%%%%%%%%%%%%%%%%%%%%
\subsection{Including Higgs decays}
\label{sec:kin_jd}

%-------------------------------------------------------
\begin{figure}[b!]
 \hspace*{0.5cm}  spin-0
 \hspace*{3.25cm} spin-1
 \hspace*{3.25cm} spin-2 
 \hspace*{3.25cm} spin-2($p_T^\text{max}$)\\
 \includegraphics[width=0.24\textwidth]{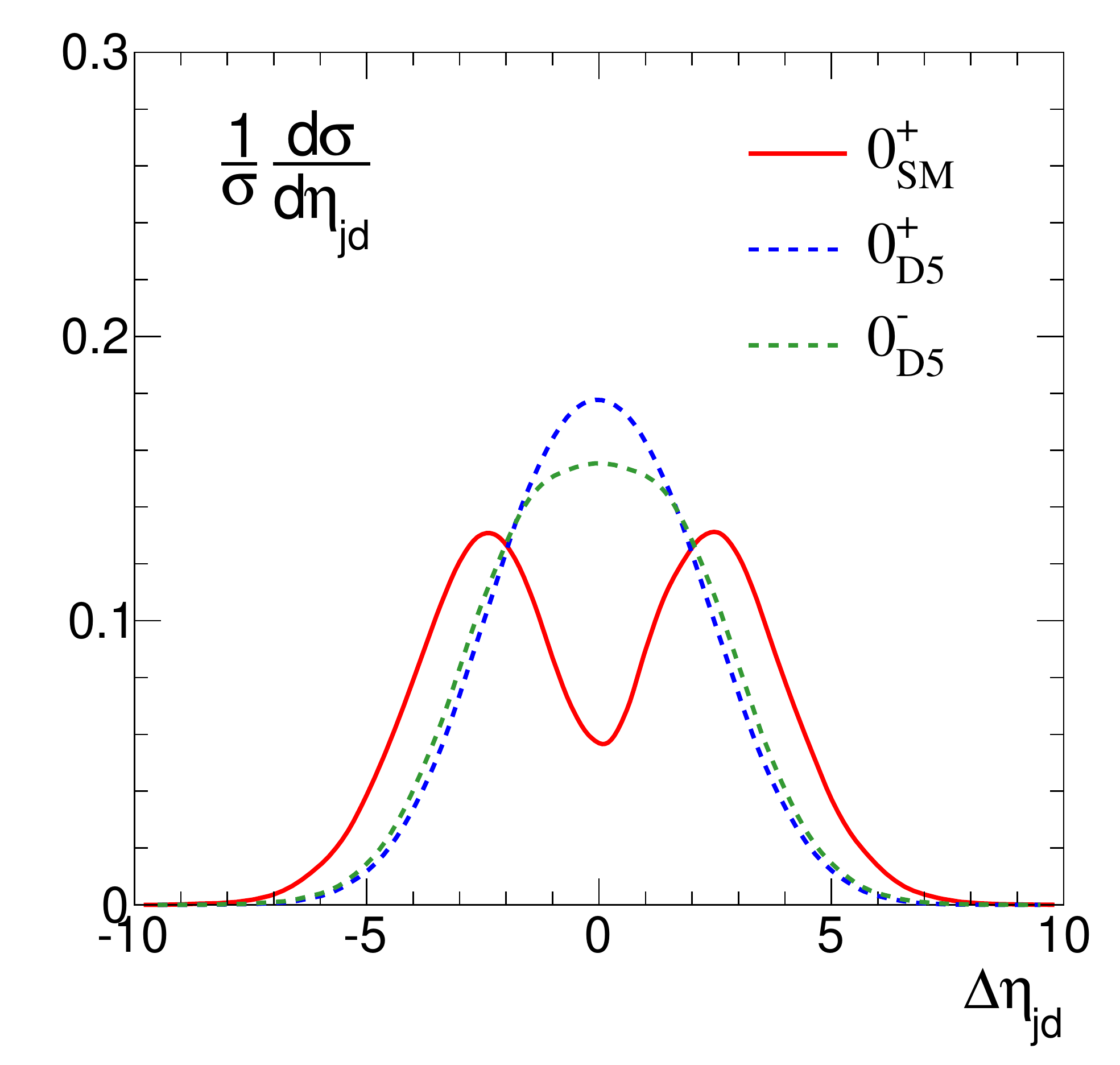}
 \hfill
 \includegraphics[width=0.24\textwidth]{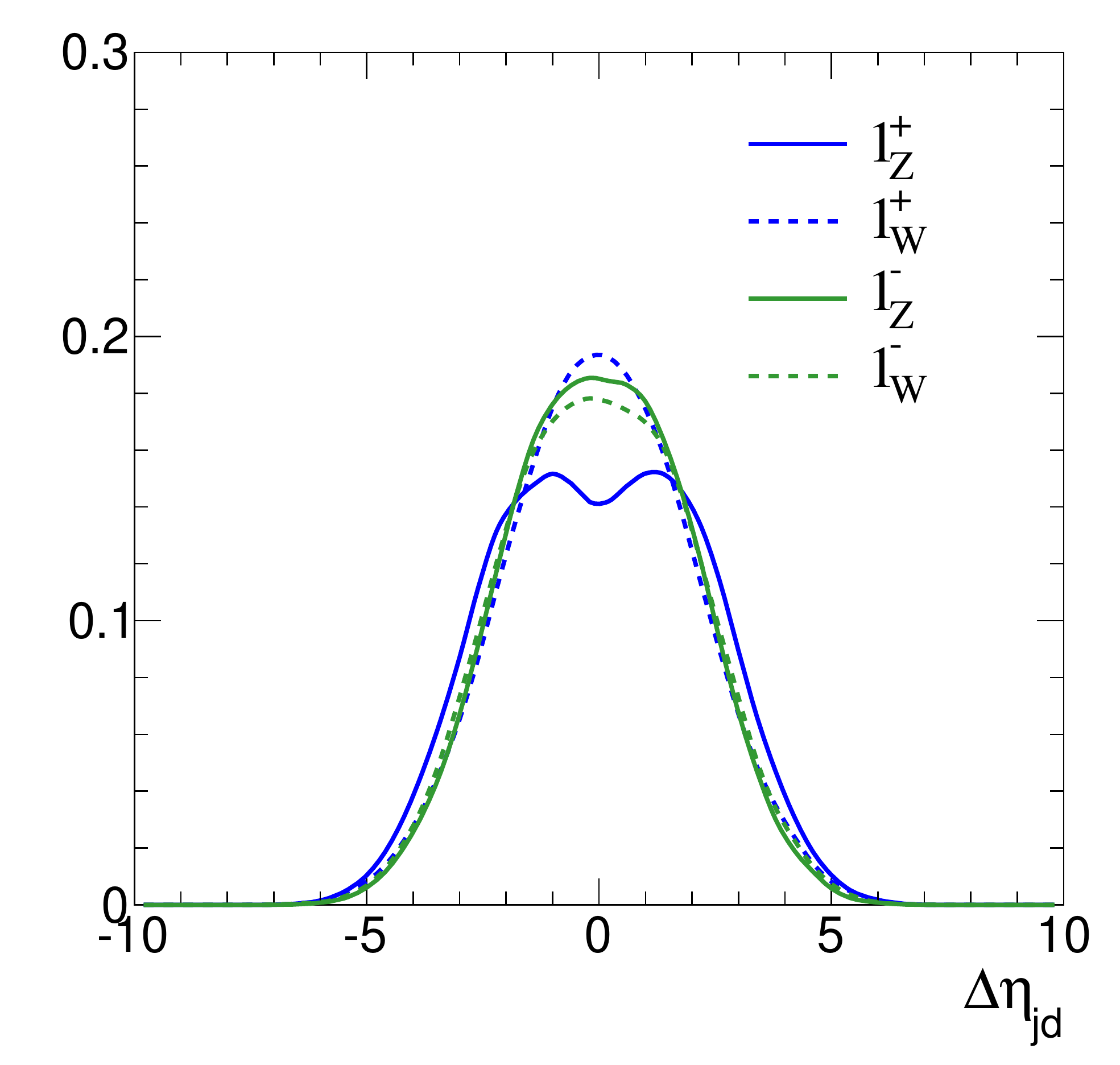}
 \hfill
 \includegraphics[width=0.24\textwidth]{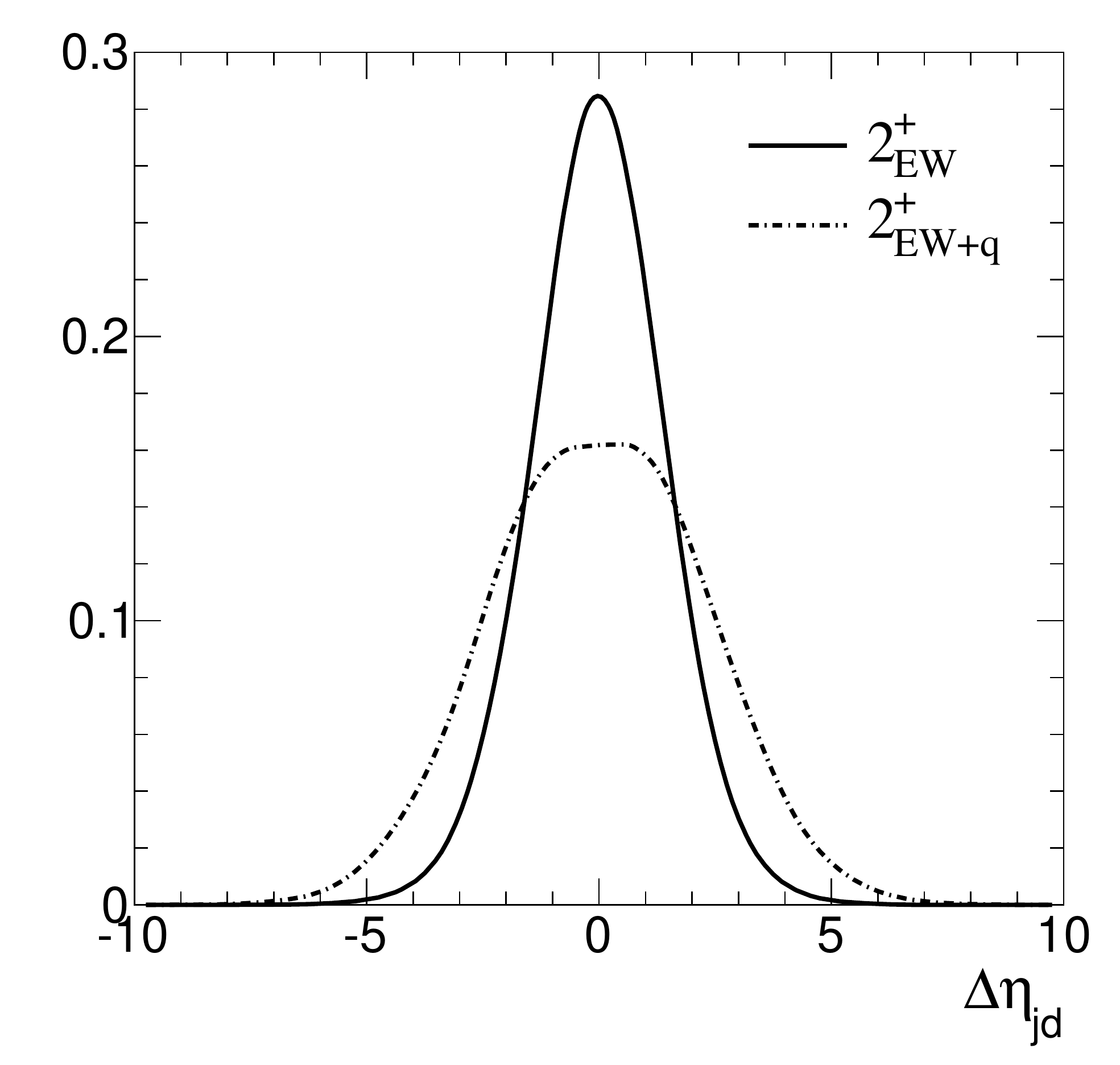}
 \hfill
 \includegraphics[width=0.24\textwidth]{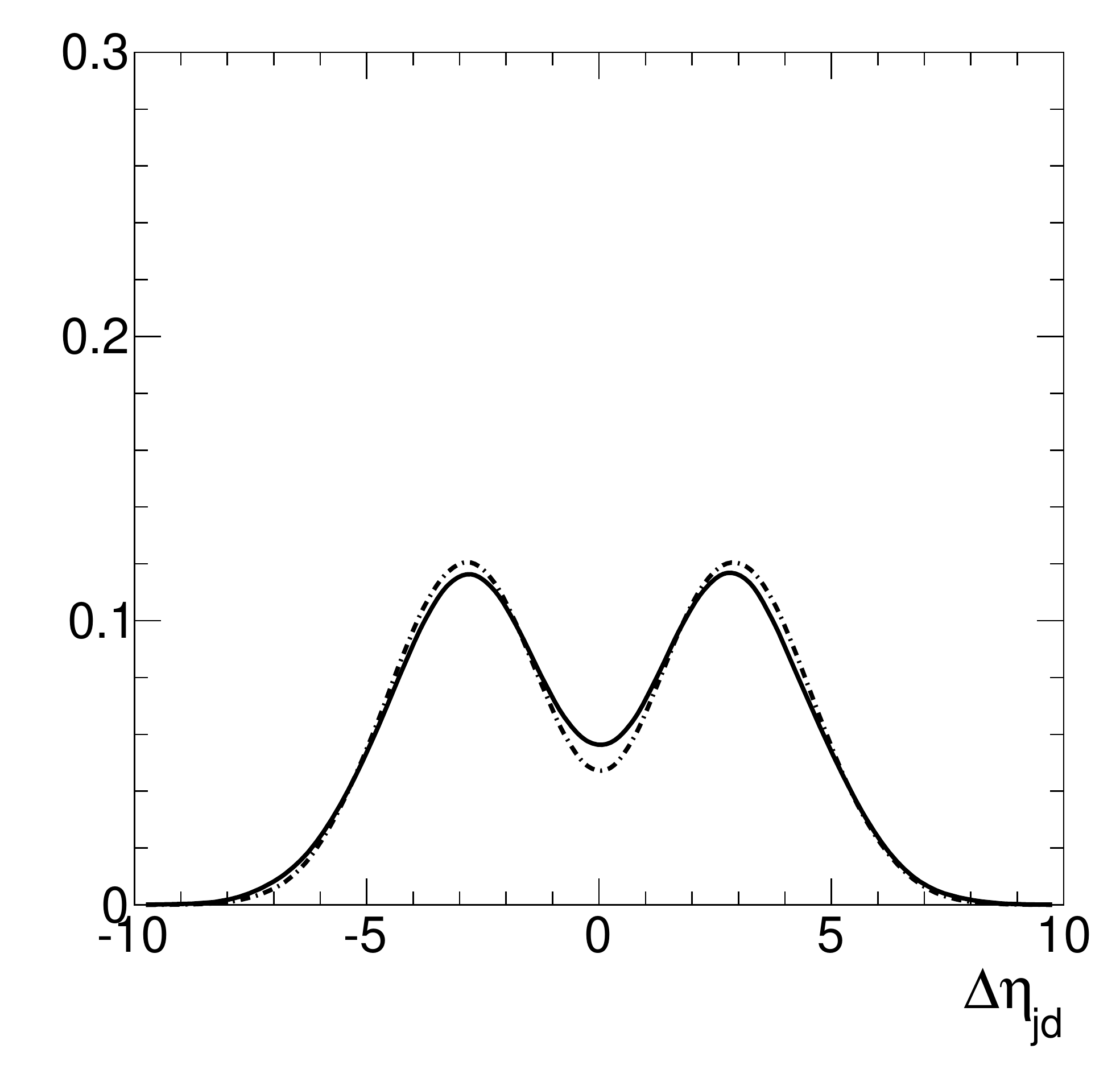}\\
 \includegraphics[width=0.24\textwidth]{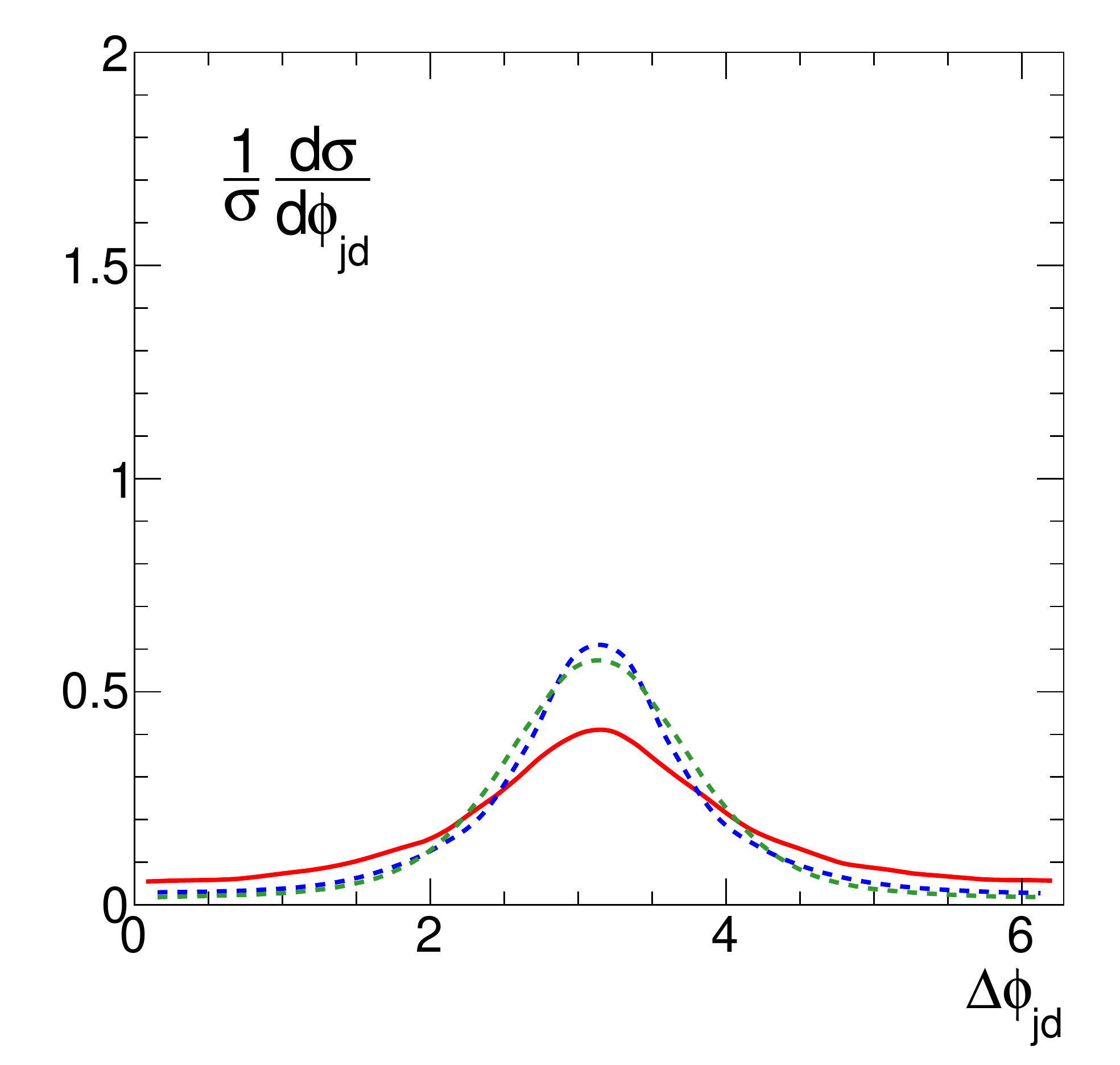}
 \hfill
 \includegraphics[width=0.24\textwidth]{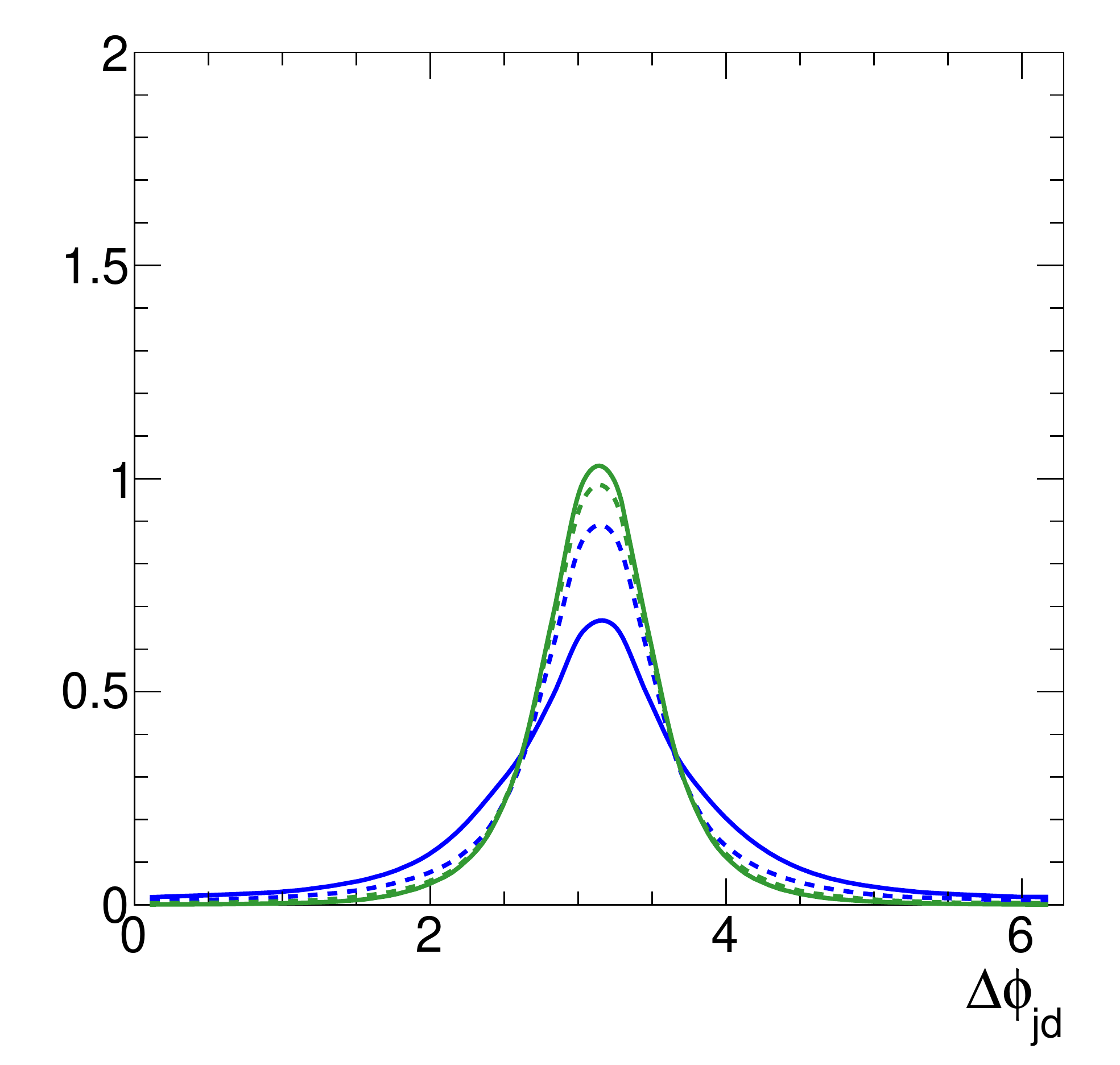}
 \hfill
 \includegraphics[width=0.24\textwidth]{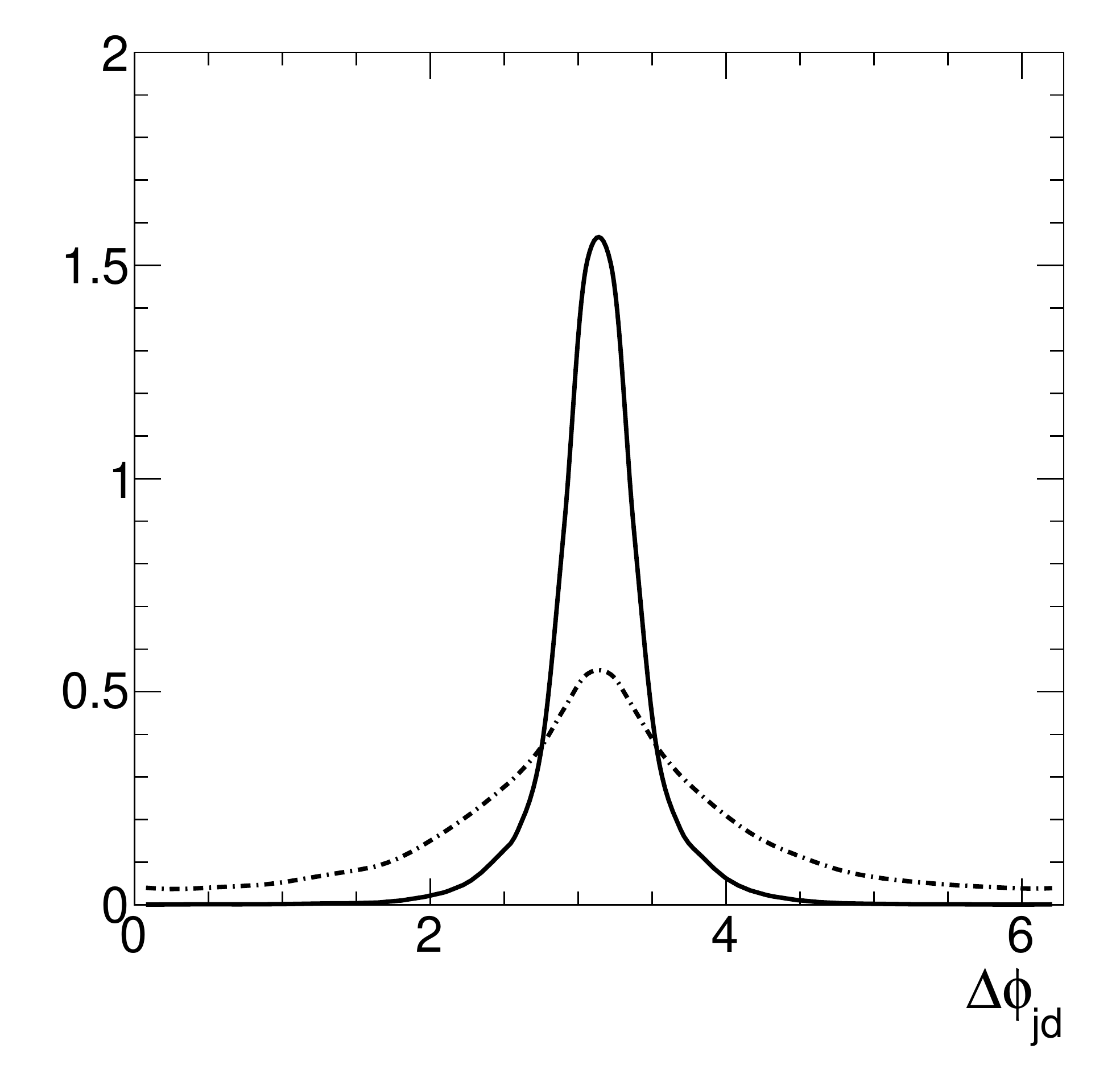}
  \hfill
 \includegraphics[width=0.24\textwidth]{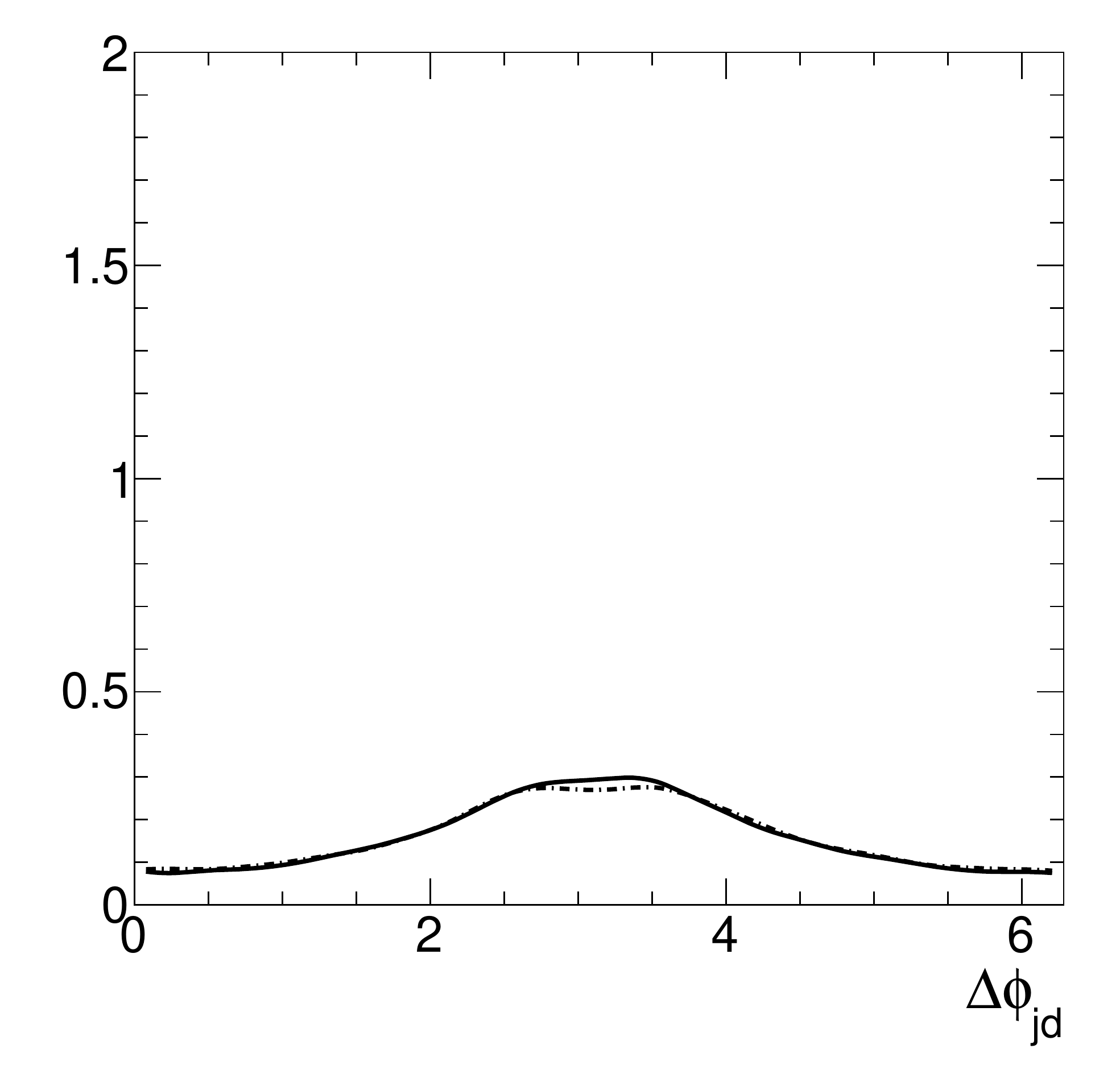} \\
 \includegraphics[width=0.24\textwidth]{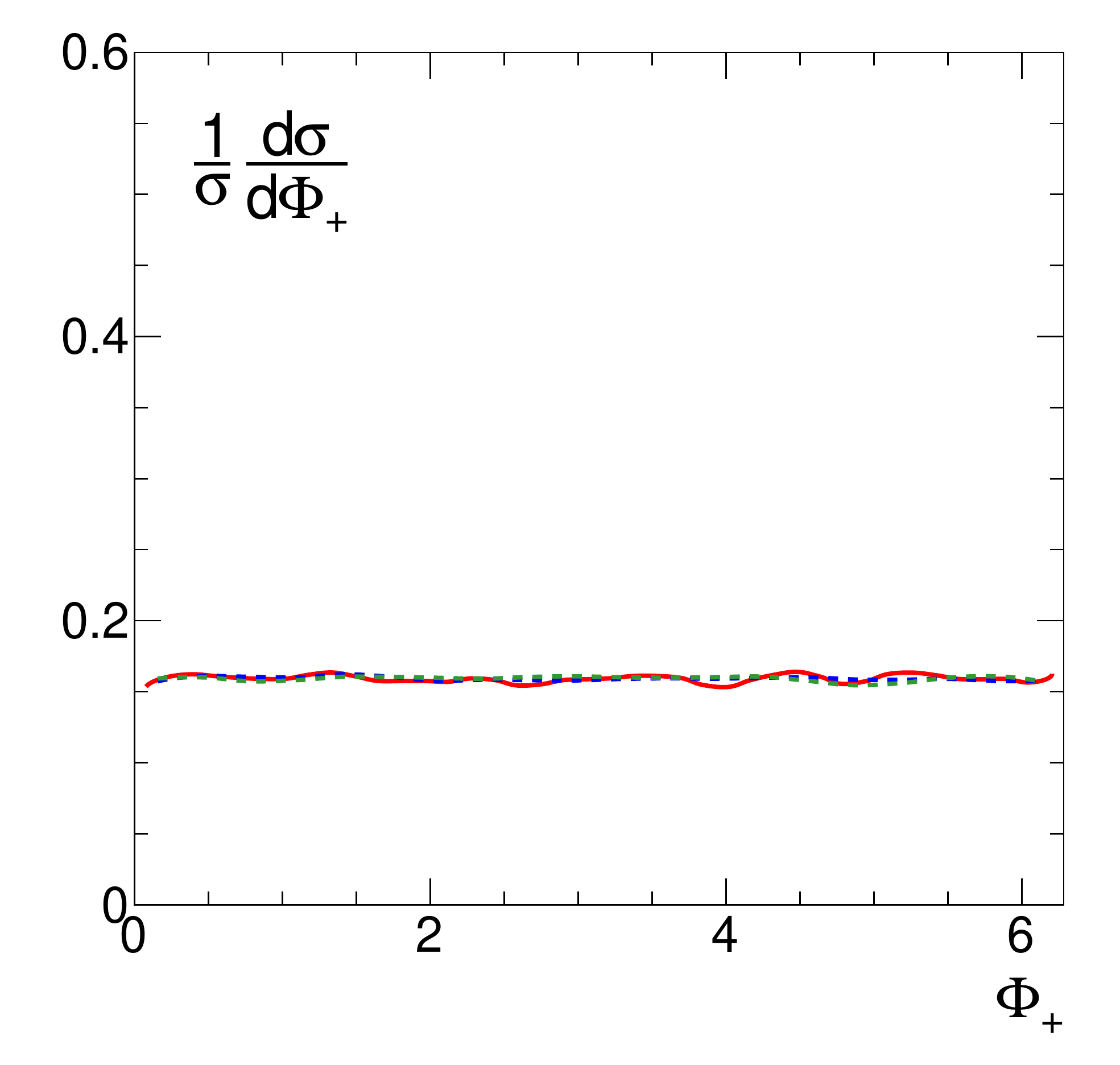}
 \hfill
 \includegraphics[width=0.24\textwidth]{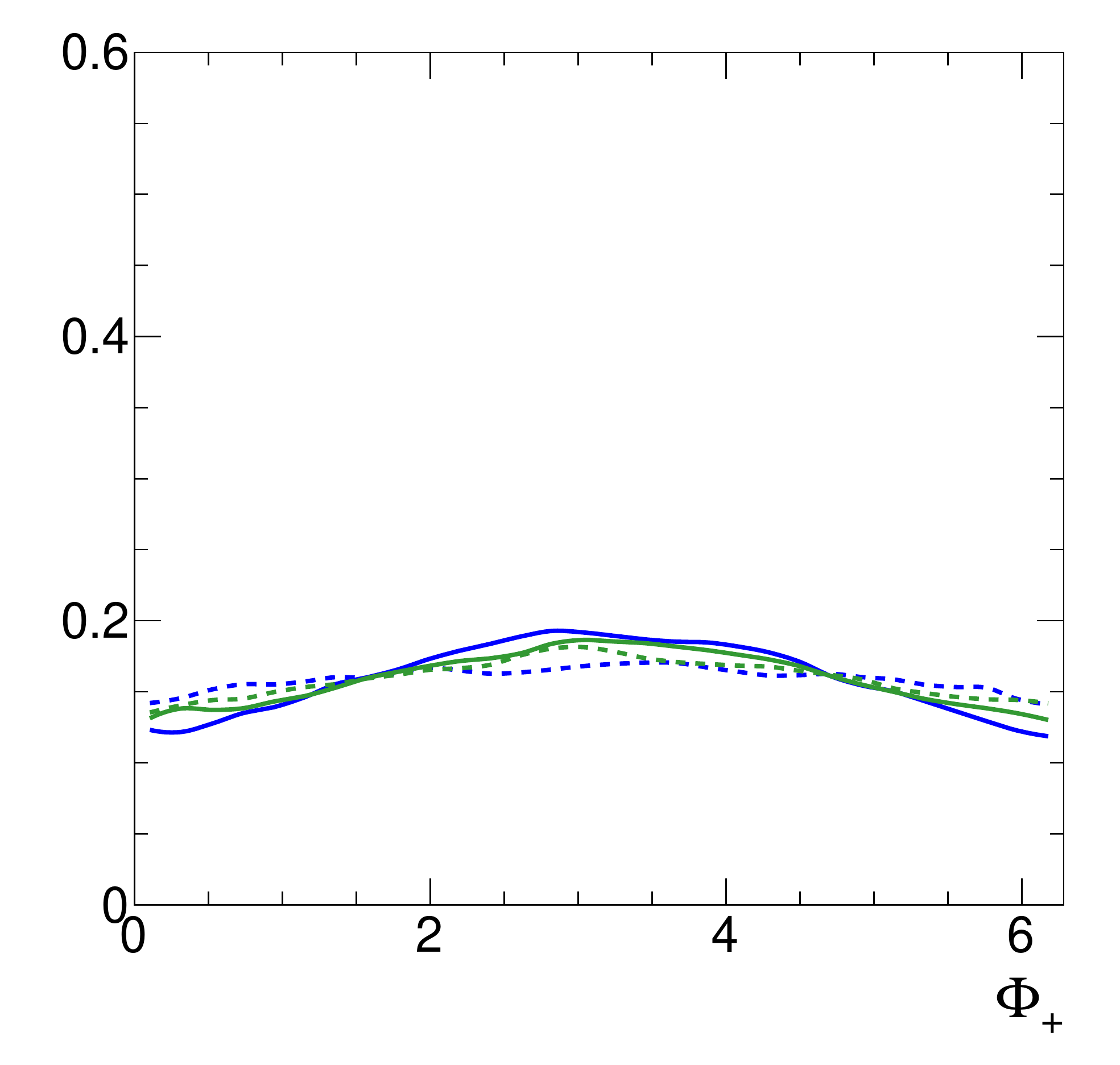}
 \hfill
 \includegraphics[width=0.24\textwidth]{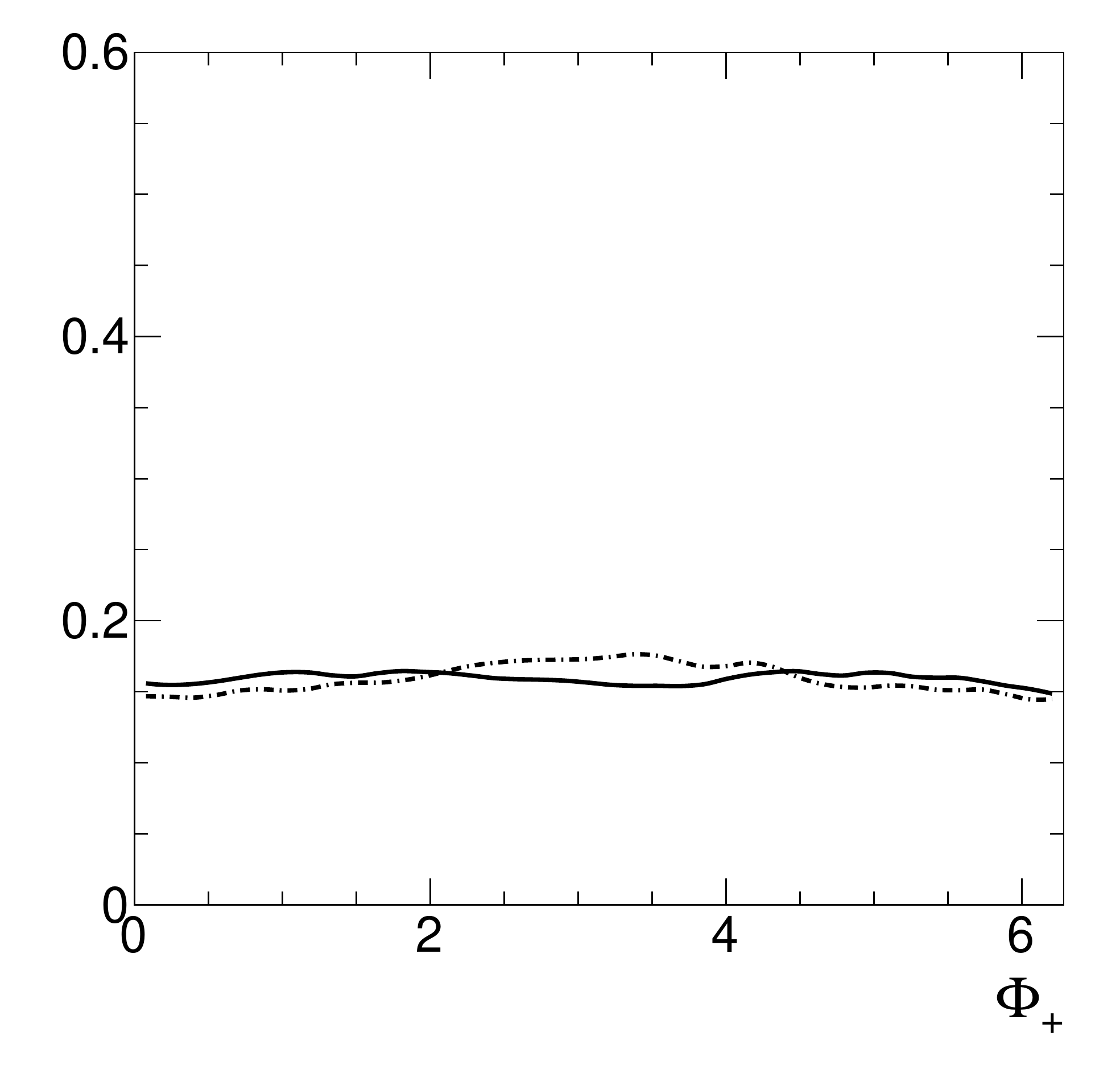}
 \hfill
 \includegraphics[width=0.24\textwidth]{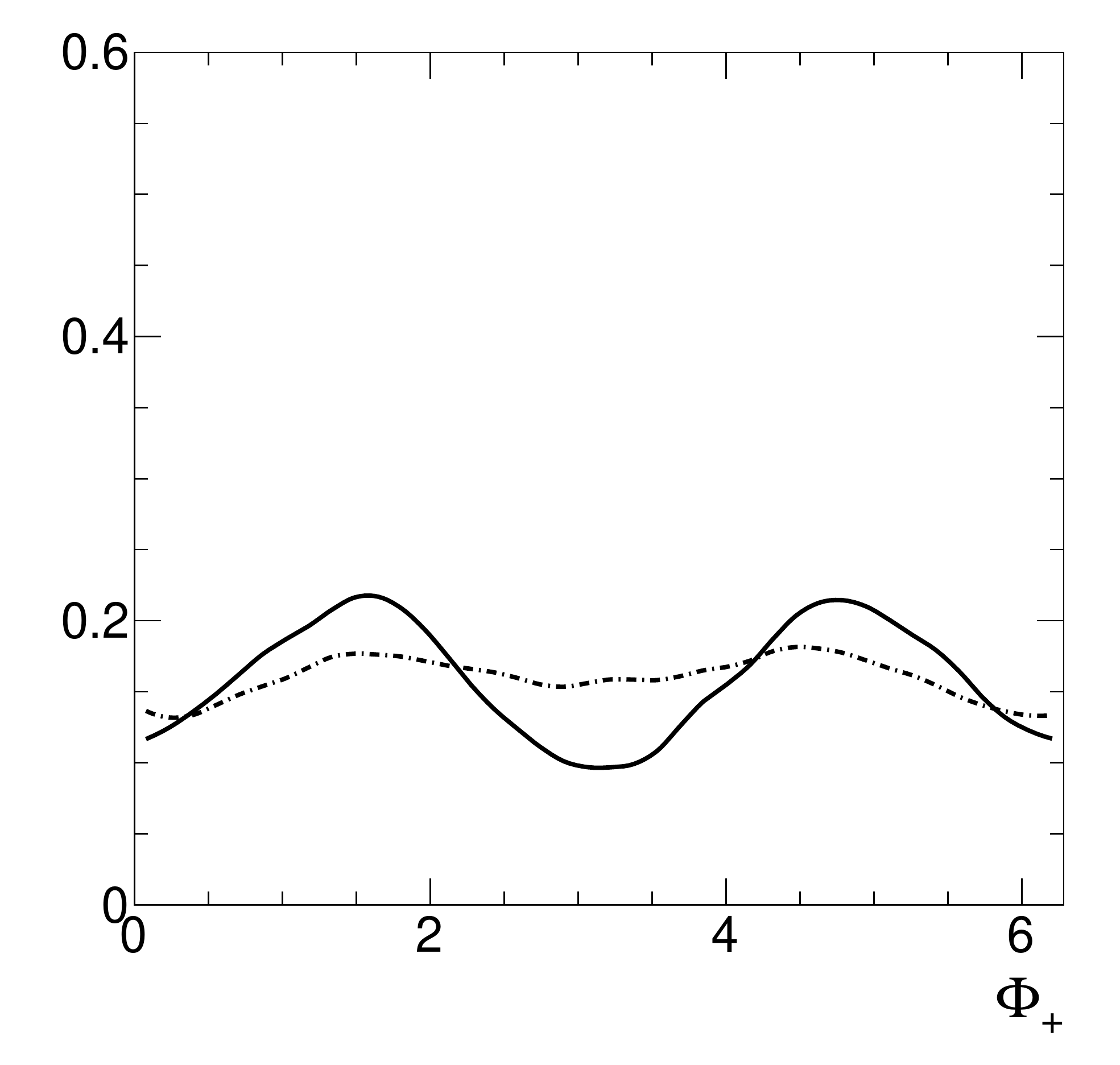}
 \caption{Normalized correlations between the leading tagging jet and
   the $X$ decay particle.  We show the difference in 
  the rapidity difference $\Delta\eta_{jd}$ as well as
 the azimuthal
   angle $\Delta \phi_{jd}$. In the bottom row we show the angle $\Phi_+$ in
   the Breit frame.}
\label{fig:kin_jd}
\end{figure}
%-------------------------------------------------------

Essentially all information on the Higgs couplings structure available
from `Higgs' production in weak boson fusion is included in the
jet-jet and jet-$X$ correlations described in the previous two
sections. The only caveat is that the resonance observed in the final
state might have a spin or polarization structure which we average
over unless we make certain requirements on the $X \to d\bar{d}$ decay
products. The price we have to pay for this additional information is
a full simulation of the $X$ decay, including its underlying coupling
structure. If we limit ourselves to $VVX$ couplings for the production
process this obviously implies additional assumptions for all decays
except for decays to massive weak gauge bosons $X \to VV$. Moreover,
the production and decay process $VV \to X \to VV$ requires a
unitarization which only the $0^+_\text{SM}$ delivers and which we
have to add for example for a spin-2 resonance. In
Fig.~\ref{fig:kin_basics} we see that this unitarization has a major
effect on all energy dependent distributions, which means we have to
take the jet-decay correlations with a grain of salt.\bigskip

Nevertheless, in Fig.\ref{fig:kin_jd} we show the usual rapidity and
azimuthal angle differences between the leading tagging jet and one of
the two $X$ decay products. For the decay channel we assume $X \to
\tau^+ \tau^-$ with the couplings described in Sec.~\ref{sec:lag}.
Comparing these distributions to the jet-$X$ distributions in
Fig.~\ref{fig:kin_jx} we see that their features are very similar, \ie
the heavy resonance decay kinematics is dominated by the jet-$X$
kinematics. Small effects come from the acceptance cuts for the decay
products. The main target for production-decay correlations is the
spin-2 hypothesis, which after the unitarization cut $p_{T,j} <
p_T^\text{max}$ in Eq.\eqref{eq:ptmax} tends to be dominated by the
helicity-2 states.

The spin-2 operator coupling to gauge bosons and to quarks is the only
scenario where $\Delta \eta_{jd}$ differs from $\Delta
\eta_{jX}$. While the latter shows a centrally flat behavior,
distinctively different from $2^+_\text{EW}$ and $0^+_\text{SM}$, the
distributions including the decay becomes much more similar to
those. In this case the jet-decay correlations are even less useful
than the jet-$X$ correlations.  The $\Delta \phi_{jd}$ distributions
also resemble their $\Delta \phi_{jX}$ counter parts closely, with a
generic smearing because the tagging jets now recoil against to decay
products instead of one resonance. Again, one change is that the two
spin-2 hypotheses are becoming indistinguishable, but possibly more
different from the spin-0 predictions. Moreover, one exception to the
general broadening of $\Delta \phi_{jd}$ is the $1^+_Z$ operator,
which becomes more similar to the other spin-1 scenarios and less
similar to the scalar couplings.\bigskip

In addition to the generic hadron collider observables there are a few
variables which are particularly well suited to identify a spin-2
resonance. One example is the Gottfried--Jackson
angle~\cite{gottfried,mrauch}, which is very closely related to $\cos
\theta^*$ in the Breit frame. We show the corresponding distributions
in the Appendix. In the bottom row of Fig.~\ref{fig:kin_jd} we show
the distribution of the angle $\Phi_+\equiv 2\phi_1 +\Delta \phi $ in the Breit
frame~\cite{kentarou}, whose angles are defined in Eq.\eqref{eq:angles_wbf}. Its
particular feature is that it has no distinguishing features unless we
unitarize the spin-2 rate, in which case it develops a clear
modulation for spin-2 couplings. In principle the same information
should be included in an appropriate combination of hadron collider
observables, but it cannot be easily observed in combinations like
$\Delta \phi_{j_1 d} +
\Delta \phi_{j_2 d}$.

%%%%%%%%%%%%%%%%%%%%%%%%%%%%%%%%%%%%%%%%%%%%%%%%%%%%%%%%%%%%%%%%%%%%%%%%
\subsection{Basic strategy}
\label{sec:strategy}

From the discussion of the different jet-jet, jet-$X$, and jet-decay
observables we know that there are a few key measurements which can
distinguish between the `Higgs' coupling structures. The hadron
collider observables $\Delta \eta$ and $\Delta \phi$ serve as
effective replacements of the
Cabibbo--Maksymowicz--Dell'Aquila--Nelson
angles~\cite{nelson,zerwas}. A particular strength of the angular
basis for $X \to ZZ$ decays is the clear link between some of some
Higgs couplings structures and particular distinctive observables. The
distribution of the azimuthal angle $\Delta\phi_{jj}$~\cite{original}
shown in Fig.~\ref{fig:dphi} constitutes such a specific link in the
WBF topology.\bigskip

Hence, the question arises if we can construct a set of measurements
which can distinguish the different `Higgs' operators shown in
Tab.~\ref{tab:model}. Such a simple flow diagram does not correspond to
an optimized experimental analysis, but it indicates what the key
observables are. For example, from Sec.~\ref{sec:analysis} we know
that we should avoid cutting on $\Delta \eta_{jj}$ before testing
`Higgs' couplings because it removes one of the key observables to
tell apart the $0^+_\text{D5}$ and $0^+_\text{SM}$ operators.
Following our earlier argument we do not include the transverse
momentum spectrum of the tagging jets in our list of observables
because the initially striking differences will be modified by form
factors for spin-1 and spin-2 resonances. For now, we consistently
assume spin-2 operators with our simplified unitarization cut
$p_T^\text{max}=100$~GeV as defined in Eq.\eqref{eq:ptmax}. Without
this cut a spin-2 could be immediately recognized from the $p_T$
spectra shown in Sec.\ref{sec:kin_jj}. \bigskip

%-------------------------------------------------------
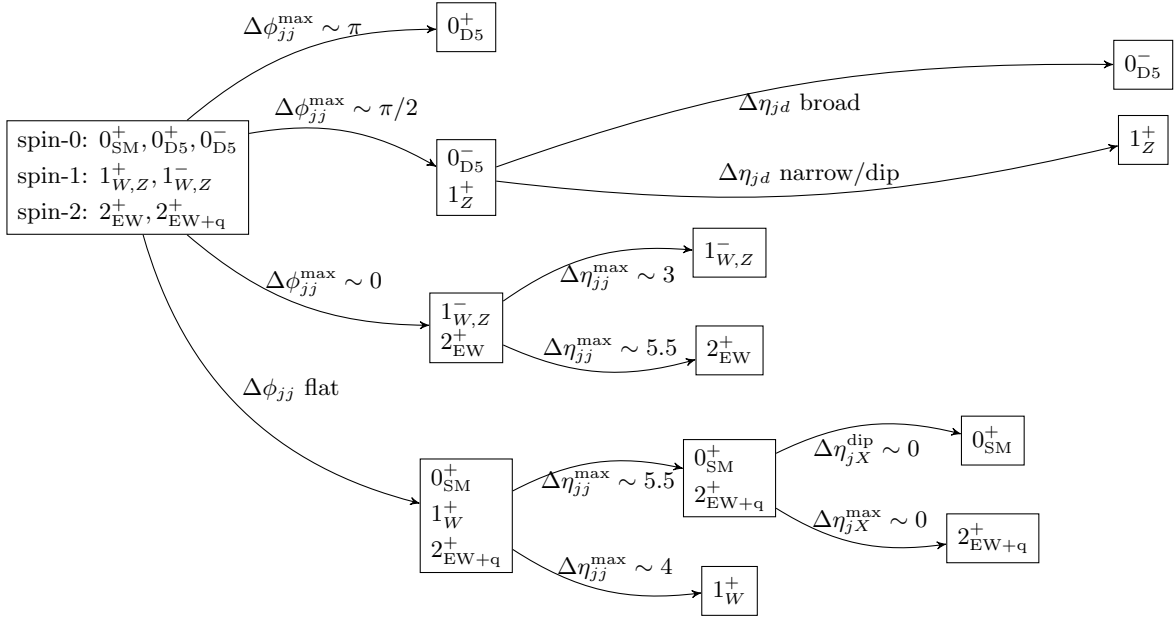
\begin{figure}[t]
\begin{tikzpicture}[->,>=stealth']

\node[state] (ORIG) 
{\begin{tabular}{l} 
  spin-0: $0^+_\text{SM},0^+_\text{D5},0^-_\text{D5}$ \\[1mm]
  spin-1: $1^+_{W,Z}, 1^-_{W,Z}$ \\[1mm]
  spin-2: $2^+_\text{EW},2^+_\text{EW+q}$
 \end{tabular}};

\node[state,yshift=2cm,right of=ORIG,node distance=4.5cm,anchor=center] (PHIJJ1)
{\begin{tabular}{l}
  $0^+_\text{D5}$ 
 \end{tabular}};
 
\node[state,yshift=1.5cm,below of=PHIJJ1,node distance=3.5cm,anchor=center] (PHIJJ2)
{\begin{tabular}{l}
  $0^-_\text{D5}$ \\[1mm]
  $1^+_Z$ 
 \end{tabular}};

\node[state,yshift=1.5cm,below of=PHIJJ2,node distance=3.5cm,anchor=center] (PHIJJ3)
{\begin{tabular}{l}
  $1^-_{W,Z}$ \\
  $2^+_\text{EW}$ 
 \end{tabular}};

\node[state,yshift=1cm,below of=PHIJJ3,node distance=3.5cm,anchor=center] (PHIJJ4)
{\begin{tabular}{l}
  $0^+_\text{SM}$ \\[1mm]
  $1^+_W$ \\[1mm]
  $2^+_\text{EW+q}$
 \end{tabular}};

%----------------------------
\node[state,yshift=1.0cm,right of=PHIJJ3,node distance=3.5cm,anchor=center] (ETAJJ3)
{\begin{tabular}{l}
  $1^-_{W,Z}$ 
 \end{tabular}};
 
\node[state,yshift=2.2cm,below of=ETAJJ3,node distance=3.5cm,anchor=center] (ETAJJ4)
{\begin{tabular}{l}
  $2^+_\text{EW}$ 
 \end{tabular}};
 
\node[state,yshift=1.8cm,below of=ETAJJ4,node distance=3.5cm,anchor=center] (ETAJJ5)
{\begin{tabular}{l}
  $0^+_\text{SM}$ \\[1mm]
  $2^+_\text{EW+q}$
 \end{tabular}};
 
\node[state,yshift=1.5cm,below of=ETAJJ5,node distance=3.0cm,anchor=center] (ETAJJ6)
{\begin{tabular}{l}
  $1^+_W$ 
 \end{tabular}};

%----------------------------
\node[state,yshift=0.5cm,right of=ETAJJ5,node distance=3.5cm,anchor=center] (ETAJX1)
{\begin{tabular}{l}
  $0^+_\text{SM}$ 
 \end{tabular}};
 
\node[state,yshift=2.2cm,below of=ETAJX1,node distance=3.5cm,anchor=center] (ETAJX2)
{\begin{tabular}{l}
  $2^+_\text{EW+q}$
 \end{tabular}};

%----------------------------
\node[state,yshift=5cm,right of=ETAJX1,node distance=2.0cm,anchor=center] (ETAJD1)
{\begin{tabular}{l}
  $0^-_\text{D5}$ 
 \end{tabular}};

\node[state,yshift=2.5cm,below of=ETAJD1,node distance=3.5cm,anchor=center] (ETAJD2)
{\begin{tabular}{l}
  $1^+_Z$ 
 \end{tabular}};

\path (ORIG) edge[bend left=20]  node[anchor=south,above]{$\Delta \phi_{jj}^\text{max} \sim\pi$} (PHIJJ1);
\path (ORIG) edge[bend left=20] node[anchor=south,above]{$\Delta \phi_{jj}^\text{max} \sim \pi/2$} (PHIJJ2);
\path (ORIG) edge[bend right=20] node[anchor=south,above]{$\qquad \Delta \phi_{jj}^\text{max} \sim 0$} (PHIJJ3);
\path (ORIG) edge[bend right=30] node[anchor=south,above]{$\qqquad \Delta \phi_{jj}$ flat} (PHIJJ4);
\path (PHIJJ3) edge[bend left=20] node[anchor=south,below]{$\qquad \Delta \eta_{jj}^\text{max} \sim 3$} (ETAJJ3);
\path (PHIJJ3) edge[bend right=20] node[anchor=south,above]{$\quad \Delta \eta_{jj}^\text{max} \sim 5.5$} (ETAJJ4);
\path (PHIJJ4) edge[bend left=20] node[anchor=south,below]{$\quad \Delta \eta_{jj}^\text{max} \sim 5.5$} (ETAJJ5);
\path (PHIJJ4) edge[bend right=20] node[anchor=south,above]{$\quad \Delta \eta_{jj}^\text{max} \sim 4$} (ETAJJ6);
\path (ETAJJ5) edge[bend left=20] node[anchor=south,below]{$\Delta \eta_{jX}^\text{dip} \sim 0$} (ETAJX1);
\path (ETAJJ5) edge[bend right=20] node[anchor=south,above]{$\quad \Delta \eta_{jX}^\text{max} \sim 0$} (ETAJX2);
\path (PHIJJ2) edge[bend left=10] node[anchor=south,below]{$\Delta \eta_{jd}$ broad} (ETAJD1);
\path (PHIJJ2) edge[bend right=10] node[anchor=south,above]{$\Delta \eta_{jd}$ narrow/dip} (ETAJD2);

\end{tikzpicture}
\caption{Flow diagram for testing Higgs coupling structure in WBF
  production. All spin-2 models are defined including the
  unitarization cutoff Eq.(\ref{eq:ptmax}).}
\label{fig:flow}
\end{figure}
%-------------------------------------------------------

In Fig.~\ref{fig:flow} we show some of the most distinctive
measurements reflecting the couplings of a heavy resonance. The first
jet-jet variable, $\Delta \phi_{jj}$, is well suited to clearly
separate the spin-0 structures. As we will see in the following
section the $\Delta \eta_{jj}$ distribution might be even more
powerful than $\Delta \phi_{jj}$, so there should be no problem in
distinguishing the different spin-0 structures just based on the
tagging jets.

The problem is to tell apart some of the spin-1 and spin-2 hypotheses
from their closest spin-0 models in the jet-jet correlations. For
example, to tell apart the Standard Model Higgs boson $0^+_\text{SM}$
and the graviton $2^+_\text{EW+q}$ we have to resort to the jet-$X$
correlations $\Delta \eta_{jX}$. From Sec.~\ref{sec:kin_jd} we know
that Breit frame angles would also be well suited for this
distinction, most notably $\Phi_+$~\cite{kentarou}.

Similarly, telling apart $0^-_\text{D5}$ and $1^+_Z$ we requires
jet-decay correlations like $\Delta \eta_{jd}$. Alternatively, we
could use $\Phi_+$ shown in Fig.~\ref{fig:kin_jd}, the
Breit-frame angle $\theta^*$, or the Gottfried--Jackson angle
$\theta_\text{GJ}$. The two operators $1^-_{W,Z}$ appear
indistinguishable using the set of observables at hand.\bigskip

The qualitative bottom line of this section is that clearly the
jet-jet correlations are the most decisive, in particular to separate
the different scalar coupling structures~\cite{original}. For some
spin-1 and (unitarized) spin-2 models we need to add information from
the heavy resonance or its decays~\cite{kentarou}.

%%%%%%%%%%%%%%%%%%%%%%%%%%%%%%%%%%%%%%%%%%%%%%%%%%%%%%%%%%%%%%%%%%%%%%%%
\subsection{Comparison of observables}
\label{sec:stats}

Following Sec.~\ref{sec:strategy} we know what the candidate
observables for the distinction of model hypotheses are. However, for
example for the distinction of the $0^+_\text{SM}$ and
$0^+_\text{D5}$ coupling structures there are several such
variables. Only a complete shape analysis can tell us how promising
these distributions really are.\bigskip

The statistical power of a well-defined set of observables we compute
based on statistical errors, ignoring systematics as well a
theoretical uncertainties and errors. This approximation is justified
for relatively low luminosities --- for example asking the question
which of the distributions would first cross a given confidence level
in the comparison between two hypotheses.

To asses such a minimal luminosity we perform a binned log-likelihood
ratio hypothesis test~\cite{llhr}.  Our zero hypothesis or background
hypothesis is the Standard Model coupling structure
$0^+_\text{SM}$. This hypothesis also fixes the number of events in
the distribution. The exotic signal we would like to discover or
reject are the alternative spin-0 or spin-2 coupling structures.
Backgrounds we ignore in this computation, because they mostly affect
the exact value of the luminosity required for a 95\% exclusion, but
not the relative strength of the different observables. The binned
likelihood is defined in terms of a Poisson likelihood $L$,
\begin{alignat}{5}
L( \text{data}\, |\,  {\text{hypothesis $X$}}) 
& = \frac{N^{n}(X) e^{-N(X)}}{n!} 
\notag \\
\mathcal{Q}
& = -2\log \frac{L( \text{data}\, |\, {\text{hypothesis 1}} ) } 
  { L( \text{data}\,| \, {\text{SM}} ) } \; .
\label{eq:loglike2}
\end{alignat}
The expected number of signal events (per bin) given the luminosity
$\mathcal{L}$ is $N(X)=\sigma(X){\cal{L}}$, while $n$ is the observed
number of signal events. The logarithm is additive, so the
generalization of Eq.\eqref{eq:loglike2} to binned distributions with
(in our case) 20 bins per distribution is straightforward. The
Neyman--Pearson lemma states that this is the statistically `best'
discriminator.  The different hypotheses in Eq.\eqref{eq:loglike2} are
sampled with randomly generated pseudo-data, giving us probability
density functions
% ${{Q}}^{\text{hyp 1}}({\cal{Q}})$ and ${{Q}}^{\text{hyp 2}}
%({\cal{Q}})$. 
whose overlap define the 95\%~CL curves. As alluded to above, the
absolute value of the luminosity required for a 95\% CL exclusion of
alternative coupling structure should be taken with a grain of
salt. What we aim at is the relative strength of the different
observables, which might be hard to judge by eye.\bigskip

%-------------------------------------------------------
\begin{figure}[t]
\includegraphics[width=0.32\textwidth]{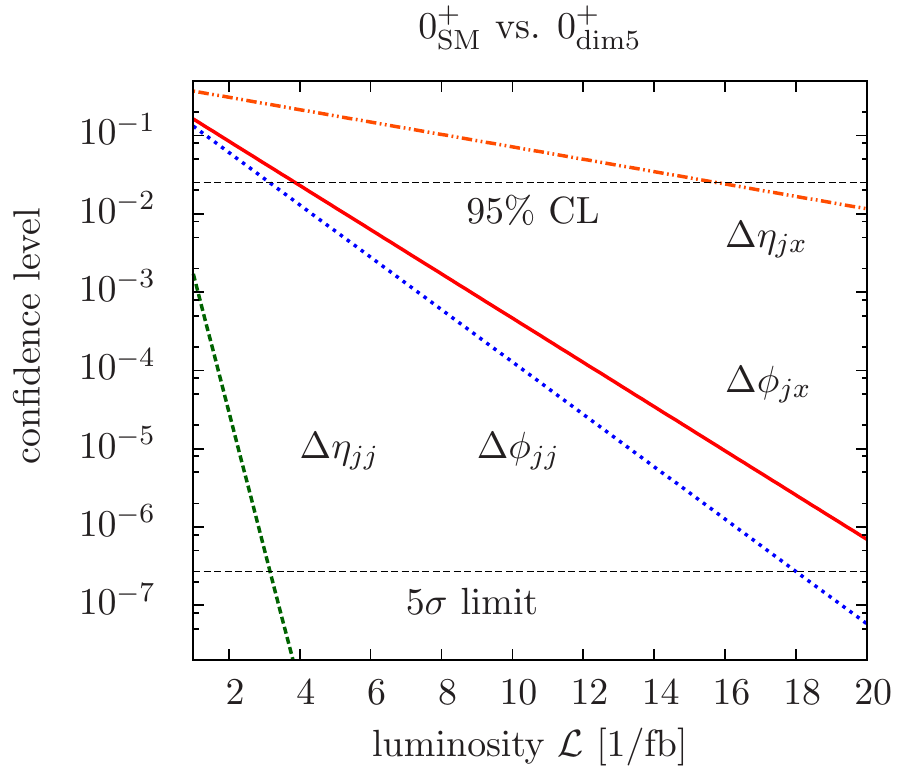}
\hfill
\includegraphics[width=0.32\textwidth]{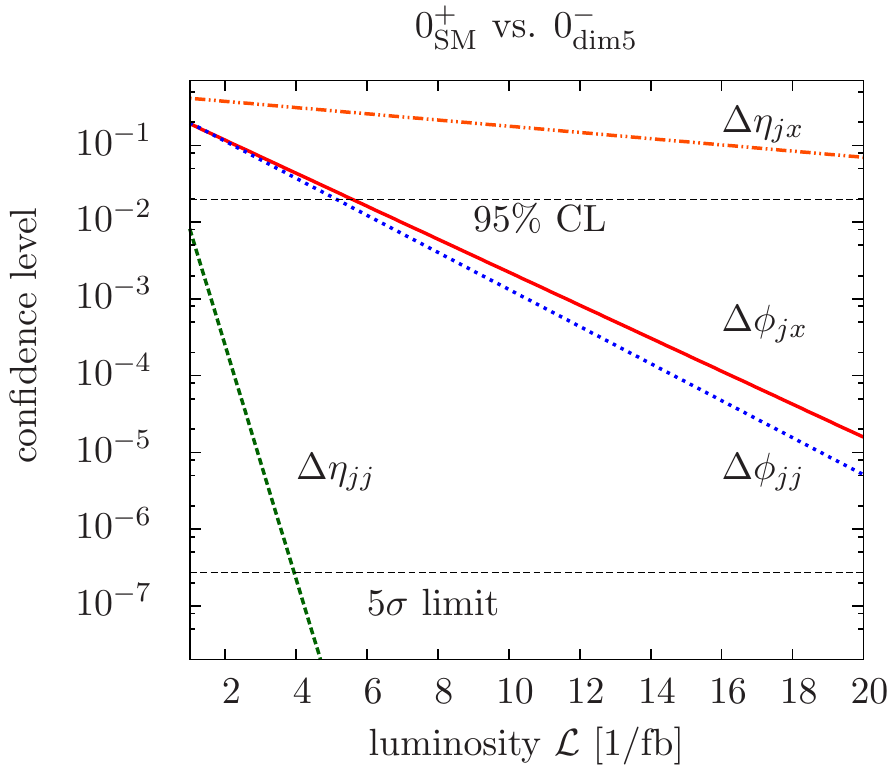}
\hfill
\includegraphics[width=0.32\textwidth]{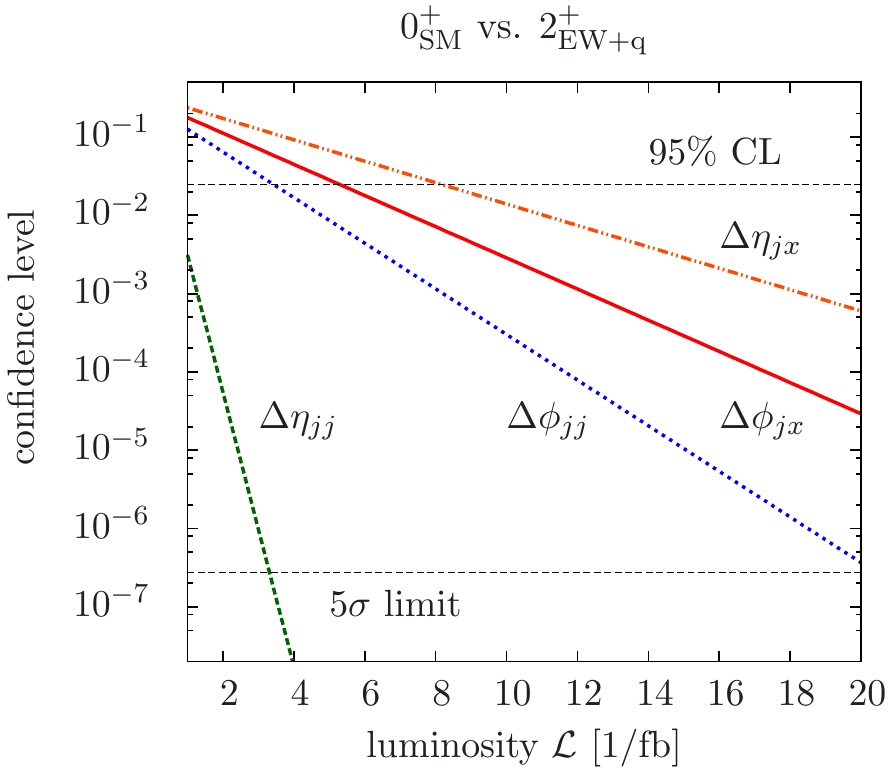}
\caption{Confidence level for the distinction of different spin-0
  coupling structures as a function of the integrated luminosity. We
  assume statistical uncertainties only for the WBF signal in the
  absence of backgrounds. The plotted confidence level refers to
  exclusion of the SM WBF component. The cross section normalization
  is 2.8 fb which corresponds to a good $\tau$ reconstruction in the
  purely leptonic channels (see \cite{spin_fatjets2}).}
\label{fig:likelihood}
\end{figure}
%-------------------------------------------------------

In Fig.~\ref{fig:likelihood} we compare the statistical power of
different observables in WBF kinematics. In the left panel we show how
well the different observables can distinguish the $0^+_\text{D5}$
operator from the Standard Model coupling structure. The two most
promising observables are the rapidity difference between the two
tagging jets and their azimuthal angle difference, \ie observables
which are independent of the Higgs decay. All that is required is that
the event sample has a good signal-to-background ratio, such that such
an analysis is not dominated by the systematics of the background
subtraction. The necessary clean-up of the event sample can be
achieved by central jet vetos~\cite{jetveto} and/or the matrix element
method~\cite{wbf_gamma}.  We see that the rapidity difference turns
out more powerful than the azimuthal angle.

In the second panel we show the same set of observables for a
discovery of the CP-odd $0^-_\text{D5}$ coupling structure, a
pseudo-scalar Higgs boson. Again, the azimuthal angle $\Delta
\phi_{jj}$ is somewhat less sensitive compared to the rapidity
difference. Finally, we study the most relevant observables to
identify a $2^+$ resonance without a constraint on the high-energy
tails of the jet momenta. As before, the rapidity difference of the
tagging jets is the most powerful. These results are a direct
consequence of the high discriminative power of the $\Delta \eta_{jj}$
distribution shown in Fig.~\ref{fig:kin_jj}, where all the alternative
couplings tested in this section peak around $\Delta \eta_{jj} \sim
4$, compared to the Standard Model $0^+_\text{SM}$ case with $\Delta
\eta_{jj} \sim 5.5$.  For spin-2 identification adding information
from the heavy resonance or its decays implies a significant
improvement~\cite{kentarou}. Our results for the best-suited
distribution reflects the setup of the likelihood test: the required
luminosity is dominated by those bins which show a large ratio of
events between the two different hypotheses.\bigskip

Going beyond our idealized setup, these jet-jet observables differ
when it comes to detector and QCD effects, \eg pile-up suppression is
a function of rapidity. Furthermore it is well-known that $\Delta
\phi_{jj}$ acts as good measurement quantity to extract mixed CP
properties~\cite{klamke}. Such an angular correlation is
insufficiently reflected in the rapidity difference. This way $\Delta
\phi_{jj}$ remains a well-suited observable to test more complex CP
properties in WBF Higgs production, while a wide operator spectroscopy
preferably relies on the leading observable $\Delta \eta_{jj}$.

%%%%%%%%%%%%%%%%%%%%%%%%%%%%%%%%%%%%%%%%%%%%%%%%%%%%%%%%%%%%%%%%%%%%%%%%
\section{ZH production}

The study of WBF kinematics with the different decay channels gives us
a wide array of tests of the Higgs coupling structures. However,
fermions still only feature in one channel, namely $X \to \tau^+
\tau^-$ decays. Once the LHC runs at 14~TeV, another channel will
become available to test the `Higgs' couplings to fermions,
\begin{alignat}{5}
 pp\to (Z \to \ell^+\ell^-) (X \to b\bar{b} ) \; .
\end{alignat}
Independent of the use of modern Higgs finding methods~\cite{bdrs}
this channel will only be observable for boosted $Z$ and $X$ decays.
The kinematics of this process can be directly mapped onto the decay
$X\to ZZ$, so the Cabibbo--Maksymowicz--Dell'Aquila--Nelson
angles~\cite{nelson} are a natural choice for spin-sensitive
observables.  In complete analogy to Eq.\eqref{eq:angles_zz} we define
\begin{alignat}{5}
p_{Z^*} = p_Z + p_X \; ,
\qqquad 
p_X = p_b + p_{\bar{b}} \; ,
\qqquad 
p_Z = p_{\ell^-} + p_{\ell^+} \; .
\end{alignat}
as well as the corresponding unit three-momenta $\hat{p}_i$ in the
$Z^*$ and the $Z,X$ rest frames. Similar to the original angles
defined in Sec.~\ref{sec:angles} the $Z^*$ in the $s$-channel will be
off-shell, which does not prevent us from using it as a reference
frame. The basis of angles is then defined as
\begin{alignat}{5}
&\cos \theta_b =   
  \hat{p}_{b_1} \cdot\hat{p}_Z \Big|_X 
&\qquad 
&\cos \theta_\ell =   
  \hat{p}_{\ell^-} \cdot\hat{p}_X \Big|_Z 
\qqqquad 
\cos \theta^* = 
  \hat{p}_X \cdot \hat{p}_\text{beam} \Big|_{Z^*}  \notag \\
&\cos \phi_b = 
  (\hat{p}_\text{beam} \times \hat{p}_Z) \cdot (\hat{p}_Z \times \hat{p}_{b_1}) \Big|_X 
  &\qquad
&\cos \Delta \phi =
(\hat{p}_b \times \hat{p}_{\bar{b}}) \cdot (\hat{p}_{\ell^-} \times \hat{p}_{\ell^+}) \Big|_{Z^*} \; .
\label{eq:angles_zh}
\end{alignat}
The signal distributions for the different $X$ hypotheses are
generated with the rough acceptance cuts
\begin{alignat}{5}
p_{T,b} \ge 20~\gev \qqquad
p_{T,\ell} \ge 10~\gev \qqquad
\Delta R \ge 0.4 \qqquad
|\eta_b|\le 5 \qqquad
|\eta_{\ell}| \le 2.5 \;,
\label{eq:cuts_zh}
\end{alignat}
where we assume the same value for all $\Delta R$ value for $b$-jets
and leptons. In addition, we require
\begin{alignat}{5}
p_{T,(b\bar{b})} > 100~\gev \; ,
\end{alignat}
to be consistent with the proposed LHC analyses.\bigskip

%-------------------------------------------------------
\begin{figure}[t]
 \includegraphics[width=0.24\textwidth]{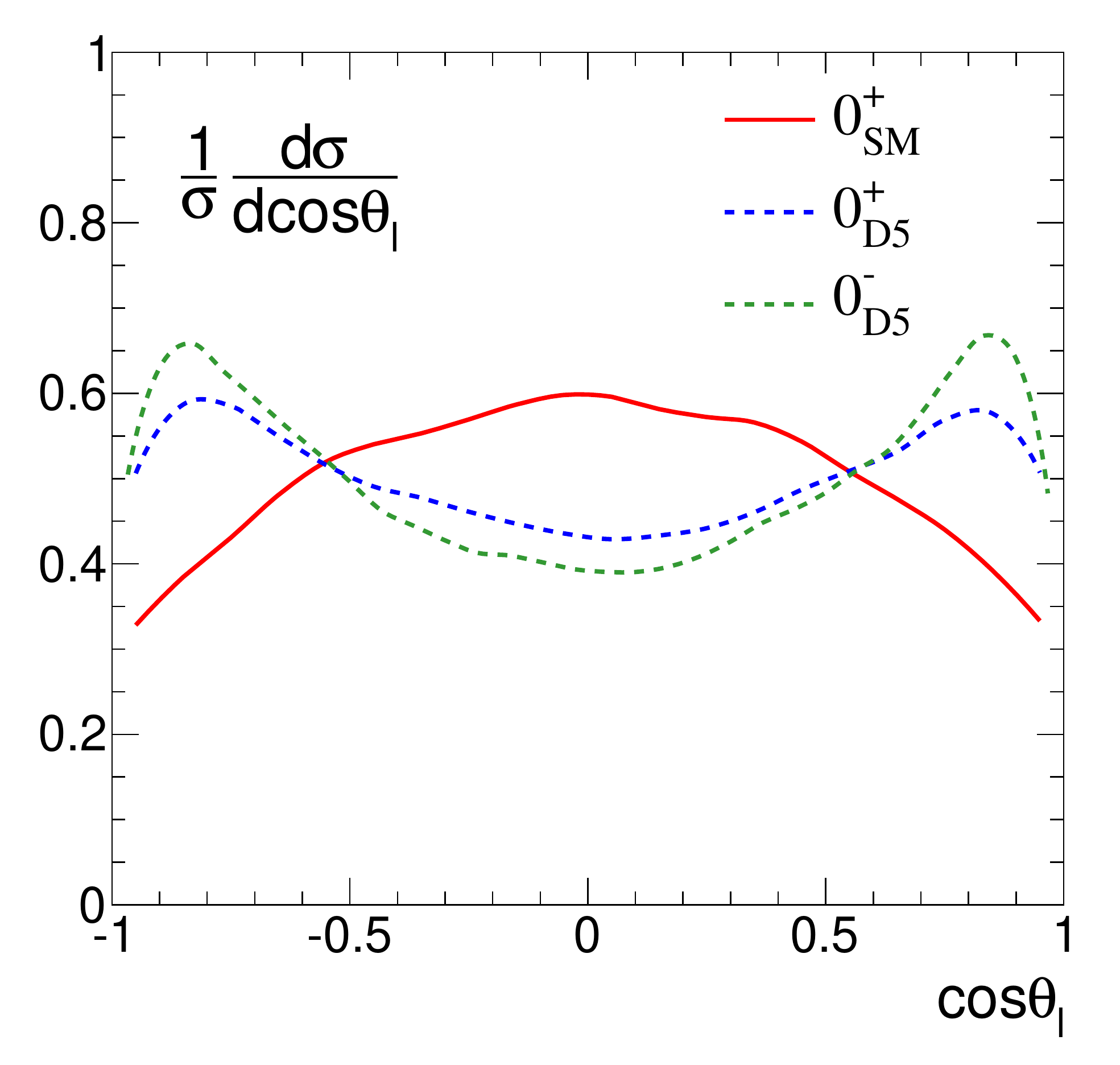}
 \hfil
 \includegraphics[width=0.24\textwidth]{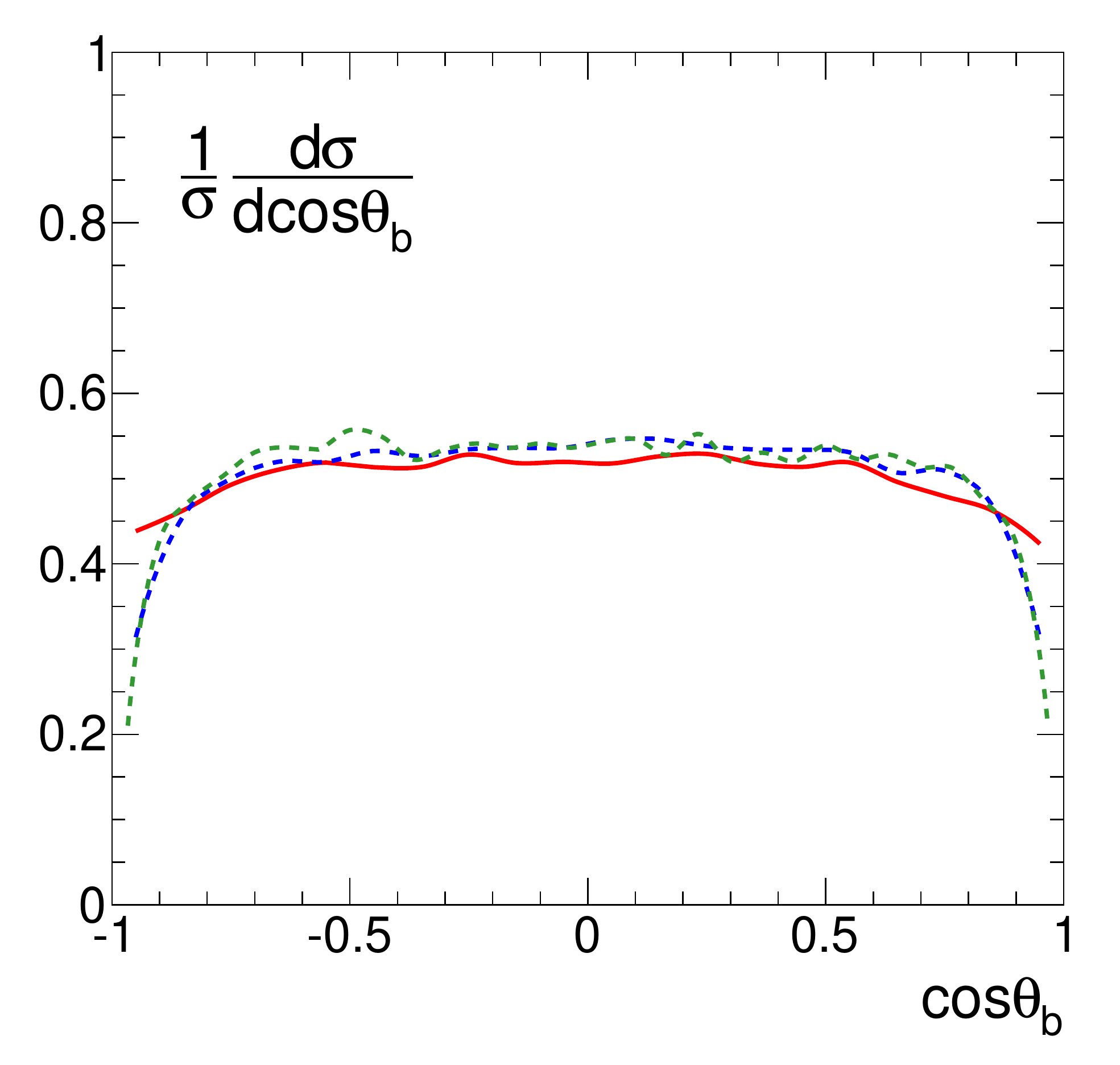}
 \hfil
 \includegraphics[width=0.24\textwidth]{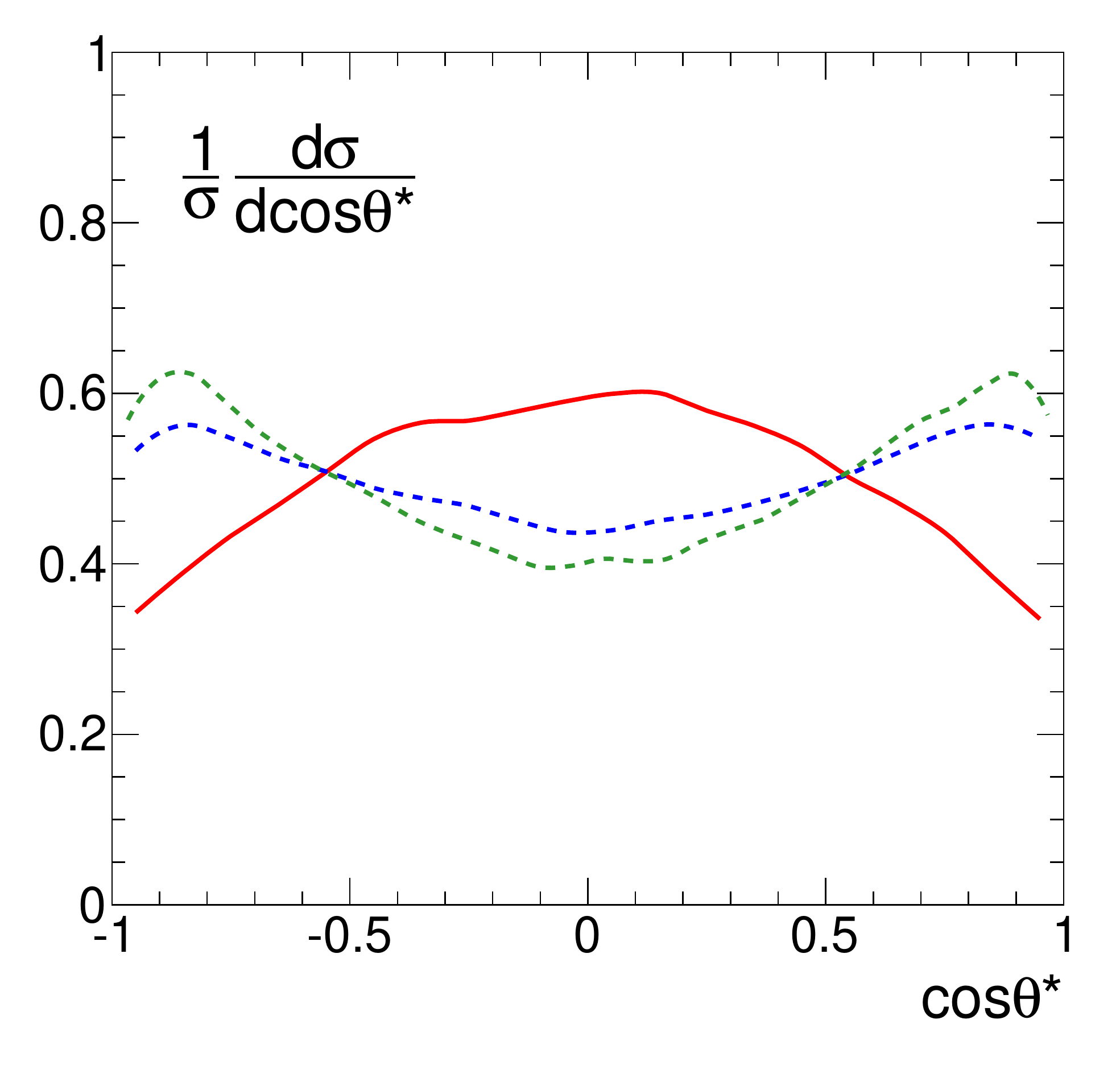}
 \hfil
 \includegraphics[width=0.24\textwidth]{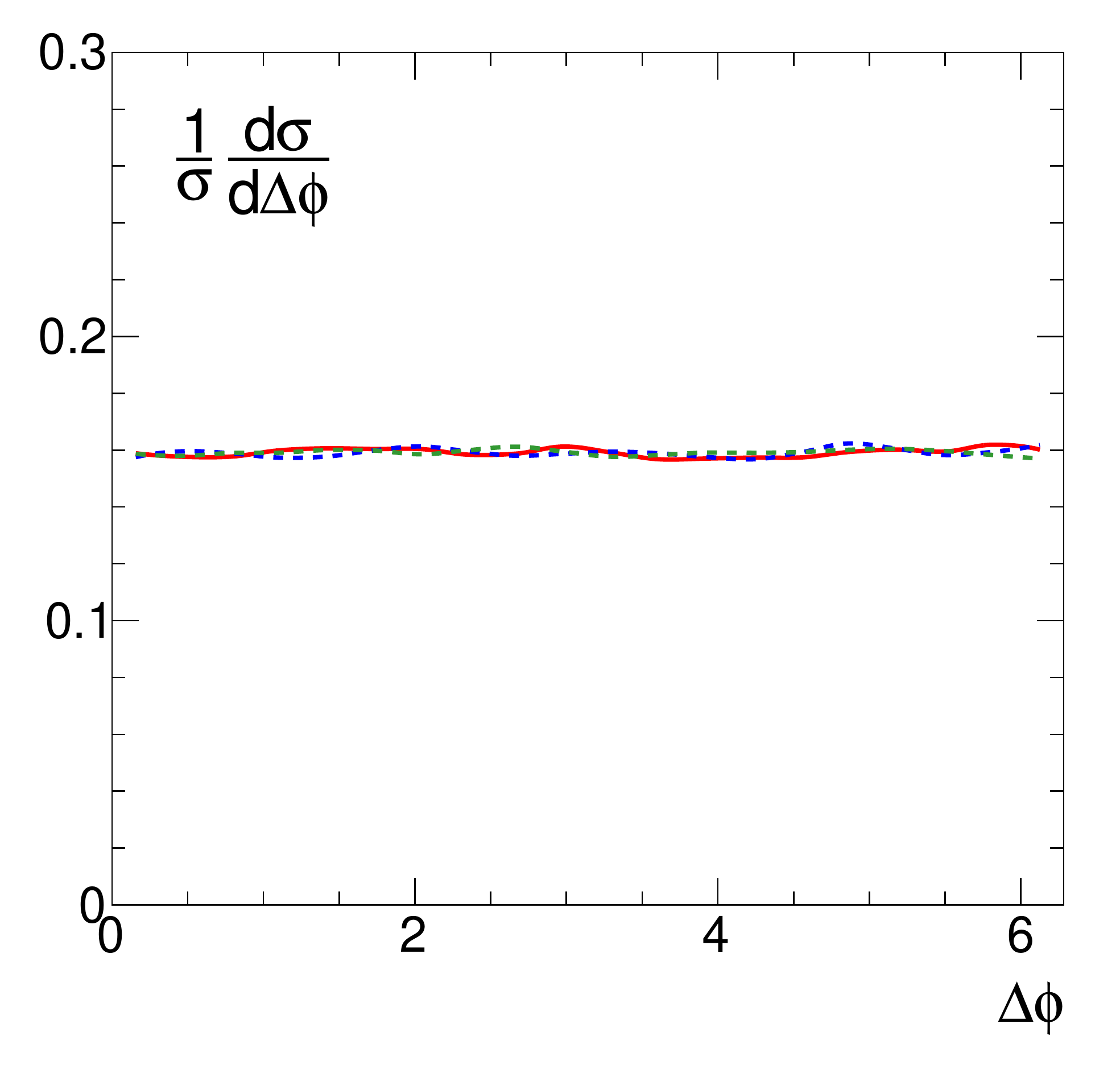} \\
 \includegraphics[width=0.24\textwidth]{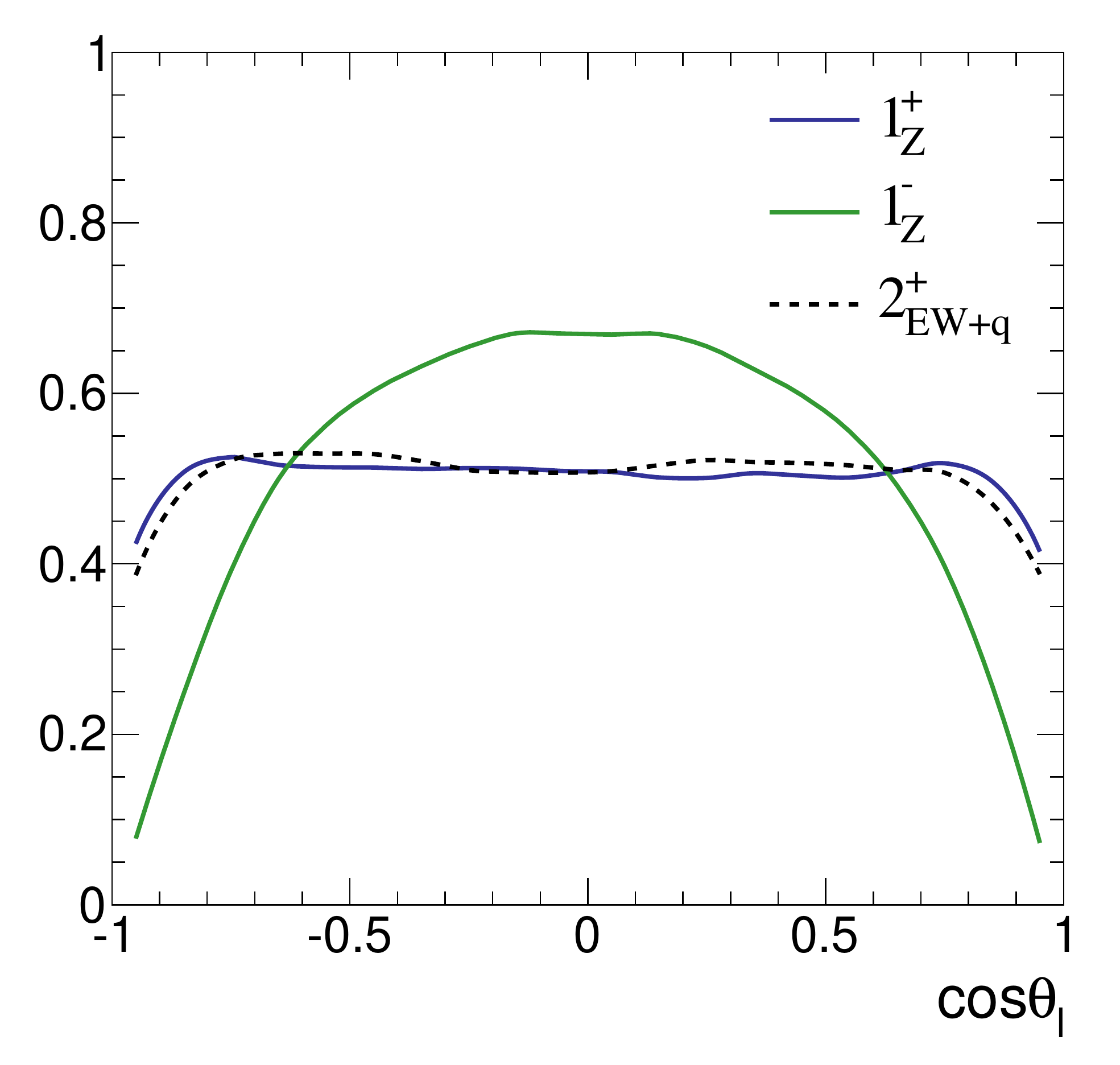}
 \hfil
 \includegraphics[width=0.24\textwidth]{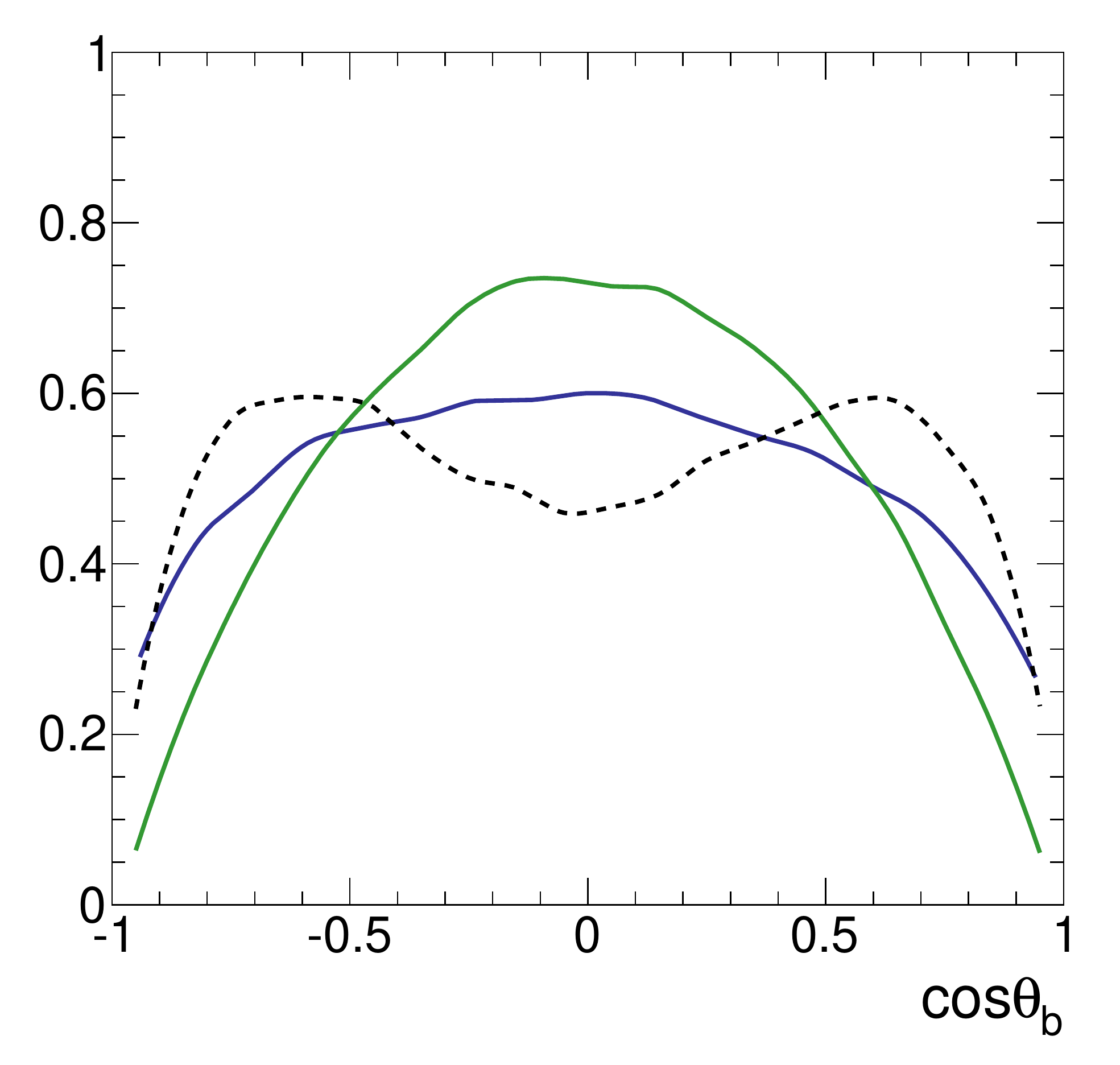}
 \hfil
 \includegraphics[width=0.24\textwidth]{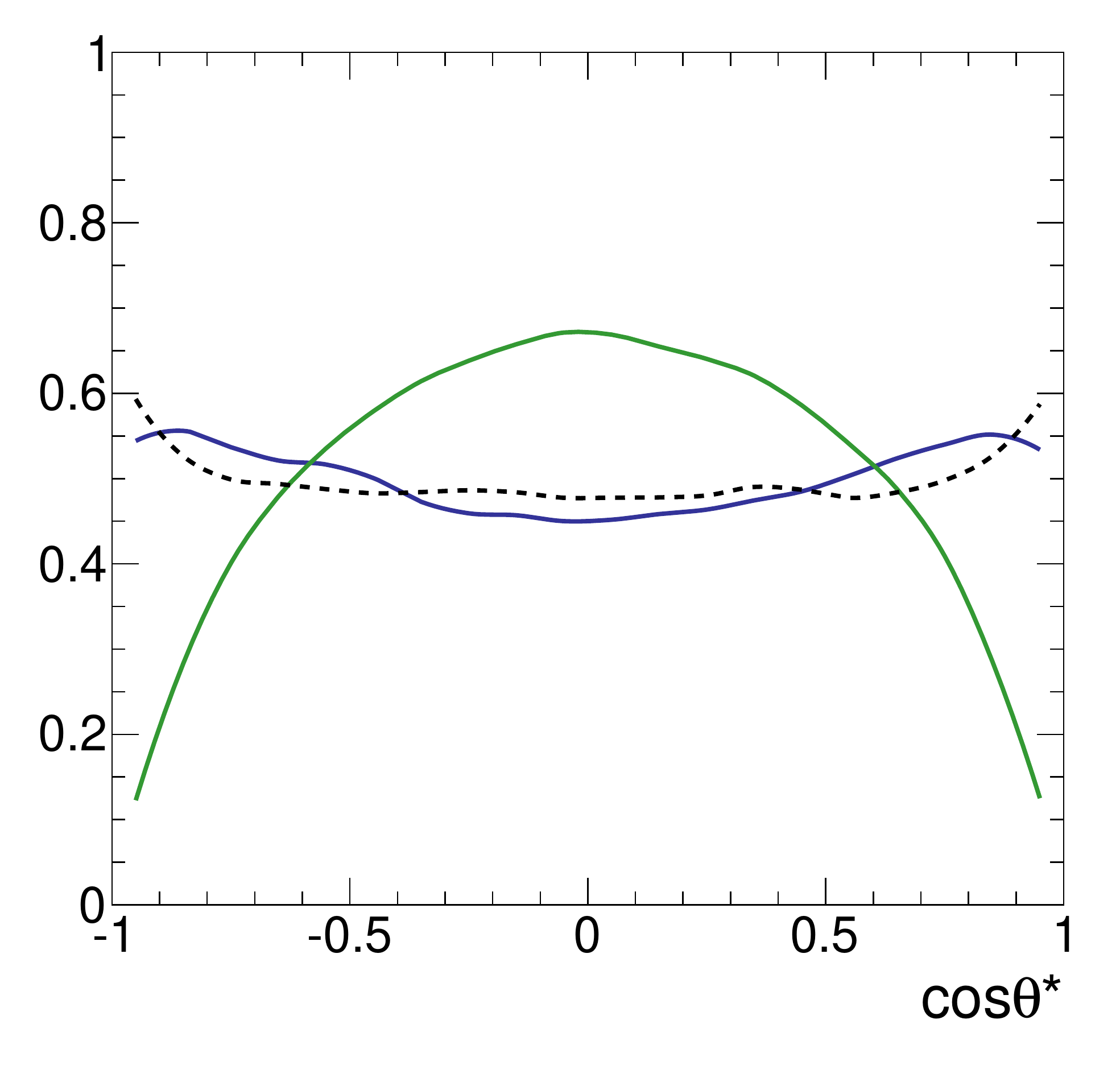}
 \hfil
 \includegraphics[width=0.24\textwidth]{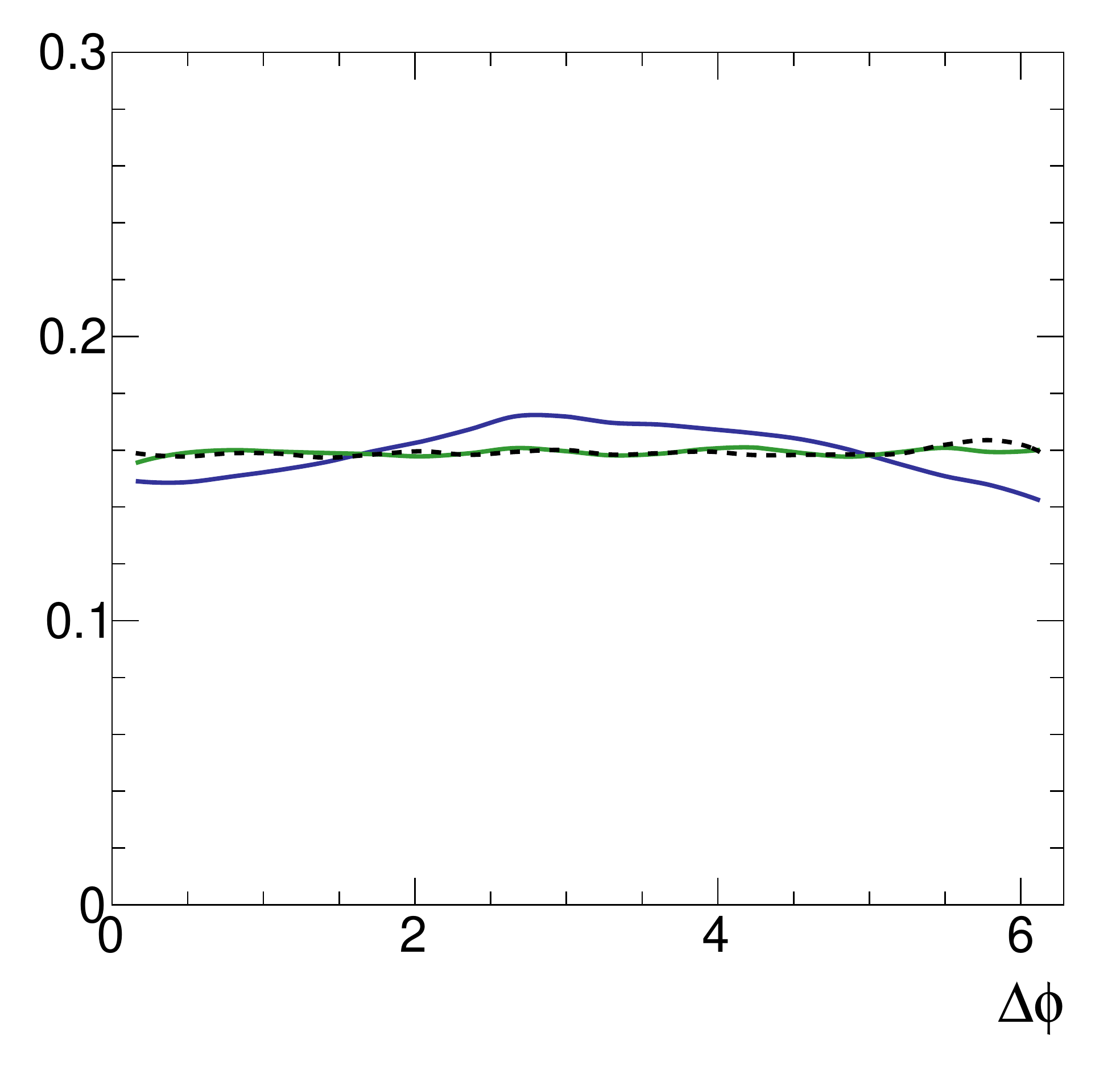}
\caption{Angular correlations for $(Z \to \ell^+\ell^-) (X \to b\bar{b}
  )$ production. We show $0_\text{SM}^+$ (red), $0_\text{D5}^+$
  (blue), $0_\text{D5}^-$ (green), $1_Z^+$ (blue) in the upper panels
  and $1_Z^-$ (green), and $2^+_\text{EW+q}$ (black dashed) in the
  lower panels. All operators are defined in Tab.~\ref{tab:model}.}
   \label{fig:angles_zh}
\end{figure}
%-------------------------------------------------------

Because this analysis requires reconstructed $Z \to \ell^+\ell^-$ and
the $H \to b\bar{b}$ decays we limit our model hypotheses from
Tab.~\ref{tab:model} to those which predict a $XZZ$ coupling as well
as a $Xb\bar{b}$ coupling.  For this sub-set of spin-0, spin-1, and
spin-2 models we should the most discriminating distributions in
Fig.~\ref{fig:angles_zh}. The three scalar operators can be identified
by a flat $\cos \theta_b$ distribution. The SM and the two
higher-dimensional scalar couplings then show a distinctively
different behavior in $\cos \theta_\ell$, which carries the same physical 
information as the $\cos \theta_e$ distribution presented in
 Fig.\ref{fig:costheta}  for the $X\to ZZ$ decay.  The challenge in the scalar
sector will be the measurement of the CP property of the
$0_\text{D5}^\pm$ operators. The problem with these two states is that
they are both scalars, washing out any angular correlation between the
$X \to b\bar{b}$ and the $Z \to \ell^+\ell^-$ decays. If their
dimension-5 origin gives them a similar energy behavior we would need
a dedicated CP asymmetry to distinguish them.

Comparing the different spin-1 and spin-2 hypotheses is made easy by
the distinctive forward-backward peaks in the spin-2 prediction for
$\cos \theta_b$. For $\theta_\ell$ and $\theta^*$ the $1^-_Z$
distribution has a significantly larger preference for $\cos \theta =
0$. While the boosted kinematics might well start to wash out the
angular correlations of the $Z,X$ decay products, this simple set of
predictions indicates that the determination of Higgs coupling
structures in associated $ZH$ production is promising.

%%%%%%%%%%%%%%%%%%%%%%%%%%%%%%%%%%%%%%%%%%%%%%%%%%%%%%%%%%%%%%%%%%%%%%%%
\section{Outlook}

After the discovery of a Higgs-like resonance by ATLAS and CMS the
main focus of Higgs analyses at the LHC will be the determination of
the Higgs Lagrangian. This includes the structure of the operators
(linked to the spin and CP quantum numbers of the `Higgs' boson) as
well as an independent measurement of the coupling strength. Similar
to the hugely successful electroweak precision program centered around
LEP this should give us hints about the embedding of the Higgs
mechanism in an ultraviolet completion of the Standard Model.

We present a comprehensive study for the determination of the Higgs
coupling structure in weak boson fusion.  Starting from the
Cabibbo--Maksymowicz--Dell'Aquila--Nelson angles in $X \to ZZ$ decays
we show how the WBF kinematics can be described either using Breit
frame angles or rapidity and azimuthal angles.\bigskip

The most distinctive observables in the WBF topology are jet-jet
correlations. It is well known that the azimuthal angle $\Delta
\phi_{jj}$ can distinguish between different spin-0 couplings without
ever requiring a reconstructed resonance
momentum~\cite{original,klamke}. Based on a signal-only and parton
level statistical analysis we expect the rapidity difference $\Delta
\eta_{jj}$ to be at least as effective in distinguishing different
spin-0 and spin-1 hypotheses. The advantage of jet-jet correlations is
that we do not have to model `Higgs' decays based on different models,
where the production in weak boson fusion and a decay to photons or
fermions might probe very different underlying assumptions on the
production and decay sides.

Generalizing our analysis to spin-1 and spin-2 hypotheses requires
additional information from the heavy resonance
decays~\cite{kentarou}. Again, the more general approach is to
reconstruct the heavy resonance momentum without any assumption on the
decay vertex and test angular jet-$X$ correlations. We find that they
indeed allow for a distinction of spin-0 and spin-1 or spin-2
models. Adding decays of the heavy resonances and computing jet-decay
correlations adds only little in terms of the hadron collider
observables $\Delta \eta$ and $\Delta \phi$.

Nevertheless, it might be useful to add dedicated angular correlations
defined differently. This includes an analytically continuation of the
$X \to ZZ$ angles, or some Breit frame angles like $\theta^*$ or
$\Phi_+$.\bigskip

For associated $ZH$ production with fully reconstructed decays $Z \to
\ell^+ \ell^-$ and $X \to b\bar{b}$ we can immediately employ the $X
\to ZZ$ angles. It should be possible to confirm the $0^+_\text{SM}$
coupling structure also in this channel without any problems.

\acknowledgments

We would like to thank Mario Campanelli for suggesting the analysis of
angular correlations in the $ZH$ channel.  CE acknowledges funding by
the Durham International Junior Research Fellowship scheme and thanks
Michael Spannowsky for helpful discussions. KM has been in part supported by the Concerted
 Research action ``Supersymmetric Models and their Signatures at the Large Hadron
 Collider'' and the Research Council of the Vrije Universiteit Brussel, and 
in part by the Belgian Federal Science Policy Office through the Interuniversity 
Attraction Pole P7/37.

%%%%%%%%%%%%%%%%%%%%%%%%%%%%%%%%%%%%%%%%%%%%%%%%%%%%%%%%%%%%%%%%%%%%%%%%
\newpage
\appendix
\section*{Complete set of observables}

In Sections~\ref{sec:kin_jj}, \ref{sec:kin_jx}, and \ref{sec:kin_jd}
we show a selected set of observables which should allow us to
determine the coupling structure of a heavy Higgs-like resonance in
weak boson fusion. We largely rely on the hadron collider observables
$\Delta \eta$ and $\Delta \phi$, even though in Sec.\ref{sec:angles}
we introduce Breit frame observables as well as dedicated angles like
the Gottfried-Jackson angle. In this appendix we give a complete set
of distributions for all model hypotheses listed in
Tab~\ref{tab:model}.\bigskip

%-------------------------------------------------------
\begin{figure}[!b]
spin-0 \hspace*{3.cm} spin-1 \hspace*{3.cm} spin-2 \hspace*{3.cm} spin-2($p_T^\text{max}$) \\
 \includegraphics[width=0.24\textwidth]{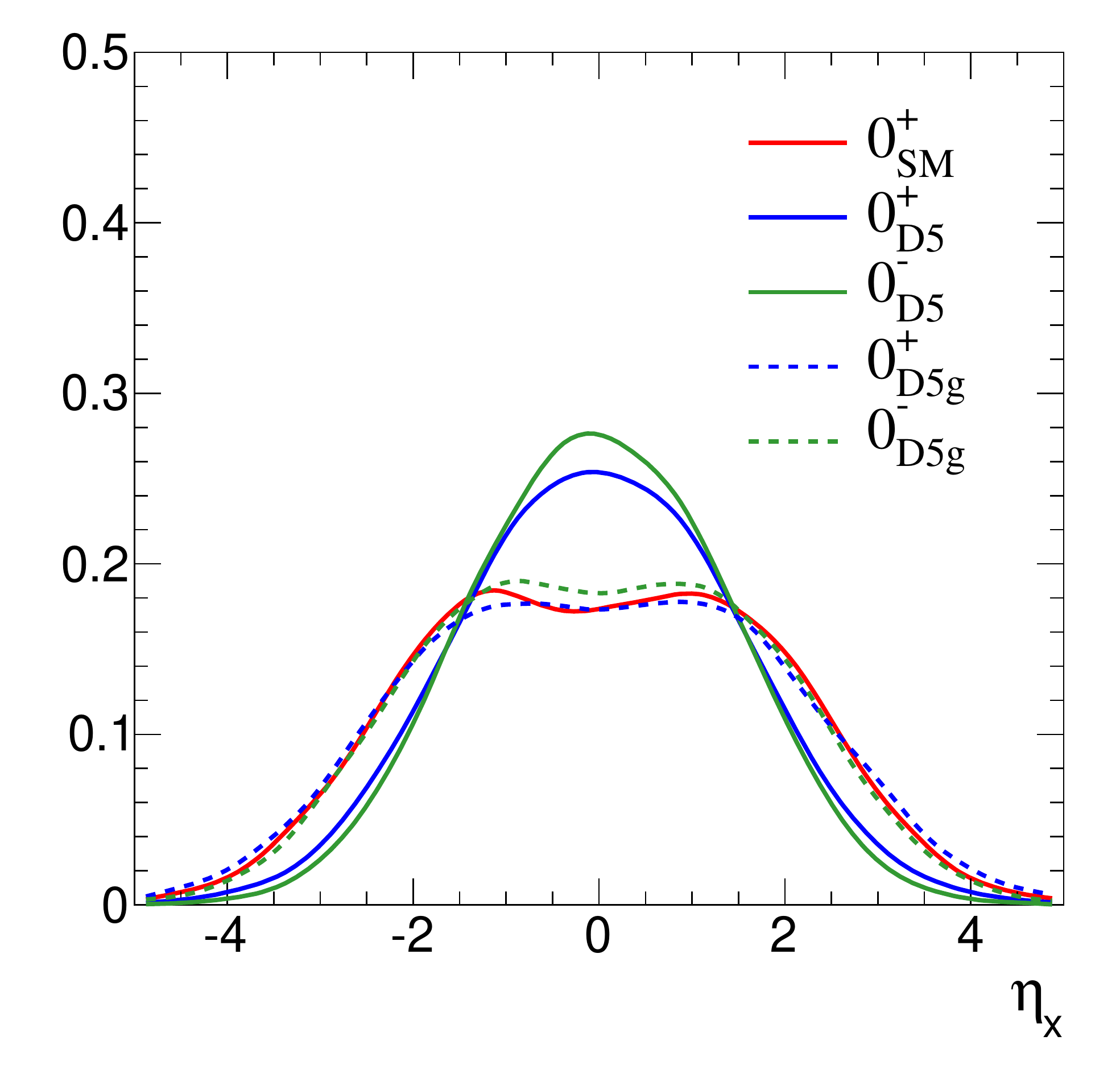}
 \includegraphics[width=0.24\textwidth]{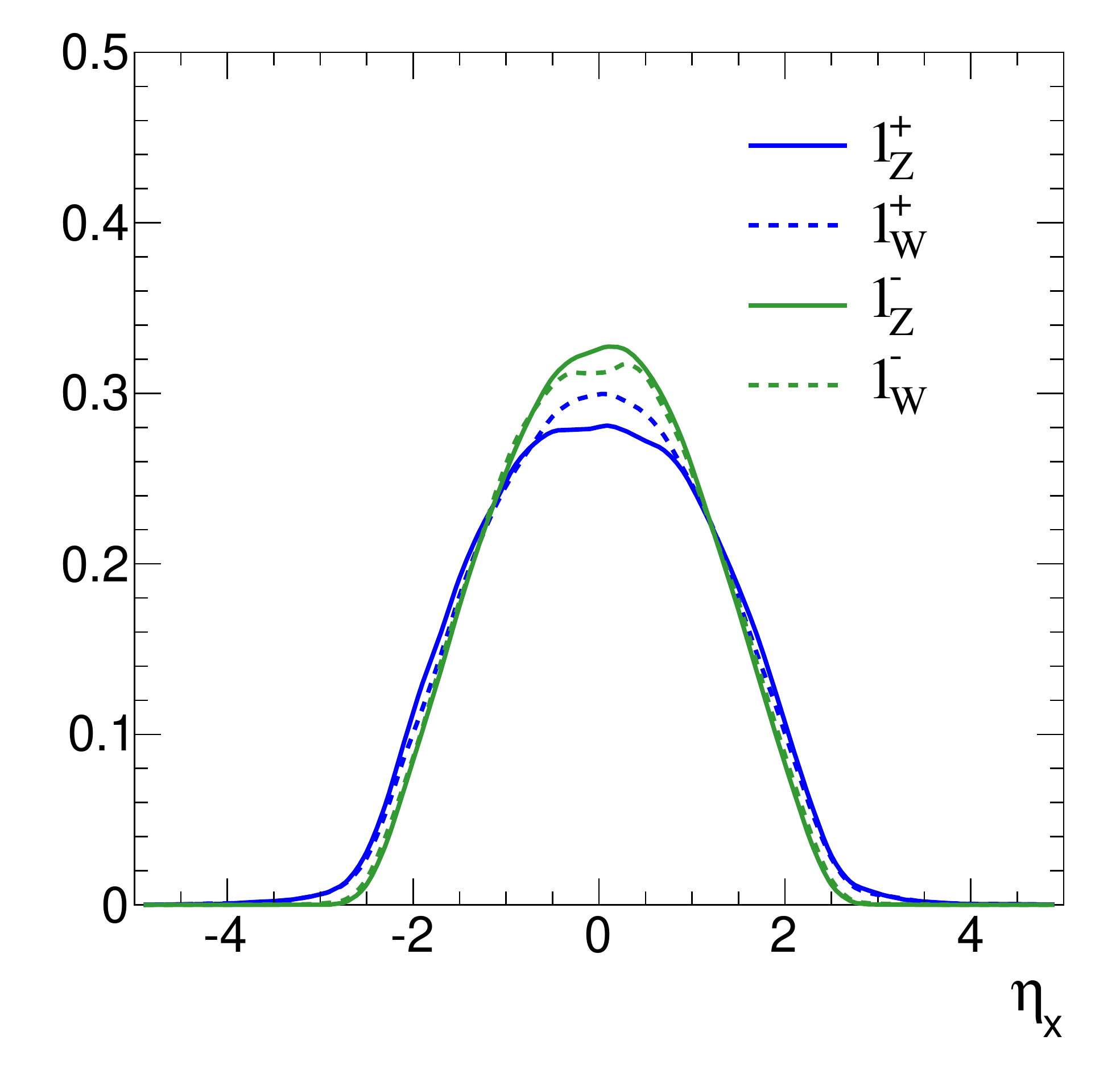}
 \includegraphics[width=0.24\textwidth]{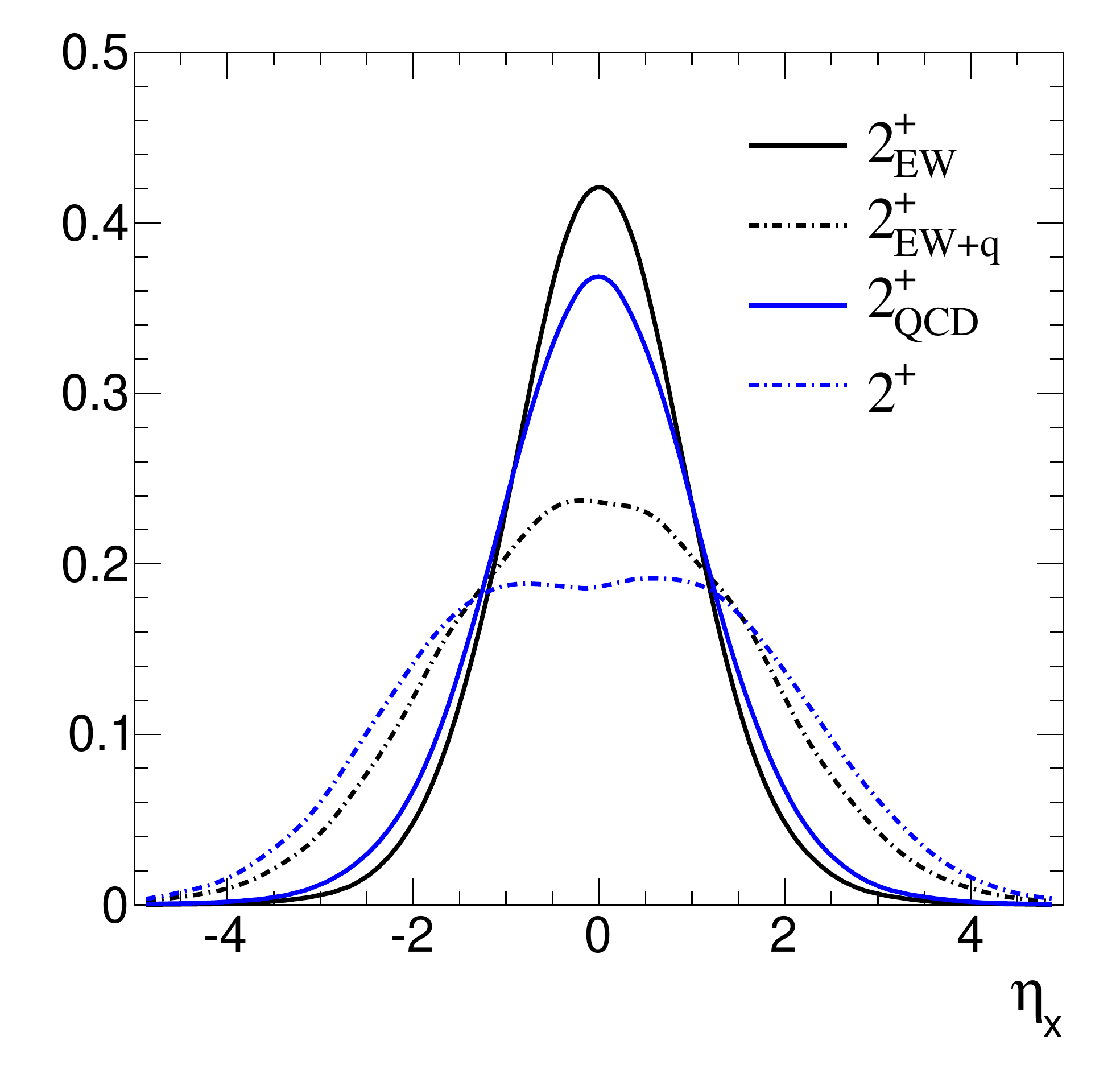}
 \includegraphics[width=0.24\textwidth]{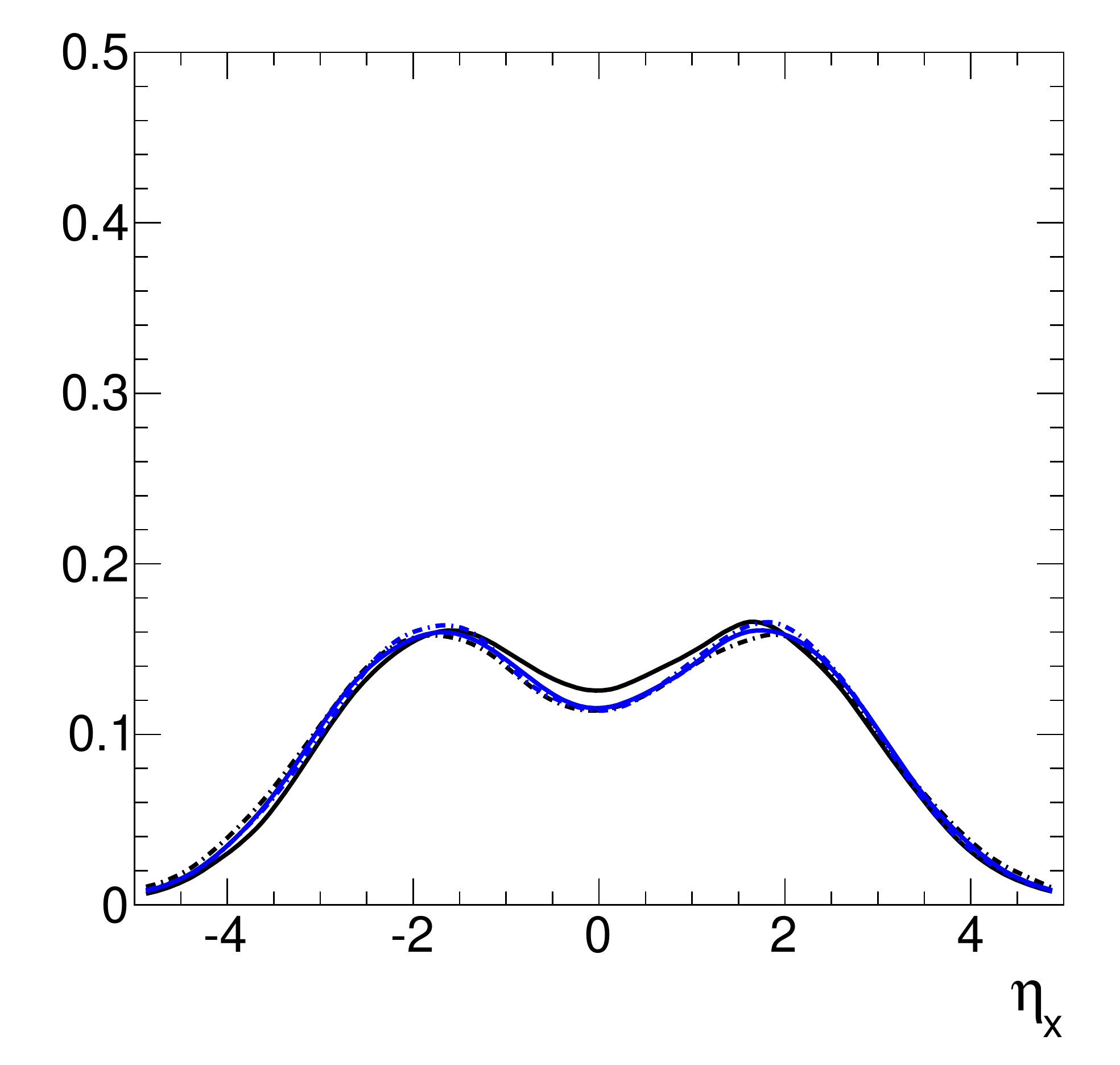}\\
 \includegraphics[width=0.24\textwidth]{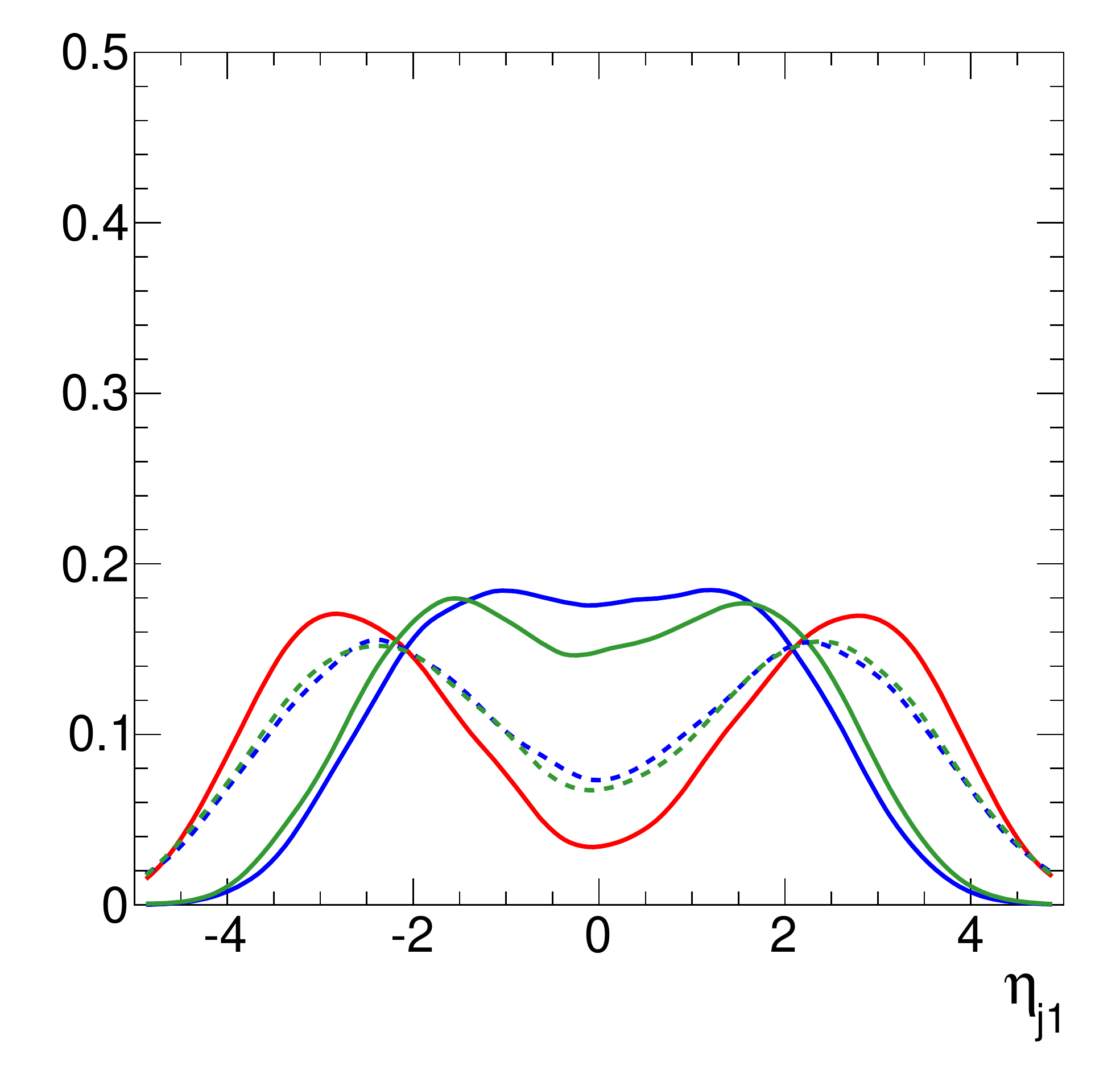}
 \includegraphics[width=0.24\textwidth]{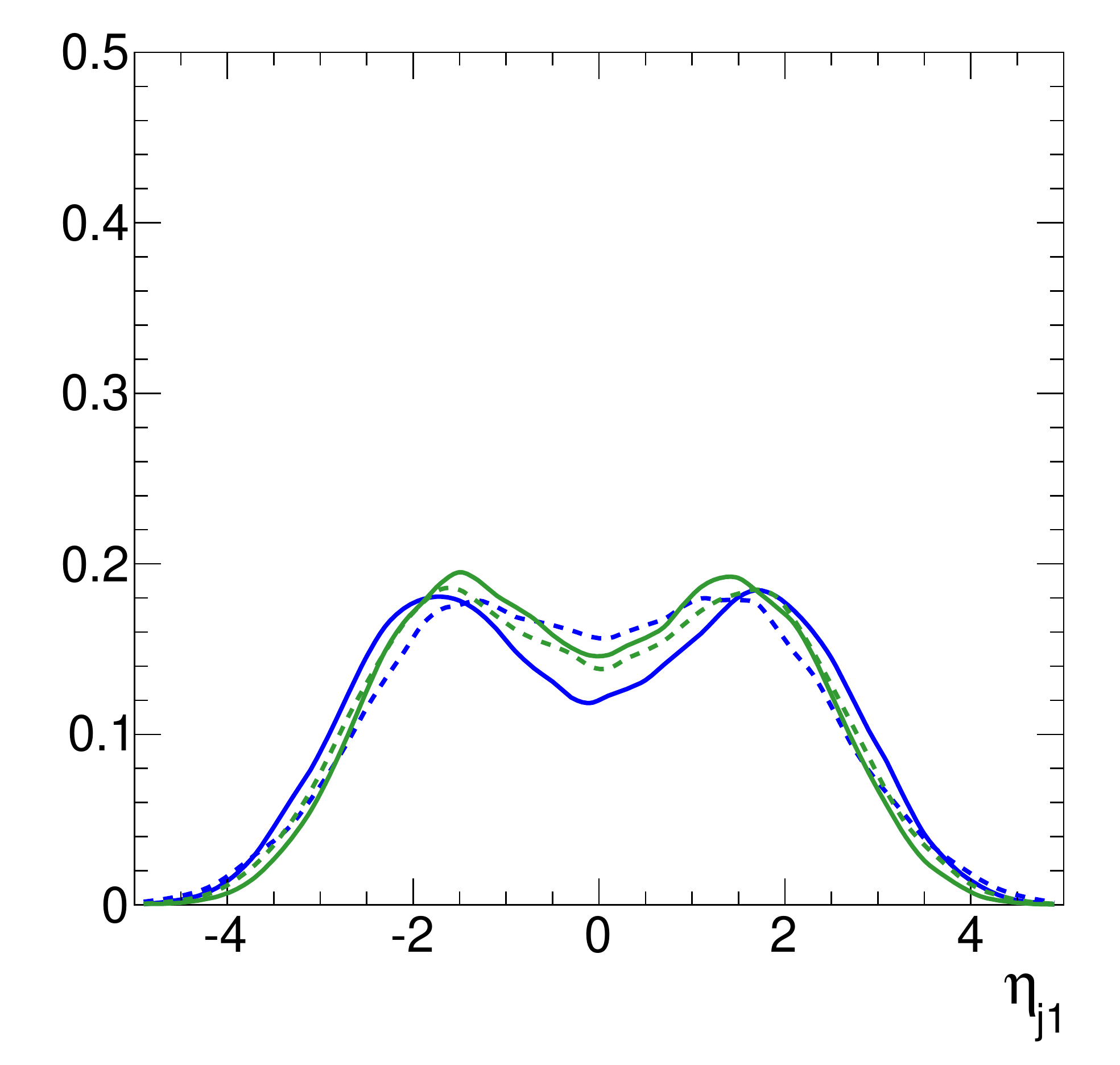}
 \includegraphics[width=0.24\textwidth]{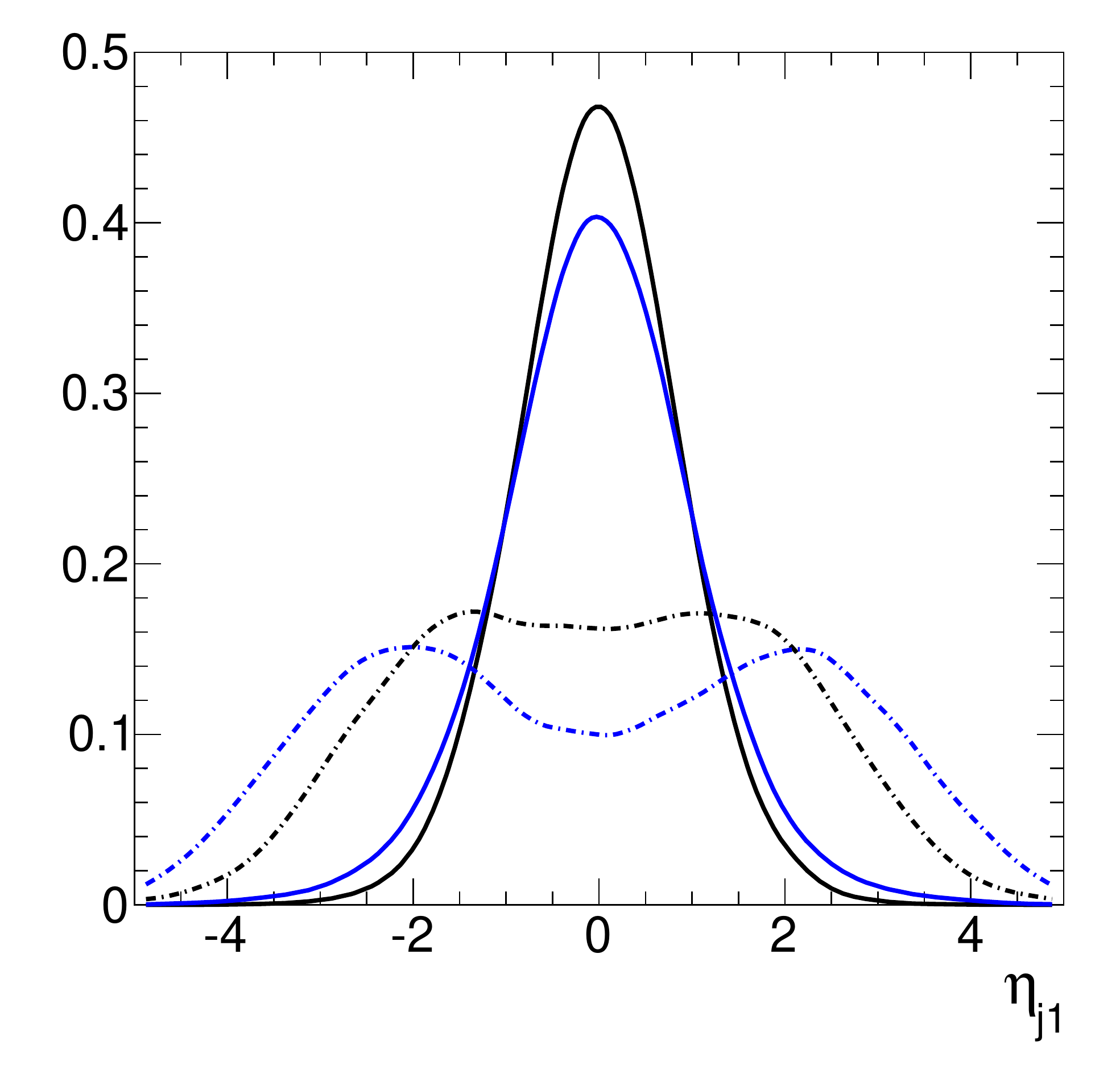}
 \includegraphics[width=0.24\textwidth]{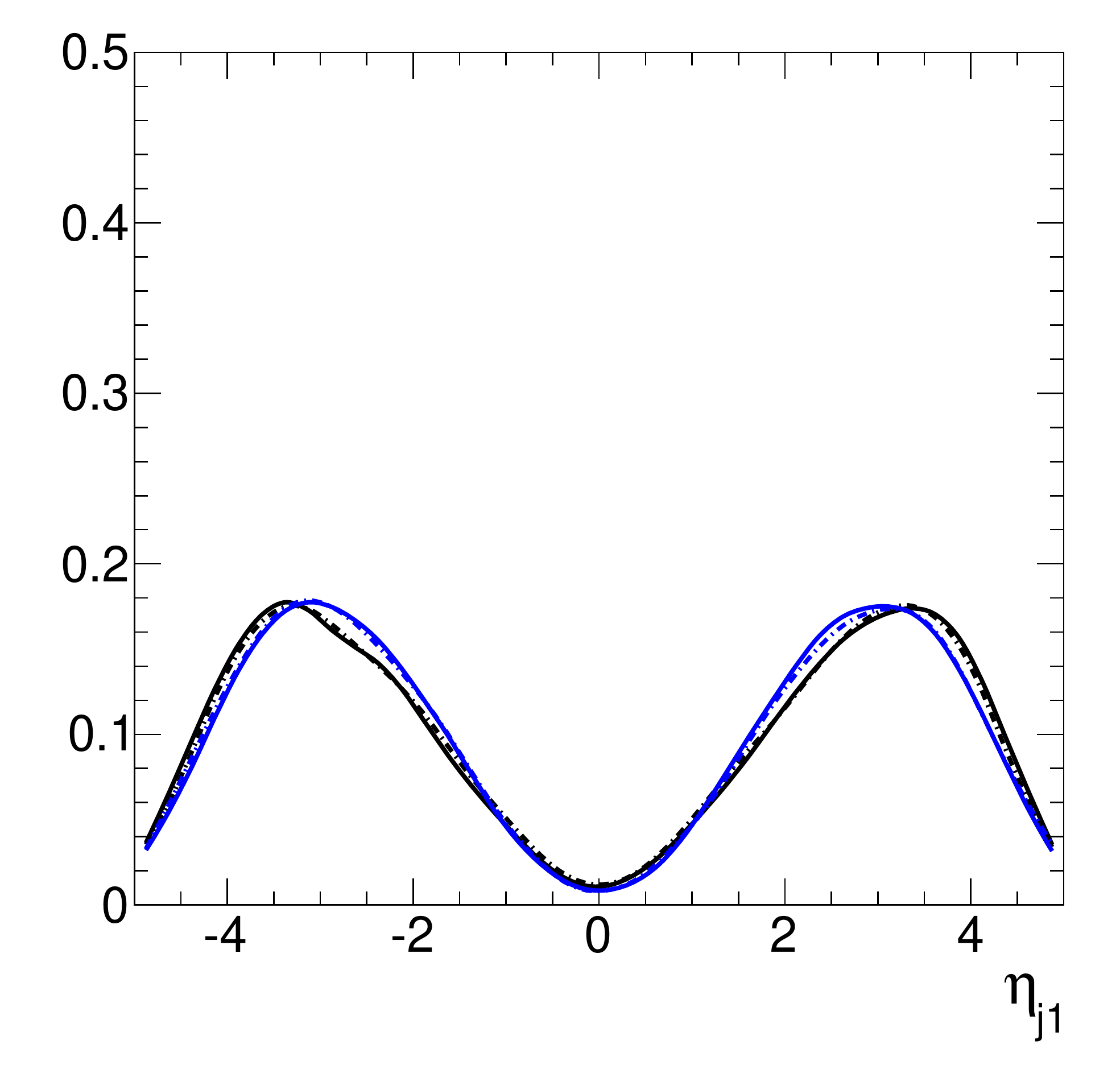}\\
 \includegraphics[width=0.24\textwidth]{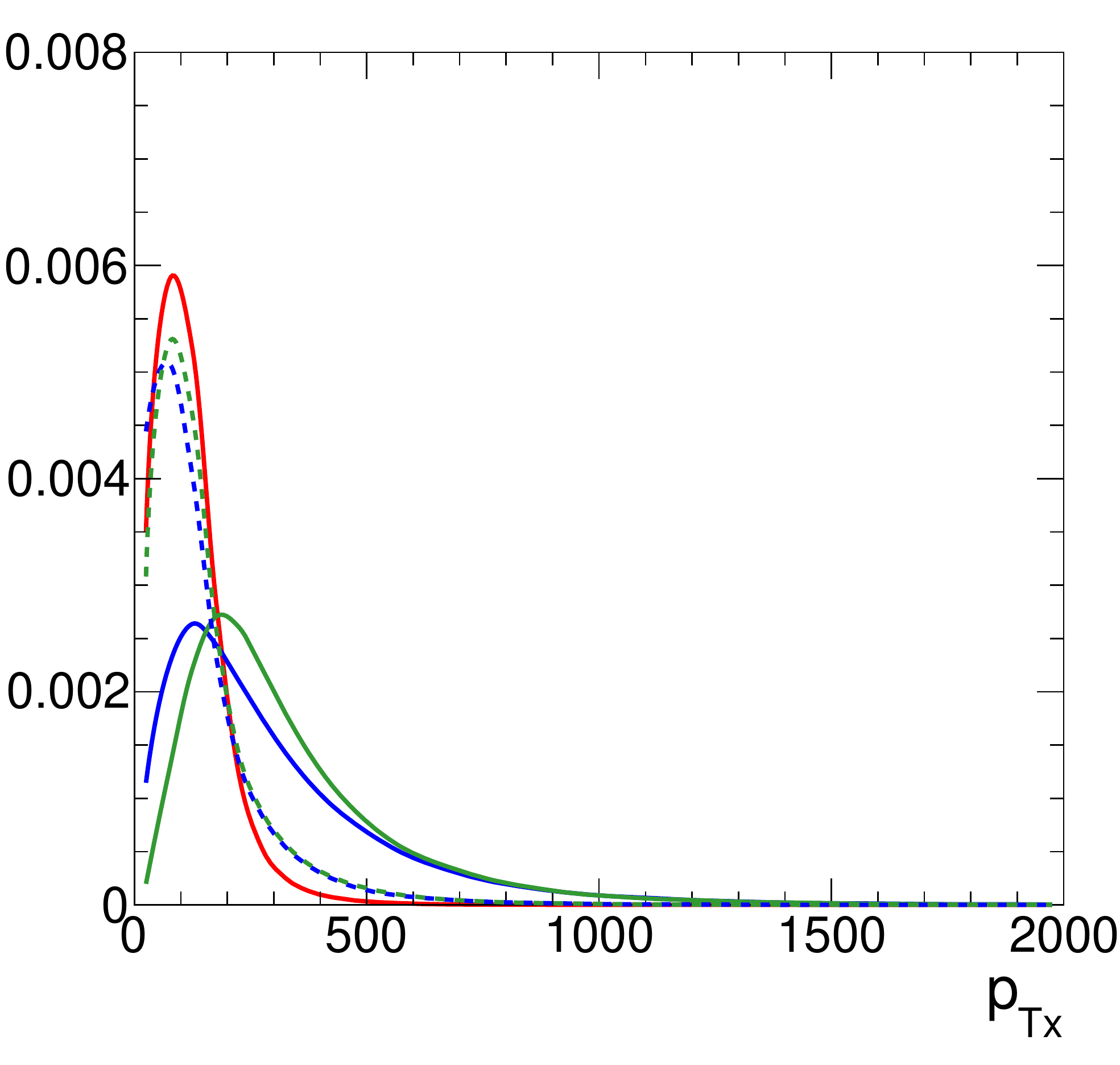}
 \includegraphics[width=0.24\textwidth]{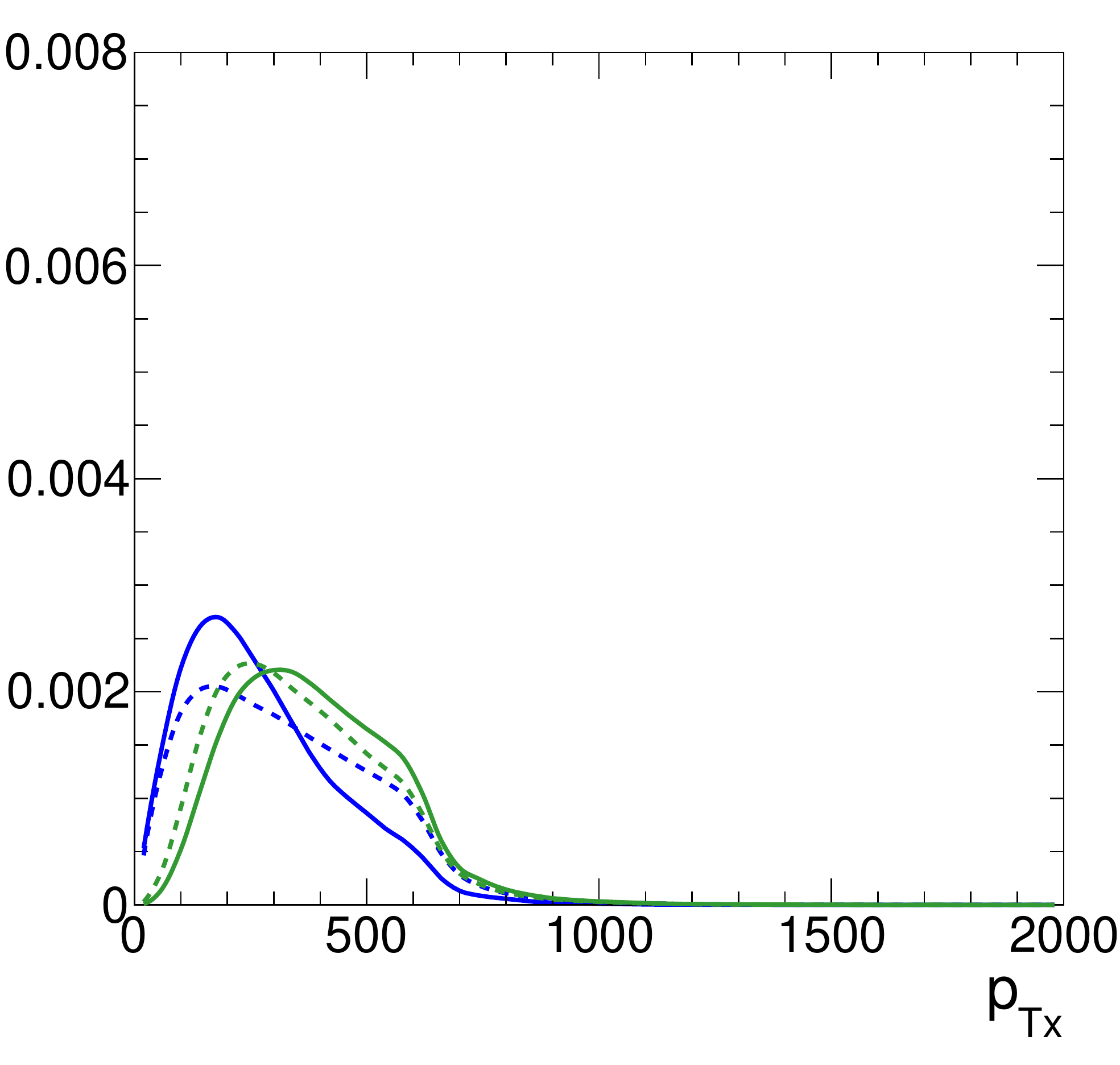}
 \includegraphics[width=0.24\textwidth]{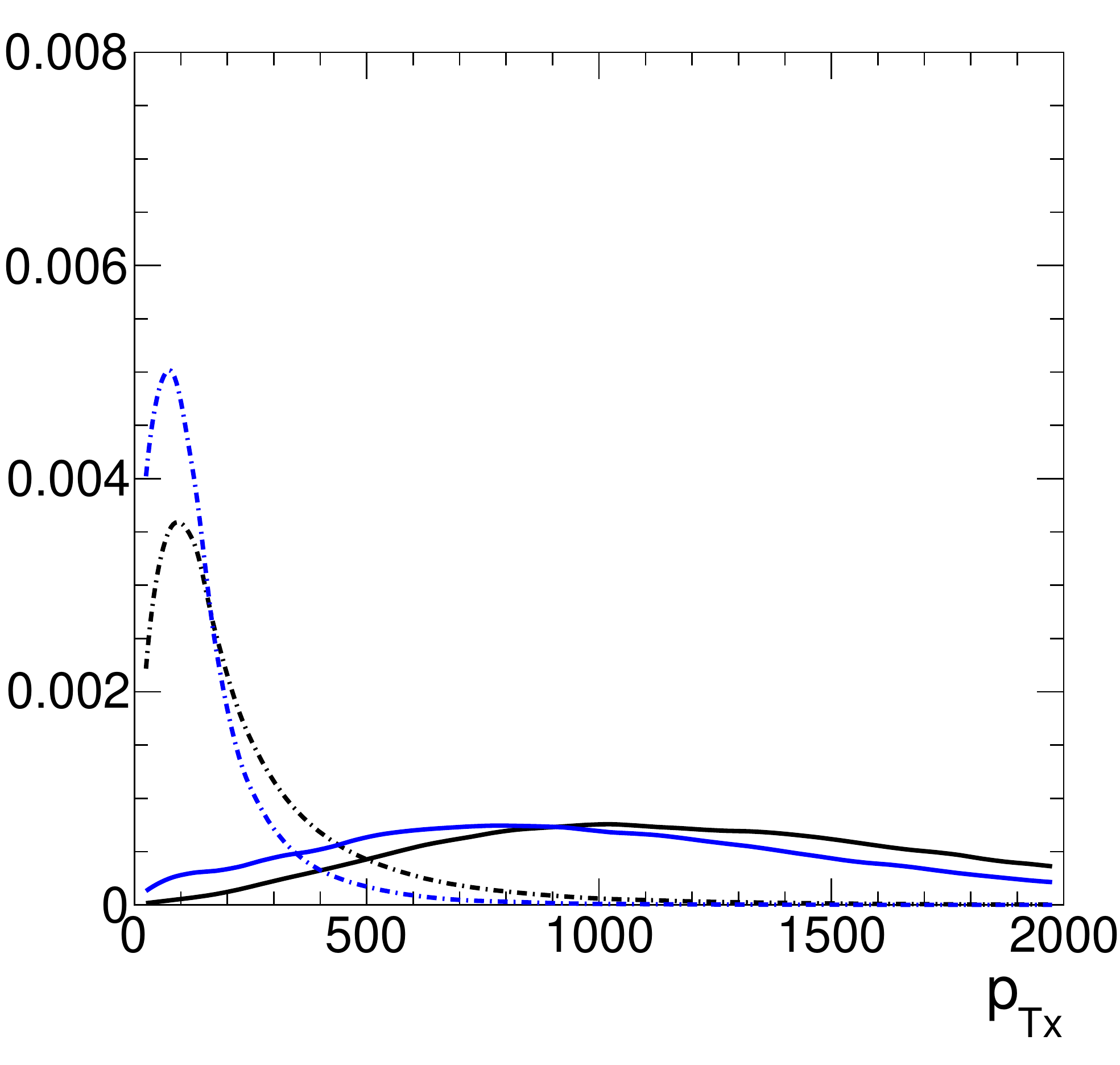}
 \includegraphics[width=0.24\textwidth]{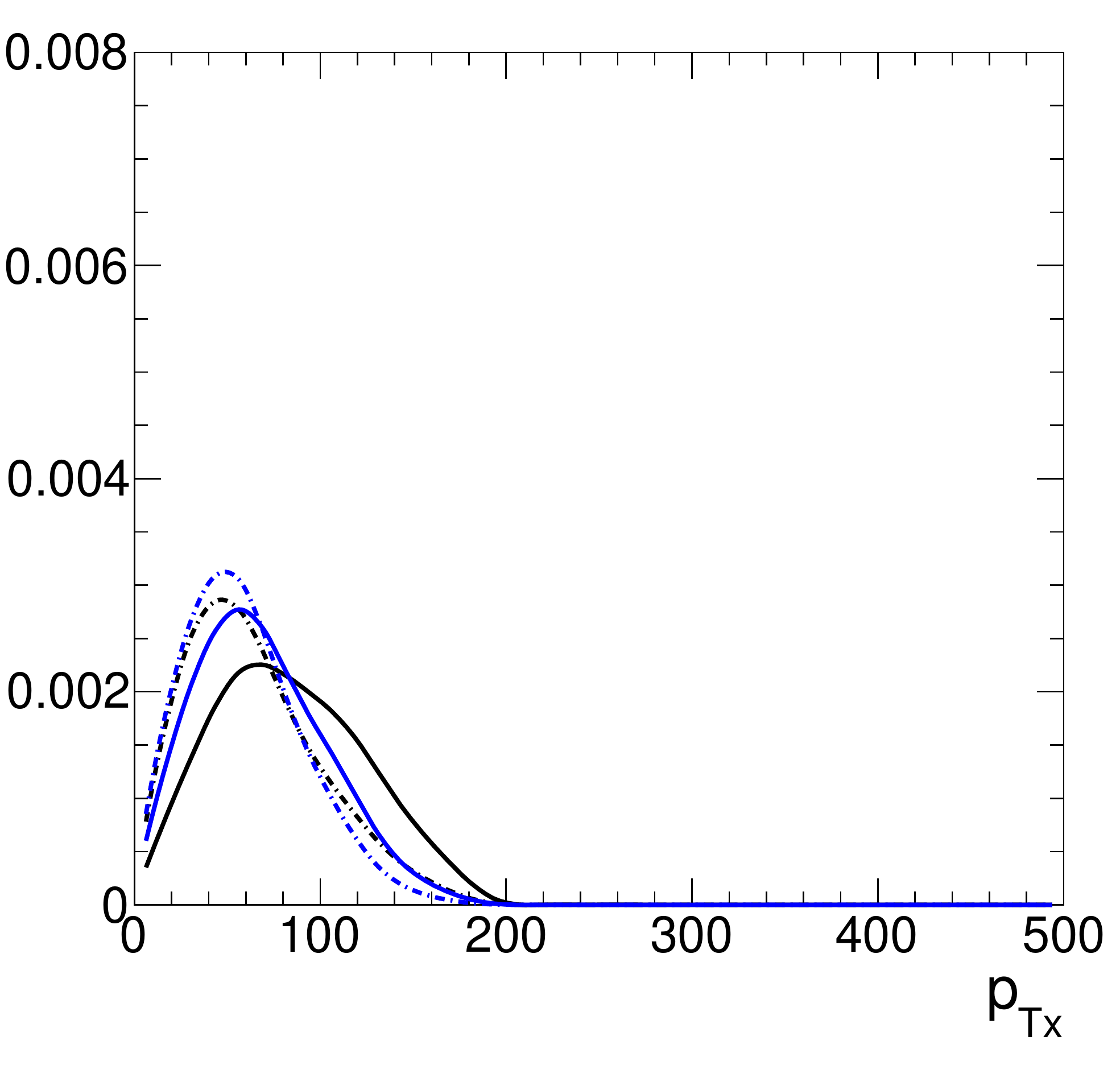} \\
 \includegraphics[width=0.24\textwidth]{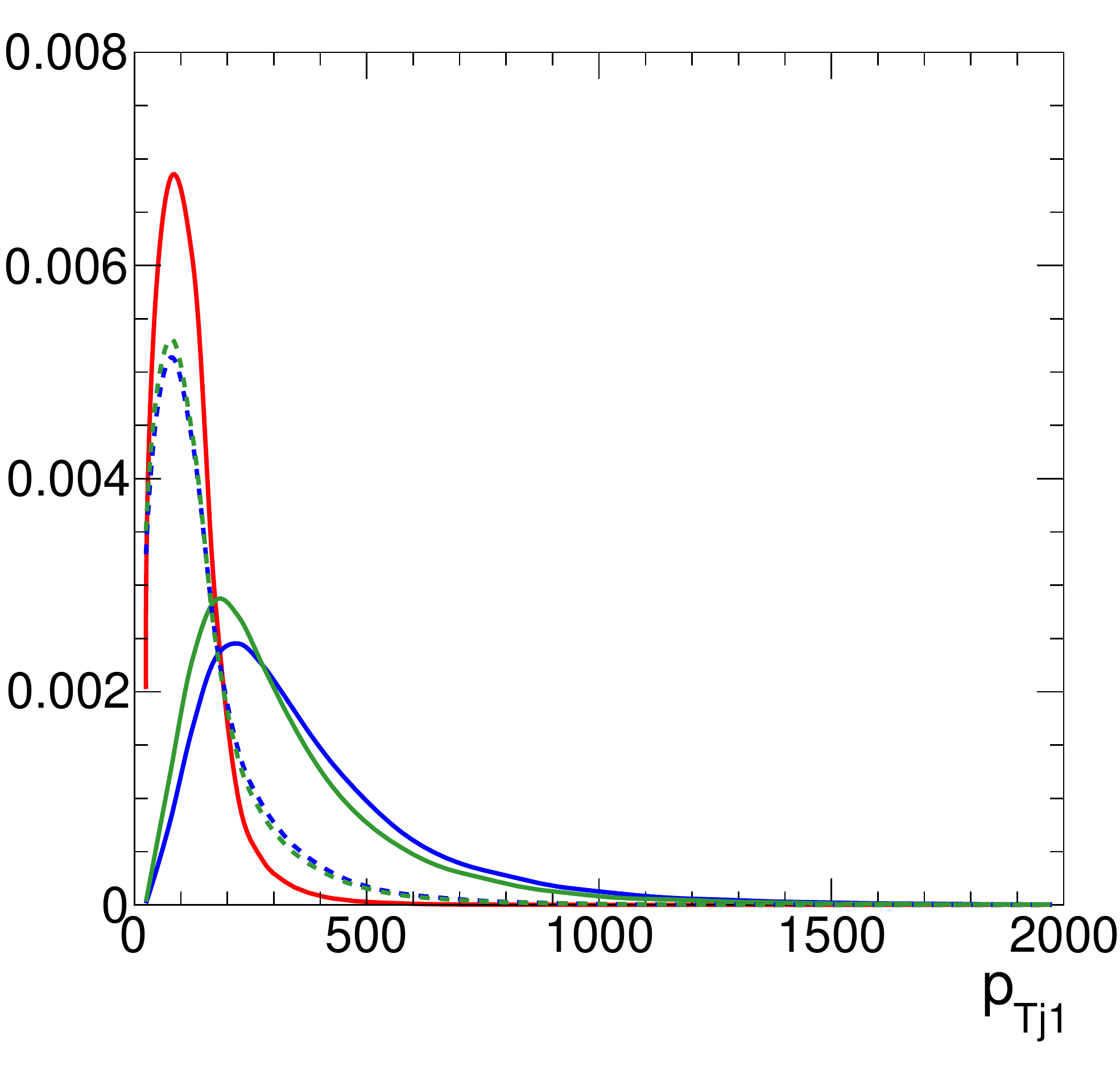}
 \includegraphics[width=0.24\textwidth]{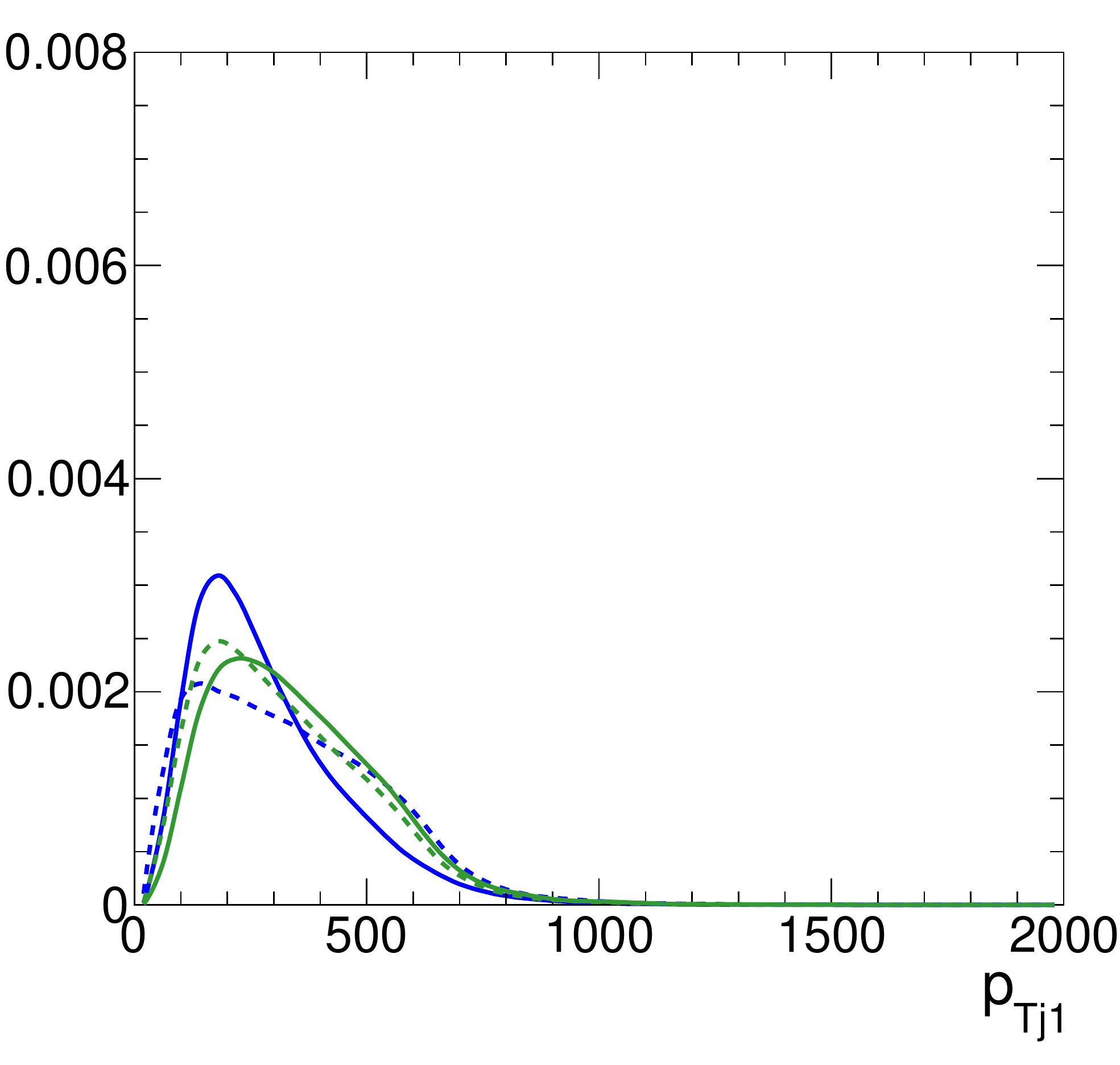}
 \includegraphics[width=0.24\textwidth]{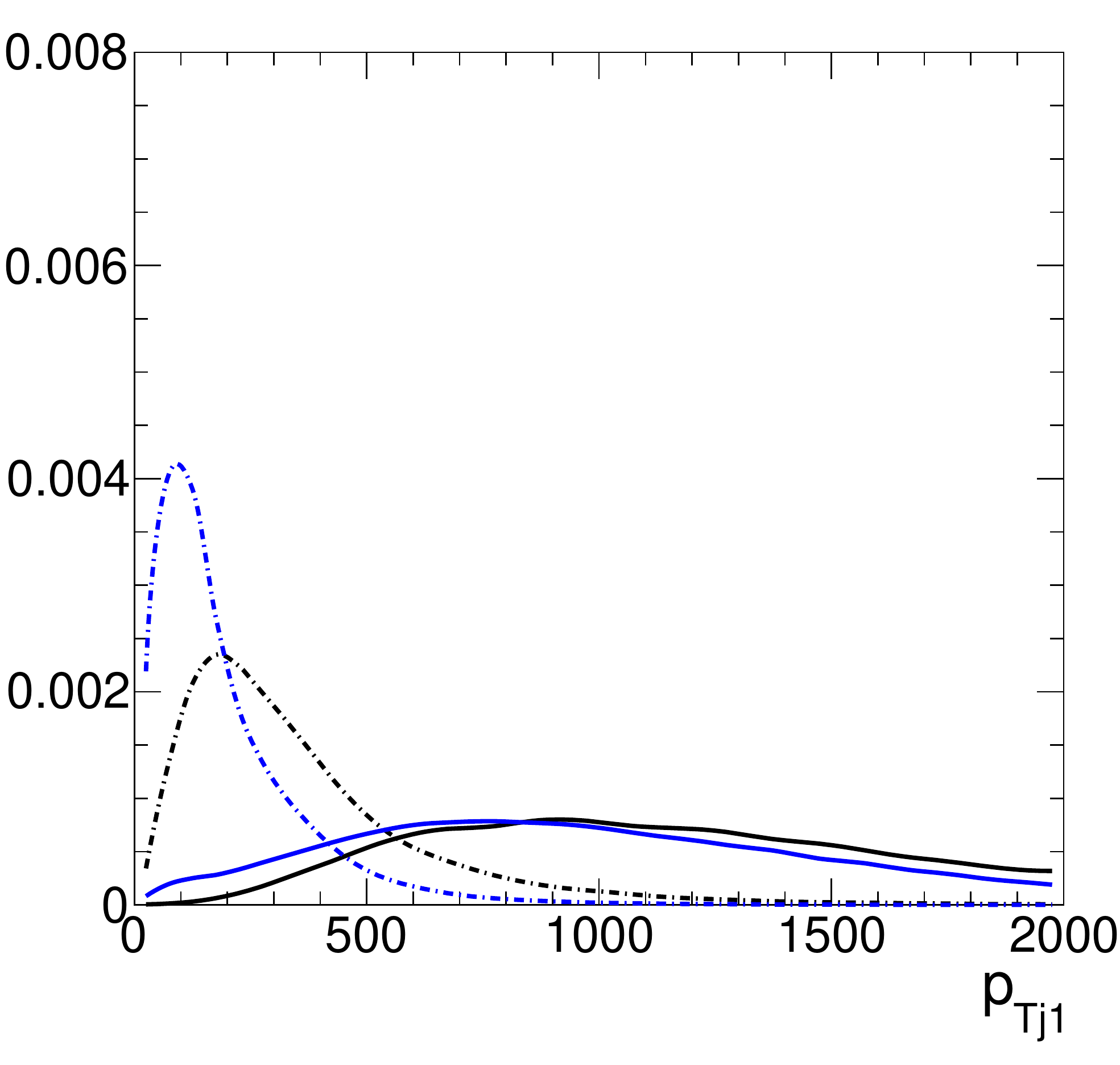}
 \hspace*{0.24\textwidth} 
\caption{Normalized distributions of the heavy resonance $X$ and the
  leading tagging jet after the minimal kinematical cuts of
  Eq.\eqref{eq:general_cuts} and $m_{jj}>600$~GeV in Eq.\eqref{eq:wbf_cuts1}. The hypotheses are
  $0_\text{SM}^+$ (red), $0_\text{D5}^+$ (blue), $0_\text{D5}^-$
  (green), $0_\text{D5(g)}^+$ (blue dashed), $0_\text{D5(g)}^-$ (green
  dashed); $1^+_Z$ (blue), $1^+_W$ (blue dashed), $1^-_Z$ (green),
  $1^-_W$ (green dashed); $2^+_\text{EW}$ (black), $2^+_\text{EW+q}$
  (black dashed), $2^+_\text{QCD}$ (blue), $2^+$ (blue dashed), all
  defined in Tab.~\ref{tab:model}.}
\label{fig:app1}
\end{figure}
%-------------------------------------------------------

In Fig.~\ref{fig:app1} we start by giving the complete set of rapidity
and transverse momentum distributions for the heavy resonance
$X$ and the leading tagging jet. In addition to the curves shown in
Fig.~\ref{fig:kin_basics} we also show the scalar and spin-2 couplings
to incoming gluons, as defined in Tab.~\ref{tab:model}.\bigskip

%-------------------------------------------------------
\begin{figure}[!t]
spin-0 \hspace*{3.cm} spin-1 \hspace*{3.cm} spin-2 \hspace*{3.cm} spin-2($p_T^\text{max}$) \\
 \includegraphics[width=0.24\textwidth]{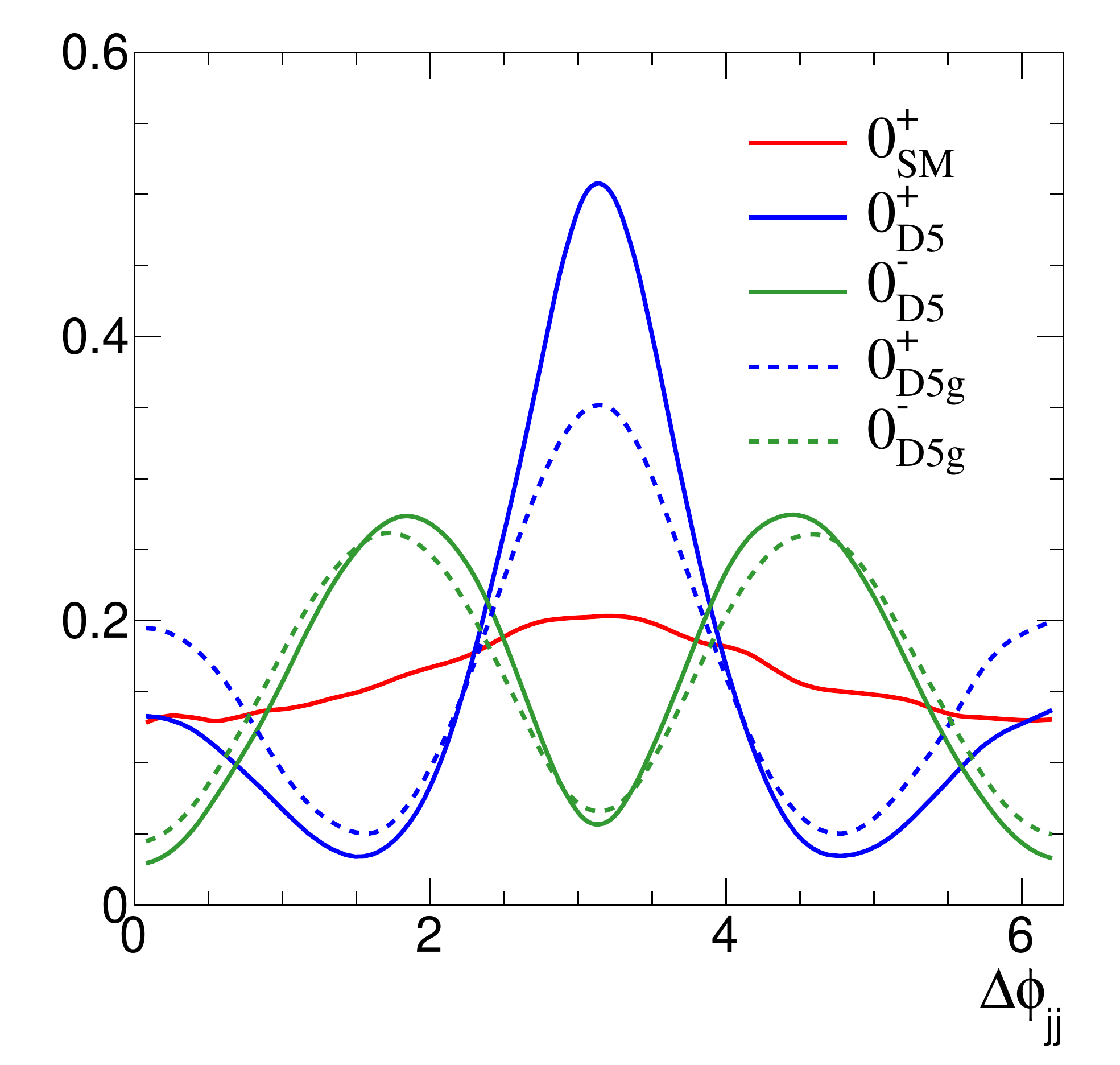}
 \hfill
 \includegraphics[width=0.24\textwidth]{spin1_dphijj}
 \hfill
 \includegraphics[width=0.24\textwidth]{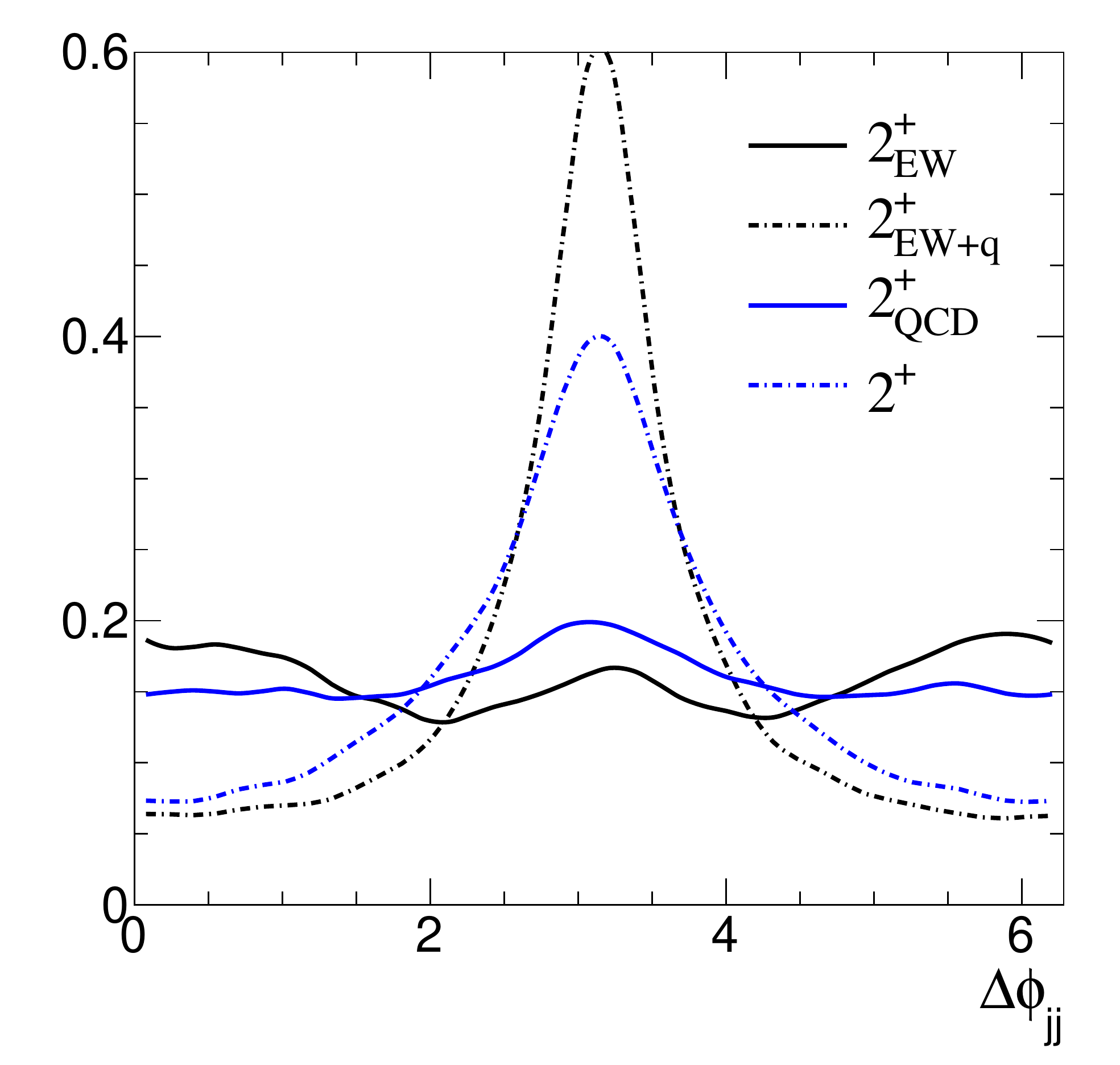}
 \hfill
 \includegraphics[width=0.24\textwidth]{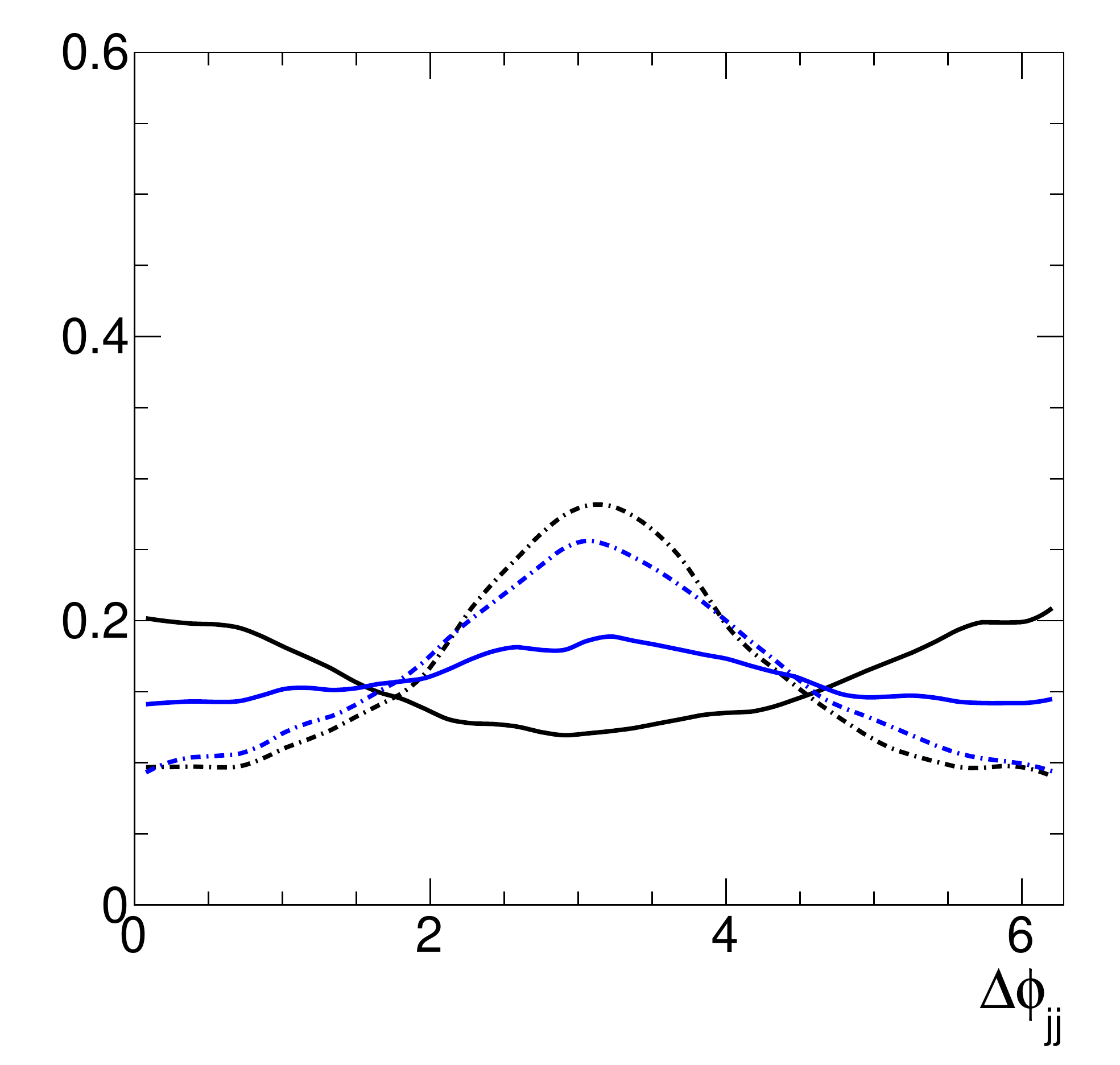} \\
 \hfill
 \includegraphics[width=0.24\textwidth]{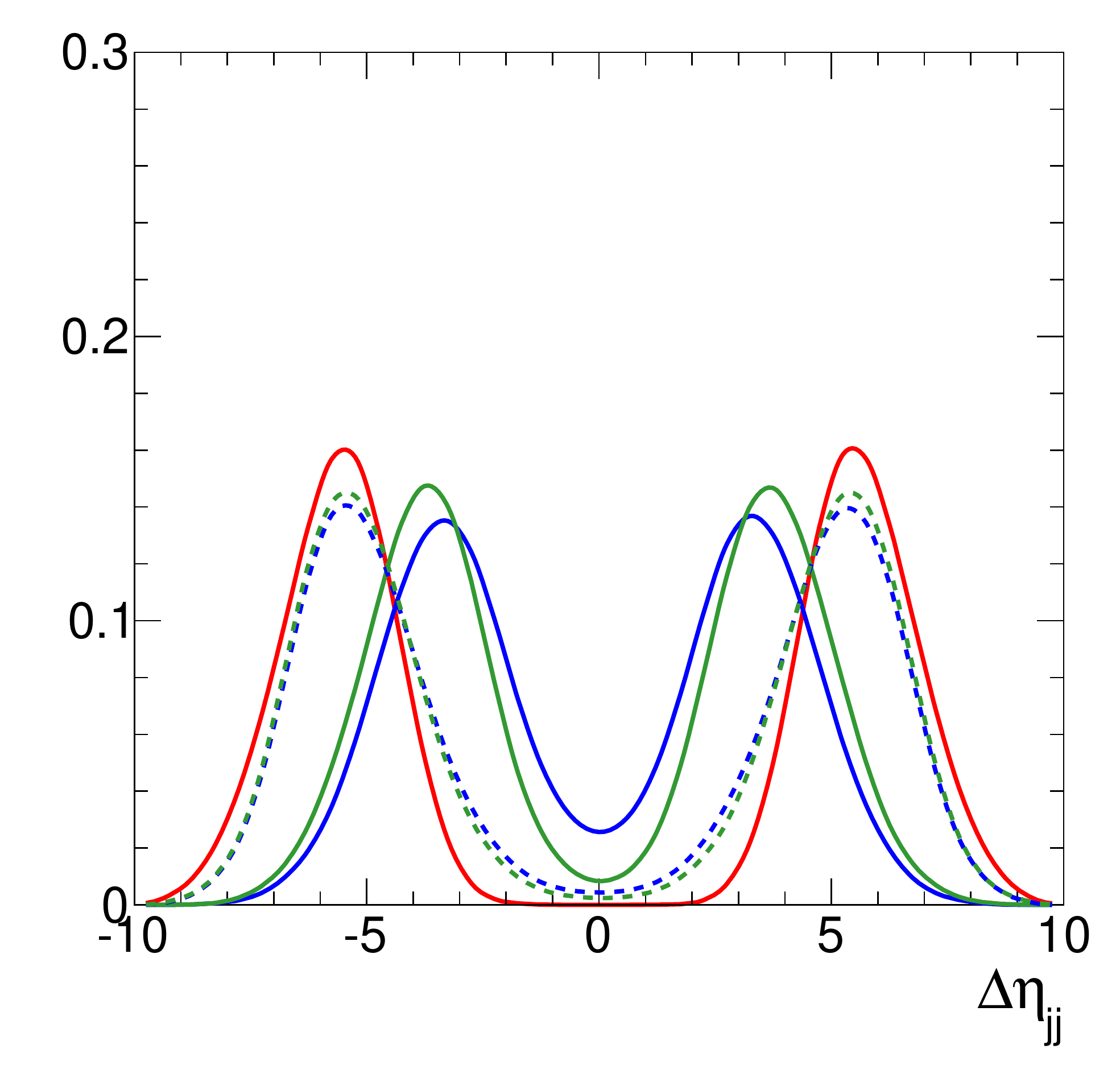}
 \hfill
 \includegraphics[width=0.24\textwidth]{spin1_detajj}
 \hfill
 \includegraphics[width=0.24\textwidth]{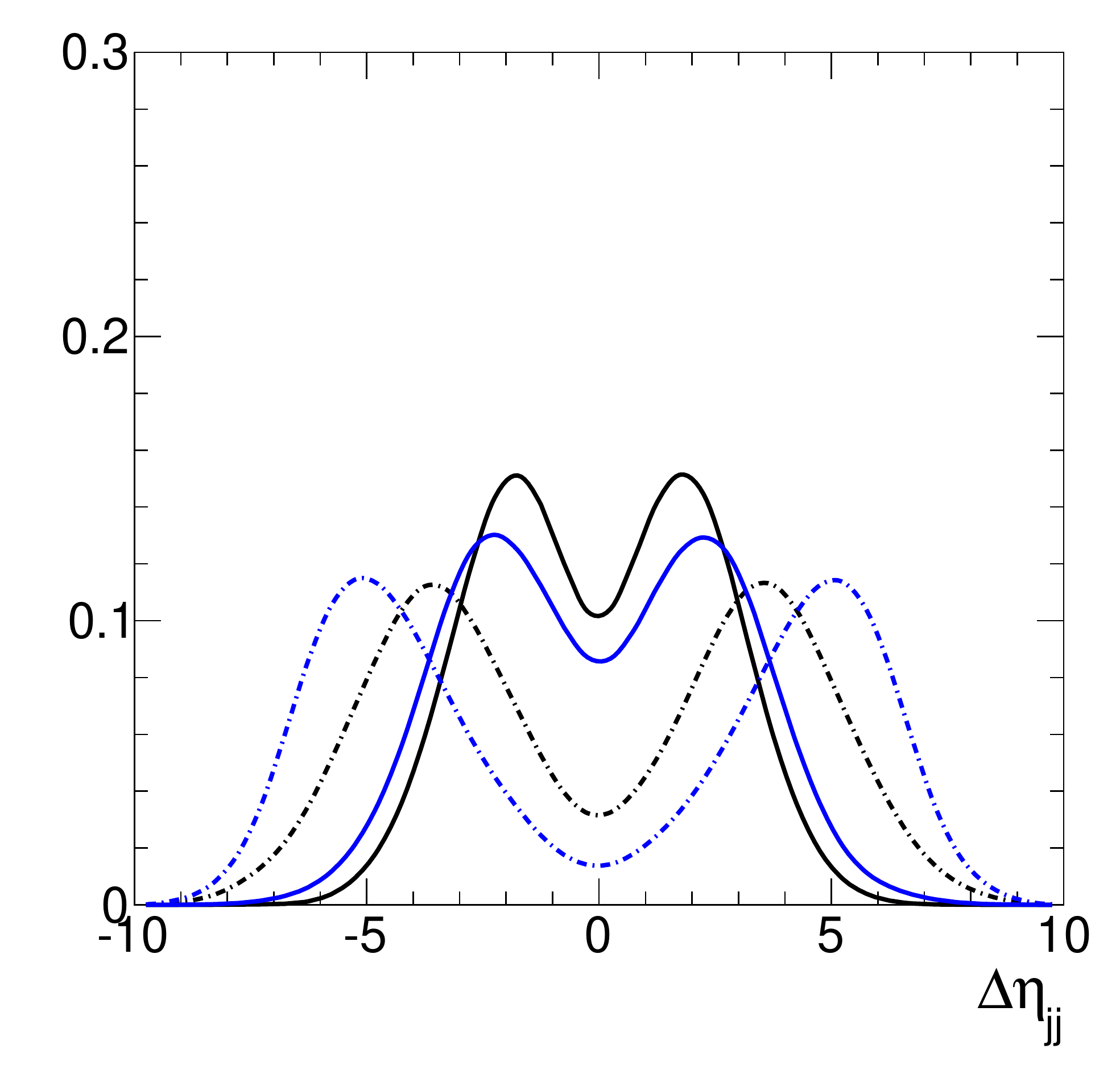}
 \hfill
 \includegraphics[width=0.24\textwidth]{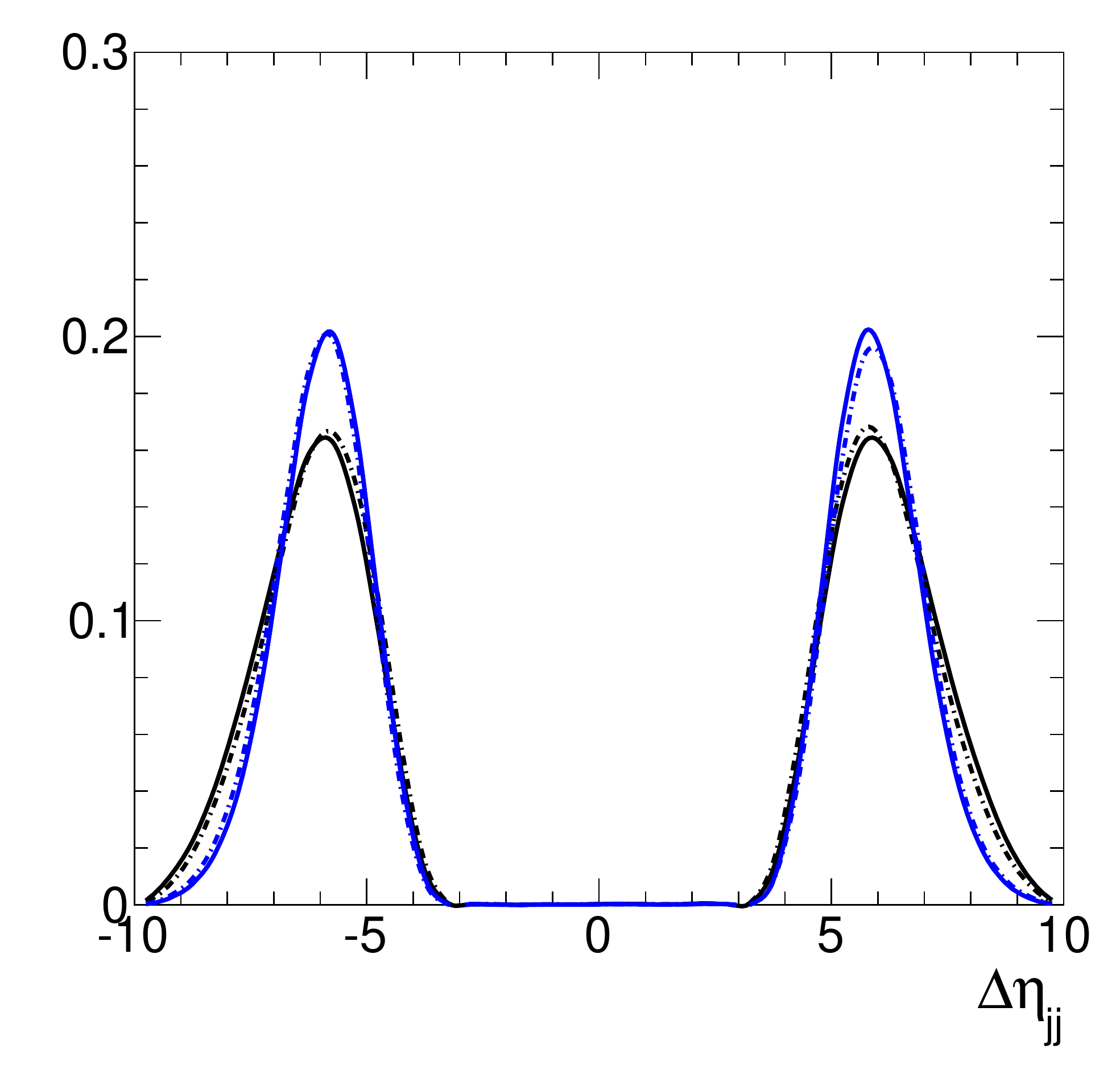}
 \caption{Correlations between the tagging jets.
 The different
   curves are the same as in Fig.~\ref{fig:app1}.}
\label{fig:app2a}
\end{figure}
%-------------------------------------------------------

%-------------------------------------------------------
\begin{figure}[!b]
spin-0 \hspace*{3.cm} spin-1 \hspace*{3.cm} spin-2 \hspace*{3.cm} spin-2($p_T^\text{max}$) \\
 \includegraphics[width=0.24\textwidth]{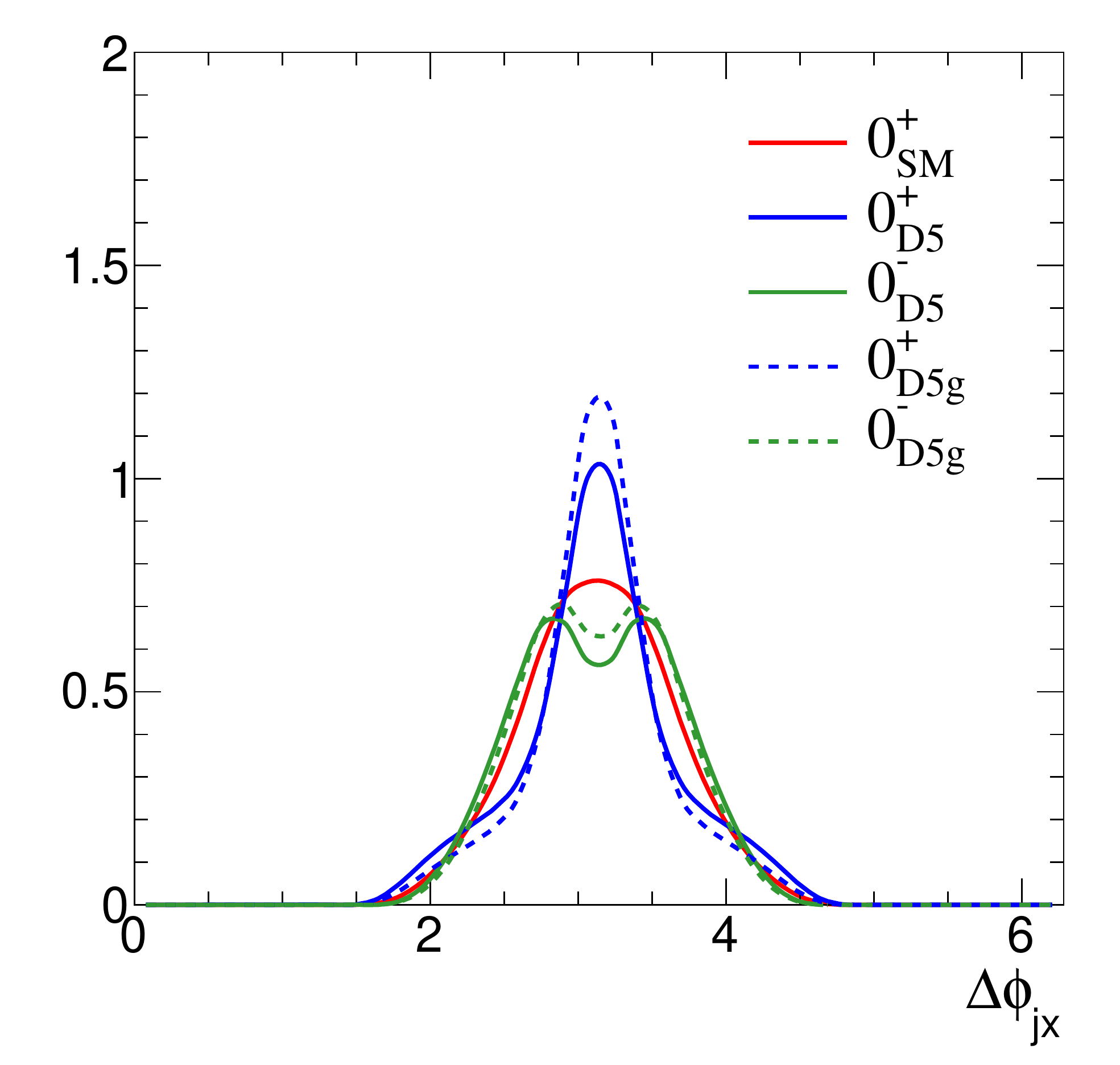}
 \hfill
 \includegraphics[width=0.24\textwidth]{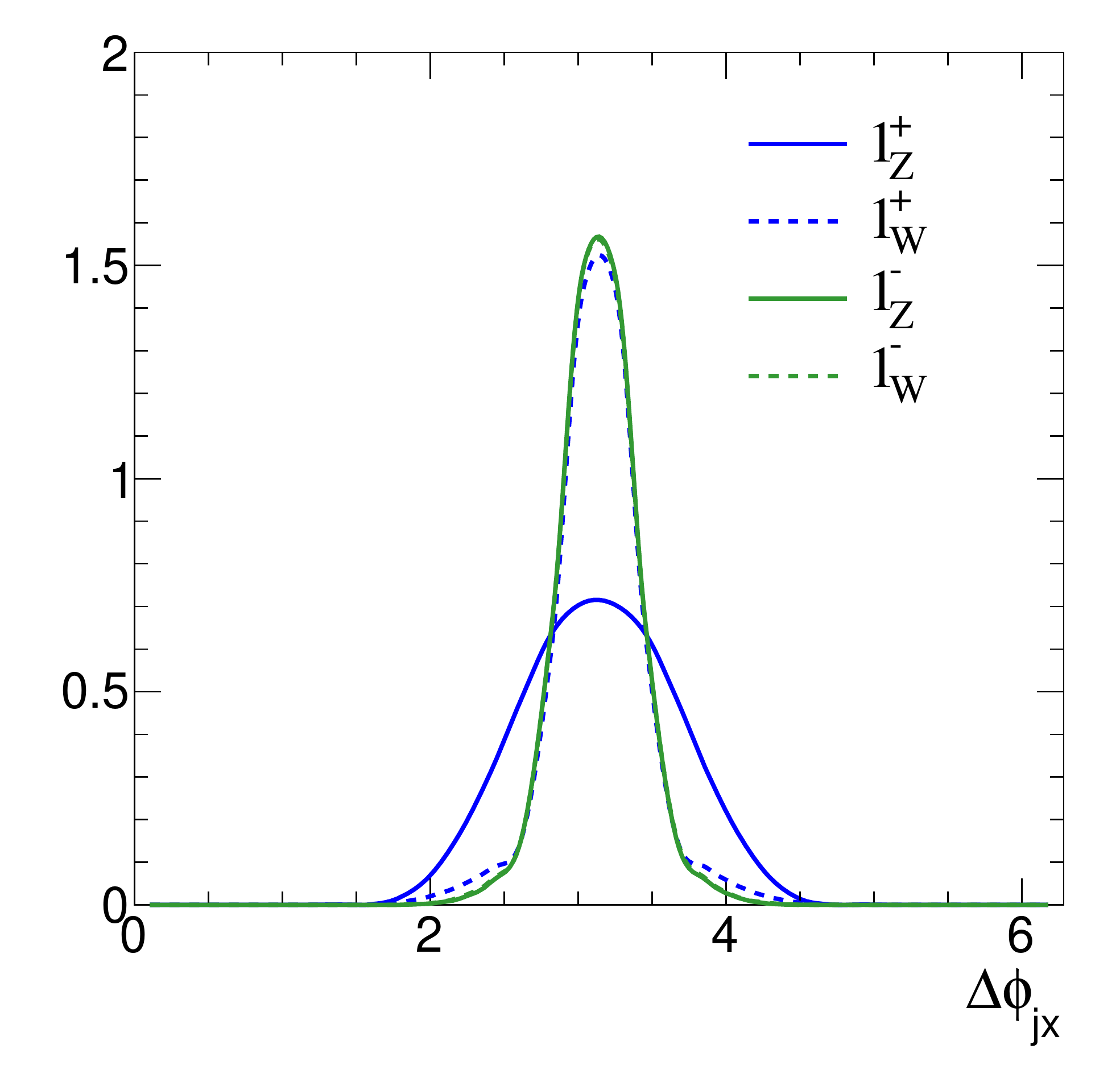}
 \hfill
 \includegraphics[width=0.24\textwidth]{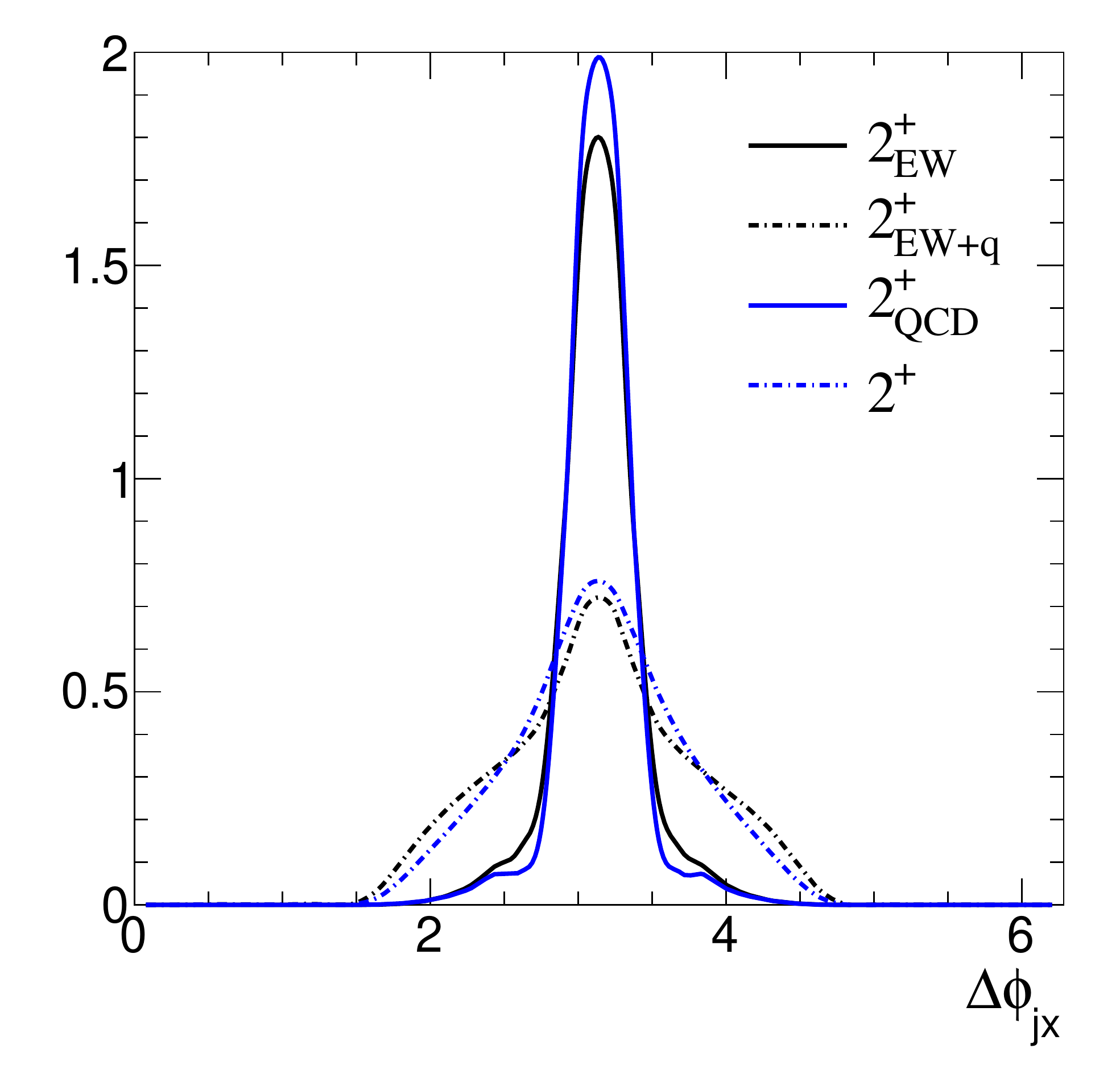}
 \hfill
 \includegraphics[width=0.24\textwidth]{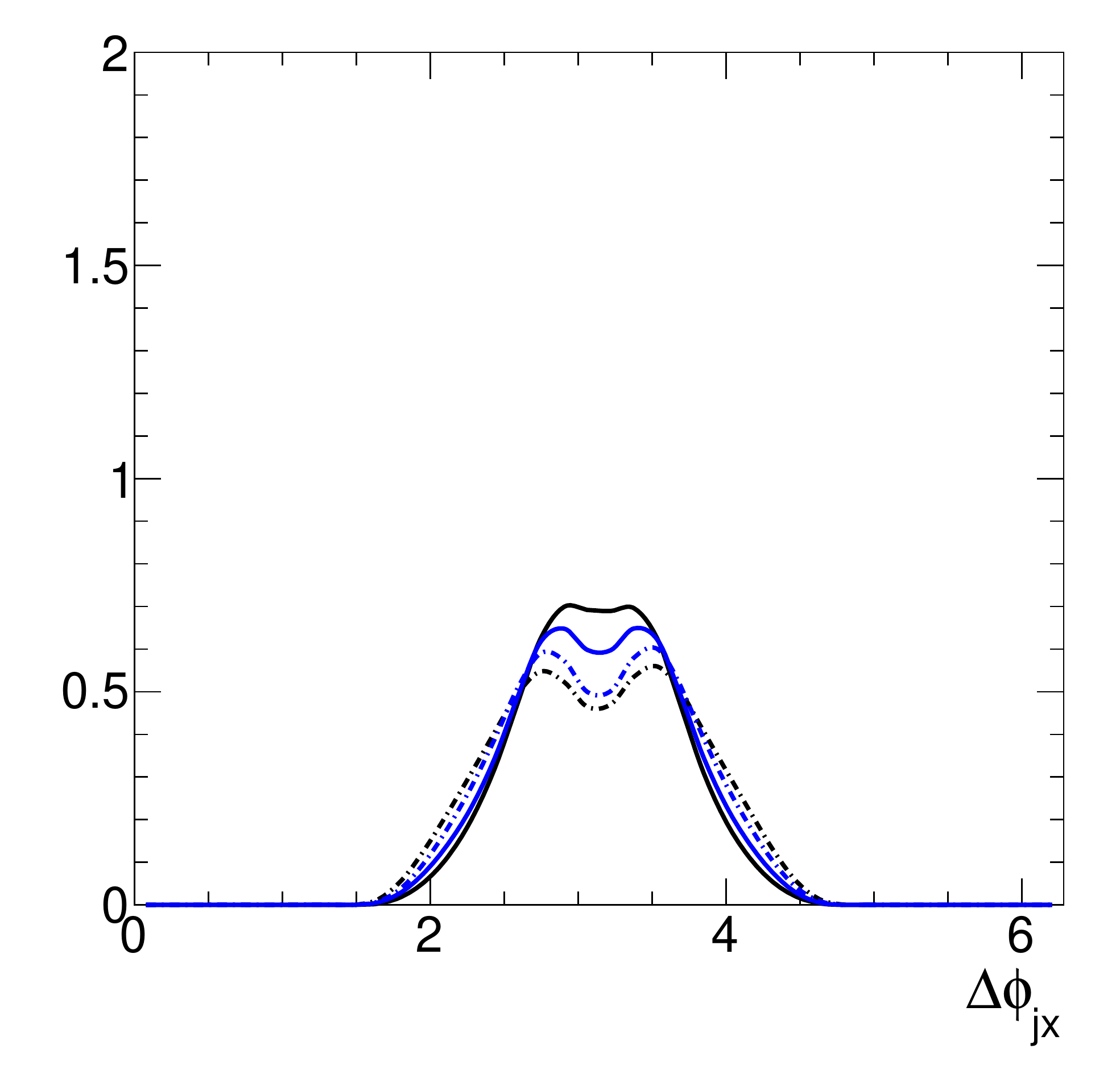}\\
 \hfill
 \includegraphics[width=0.24\textwidth]{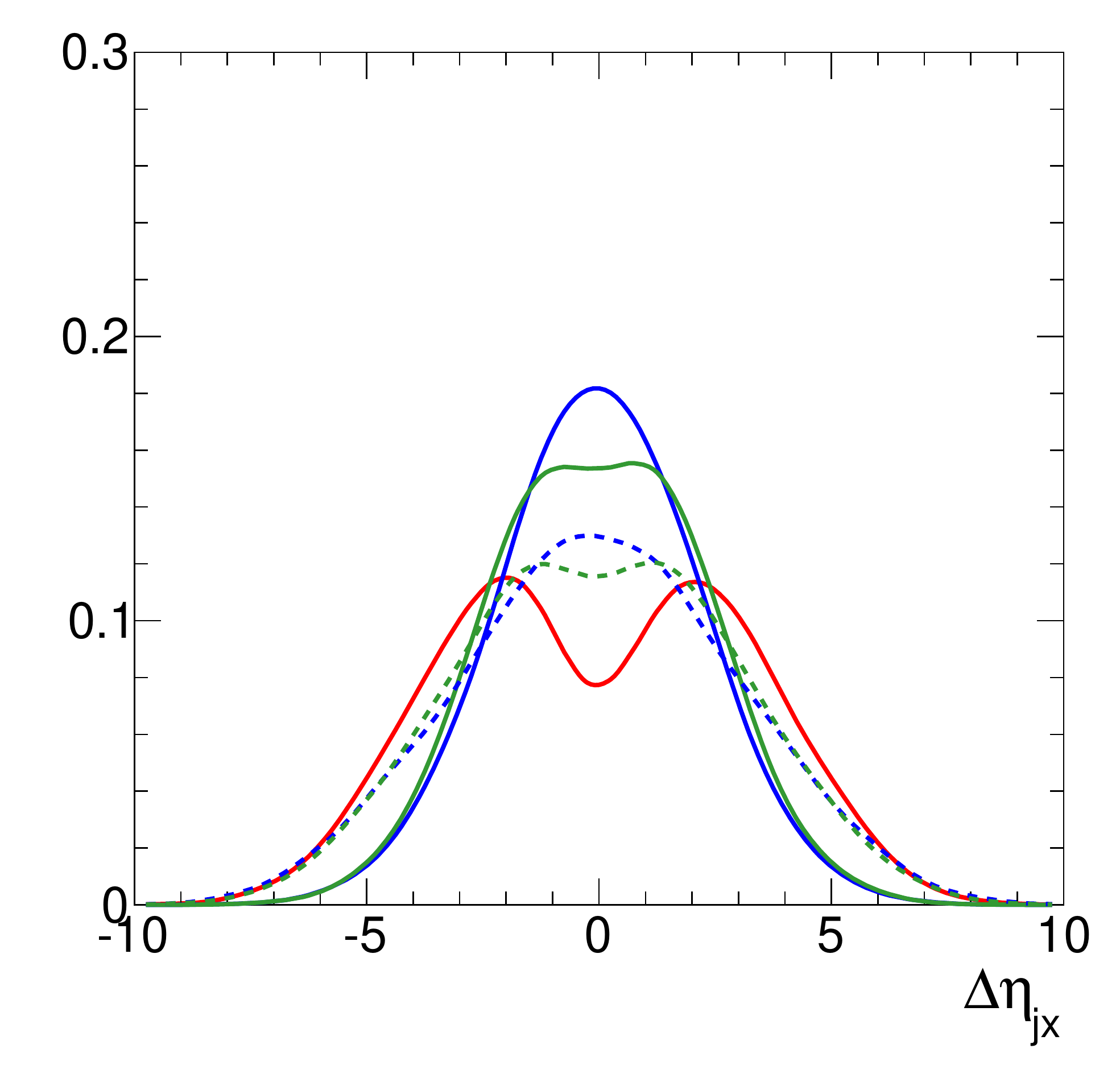}
 \hfill
 \includegraphics[width=0.24\textwidth]{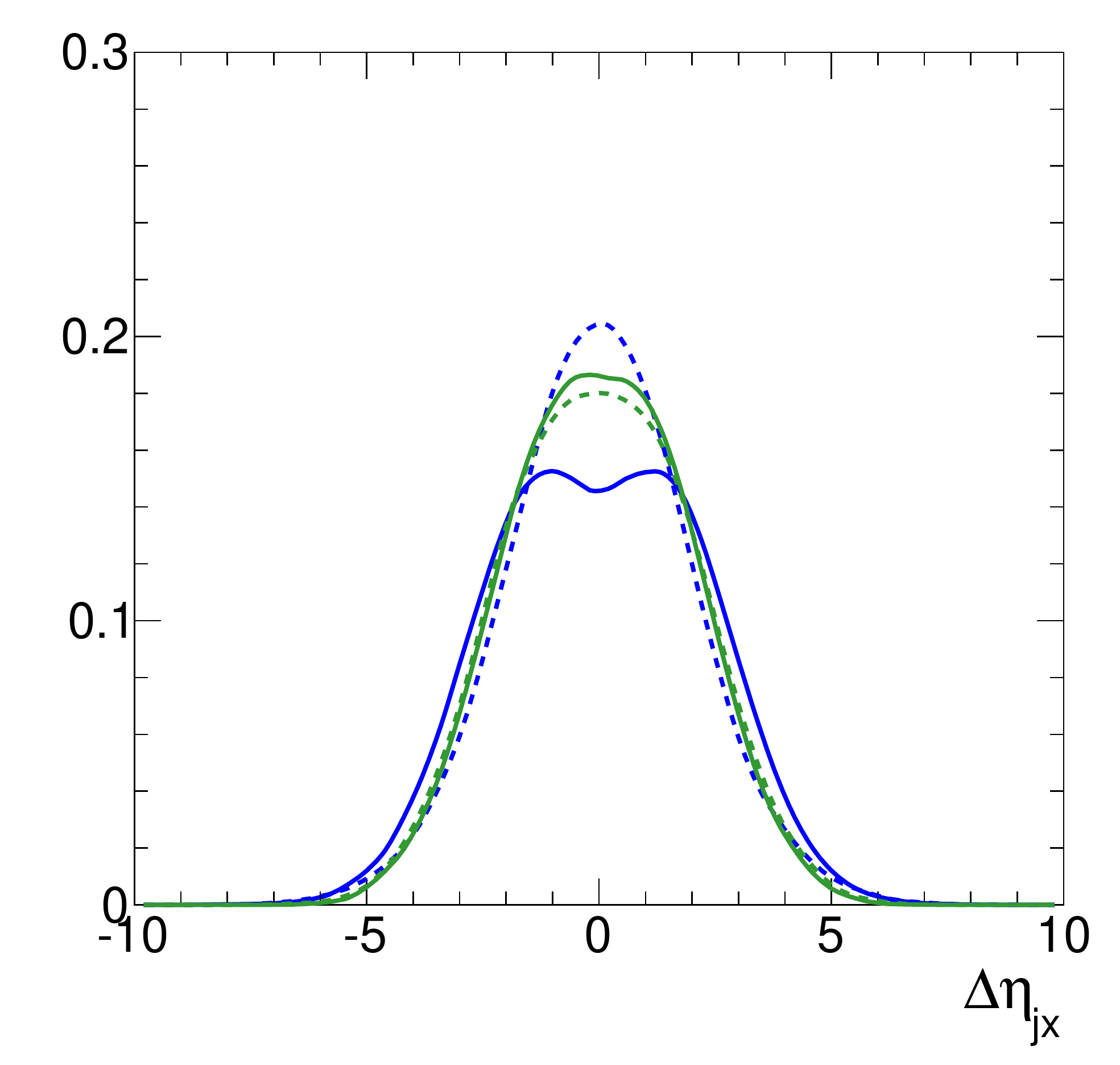}
 \hfill
 \includegraphics[width=0.24\textwidth]{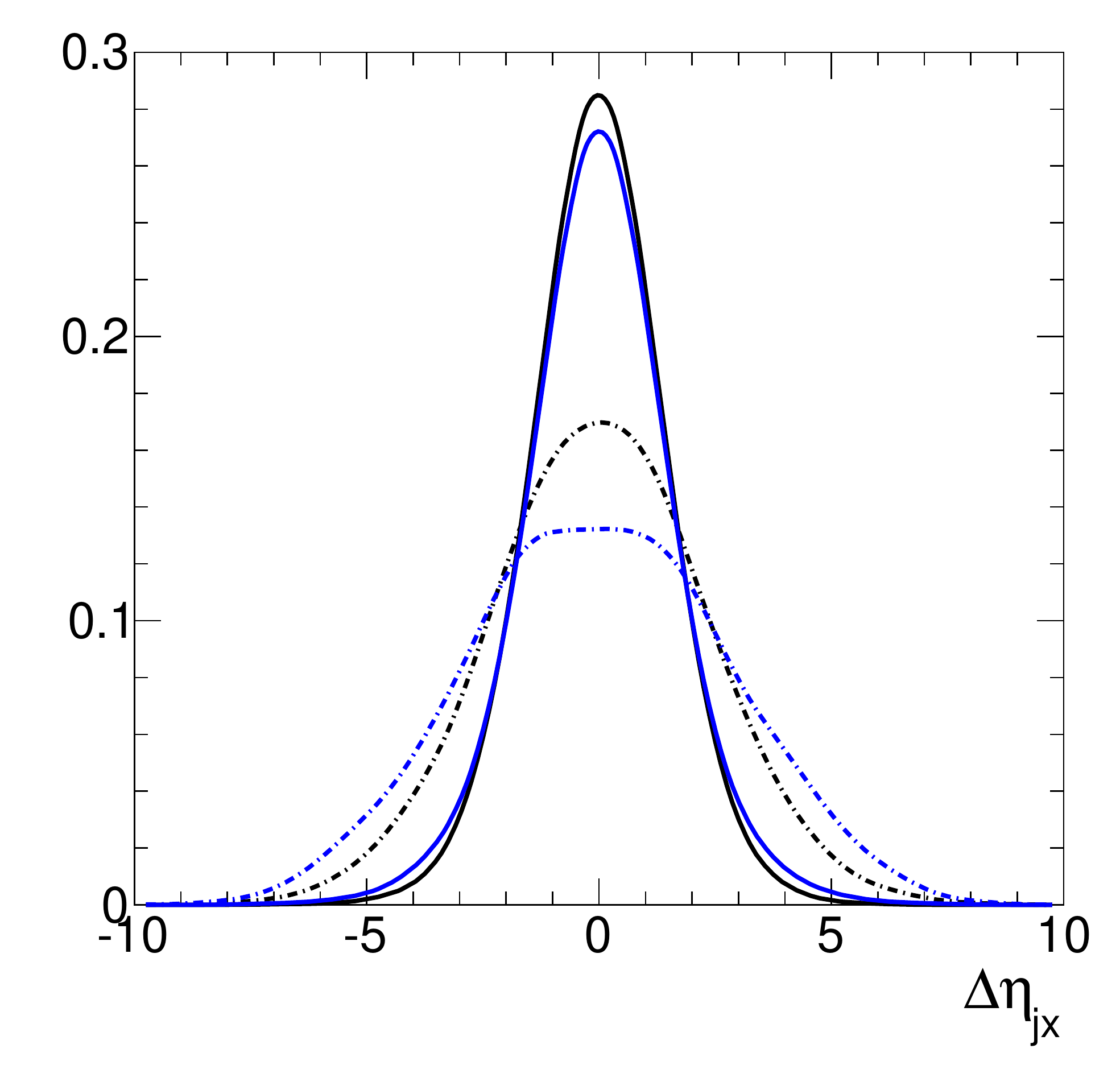}
 \hfill
 \includegraphics[width=0.24\textwidth]{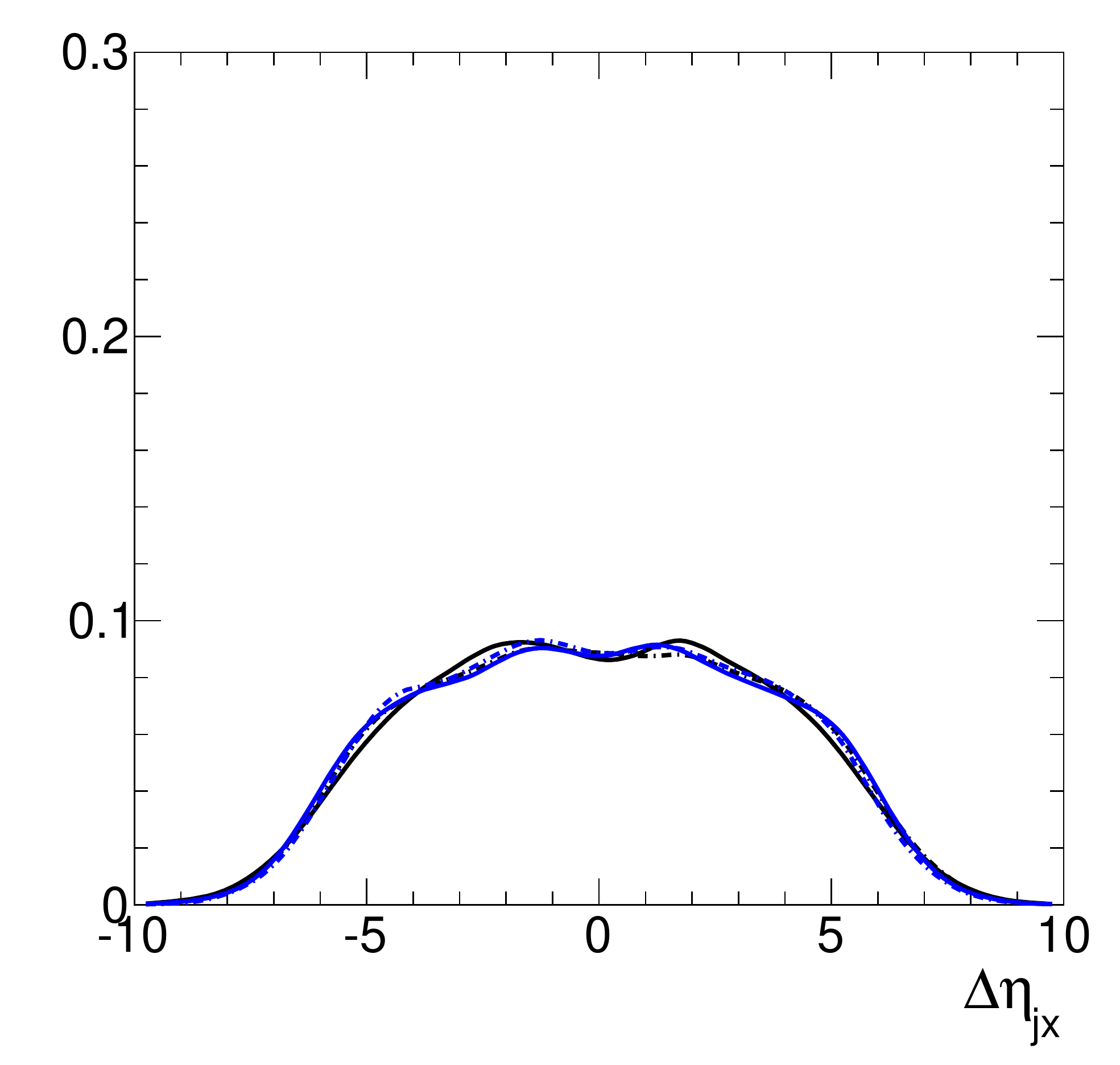}
 \caption{Same as Fig.~\ref{fig:app2a}, but among the heavy resonance
   $X$ and the leading tagging jet.}
\label{fig:app2b}
\end{figure}
%-------------------------------------------------------

\begin{figure}[!t]
spin-0 \hspace*{3.cm} spin-1 \hspace*{3.cm} spin-2 \hspace*{3.cm} spin-2($p_T^\text{max}$) \\
 \includegraphics[width=0.24\textwidth]{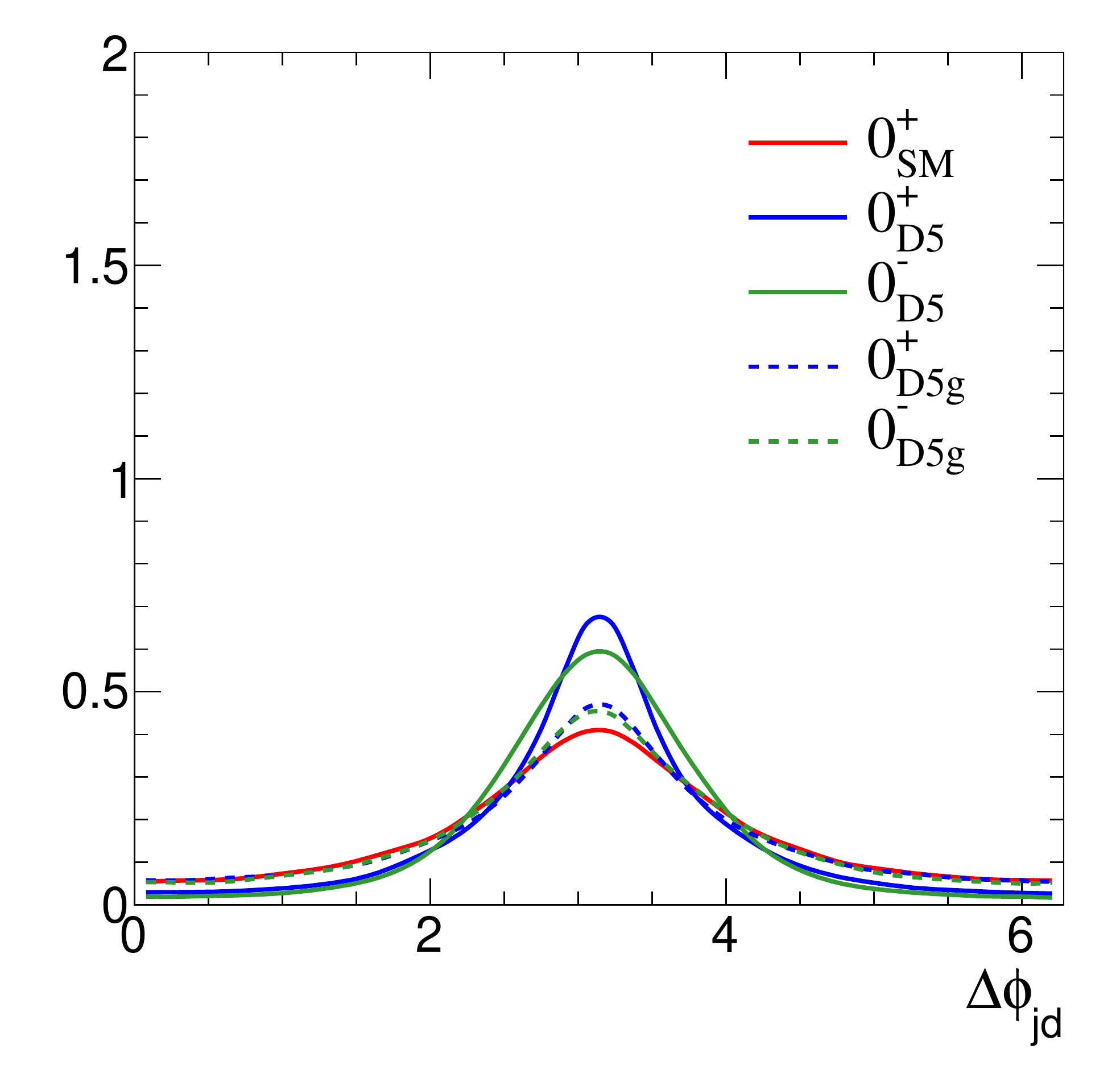}
 \hfill
 \includegraphics[width=0.24\textwidth]{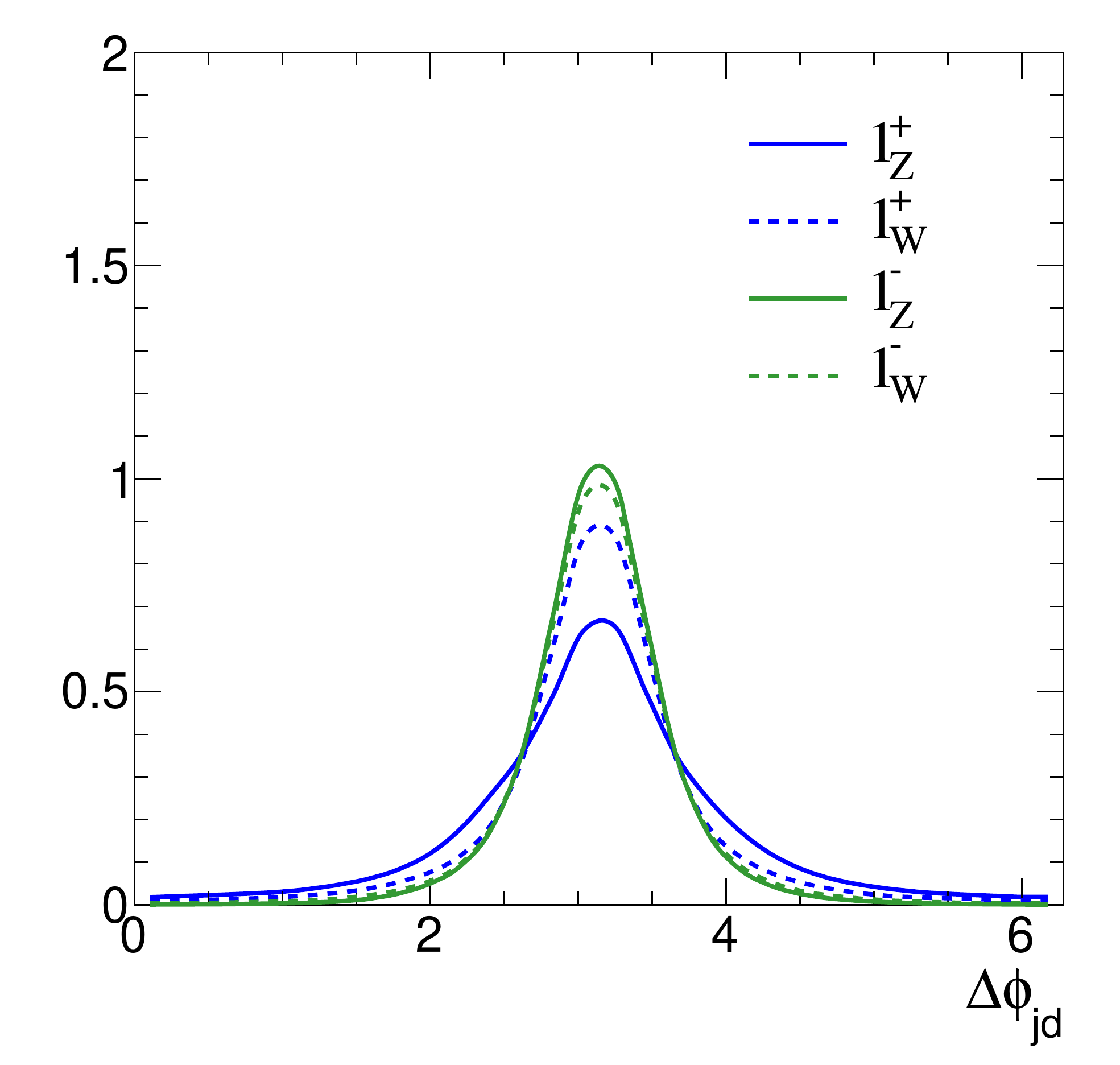}
 \hfill
 \includegraphics[width=0.24\textwidth]{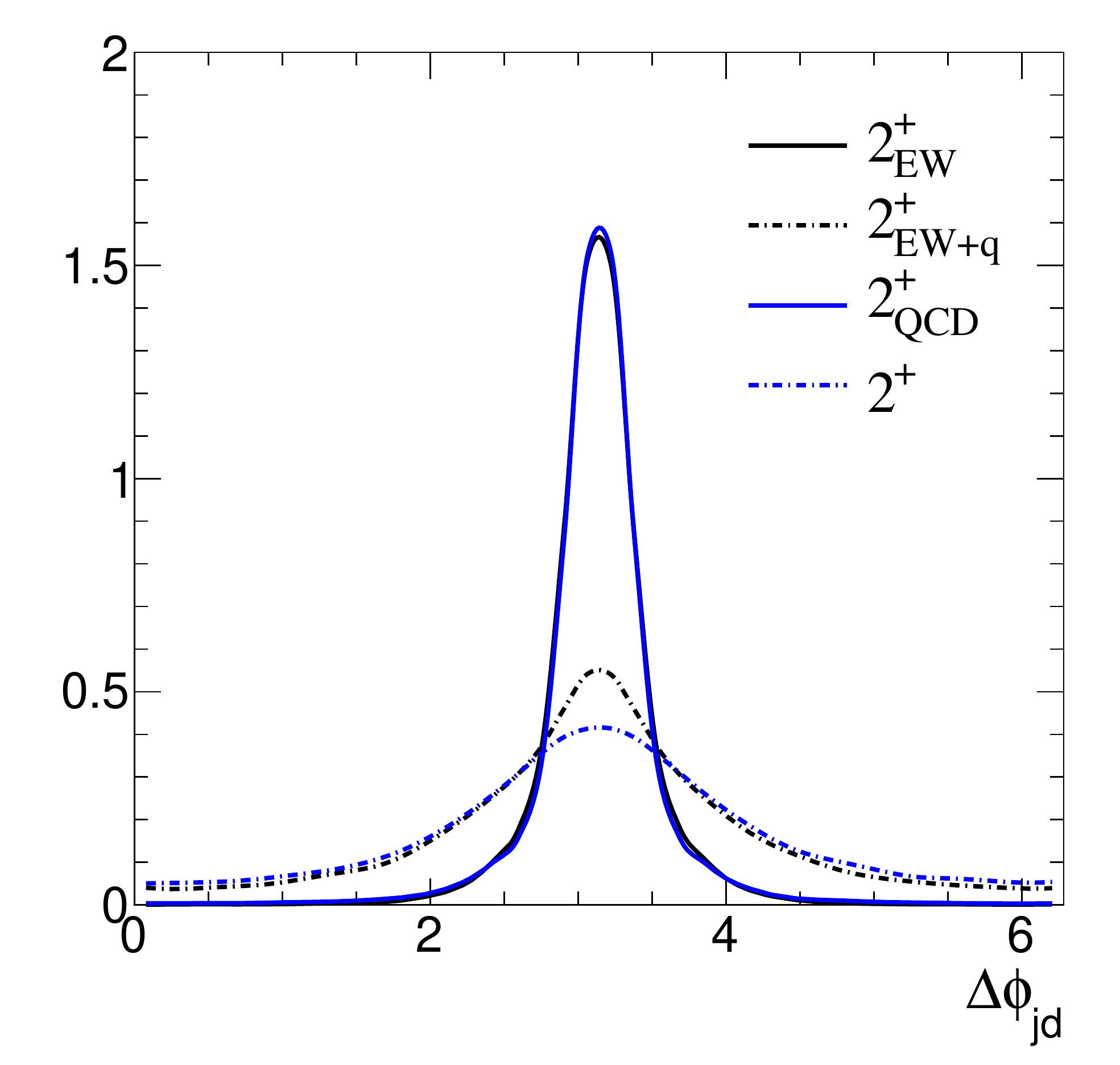}
 \hfill
 \includegraphics[width=0.24\textwidth]{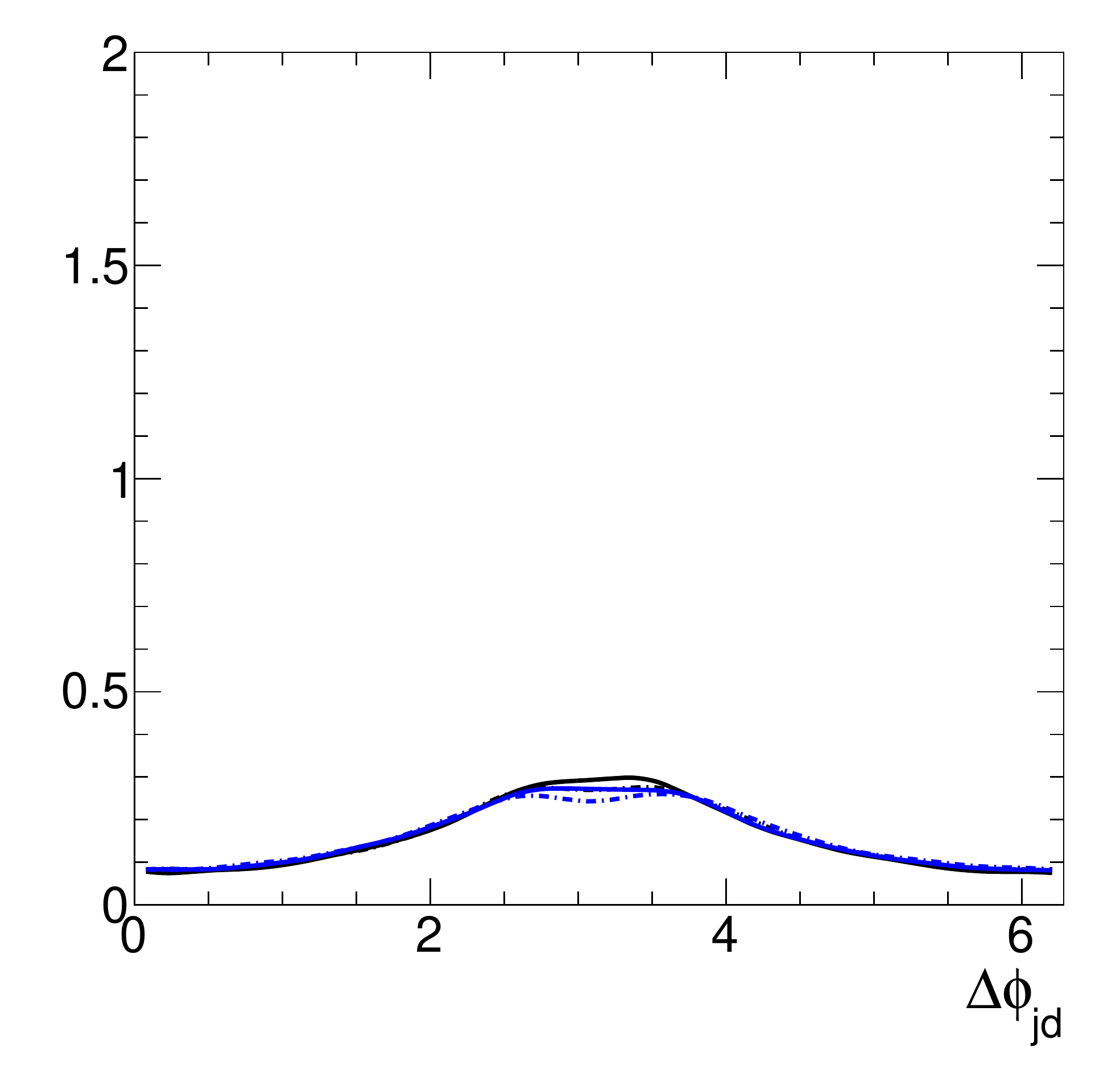} \\
 \hfill
 \includegraphics[width=0.24\textwidth]{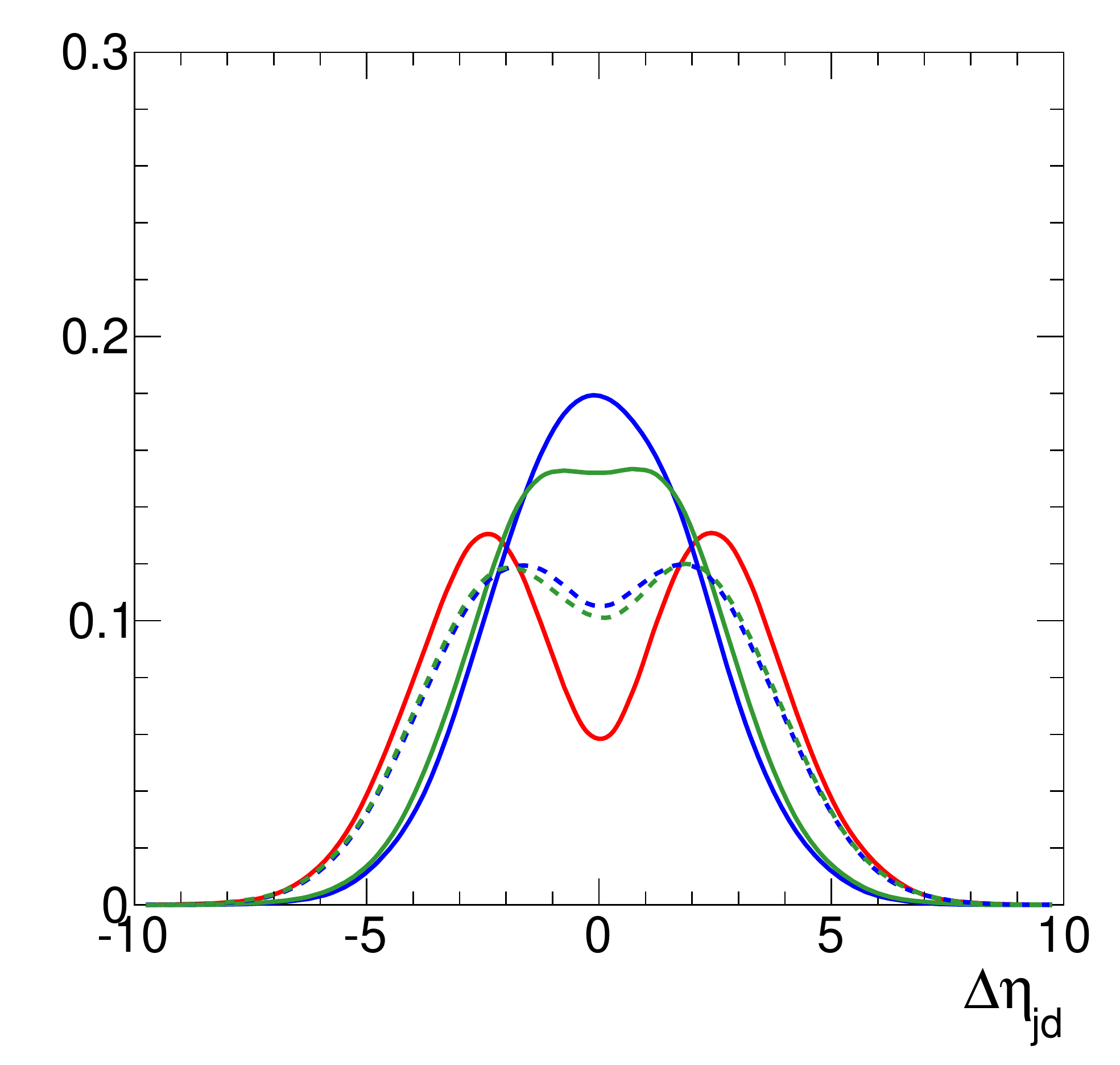}
 \hfill
 \includegraphics[width=0.24\textwidth]{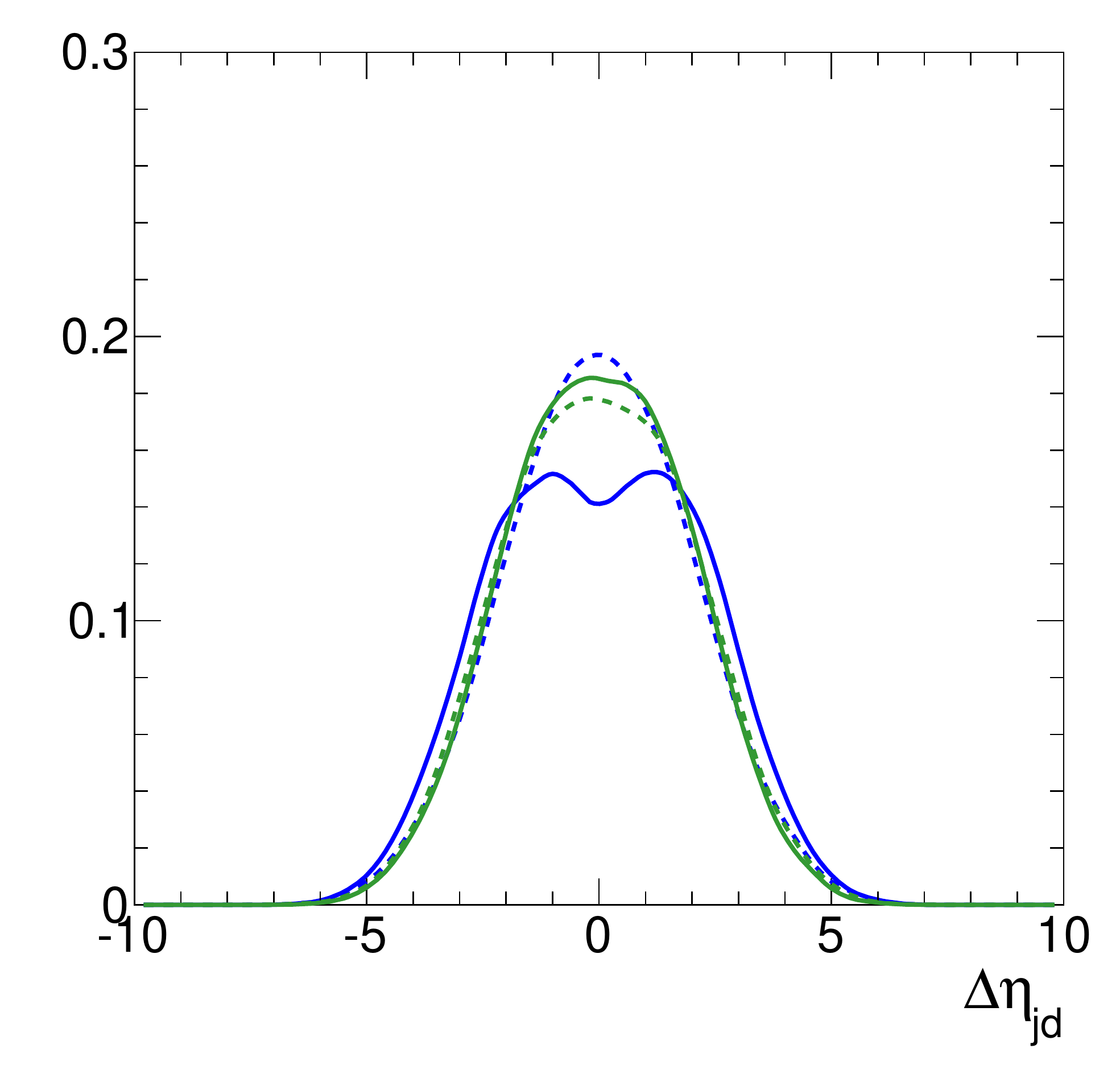}
 \hfill
 \includegraphics[width=0.24\textwidth]{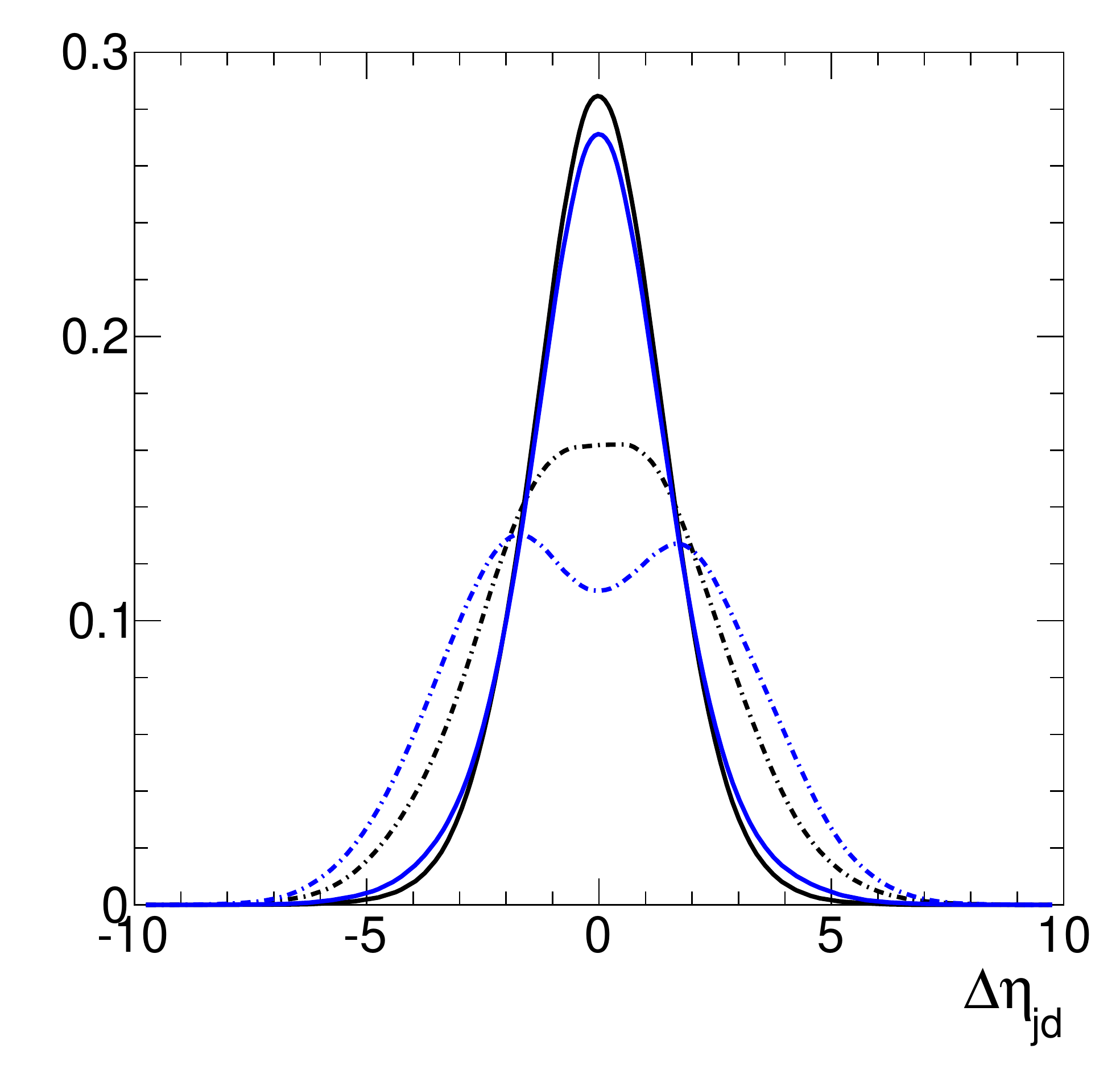}
 \hfill
 \includegraphics[width=0.24\textwidth]{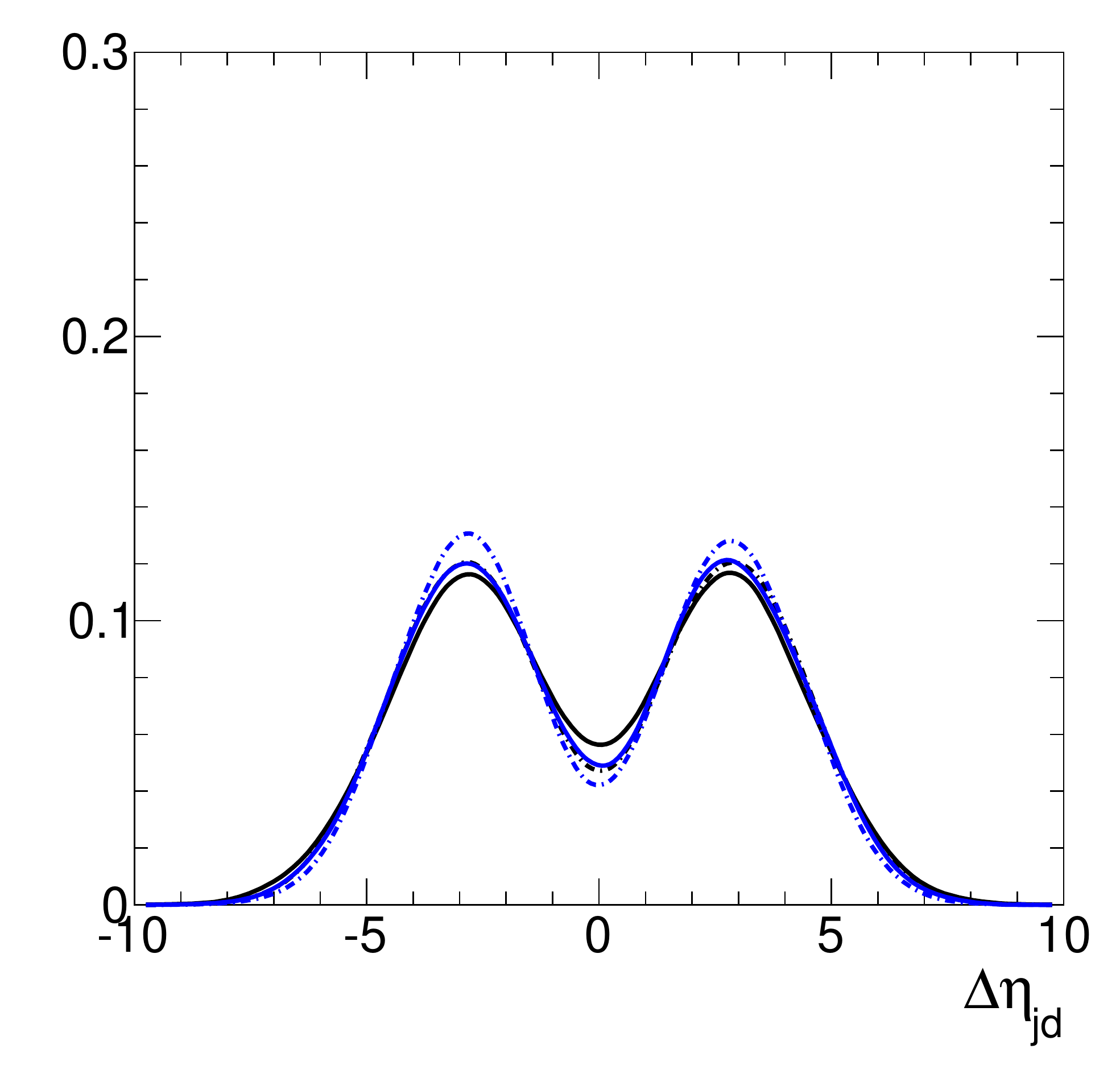}
 \caption{Same as Fig.~\ref{fig:app2a}, but among the leading tagging
   jet and the $X$ decay products.}
\label{fig:app2c}
\end{figure}

In Figs.~\ref{fig:app2a}, \ref{fig:app2b}, and \ref{fig:app2c} we show
the jet-jet, jet-$X$, and jet-decay correlations for the same full set
of operators. The confirm the picture that comes out of
Sec.~\ref{sec:analysis}, that for the WBF topology the jet-jet
correlations are the most sensitive because they show the most
distinctive kinematic patterns. Moreover, Fig.~\ref{fig:app2a}
confirms that adding a small fraction of gluon fusion events hardly
changes or even dilutes the general features of the WBF
kinematics. For all distributions we also observe a qualitative
difference between the generic spin-2 couplings and the spin-2
couplings after a unitarization cut $p_{T,j} = p_T^\text{max}$. A
completely model independent test of spin-2 couplings is unfortunately
not possible.\bigskip

\begin{figure}[!b]
spin-0 \hspace*{3.cm} spin-1 \hspace*{3.cm} spin-2 \hspace*{3.cm} spin-2($p_T^\text{max}$) \\
 \includegraphics[width=0.24\textwidth]{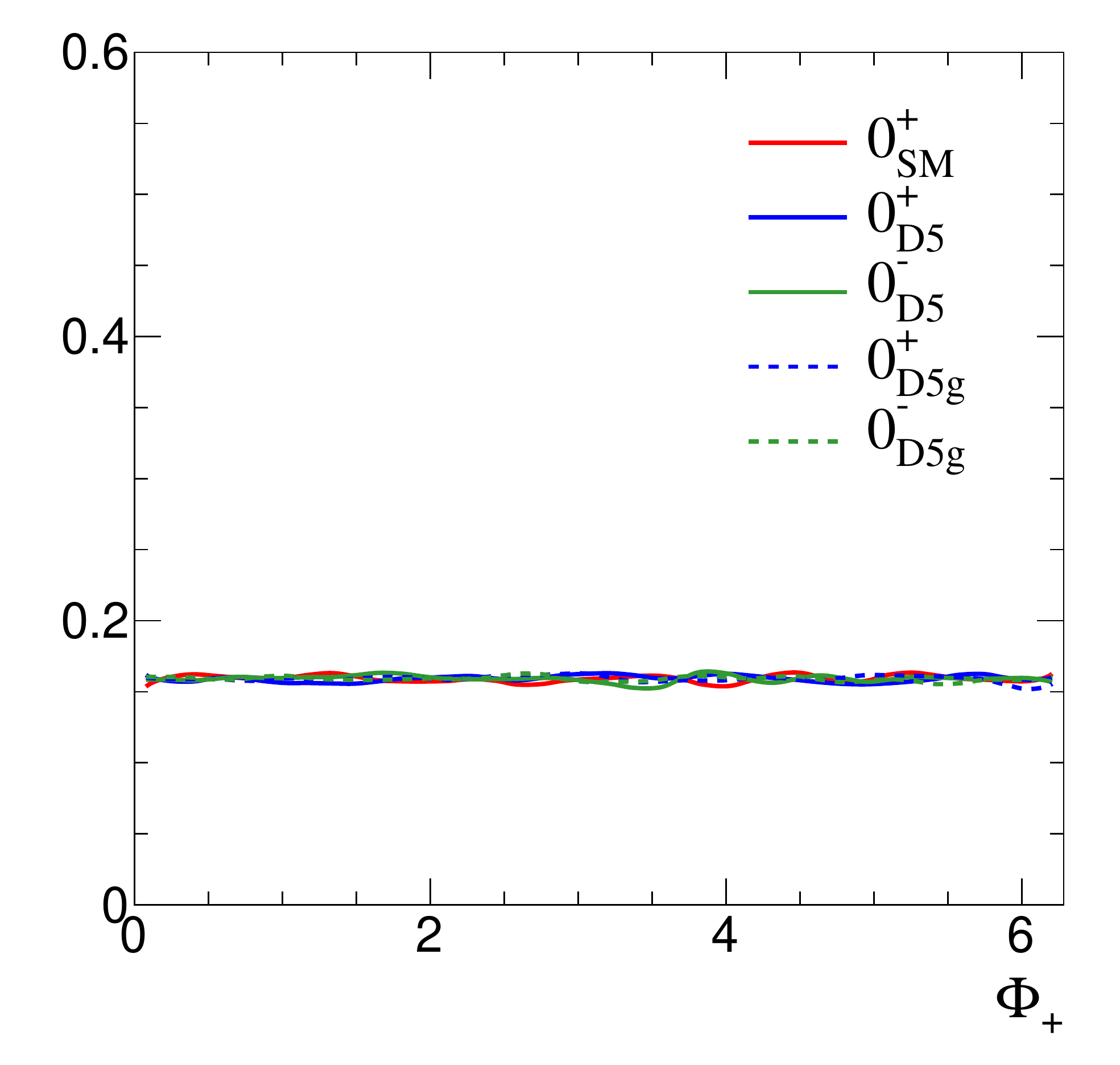}
 \hfill
 \includegraphics[width=0.24\textwidth]{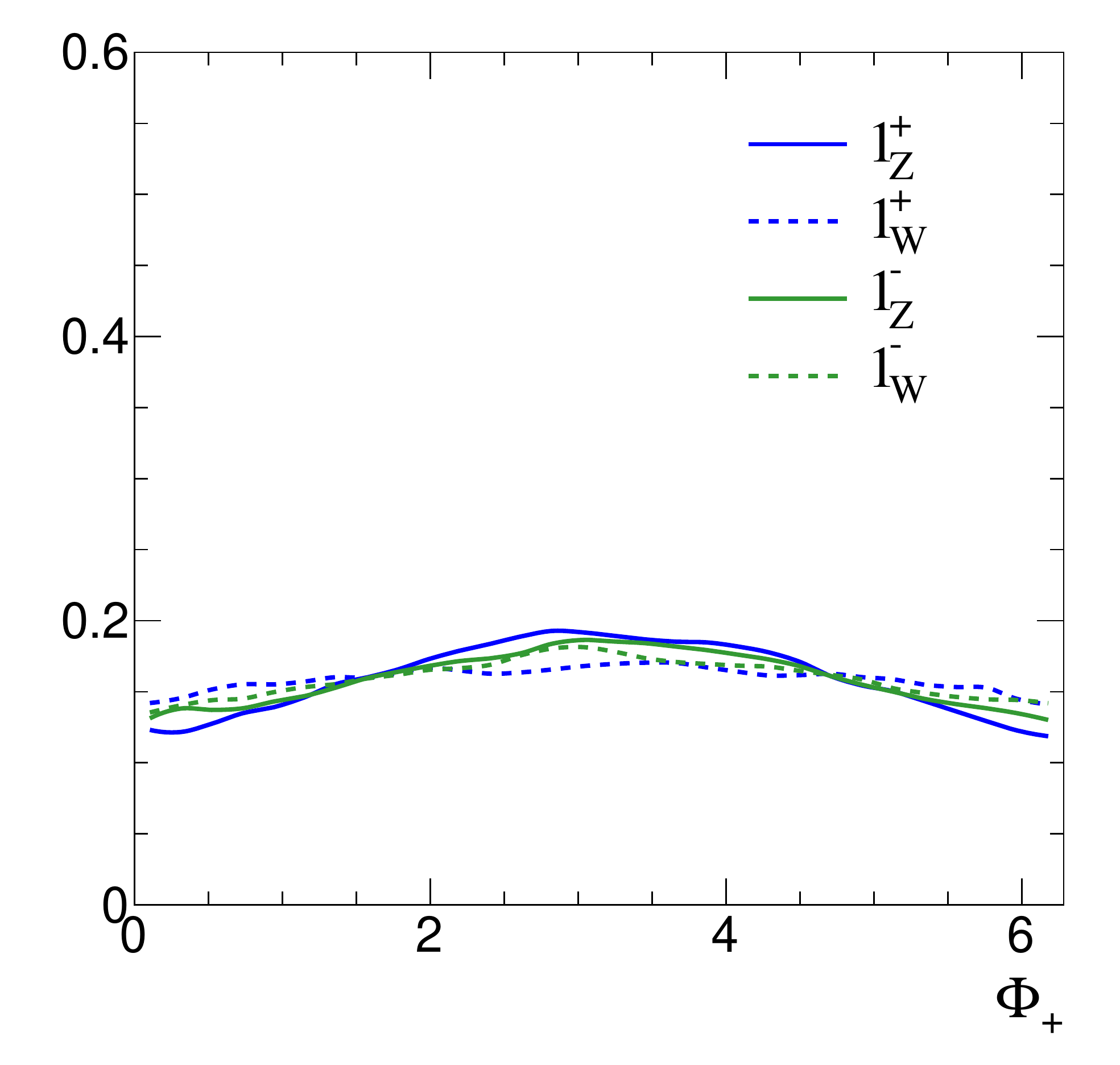}
 \hfill
 \includegraphics[width=0.24\textwidth]{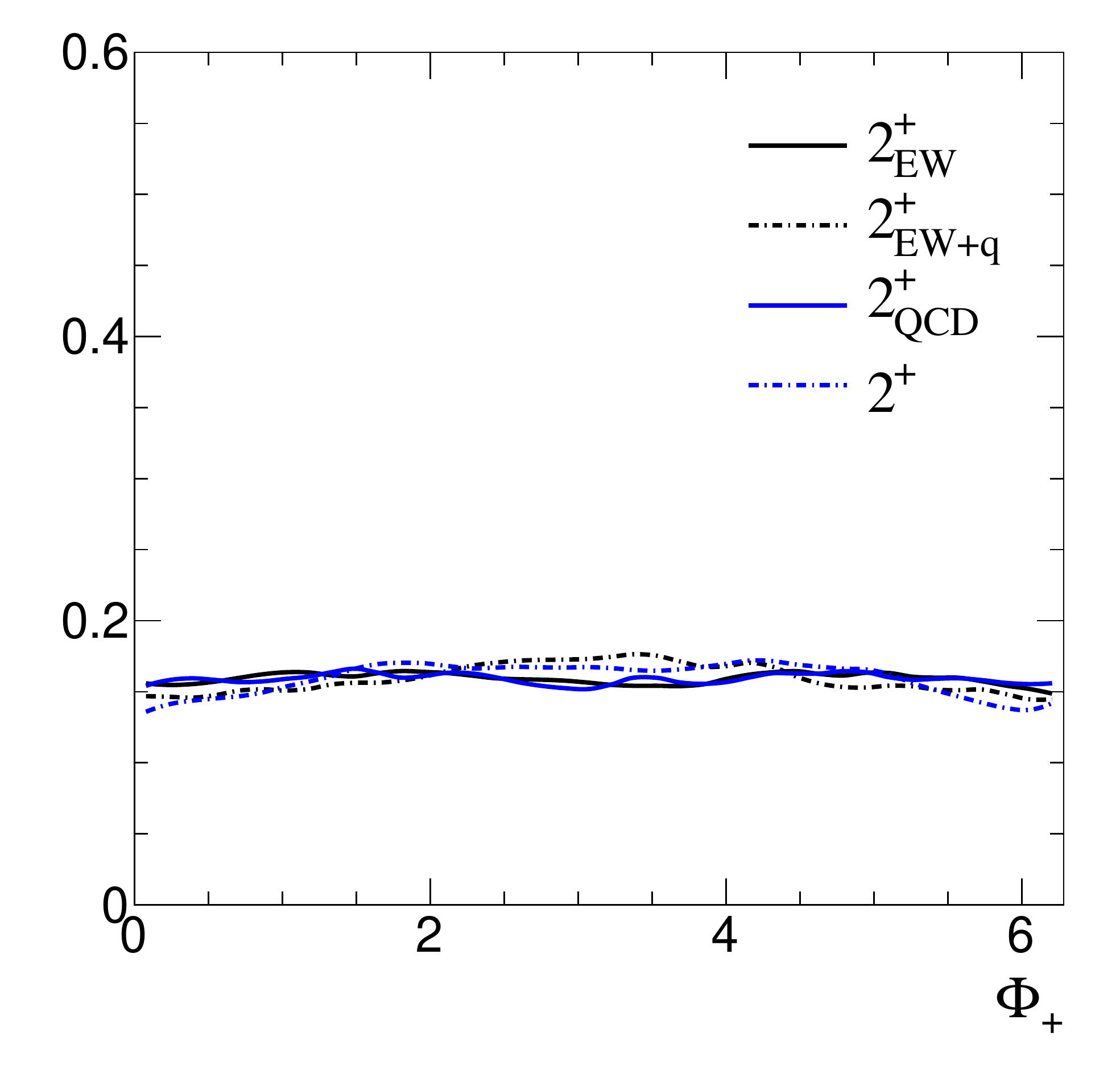}
 \hfill
 \includegraphics[width=0.24\textwidth]{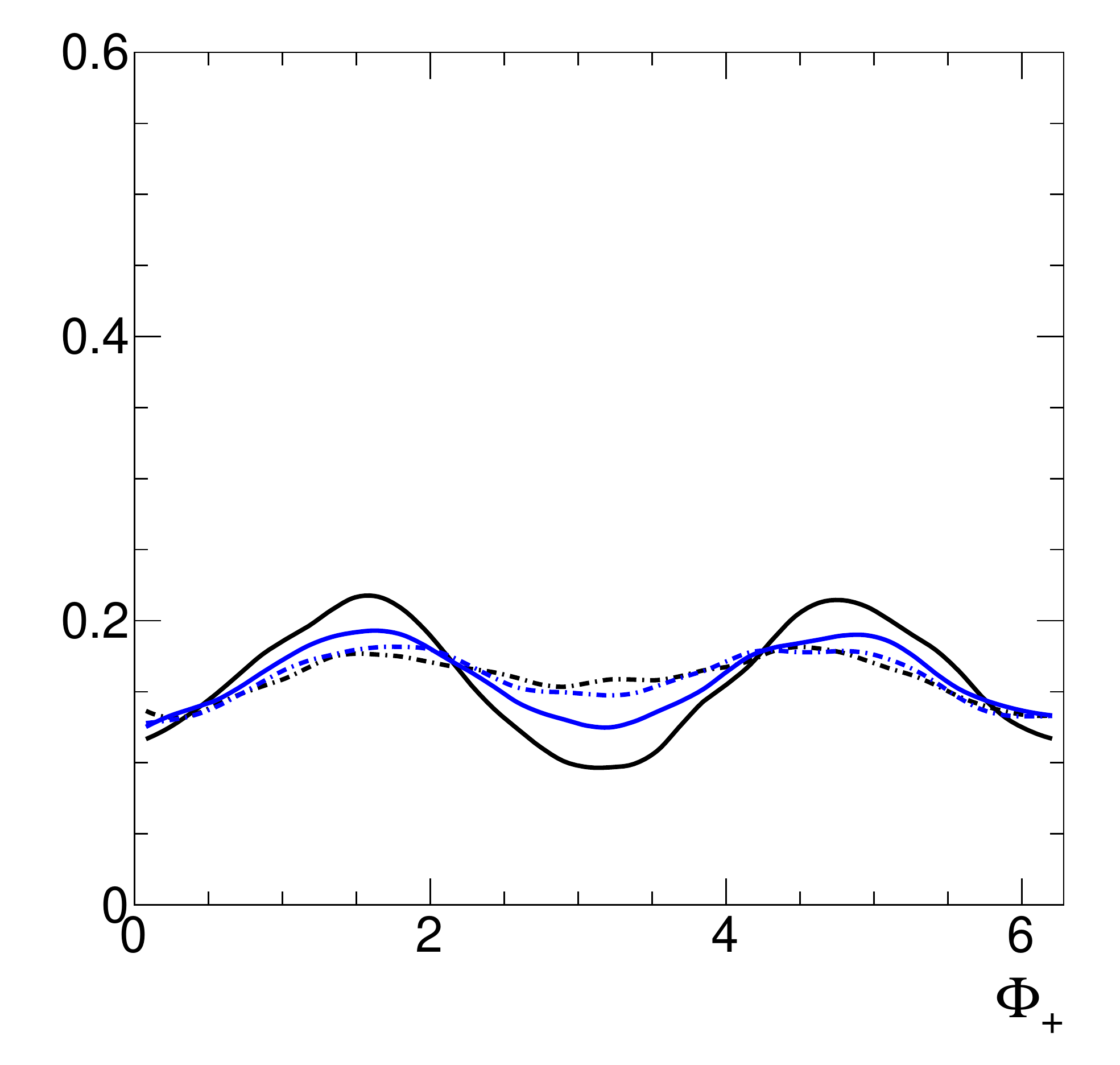}\\
 \includegraphics[width=0.24\textwidth]{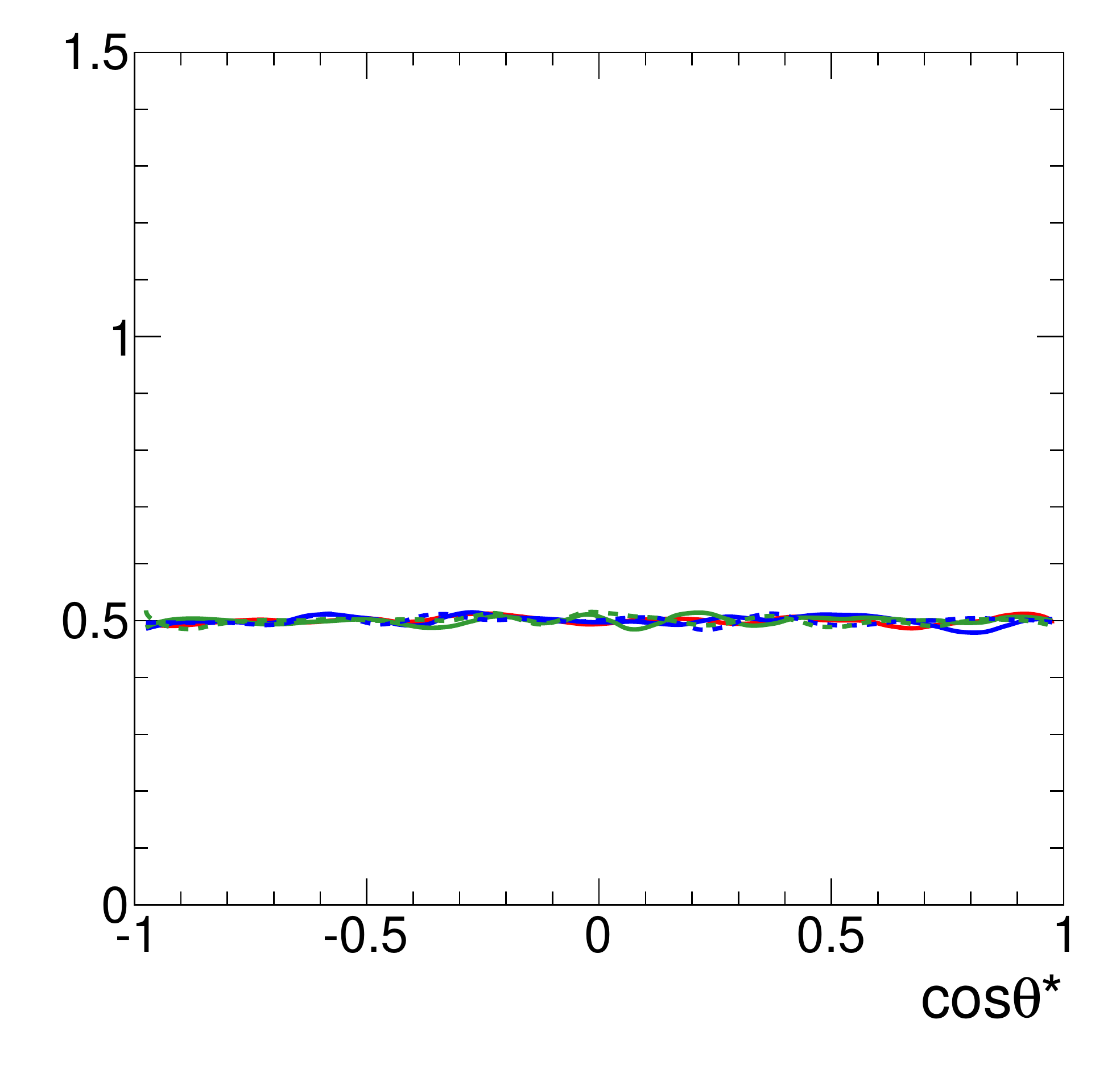}
 \hfill
 \includegraphics[width=0.24\textwidth]{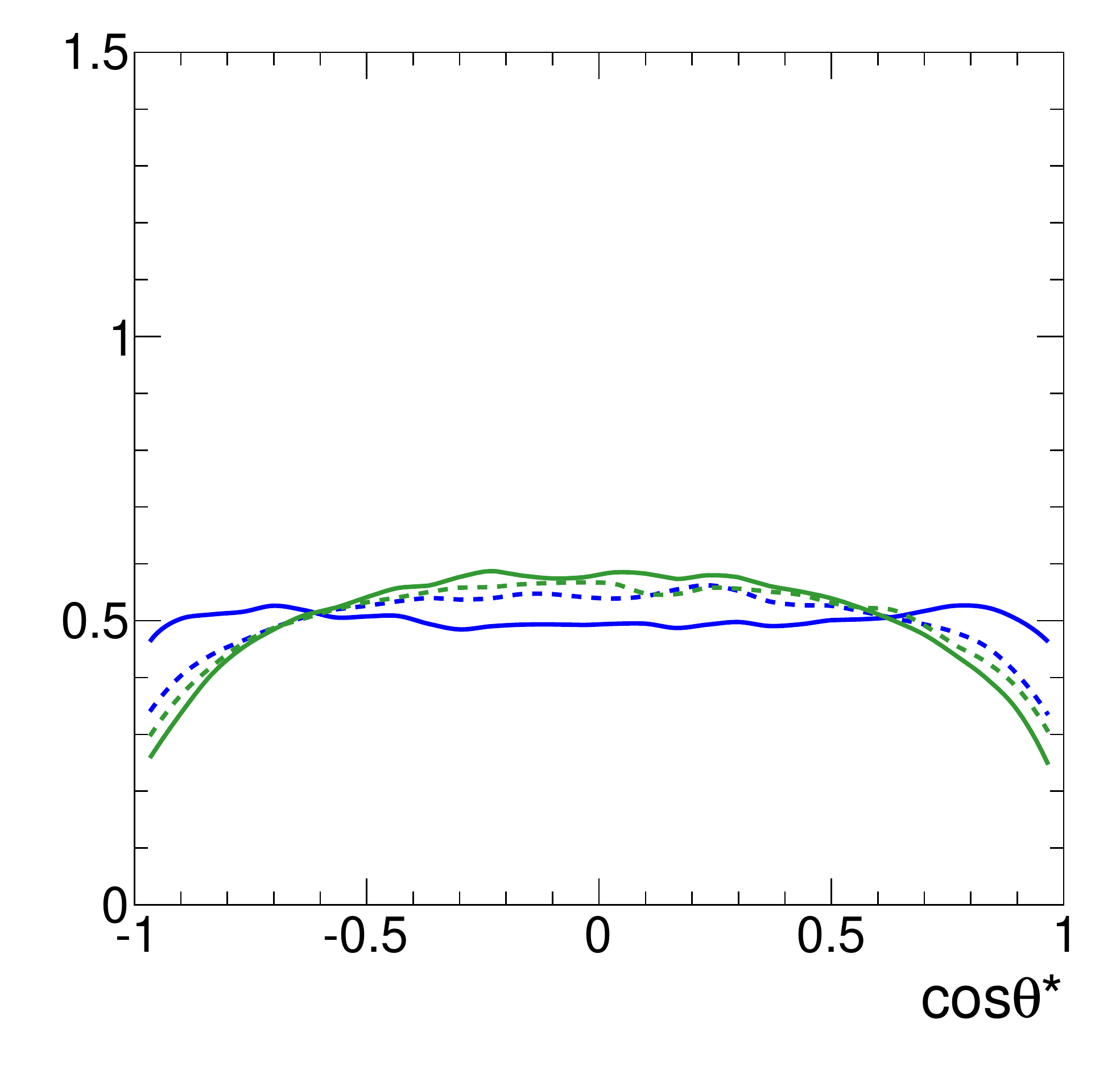}
 \hfill
 \includegraphics[width=0.24\textwidth]{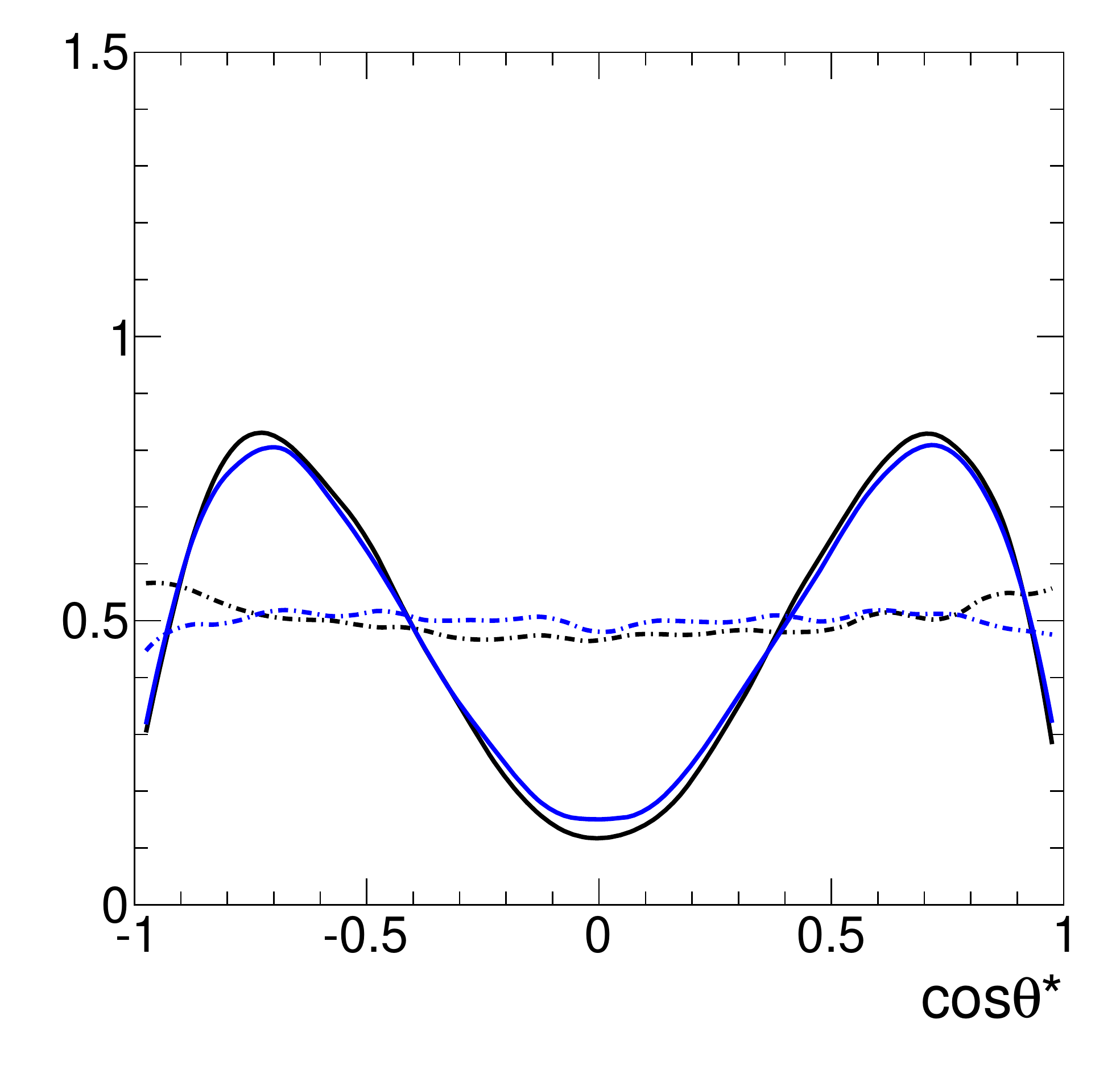}
 \hfill
 \includegraphics[width=0.24\textwidth]{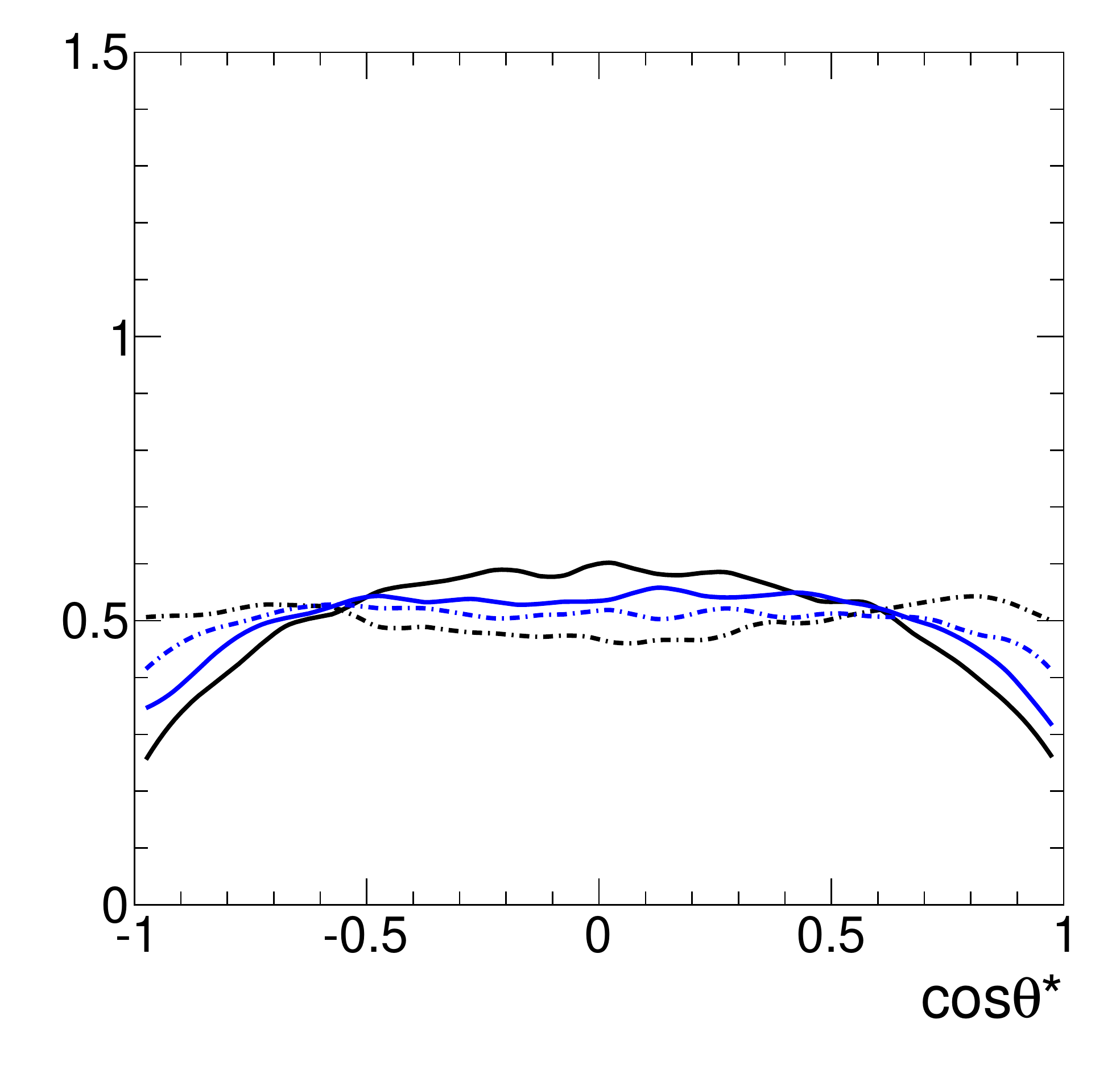}\\
 \includegraphics[width=0.24\textwidth]{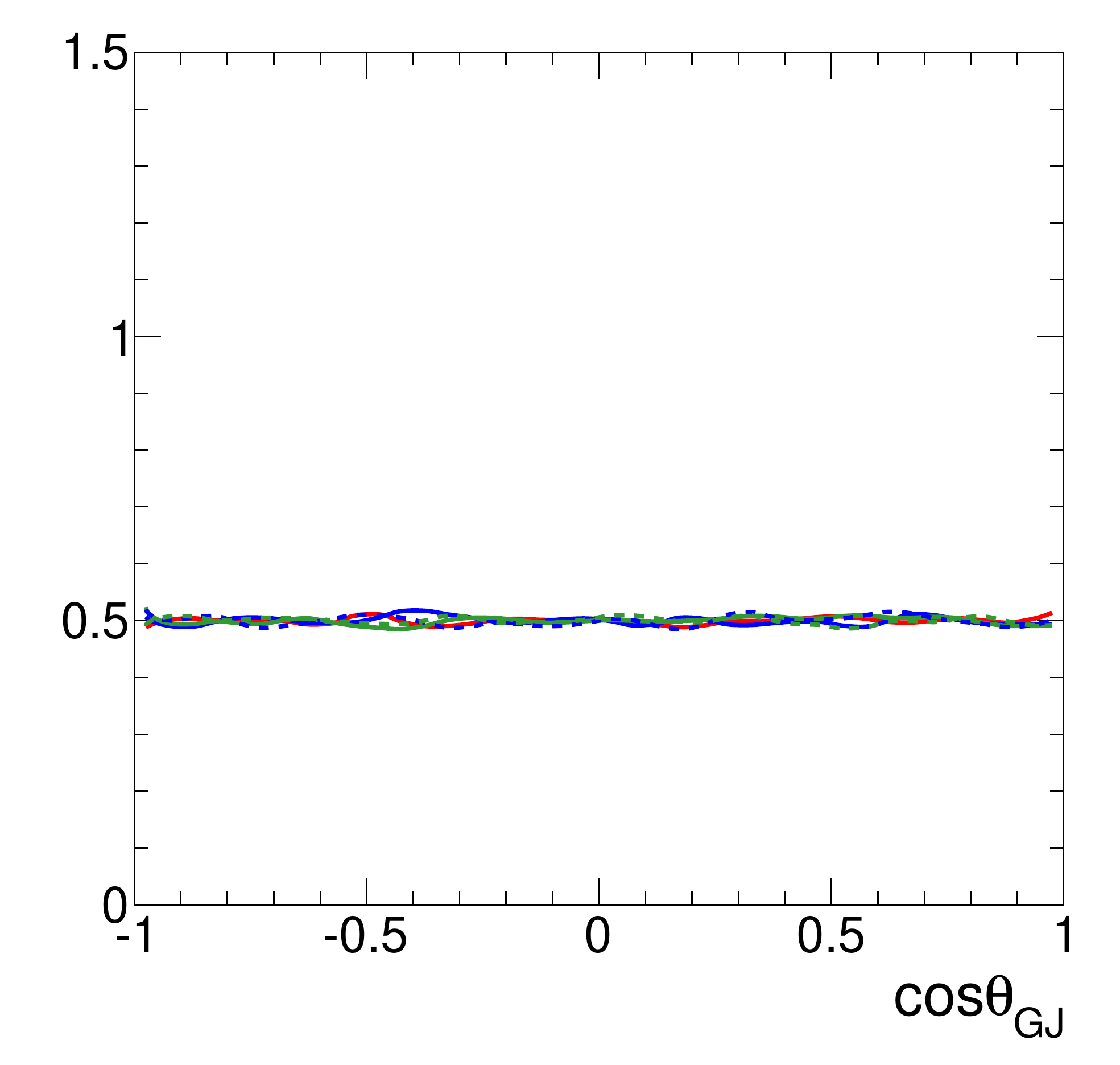}
 \hfill
 \includegraphics[width=0.24\textwidth]{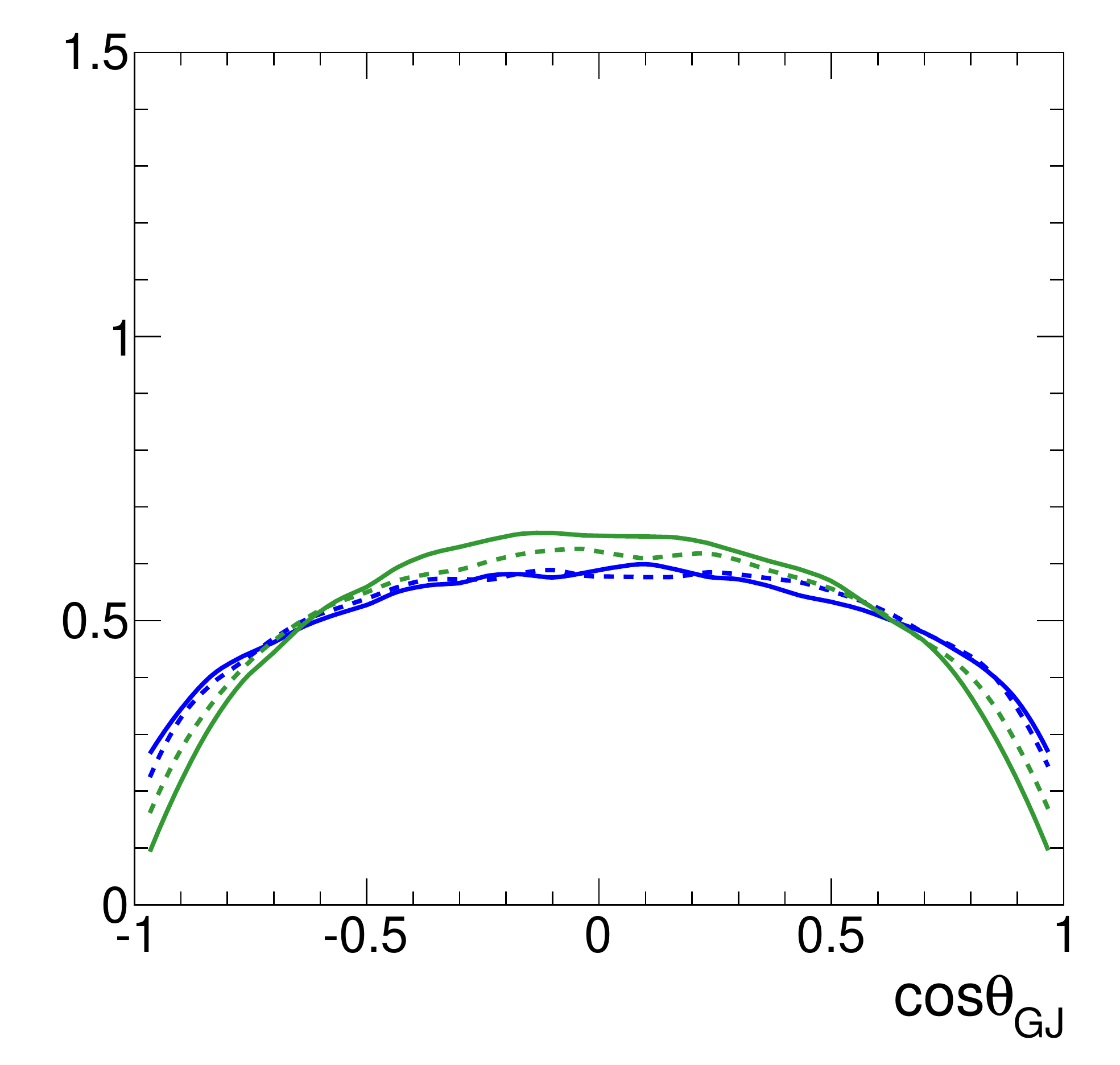}
 \hfill
 \includegraphics[width=0.24\textwidth]{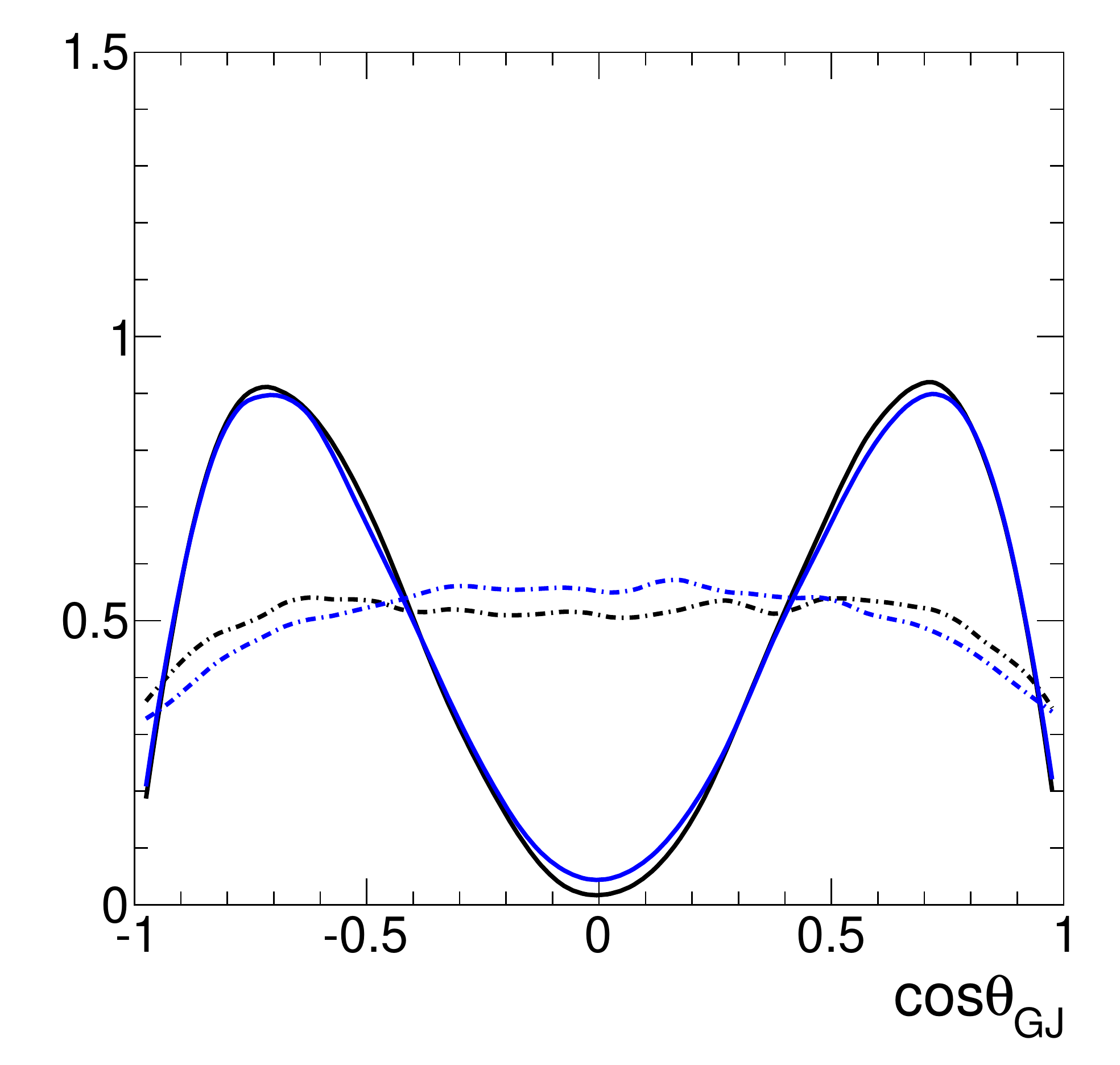}
 \hfill
 \includegraphics[width=0.24\textwidth]{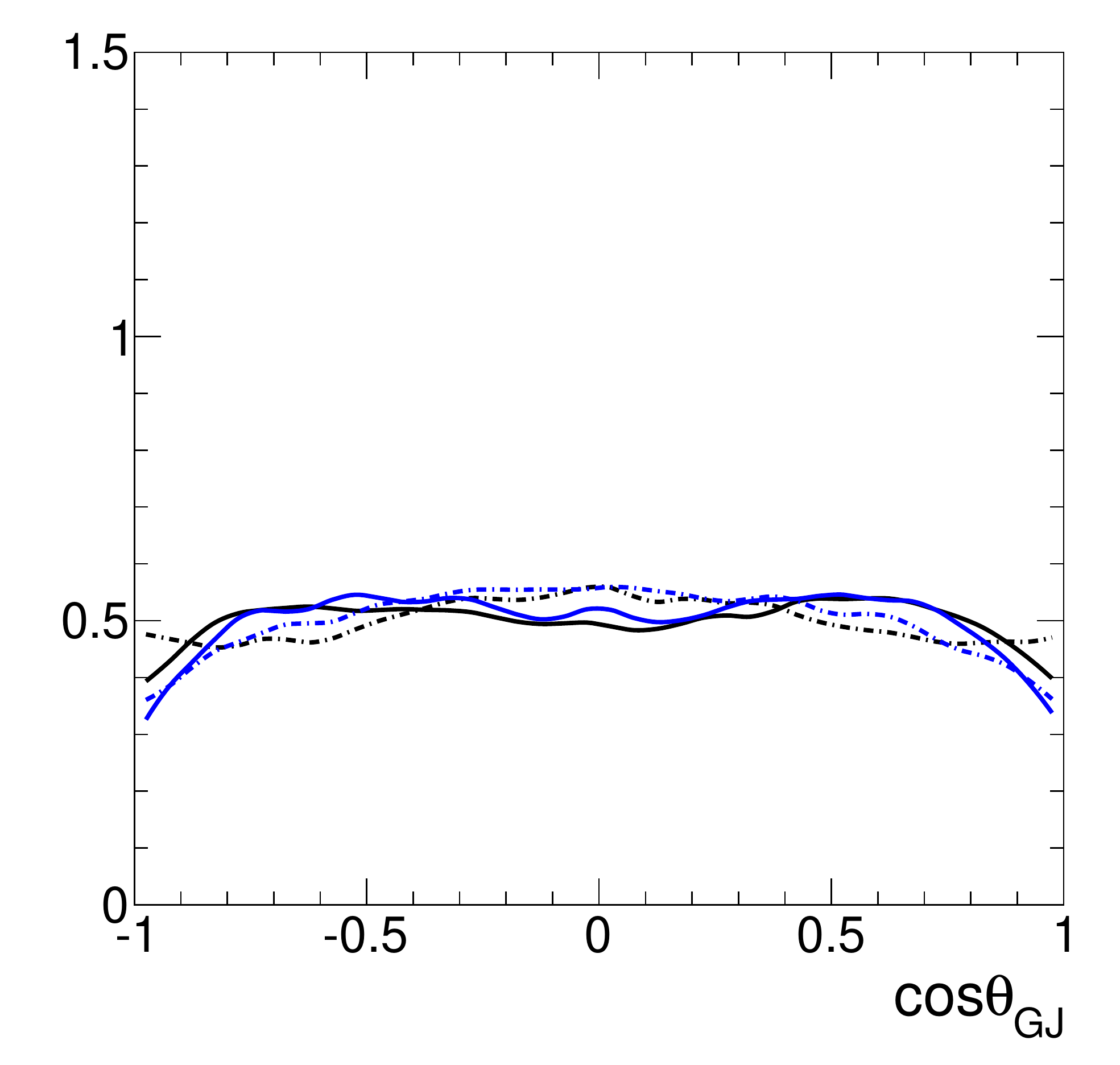}
 \caption{Specific angles targeted at identifying a spin-2 resonance.
   The different curves are the same as in Fig.~\ref{fig:app1}.}
\label{fig:app3}
\end{figure}

Finally, in Fig.~\ref{fig:app3} we show the most promising specific
angular observable for the determination of the coupling structure in
weak boson fusion. This includes the Breit-frame angles $\Phi_+$
 and $\cos \theta^*$ defined in Eq.\eqref{eq:angles_wbf} as
well as the Gottfried--Jackson angles introduced at the end of
Sec.~\ref{sec:angles}. We see that their benefit apart from a spin-2
identification is somewhat limited.

%-------------------------------------------------------
%-------------------------------------------------------

%-------------------------------------------------------
%-------------------------------------------------------

\pagebreak
%%%%%%%%%%%%%%%%%%%%%%%%%%%%%%%%%%%%%%%%%%%%%%%%%%%%%%%%%%%%%%%%%%%%%%%%
%\baselineskip15pt

\end{document}